\documentclass{article}
\pdfoutput=1

\usepackage{epubtk}    
\usepackage{graphicx}  
\usepackage{amsfonts}
\usepackage{amsmath}
\usepackage{amssymb}
\usepackage{pifont}
\usepackage{tipa}

\usepackage{dcolumn}
\usepackage{bm}
\usepackage{sectsty}
\usepackage{latexsym}
\usepackage{ifthen}
\usepackage{enumerate}
\usepackage{bbm}

\def\ba{\begin{eqnarray}}
\def\ea{\end{eqnarray}}

\def\be{\begin{eqnarray}}
\def\ee{\end{eqnarray}}

\def\A{\mathcal{A}}

\def\C{\mathcal{C}}
\def\X{\mathbb{X}}
\def\D{\mathcal{D}}
\def\U{\mathcal{U}}
\def\E{\mathcal{E}}
\def\L{\mathcal{L}}
\def\K{\mathcal{K}}

\def\nn{\nonumber}

\def\d{\,\mathrm{d}}
\def\mn{_{\mu \nu}}
\def\mupn{^\mu_{\, \, \nu}}
\def\mnu{^{\mu \nu}}
\def\({\left(}
\def\){\right)}
\def\e{\epsilon}
\def\ep{\varepsilon}
\def\ie{i.e., }
\def\eg{e.g., }

\def\mpl{M_{\rm Pl}}
\def\p{\partial}
\def\stu{St\"uckelberg }
\def\Ein{\hat{\mathcal{E}}}
\def\ab{_{\alpha\beta}}
\def\abu{^{\alpha\beta}}
\def\etat{\tilde{\eta}}
\def\Lc{\Lambda_3}
\def\R{{\cal R}}
\def\wed{\wedge}
\def\g{\gamma}
\def\Nf{\bar{\mathcal{N}}}
\def\uf{{}^{(5)}\hspace{-1.5pt}}

\def\napprox{\approx\hspace{-11pt}/ \hspace{5pt}}

\numberwithin{equation}{section}

\showlistoftablesfalse

\begin{document}

\title{Massive Gravity}

\author{\epubtkAuthorData{Claudia de Rham}{%
CERCA \& Physics Department \\
Case Western Reserve University \\
10900 Euclid Ave, Cleveland, OH 44106, USA}{%
Claudia.deRham@case.edu}{%
http://www.phys.cwru.edu/~claudia/}%
}

\date{}
\maketitle

\begin{abstract}
We review recent progress in massive gravity. We start by showing how different theories of massive gravity emerge from a higher-dimensional theory of general relativity, leading to the Dvali--Gabadadze--Porrati model (DGP), cascading gravity and ghost-free massive gravity. We then explore their theoretical and phenomenological consistency, proving the absence of Boulware--Deser ghosts and reviewing the Vainshtein mechanism and the cosmological solutions in these models. Finally we present alternative and related models of massive gravity such as new massive gravity, Lorentz-violating massive gravity and non-local massive gravity.
\end{abstract}

\epubtkKeywords{General relativity, Gravity, Alternative theories of
  gravity, Modified theories of gravity, Massive gravity}

\newpage

\tableofcontents

\newpage

\section{Introduction}

For almost a century the theory of general relativity (GR) has been known to describe the force of gravity with impeccable agreement with observations. Despite all the successes of GR the search for alternatives has been an ingoing challenge since its formulation. Far from a purely academic exercise, the existence of consistent alternatives to describe the theory of gravitation is actually essential to test the theory of GR. Furthermore the open questions that remain behind the puzzles at the interface between gravity/cosmology and particle physics such as the hierarchy problem, the old cosmological constant problem and the origin of the late-time acceleration of the Universe have pushed the search for alternatives to GR.

While it was not formulated in this language at the time, from a more modern particle physics perspective GR can the thought of as the unique theory of a massless spin-2 particle~\cite{Gupta:1954zz,Weinberg:1965rz,Deser:1969wk,Feynman:1996kb,Boulware:1974sr}, and so in order to find alternatives to GR one should break one of the underlying assumptions behind this uniqueness theorem. Breaking Lorentz invariance and the notion of spin along with it is probably the most straight forward since non-Lorentz invariant theories include a great amount of additional freedom. This possibility has been explored in length in the literature see for instance~\cite{Mattingly:2005re} for a review. Nevertheless, Lorentz invariance is observationally well constrained by both particle and astrophysics. Another possibility is to maintain Lorentz invariance and the notion of spin that goes with it but to consider gravity as being the representation of a higher spin. This idea has also been explored see for instance~\cite{Vasiliev:1995dn,Bekaert:2005vh} for further details. In this review we shall explore yet another alternative: Maintaining the notion that gravity is propagated by a spin-2 particle but considering this particle to be massive. From the particle physics perspective, this extension seems most natural since we know that the particles carrier of the electroweak forces have to acquire a mass through the Higgs mechanism.

Giving a mass to a spin-2 (and spin-1) field is an old idea and in this review we shall summarize the approach of Fierz and Pauli which dates back to 1939~\cite{Fierz:1939ix}. While the theory of a massive spin-2 field is in principle simple to derive, complications arise when we include interactions between this spin-2 particle and other particles as should be the case if the spin-2 field is to describe the graviton.

At the linear level, the theory of a massless spin-2 field enjoys a linearized diffeomorphism (diff) symmetry, just as a photon enjoys a $U(1)$ gauge symmetry. But unlike for a photon, coupling the spin-2 field with external matter forces this symmetry to be realized in a different way non-linearly. As a result GR is a fully non-linear theory which enjoys non-linear diffeomorphism invariance (also known as general covariance or coordinate invariance). Even though this symmetry is broken when dealing with a massive spin-2 field, the non-linearities are inherited by the field. So unlike a single isolated massive spin-2 field, a theory of massive gravity is always fully non-linear (and as a consequence non-renormalizable) just as for GR. The fully non-linear equivalent to GR for massive gravity has been a much more challenging theory to obtain. In this review we shall summarize a few different approaches in deriving consistent theories of massive gravity and shall focus on recent progress. See Ref.~\cite{Hinterbichler:2011tt} for an earlier review on massive gravity, as well as Refs.~\cite{deRham:2012az} and \cite{Khoury:2013tda} for other reviews relating Galileons and massive gravity.

When dealing with a theory of massive gravity two elements have been known to be problematic since the seventies. First, a massive spin-2 field propagates five degrees of freedom no matter how small its mass is. At first sight this seems to suggest that even in the massless limit, a theory of massive gravity could never resemble GR, \ie a theory of a massless spin-2 field with only two propagating degrees of freedom. This subtlety is at the origin of the vDVZ discontinuity (van Dam--Veltman--Zakharov~\cite{vanDam:1970vg,Zakharov:1970cC}). The resolution behind that puzzle was provided by Vainshtein two years later and lies in the fact the extra degree of freedom responsible for the vDVZ discontinuity gets screened by its own interactions which dominate over the linear terms in the massless limit. This process is now relatively well understood~\cite{Vainshtein:1972sx} (see also Ref.~\cite{Babichev:2013usa} for a recent review). The Vainshtein mechanism also comes hand in hand with its own set of peculiarities like strong coupling and superluminalities which we shall discuss in this review.

A second element of concern in dealing with a theory of massive gravity is the realization that most non-linear extensions of Fierz--Pauli massive gravity are plagued with a ghost, now known as the Boulware--Deser (BD) ghost~\cite{Boulware:1973my}. The past decade has seen a revival of interest in massive gravity with the realization that this BD ghost could be avoided either in a model of soft massive gravity (not a single massive pole for the graviton but rather a resonance) as in the DGP (Dvali--Gabadadze--Porrati) model or its extensions~\cite{Dvali:2000hr,Dvali:2000rv,Dvali:2000xg}, or in a three-dimensional model of massive gravity as in `new massive gravity' (NMG)~\cite{Bergshoeff:2009hq} or more recently in a specific ghost-free realization of massive gravity (also known as dRGT in the literature)~\cite{deRham:2010kj}.

With these developments the possibility to test massive gravity as an alternative to GR has become a reality. We will summarize the different phenomenologies of these models and their theoretical as well as observational bounds through this review. Except in specific cases, the graviton mass is typically bounded to be a few times the Hubble parameter today, that is $m\lesssim 10^{-30}-10^{-33}\mathrm{\ eV}$ depending on the exact models. In all of these models, if the graviton had a mass much smaller than $10^{-33}\mathrm{\ eV}$, its effect would be unseen in the observable Universe and such a mass would thus be irrelevant. Fortunately there is still to date an open window of opportunity for the graviton mass to be within an interesting range and providing potentially new observational signatures. Independently of this, developments in massive gravity have also opened new theoretical avenues, which we will summarize, and these remain very much an active area of progress.

This review is organized as follows: We start by setting the formalism for massive and massless spin-1 and -2 fields in Section~\ref{sec:MassiveFields} and emphasize the \stu language both for the Proca and the Fierz--Pauli fields. In Part~\ref{part:extradim} we then derive consistent theories using a higher-dimensional framework, either using a braneworld scenario \`a la DGP in Section~\ref{sec:DGP}, or via a discretization (or Kaluza--Klein reduction) of the extra dimension in Section~\ref{sec:deconstruction}. This second approaches leads to the theory of ghost-free massive gravity (also known as dRGT) which we review in more depth in Part~\ref{sec:ghost-free MG}. Its formulation is summarized in Section~\ref{sec:dRGTsummary}, before tackling other interesting aspects such as the fate of the BD ghost in Section~\ref{sec:Abscence of ghost proof}, deriving its decoupling limit in Section~\ref{sec:DL section}, and various extensions in section~{sec:Extensions}. The Vainshtein mechanism and other related aspects are discussed in Section~\ref{sec:MGFT}.
The phenomenology of ghost-free massive gravity is then reviewed in Part~\ref{part:pheno} including a discussion on solar system tests, gravitational waves, weak lensing, pulsars, black holes and cosmology. We then conclude with other related theories of massive gravity in Part~\ref{Part:OtherMG} including new massive gravity, Lorentz breaking theories of massive gravity and non-local versions.

\paragraph*{Notations and conventions:}

Throughout this review we work in units where the reduced Planck constant $\hbar$ and the speed of light $c$ are set to unity. The gravitational Newton constant $G_N$ is related to the Planck scale by $8 \pi G_N = \mpl^{-2}$. Unless specified otherwise $d$ represents the number of spacetime dimensions.
We use the mainly $+$ convention $(-+\cdots+)$ and space indices are denoted by $i,j,\cdots=1,\cdots, d-1$ while $0$ represents the time-like direction, $x^0=t$.

We also use the symmetric convention: $(a,b)=\frac 12 (ab+ba)$ and $[a,b]=\frac 12 (ab-ba)$. Throughout this review square brackets of a tensor indicates the trace of tensor, for instance $[\mathbb{X}]=\mathbb{X}^\mu_{\, \mu}$, $[\mathbb{X}^2]=\mathbb{X}^\mu_{\, \nu}\mathbb{X}^\nu_{\, \mu}$, etc\ldots. We also use the notation $\Pi\mn=\p_\mu \p_\nu \pi$, and $\mathbb{I}=\delta\mupn$.
$\ep_{\mu\nu\alpha\beta}$ and $\ep_{ABCDE}$ represent the Levi-Cevita symbol in respectively four and five dimensions, $\ep_{0123}=\ep_{01234}=1=\ep^{0123}$.

\newpage


\section{Massive and Interacting Fields}
\label{sec:MassiveFields}

\subsection{Proca field}

\subsubsection{Maxwell kinetic term}
\label{sec:Maxwell}

Before jumping into the subtleties of massive spin-2 field and gravity in general, we start this review with massless and massive spin-1 fields as a warm up. Consider a Lorentz vector field $A_\mu$ living on a four-dimensional Minkowski manifold. We focus this discussion to four dimensions and the extension to $d$ dimensions is straightforward. Restricting ourselves to Lorentz invariant and local actions for now, the kinetic term can be decomposed into three possible contributions:
\ba
\label{VecKin1}
\L_{\rm kin}^{\rm spin-1}= a_1 \L_1 + a_2 \L_2+a_3 \L_3\,,
\ea
where $a_{1,2,3}$ are so far arbitrary dimensionless coefficients and the possible kinetic terms are given by
\ba
\L_1&=&\p_\mu A^\nu \p^\mu A_\nu\\
\L_2&=&\p_\mu A^\mu \p_\nu A^\nu \\
\L_3&=&\p_\mu A^\nu \p_\nu A^\mu\,,
\ea
where in this section, indices are raised and lowered with respect to the flat Minkowski metric. The first and third contributions are equivalent up to a boundary term, so we set $a_3=0$ without loss of generality.

We now proceed to establish the behaviour of the different degrees of freedom (dofs) present in this theory. \textit{A priori} a Lorentz vector field $A_\mu$ in four dimensions could have up to four dofs, which we can split as a transverse contribution $A_\mu^\bot$ satisfying $\p^\mu A^\bot_\mu=0$ bearing a priori three dofs and a longitudinal mode $\chi$ with $A_\mu=A_\mu^\bot+\p_\mu \chi$.

\subsubsection*{Helicity-0 Mode}

Focusing on the longitudinal (or helicity-0) mode $\chi$, the kinetic term takes the form
\ba
\label{VecKin2}
\L_{\rm kin}^{\chi}=(a_1+a_2) \p_\mu\p_\nu \chi \p^\mu\p^\nu\chi= (a_1+a_2)(\Box \chi)^2\,,
\ea
where $\Box=\eta^{\mu\nu}\p_\mu\p_\nu$ represents the d'Alembertian in flat Minkowski space and the second equality holds after integrations by parts. We directly see that unless $a_1=-a_2$, the kinetic term for the field $\chi$ bears higher time (and space) derivatives. As a well known consequence of Ostrogradsky's theorem~\cite{Ostrogradsky}, two dofs are actually hidden in $\chi$ with an opposite sign kinetic term. This can be seen by expressing the propagator $\Box^{-2}$ as the sum of two propagators with opposite signs:
\ba
\frac{1}{\Box^2}=\lim_{m\to 0} \frac{1}{2m^2}\(\frac{1}{\Box-m^2}-\frac{1}{\Box+m^2}\)\,,
\ea
signaling that one of the modes always couples the wrong way to external sources. The mass $m$ of this mode is arbitrarily low which implies that the theory~\eqref{VecKin1} with $a_3=0$ and $a_1+a_2\ne 0$ is always sick. Alternatively, one can see the appearance of the Ostrogradsky instability by introducing a Lagrange multiplier $\tilde \chi(x)$, so that the kinetic action~\eqref{VecKin2} for $\chi$ is equivalent to
\ba
\L_{\rm kin}^{\chi}=(a_1+a_2)\(\tilde\chi \Box \chi -\frac 14 \tilde \chi^2\)\,,
\ea
after integrating out the Lagrange multiplier\epubtkFootnote{The equation of motion with respect to $\chi$ gives $\Box \tilde \chi=0$, however this should be viewed as a dynamical relation for $\tilde \chi$, which should not be plugged back into the action. On the other hand, when deriving the equation of motion with respect to $\tilde \chi$, we obtain a constraint equation for $\tilde \chi$: $\tilde \chi=2 \Box \chi$ which can be plugged back into the action (and $\chi$ is then treated as the dynamical field).} $\tilde \chi \equiv 2 \Box \chi$. We can now perform the change of variables $\chi=\phi_1+\phi_2$ and $\tilde \chi=\phi_1-\phi_2$ giving the resulting Lagrangian for the two scalar fields $\phi_{1,2}$
\ba
\L_{\rm kin}^{\chi}=(a_1+a_2)\(\phi_1 \Box \phi_1-\phi_2 \Box \phi_2-\frac 14 (\phi_1-\phi_2)^2\)\,.
\ea
As a result, the two scalar fields $\phi_{1,2}$ always enter with opposite kinetic terms, signaling that one of them is always a ghost\epubtkFootnote{This is already a problem at the classical level, well before the notion of particle needs to be defined, since classical configurations with arbitrarily large $\phi_1$ can always be constructed by compensating with a large configuration for $\phi_2$ at no cost of energy (or classical Hamiltonian).}. The only way to prevent this generic pathology is to make the specific choice $a_1+a_2=0$, which corresponds to the well-known Maxwell kinetic term.

\subsubsection*{Helicity-1 mode and gauge symmetry}

Now that the form of the local and covariant kinetic term has been uniquely established by the requirement that no ghost rides on top of the helicity-0 mode, we focus on the remaining transverse mode $A^\bot_\mu$,
\ba
\label{Maxwell2}
\L_{\rm kin}^{\rm helicity-1}= a_1\, \(\p_\mu A^\bot_\nu\)^2\,,
\ea
which has the correct normalization if $a_1=-1/2$. As a result, the only possible local kinetic term for a spin-1 field is the Maxwell one:
\ba
\label{Maxwell}
\L_{\rm kin}^{\rm spin-1}=-\frac{1}{4}F\mn^2
\ea
with $F\mn=\p_\mu A_\nu-\p_\nu A_\mu$.  Restricting ourselves to a massless spin-1 field, (with no potential and other interactions), the resulting Maxwell theory satisfies the following $U(1)$ gauge symmetry:
\ba
\label{U(1)gauge}
A_\mu \to A_\mu +\p_\mu \xi\,.
\ea
This gauge symmetry projects out two of the naive four degrees of freedom.
This can be seen at the level of the Lagrangian directly, where the gauge symmetry~\eqref{U(1)gauge} allows us to fix the gauge of our choice. For convenience, we perform a $(3+1)$-split and choose Coulomb gauge $\p_i A^i=0$, so that only two dofs are present in $A_i$, \ie $A_i$ contains no longitudinal mode, $A_i=A_i^t+\p_i A^l$, with $\p^i A_i^t=0$ and the Coulomb gauge sets the longitudinal mode $A^l=0$. The time-component $A_0$ does not exhibit a kinetic term,
\ba
\L_{\rm kin}^{\rm spin-1}=\frac 12 (\p_t A_i)^2-\frac 12 (\p_i A^0)^2-\frac 14 (\p_i A_j)^2\,,
\ea
and appears instead as a Lagrange multiplier imposing the constraint
\ba
\p_i\p^i A_0\equiv 0\,.
\ea
The Maxwell action has therefore only two propagating dofs in $A_i^t$,
\ba
\L_{\rm kin}^{\rm spin-1}=-\frac 12 (\p_\mu A^t_i)^2\,.
\ea
To summarize, the Maxwell kinetic term for a vector field and the fact that a massless vector field in four dimensions only propagates 2 dofs is not a choice but has been imposed upon us by the requirement that no ghost rides along with the helicity-0 mode. The resulting theory is enriched by a $U(1)$ gauge symmetry which in turn freezes the helicity-0 mode when no mass term is present. We now `promote' the theory to a massive vector field.

\subsubsection{Proca mass term}
\label{sec:ProcaMass}

Starting with the Maxwell action, we consider a covariant mass term $A_\mu A^\mu$ corresponding to the Proca action
\ba
\L_{\rm Proca}=-\frac 14 F\mn^2 -\frac 12 m^2 A_\mu A^\mu\,,
\ea
and emphasize that the presence of a mass term does not change the fact that the kinetic has been uniquely fixed by the requirement of the absence of ghost. An immediate consequence of the Proca mass term is the breaking of the $U(1)$ gauge symmetry~\eqref{U(1)gauge}, so that the Coulomb gauge can no longer be chosen and the longitudinal mode is now dynamical. To see this, let us use the previous decomposition $A_\mu = A^\bot_\mu+\p_\mu \hat \chi$ and notice that the mass term now introduces a kinetic term for the helicity-0 mode $\chi=m \hat \chi$,
\ba
\label{Proca}
\L_{\rm Proca}=-\frac 12 (\p_\mu A^\bot_\nu)^2 - \frac 12 m^2 (A^\bot_\mu)^2-\frac {1}2 (\p_\mu \chi)^2\,.
\ea
A massive vector field thus propagates three dofs, namely two in the transverse modes $A^\bot_\mu$ and one in the longitudinal mode $\chi$. Physically, this can be understood by the fact that a massive vector field does not propagate along the light-cone, and the fluctuations along the line of propagation correspond to an additional physical dof.

Before moving to the Abelian Higgs mechanism which provides a dynamical way to give a mass to bosons, we first comment on the discontinuity in number of dofs between the massive and massless case. When considering the Proca action~\eqref{Proca} with the properly normalized fields $A^\bot_\mu$ and $\chi$, one does not recover the massless Maxwell action~\eqref{Maxwell2} or~\eqref{Maxwell} when sending the boson mass $m\to 0$. A priori this seems to signal the presence of a discontinuity which would allow us to distinguish between for instance a massless photon and a massive one no matter how tiny the mass. In practise however, the difference is physically indistinguishable so long as the photon couples to external sources in a way which respects the $U(1)$ symmetry. Note however that quantum anomalies remain sensitive to the mass of the field so the discontinuity is still present at this level, see Refs.~\cite{Dilkes:2001av,Duff:2002sm}.

To physically tell the difference between a massless vector field and a massive one with tiny mass, one has to probe the system, or in other words include interactions with external sources
\ba
\L_{\rm sources}=-A_\mu J^\mu\,.
\ea
The $U(1)$ symmetry present in the massless case is preserved only if the external sources are conserved, $\p_\mu J^\mu=0$. Such a source produces a vector field which satisfies
\ba
\Box A_\mu^\bot=J_\mu
\ea
in the massless case. The exchange amplitude between two conserved sources $J_\mu$ and $J'_\mu$ mediated by a massless vector field is given by
\ba
\mathcal{A}^{\rm massless}_{JJ'}=\int \d^4x A^\bot_\mu J'{}^\mu=\int\d ^4 x J'{}^\mu \frac{1}{\Box}J_\mu\,.
\ea
On the other hand, if the vector field is massive, its response to the source $J^\mu$ is instead
\ba
(\Box-m^2) A_\mu^\bot=J_\mu \qquad \text{and}\qquad \Box \chi=0\,.
\ea
In that case one needs to consider both the transverse and the longitudinal modes of the vector field in the exchange amplitude between the two sources $J_\mu$ and $J'_\mu$. Fortunately, a conserved source does not excite the longitudinal mode and the exchange amplitude is uniquely given by the transverse mode,
\ba
\label{exchance vector massive}
\mathcal{A}^{\rm massive}_{JJ'}=\int \d^4x \(A^\bot_\mu+\p_\mu \chi\) J'{}^\mu=\int\d ^4 x J'{}^\mu \frac{1}{\Box-m^2}J_\mu\,.
\ea
As a result, the exchange amplitude between two conserved sources is the same in the limit $m\to 0$ no matter whether the vector field is intrinsically massive and propagates 3 dofs or if it is massless and only propagates 2 modes. It is therefore impossible to probe the difference between an exactly massive vector field and a massive one with arbitrarily small mass.

Notice that in the massive case no $U(1)$ symmetry is present and the source needs not be conserved. However the previous argument remains unchanged so long as $\p_\mu J^\mu$ goes to zero in the massless limit at least as quickly as the mass itself. If this condition is violated, then the helicity-0 mode ought to be included in the exchange amplitude~\eqref{exchance vector massive}. In parallel, in the massless case the non-conserved source provides a new kinetic term for the longitudinal mode which then becomes dynamical.

\subsubsection{Abelian Higgs mechanism for electromagnetism}

Associated with the absence of an intrinsic discontinuity in the massless limit is the existence of a Higgs mechanism for the vector field whereby the vector field acquires a mass dynamically. As we shall see later, the situation is different for gravity where no equivalent dynamical Higgs mechanism has been discovered to date. Nevertheless, the tools used to describe the Abelian Higgs mechanism and in particular the introduction of a \stu field will prove useful in the gravitational case as well.

To describe the Abelian Higgs mechanism we start with a vector field $A_\mu$ with associated Maxwell tensor $F\mn$ and a complex scalar field $\phi$ with quartic potential
\ba
\L_{\rm AH}=-\frac 14 F\mn^2 -\frac 12 \(\mathcal{D}_\mu \phi\)\(\mathcal{D}^\mu \phi\)^*-\lambda\(\phi \phi^*-\Phi_0^2\)^2\,.
\ea
The covariant derivative, $\mathcal{D}_\mu=\p_\mu-i q A_\mu$ ensures the existence of the $U(1)$ symmetry, which in addition to~\eqref{U(1)gauge} shifts the scalar field as
\ba
\phi \to \phi e^{i q \xi}\,.
\ea
Splitting the complex scalar field $\phi$ into its norm and phase $\phi=\varphi e^{i \chi}$, we see that the covariant derivative plays the role of the mass term for the vector field, when scalar field acquires a non-vanishing vacuum expectation value (vev),
\ba
\L_{\rm AH}=-\frac 14 F\mn^2 -\frac 12 \varphi^2 \(q A_\mu -\p_\mu \chi\)^2 -\frac 12 (\p_\mu \varphi)^2
-\lambda\(\varphi ^2-\Phi_0^2\)^2\,.
\ea
The Higgs field $\varphi$ can be made arbitrarily massive by setting $\lambda\gg1$ in such a way that its dynamics may be neglected and the field can be treated as frozen at $\varphi\equiv \Phi_0=$const. The resulting theory is that of a massive vector field,
\ba
\L_{\rm AH}=-\frac 14 F\mn^2 -\frac 12 \Phi_0^2 \(q A_\mu -\p_\mu \chi\)^2 \,,
\ea
where the phase $\chi$ of the complex scalar field plays the role of a \stu which restores the $U(1)$ gauge symmetry in the massive case,
\ba
A_\mu &\to& A_\mu + \p_\mu \xi(x)\\
\chi & \to & \chi+q\, \xi(x)\,.
\ea
In this formalism, the $U(1)$ gauge symmetry is restored at the price of introducing explicitly a \stu field which transforms in such a way so as to make the mass term invariant. The symmetry ensures that the vector field $A_\mu$ propagates only 2 dofs, while the \stu $\chi$ propagates the third dof. While no equivalent to the Higgs mechanism exists for gravity, the same \stu trick to restore the symmetry can be used in that case. Since the in that context the symmetry broken is coordinate transformation invariance, (full diffeomorphism invariance or covariance), four \stu fields should in principle be included in the context of massive gravity, as we shall see below.

\subsubsection{Interacting spin-1 fields}

Now that we have introduced the notion of a massless and a massive spin-1 field, let us look at $N$ interacting spin-1 fields. We start with $N$ free and massless gauge fields, $A^{(a)}_\mu$, with $a=1, \cdots, N$, and respective Maxwell tensors $F^{(a)}\mn=\p_\mu A^{(a)}-\p_\nu A^{(a)}_\mu$,
\ba
\label{Lkin N spin1}
\L^{\rm N \, spin-1}_{\rm kin}= -\frac 14 \sum_{a=1}^N\(F^{(a)}\mn\)^2\,.
\ea
The theory is then manifestly Abelian and invariant under $N$ copies of $U(1)$, (\ie the symmetry group is $U(1)^N$ which is Abelian as opposed to $U(N)$ which would correspond to a Yang--Mills theory and would not be Abelian).

However, in addition to these $N$ gauge invariances, the kinetic term is invariant under global rotations in field space,
\ba
\label{rot N spin 1}
A^{(a)}_\mu \ \longrightarrow \ \tilde A^{(a)}_\mu = O^{a}_{\ b} A^{(b)}_\mu\,,
\ea
where $O^a_{\ b}$ is a (global) rotation matrix.
Now let us consider some interactions between these different fields. At the linear level (quadratic level in the action), the most general set of interactions is
\ba
\L_{\rm int}= -\frac 12 \sum_{a,b}\mathcal{I}_{ab}\, A^{(a)}_\mu A^{(b)}_\nu \eta^{\mu\nu}\,,
\ea
where $\mathcal{I}_{ab}$ is an arbitrary symmetric matrix with constant coefficients. For an arbitrary rank-N matrix, all $N$ copies of $U(1)$ are broken, and the theory then propagates $N$ additional helicity-0 modes, for a total of $3N$ independent polarizations in four spacetime dimensions. However if the rank $r$ of $\mathcal{I}$ is $r< N$, \ie if some of the eigenvalues of $\mathcal I$ vanish, then there are $N-r$ special directions in field space which receive no interactions, and the theory thus keeps $N-r$ independent copies of $U(1)$. The theory then propagates $r$ massive spin-1 fields and $N-r$ massless spin-2 fields, for a total of $3N-r$ independent polarizations in four dimensions.

We can see this statement more explicitly in the case of $N$ spin-1 fields by diagonalizing the mass matrix $\mathcal{I}$. A mentioned previously, the kinetic term is invariant under field space rotations,~\eqref{rot N spin 1}, so one can use this freedom to work in a field representation where the mass matrix $I$ is diagonal,
\ba
\mathcal{I}_{ab}={\rm diag}\(m_1^2, \cdots, m_N^2\)\,.
\ea
In this representation the gauge fields are the mass eigenstates and the mass spectrum is simply given by the eigenvalues of $\mathcal{I}_{ab}$.

\subsection{Spin-2 field}

As we have seen in the case of a vector field, as long as it is local and Lorentz-invariant, the kinetic term is uniquely fixed by the requirement that no ghost be present. Moving now to a spin-2 field, the same argument applies exactly and the Einstein--Hilbert term appears naturally as the unique kinetic term free of any ghost-like instability. This is possible thanks to a symmetry which projects out all unwanted dofs, namely diffeomorphism invariance (linear diffs at the linearized level, and non-linear diffs/general covariance at the non-linear level).

\subsubsection{Einstein--Hilbert kinetic term}
\label{subsec:EHkinetic}

We consider a symmetric Lorentz tensor field $h\mn$. The kinetic term can be decomposed into four possible local contributions (assuming Lorentz invariance and ignoring terms which are equivalent upon integration by parts):
\ba
\label{TenKin1}
\L_{\rm kin}^{\rm spin-2}= \frac 12 \p^\alpha h^{\mu\nu} \(b_1 \p_\alpha h\mn+2 b_2 \p_{(\mu}h_{\nu) \alpha}
+b_3 \p_\alpha h \eta\mn+ 2 b_4 \p_{(\mu}h \eta_{\nu) \alpha}\)\,,
\ea
where $b_{1,2,3,4}$ are dimensionless coefficients which are to be determined in the same way as for the vector field. We split the 10 components of the symmetric tensor field $h\mn$ into a transverse tensor $h\mn^T$ (which carries 6 components) and a vector field $\chi_\mu$ (which carries 4 components),
\ba
\label{Tensor split}
h\mn=h^T\mn+ 2\p_{(\mu}\chi_{\nu)}\,.
\ea
Just as in the case of the spin-1 field, an arbitrary kinetic term of the form~\eqref{TenKin1} with untuned coefficients $b_i$ would contain higher derivatives for $\chi_\mu$ which in turn would imply a ghost. As we shall see below, avoiding a ghost within the kinetic term automatically leads to gauge-invariance. After substitution of $h\mn$ in terms of $h^T\mn$ and $\chi_\mu$, the potentially dangerous parts are
\ba
\L_{\rm kin}^{\rm spin-2} &\supset& (b_1+b_2)\chi^\mu \Box^2 \chi_\mu + (b_1+3b_2+2 b_3+4 b_4)\chi^\mu \Box \p_\mu\p_\nu \chi^\nu\\
&&-2 h^{T \mu\nu}\big((b_2+b_4)\p_\mu\p_\nu \p_\alpha \chi^\alpha+(b_1+b_2)\p_\mu \Box \chi_\mu\nn\\
&&\phantom{-2 h^{T \mu\nu}}+ (b_3+b_4)\Box \p_\alpha \chi^\alpha\, \eta\mn\big)\nn\,.
\ea
Preventing these higher derivative terms from arising sets
\ba
b_4=-b_3=-b_2=b_1\,,
\ea
or in other words, the unique (local and Lorentz-invariant) kinetic term one can write for a spin-2 field is the Einstein--Hilbert term
\ba
\label{EH}
\L_{\rm kin}^{\rm spin-2}=-\frac 14 h^{\mu\nu} \Ein^{\alpha\beta}\mn h_{\alpha\beta}=-\frac 14 h^{T \mu\nu} \Ein^{\alpha\beta}\mn h^T_{\alpha\beta},
\ea
where $\Ein$ is the Lichnerowicz operator
\ba
\label{Lichnerowicz operator}
\Ein^{\alpha\beta}\mn h_{\alpha\beta}=-\frac 12 \(\Box h\mn-2\p_{(\mu}\p_\alpha  h^\alpha_{\nu)}+\p_\mu\p_\nu h-\eta\mn (\Box h-\p_\alpha\p_\beta h^{\alpha\beta})\)\,,
\ea
and we have set $b_1=-1/4$ to follow standard conventions.
As a result, the kinetic term for the tensor field $h\mn$ is invariant under the following gauge transformation,
\ba
\label{diff}
h\mn \to h\mn + \p_{(\mu}\xi_{\nu)}\,.
\ea
We emphasize that the form of the kinetic term and its gauge invariance is independent on whether or not the tensor field has a mass, (as long as we restrict ourselves to a local and Lorentz-invariant kinetic term). However just as in the case of a massive vector field, this gauge invariance cannot be maintained by a mass term or any other self-interacting potential. So only in the massless case, does this symmetry remain exact. Out of the 10 components of a tensor field, the gauge symmetry removes $2\times 4 =8$ of them, leaving a massless tensor field with only two propagating dofs as is well known from the propagation of gravitational waves in four dimensions.

In $d\ge 3$ spacetime dimensions, gravitational waves have $d(d+1)/2-2d=d(d-3)/2$ independent polarizations. This means that in three dimensions there are no gravitational waves and in five dimensions they have five independent polarizations.

\subsubsection{Fierz--Pauli mass term}
\label{sec:FP}

As seen in the previous section, for a local and Lorentz-invariant theory, the linearized kinetic term is \emph{uniquely} fixed by the requirement that longitudinal modes propagate no ghost, which in turn prevents that operator from exciting these modes altogether. Just as in the case of a massive spin-1 field, we shall see in what follows that the longitudinal modes can nevertheless be excited when including a mass term. In what follows we restrict ourselves to linear considerations and spare any non-linearity discussions for Parts~\ref{part:extradim} and \ref{sec:ghost-free MG}. See also~\cite{Jaccard:2012ut} for an analysis of the linearized Fierz--Pauli theory using Bardeen variables.

In the case of a spin-2 field $h\mn$, we are \emph{a priori} free to choose between two possible mass terms $h\mn^2$ and $h^2$, so that the generic mass term can be written as a combination of both,
\ba
\label{general mass term}
\L_{\rm mass}= -\frac 18 m^2\(h\mn^2- A h^2\)\,,
\ea
where $A$ is a dimensionless parameter. Just as in the case of the kinetic term, the stability of the theory constrains very strongly the phase space and we shall see that only for $\alpha=1$ is the theory stable at that order. The presence of this mass term breaks diffeomorphism invariance. Restoring it requires the introduction of four \stu fields $\chi_\mu$ which transform under linear diffeomorphisms in such a way as to make the mass term invariant, just as in the Abelian-Higgs mechanism for electromagnetism. Including the four linearized \stu fields, the resulting mass term
\ba
\L_{\rm mass}= -\frac 18 m^2\((h\mn+2\p_{(\mu}\chi_{\nu)})^2- A (h+2\p_\alpha \chi^\alpha)^2\) \, ,
\ea
is invariant under the simultaneous transformations:
\ba
h\mn & \to & h\mn + \p_{(\mu}\xi_{\nu)} \, , \\
\chi_\mu & \to & \chi_\mu -\frac 12 \xi_\mu\,.
\ea
This mass term then provides a kinetic term for the \stu fields
\ba
\L_{\rm kin}^{\chi}= - \frac 12 m^2 \((\p_\mu\chi_\nu)^2-A (\p_\alpha \chi^\alpha)^2\)\,,
\ea
which is precisely of the same form as the kinetic term considered for a spin-1 field~\eqref{VecKin1} in Section~\ref{sec:Maxwell} with $a_3=0$ and $a_2=A a_1$. Now the same logic as in Section~\ref{sec:Maxwell} applies and singling out the longitudinal component of these \stu fields it follows that the only combination which does not involve higher derivatives is $a_2=a_1$ or in other words $A=1$. As a result, the only possible mass term one can consider which is free from an Ostrogradsky instability is the Fierz--Pauli mass term
\ba
\label{FPmass}
\L_{\rm FP mass}= -\frac 18 m^2\((h\mn+2\p_{(\mu}\chi_{\nu)})^2- (h+2\p_\alpha \chi^\alpha)^2\)\,.
\ea
In \emph{unitary gauge}, \ie in the gauge where the \stu fields $\chi^a$ are set to zero, the Fierz--Pauli mass term simply reduces to
\ba
\label{FPmassunitary}
\L_{\rm FP\, mass}= -\frac 18 m^2\(h\mn^2- h^2\)\,,
\ea
where once again the indices are raised and lowered with respect to the Minkowski metric.

\subsubsection*{Propagating degrees of freedom}

To identify the propagating degrees of freedom we may split $\chi^a$ further into a transverse and a longitudinal mode,
\ba
\label{chi A pi}
\chi^a=\frac 1m A^a+\frac 1{m^2}\eta^{ab}\p_b \pi\,,
\ea
(where the normalization with negative factors of $m$ has been introduced for further convenience).

In terms of $h\mn$ and the \stu fields $A_\mu$ and $\pi$ the linearized Fierz--Pauli action is
\ba
\label{FP mixed}
\L_{\rm FP}&=&-\frac 14 h^{\mu\nu} \Ein^{\alpha\beta}\mn h_{\alpha\beta}
-\frac12 h^{\mu\nu} \(\Pi\mn -[\Pi]\eta\mn\)-\frac 18 F\mn^2\\
&-&\frac 18 m^2 \(h\mn^2-h^2\)-\frac 12 m \(h\mnu-h \eta\mnu\)\p_{(\mu}A_{\nu)}\,,
\nn
\ea
with $F\mn=\p_\mu A_\nu - \p_\nu A_\mu$ and $\Pi\mn=\p_\mu\p_\nu \pi$ and all the indices are raised and lowered with respect to the Minkowski metric.

Terms on the first line represent the kinetic terms for the different fields while the second line represent the mass terms and mixing.

We see that the kinetic term for the field $\pi$ is hidden in the mixing with $h\mn$. To make the field content explicit, we may diagonalize this mixing by shifting $h\mn=\tilde h\mn+\pi \eta\mn$ and the linearized Fierz--Pauli action is
\ba
\label{FP tot lin}
\L_{\rm FP}&=&-\frac 14 \tilde h^{\mu\nu} \Ein^{\alpha\beta}\mn \tilde h_{\alpha\beta}
-\frac34 (\p \pi)^2-\frac 18 F\mn^2\\
&-&\frac 18 m^2 \(\tilde h\mn^2-\tilde h^2\)
+ \frac 32 m^2 \pi^2+\frac 32 m^2 \pi \tilde h\nn \\
&-&\frac 12 m \(\tilde h\mnu-\tilde h \eta\mnu\)\p_{(\mu}A_{\nu)}
+3 m \pi \p_\alpha A^\alpha\,.
\nn
\ea
This decomposition allows us to identify the different degrees of freedom present in massive gravity (at least at the linear level): $h\mn$ represents the helicity-2 mode as already present in GR and propagates 2 dofs, $A_\mu$ represents the helicity-1 mode and propagates 2 dofs, and finally $\pi$ represents the helicity-0 mode and propagates 1 dof, leading to a total of five dofs as is to be expected for a massive spin-2 field in four dimensions.

The degrees of freedom have not yet been split into their mass eigenstates but on doing so one can easily check that all the degrees of freedom have the same positive mass square $m^2$.

Most of the phenomenology and theoretical consistency of massive gravity is related to the dynamics of the helicity-0 mode. The coupling to matter occurs via the coupling $h\mn T^{\mu\nu}=\tilde h\mn T^{\mu\nu}+\pi T$, where $T$ is the trace of the external stress-energy tensor. We see that the helicity-0 mode couples directly to conserved sources (unlike in the case of the Proca field) but the helicity-1 mode does not. In most of what follows we will thus be able to ignore the helicity-1 mode.

\subsubsection*{Higgs mechanism for gravity}

As we shall see in Section~\ref{sec:MassVarying}, the graviton mass can also be promoted to a scalar function of one or many other fields (for instance of a different scalar field), $m=m(\psi)$. We can thus wonder whether a dynamical Higgs mechanism for gravity can be considered where the field(s) $\psi$ start in a phase for which the graviton mass vanishes, $m(\psi)=0$ and dynamically evolves to acquire a non-vanishing vev for which $m(\psi)\ne 0$. Following the same logic as the Abelian Higgs for electromagnetism, this strategy can only work if the number of dofs in the massless phase $m=0$ is the same as that in the massive case $m\ne 0$. Simply promoting the mass to a function of an external field is thus not sufficient since the graviton helicity-0 and -1 modes would otherwise be infinitely strongly coupled as $m\to 0$.

To date no candidate has been proposed for which the graviton mass could dynamically evolve from a vanishing value to a finite one without falling into such strong coupling issues. This does not imply that Higgs mechanism for gravity does not exist, but as yet has not been found. For instance on AdS, there could be a Higgs mechanism as proposed in \cite{Porrati:2003sa}, where the mass term comes from integrating out some conformal fields with slightly unusual (but not unphysical) `transparent' boundary conditions. This mechanism is specific to AdS and to the existence of time-like boundary and would not apply on Minkowski or dS.

\subsubsection{Van Dam--Veltman--Zakharov discontinuity}
\label{sec:vDVZ}

As in the case of spin-1, the massive spin-2 field propagates more dofs than the massless one. Nevertheless, these new excitations bear no observational signatures for the spin-1 field when considering an arbitrarily small mass, as seen in Section~\ref{sec:ProcaMass}. The main reason for that is that the helicity-0 polarization of the photon couple only to the divergence of external sources which vanishes for conserved sources. As a result no external sources directly excite the helicity-0 mode of a massive spin-1 field. For the spin-2 field on the other hand the situation is different as the helicity-0 mode can now couple to the trace of the stress-energy tensor and so generic sources will excite not only the 2 helicity-2 polarization of the graviton but also a third helicity-0 polarization, which could in principle have dramatic consequences. To see this more explicitly, let us compute the gravitational exchange amplitude between two sources $T^{\mu\nu}$ and $T'^{\mu\nu}$ in both the massive and massless gravitational cases.

In the massless case, the theory is diffeomorphism invariant. When considering coupling to external sources, of the form $h\mn T\mnu$, we thus need to ensure that the symmetry be preserved, which implies that the stress-energy tensor $T\mnu$ should be conserved $\p_\mu T\mnu=0$. When computing the gravitational exchange amplitude between two sources we thus restrict ourselves to conserved ones. In the massive case, there is a priori no reasons to restrict ourselves to conserved sources, so long as their divergences cancel in the massless limit $m\to 0$.

\subsubsection*{Massive spin-2 field}

Let us start with the massive case, and consider the response to a conserved external source $T^{\mu\nu}$,
\ba
\L=-\frac 14 h\mnu \Ein^{\alpha\beta}\mn h_{\alpha\beta}-\frac {m^2}8 (h\mn^2-h^2)+\frac{1}{2\mpl} h\mn T\mnu\,.
\ea
The linearized Einstein equation is then
\ba
\label{Einstein eq lin massive}
\Ein\abu\mn h\ab+\frac 12 m^2 (h\mn-h \eta\mn)=\frac{1}{\mpl}T\mn\,.
\ea
To solve this modified linearized Einstein equation for $h\mn$ we consider the trace and the divergence separately,
\ba
h&=&-\frac{1}{3m^2\mpl}\(T+\frac{2}{m^2}\p_\alpha \p_\beta T\abu\)\\
\p_\mu h\mupn&=&\frac{1}{m^2 \mpl }\(\p_\mu T\mupn+\frac 13 \p_\nu T+\frac {2}{3m^2}\p_\nu \p_\alpha \p_\beta T\abu\)\,.
\ea
As is already apparent at this level, the massless limit $m\to 0$ is not smooth which is at the origin of the vDVZ discontinuity (for instance we see immediately that for a conserved source the linearized Ricci scalar vanishes $\p_\mu \p_\nu h^{\mu\nu}-\Box h=0$ see Refs.~\cite{vanDam:1970vg,Zakharov:1970cC}. This linearized vDVZ discontinuity was recently repointed out in \cite{Deser:2013rxa}.) As has been known for many decades, this discontinuity (or the fact that the Ricci scalar vanishes) is an artefact of the linearized theory and is resolved by the Vainshtein mechanism \cite{Vainshtein:1972sx} as we shall see later.

Plugging these expressions back into the modified Einstein equation, we get
\ba
\(\Box-m^2\)h\mn&=&-\frac{1}{\mpl}\Big[
T\mn-\frac 13 T\eta\mn-\frac{2}{m^2}\p_{(\mu}\p_\alpha T^\alpha_{\nu)}+\frac{1}{3m^2}\p_\mu \p_\nu T
\\
&& \phantom{-\frac{1}{\mpl}\Big[}+\frac{1}{3m^2}\p_\alpha \p_\beta T\abu \eta\mn +\frac{2}{3m^4}\p_\mu \p_\nu \p_\alpha \p_\beta T\abu\nn
\Big]\\
&=&\frac{1}{\mpl}\Big[\etat_{\mu(\alpha}\etat_{\nu \beta)}-\frac 13 \etat\mn \etat\ab\Big]T\abu\,,
\ea
with
\ba
\etat\mn=\eta\mn-\frac{1}{m^2}\p_\mu\p_\nu\,.
\ea
The propagator for a massive spin-2 field is thus given by
\ba
G^{\rm massive}_{\mu\nu\alpha\beta}(x,x')=\frac{f^{\rm massive}_{\mu\nu\alpha\beta}}{\Box-m^2} \,,
\ea
where $f^{\rm massive}_{\mu\nu\alpha\beta}$ is the polarization tensor,
\ba
\label{polarization}
f^{\rm massive}_{\mu\nu\alpha\beta}=\etat_{\mu(\alpha}\etat_{\nu \beta)}-\frac 13 \etat\mn \etat\ab\,.
\ea
In Fourier space we have
\ba
\label{polarization2}
f^{\rm massive}_{\mu\nu\alpha\beta}(p_\mu,m)&=&
\frac 2{3m^4}p_\mu p_\nu p_\alpha p_\beta+\eta_{\mu(\alpha}\eta_{\nu \beta)}-\frac 13 \eta\mn \eta\ab\\
&+&\frac{1}{m^2}\(p_{\alpha}p_{(\mu}\eta_{\nu)\beta}+p_{\beta}p_{(\mu}\eta_{\nu)\alpha}
-\frac13 p_\mu p_\nu \eta\ab-\frac 13 p_\alpha p_\beta \eta\mn\)\nn
\,.
\ea
The amplitude exchanged between two sources $T\mn$ and $T'\mn$ via a massive spin-2 field is thus given by
\ba
\mathcal{A}_{TT'}^{\rm massive}=\int \d^4x\ h\mn T'{}\mnu = \int \d^4 x \ T'{}\mnu \frac{f^{\rm massive}_{\mu\nu\alpha\beta}}{\Box-m^2}\ T^{\alpha\beta}\,.
\ea
As mentioned previously, to compare this result with the massless case, the sources ought to be conserved in the massless limit, $\p_\mu T\mupn, \p_\mu T\mupn{}' \to 0$ as $m\to 0$. The gravitational exchange amplitude in the massless limit is thus given by
\ba
\label{A m to 0}
\mathcal{A}_{TT'}^{ m\to 0} \int \d^4 x \ T'{}\mnu \frac{1}{\Box}\ \(T\mn-\frac{1}{3} T \eta\mn\)\,.
\ea
We now compare this result with the amplitude exchanged by a purely massless graviton.

\subsubsection*{Massless spin-2 field}

In the massless case, the equation of motion~\eqref{Einstein eq lin massive} reduces to the linearized Einstein equation
\ba
\Ein\abu\mn h\ab=\frac{1}{\mpl}T\mn\,,
\ea
where diffeomorphism invariance requires the stress-energy to be conserved, $\p_\mu T\mupn = 0$. In this case the transverse part of this equation is trivially satisfied (as a consequence of the Bianchi identity which follows from symmetry).
Since the theory is invariant under diffeomorphism transformations~\eqref{diff}, one can choose a gauge of our choice, for instance de Donder (or harmonic) gauge
\ba
\p_\mu h\mupn= \frac{1}{2} p_\nu \,.
\ea
In de Donder gauge, the Einstein equation then reduces to
\ba
(\Box-m^2)h\mn=-\frac{2}{\mpl}\(T\mn-\frac{1}{2} T\eta\mn\)\,.
\ea
The propagator for a massless spin-2 field is thus given by
\ba
\label{MasslessProp}
G^{\rm massless}_{\mu\nu\alpha\beta}=\frac{f^{\rm massless}_{\mu\nu\alpha\beta}}{\Box}\,,
\ea
where $f^{\rm massless}_{\mu\nu\alpha\beta}$ is the polarization tensor,
\ba
f^{\rm massless}_{\mu\nu\alpha\beta}=\eta_{\mu(\alpha}\eta_{\nu \beta)}-\frac 12 \eta\mn \eta\ab\,.
\ea
The amplitude exchanged between two sources $T\mn$ and $T'\mn$ via a genuinely massless spin-2 field is thus given by
\ba
\mathcal{A}_{TT'}^{\rm massless}= -\frac 2 \mpl \int \d^4 x \ T'{}\mnu \frac{1}{\Box}\ \(T\mn-\frac{1}{2} T \eta\mn\)\,,
\ea
and differs from the result~\eqref{A m to 0} in the small mass limit. This difference between the massless limit of the massive propagator and the massless propagator (and gravitational exchange amplitude) is a well-known fact and was first pointed out by van Dam, Veltman and Zakharov in 1970~\cite{vanDam:1970vg,Zakharov:1970cC}. The resolution to this `problem' lies within the \emph{Vainshtein} mechanism~\cite{Vainshtein:1972sx}. In 1972, Vainshtein showed that a theory of massive gravity becomes strongly coupled a low energy scale when the graviton mass is small. As a result, the linear theory is no longer appropriate to describe the theory in the limit of small mass and one should keep track of the non-linear interactions (very much as what we do when approaching the Schwarzschild radius in GR.) We shall see in Section~\ref{sec:Vainshtein} how a special set of interactions dominate in the massless limit and are responsible for the screening of the extra degrees of freedom present in massive gravity.

Another `non-GR' effect was also recently pointed out in Ref.~\cite{Gullu:2013yha} where a linear analysis showed that massive gravity predicts different spin-orientations for spinning objects.


\subsection{From linearized diffeomorphism to full diffeomorphism invariance}
\label{sec:NonLinearMG}

When considering the massless and non-interactive spin-2 field in Section~\ref{subsec:EHkinetic}, the linear gauge invariance~\eqref{diff} is exact. However if this field is to be probed and communicates with the rest of the world, the gauge symmetry is forced to include non-linear terms which in turn forces the kinetic term to become fully non-linear. The result is the well-known fully covariant Einstein--Hilbert term $\mpl^2\sqrt{-g}R$, where $R$ is the scalar curvature associated with the metric $g\mn=\eta\mn+h\mn/\mpl$.

To see this explicitly, let us start with the linearized theory and couple it to an external source $T_0^{\mu\nu}$, via the coupling
\ba
\label{linear coupling}
\L_{\rm matter}^{\rm linear}=\frac{1}{2\mpl} h\mn T_0^{\mu\nu}\,.
\ea
This coupling preserves diffeomorphism invariance if the source is conserved, $\p_\mu T_0^{\mu\nu}=0$.
To be more explicit, let us consider a massless scalar field $\varphi$ which satisfied the Klein--Gordon equation $\Box \varphi=0$. A natural choice for the stress-energy tensor $T\mnu$ is then
\ba
\label{scalar Tuv}
T\mnu_0=\p^\mu\varphi\p^\nu \varphi-\frac 12 (\p \varphi)^2 \eta\mnu\,,
\ea
so that the Klein--Gordon automatically guarantees the conservation of the stress-energy tensor on-shell at the linear level and linearized diffeomorphism invariance. However the very coupling between the scalar field and the spin-2 field affects the Klein--Gordon equation in such a way that beyond the linear order, the stress-energy tensor given in~\eqref{scalar Tuv} fails to be conserved. When considering the coupling~\eqref{linear coupling}, the Klein--Gordon equation receives corrections of the order of $h\mn/\mpl$
\ba
\Box \varphi=\frac 1\mpl \(\p^\alpha (h_{\alpha\beta} \p^\beta \varphi)-\frac 12 \p_\alpha(h^\beta_\beta\p^\alpha \varphi)\)\,,
\ea
implying a failure of conservation of $T\mnu_0$ at the same order,
\ba
\p_\mu T_0\mnu= \frac {\p^\nu\varphi}\mpl \(\p^\alpha (h_{\alpha\beta} \p^\beta \varphi)-\frac 12 \p_\alpha(h^\beta_\beta\p^\alpha \varphi)\)\,.
\ea
The resolution is of course to include non-linear corrections in $h/\mpl$ in the coupling with external matter,
\ba
\label{non-linear coupling}
\L_{\rm matter}=\frac{1}{2\mpl} h\mn T_0^{\mu\nu}+\frac{1}{2\mpl^2}h\mn h_{\alpha\beta}T_1^{\mu\nu\alpha\beta}+ \cdots\,,
\ea
and promote diffeomorphism invariance to a non-linearly realized gauge symmetry, symbolically,
\ba
h \to h +\p \xi + \frac{1}{\mpl} \p (h \xi)+ \cdots\,,
\ea
so this gauge invariance is automatically satisfied on-shell order by order in $h/\mpl$, \ie the scalar field (or general matter field) equations of motion automatically imply the appropriate relation for the stress-energy tensor to all orders in $h/\mpl$. The resulting symmetry is the well-known fully non-linear coordinate transformation invariance (or covariance), which requires the stress-energy tensor to be covariantly conserved. To satisfy this symmetry, the kinetic term~\eqref{EH} should then be promoted to a fully non-linear contribution,
\ba
\L_{\rm kin\ linear}^{\rm spin-2}= -\frac 14 h^{\mu\nu} \Ein^{\alpha\beta}\mn h_{\alpha\beta}
\qquad\longrightarrow \qquad
\L_{\rm kin\ covariant}^{\rm spin-2}= \frac {\mpl^2}2 \sqrt{-g}R[g]\, .
\ea
Just as the linearized version $h^{\mu\nu} \Ein^{\alpha\beta}\mn h_{\alpha\beta}$ was unique, the non-linear realization $\sqrt{-g}R$ is also unique\epubtkFootnote{Up to other Lovelock invariants. Note however that $f(R)$ theories are not exceptions, as the kinetic term for the spin-2 field is still given by $\sqrt{-g}R$. See Section~\ref{sec:NoNewKin} for more a more detailed discussion in the case of massive gravity.}.
As a result, any theory of an interacting spin-2 field is necessarily fully non-linear and leads to the theory of gravity where non-linear diffeomorphism invariance (or covariance) plays the role of the local gauge symmetry that projects out four out of the potential six degrees of freedom of the graviton and prevents the excitation of any ghost by the kinetic term.

The situation is very different from that of a spin-1 field as seen earlier, where coupling with other fields can be implemented at the linear order without affecting the $U(1)$ gauge symmetry. The difference is that in the case of a $U(1)$ symmetry, there is a unique nonlinear completion of that symmetry, \ie the unique nonlinear completion of a $U(1)$ is nothing else but a $U(1)$. Thus any nonlinear Lagrangian which preserves the full $U(1)$ symmetry will be a consistent interacting theory. On the other hand for spin-2 fields, there are two, and only two ways to nonlinearly complete linear diffs, one as linear diffs in the full theory and the other as full non-linear diffs. While it is possible to write self-interactions which preserve linear diffs, there are no interactions between matter and $h_{\mu\nu}$ which preserve linear diffs. Thus any theory of gravity must exhibit full nonlinear diffs and is in this sense what leads us to GR.

\subsection{Non-linear \stu decomposition}
\label{sec:nonlinearStuck}

\subsubsection*{On the need for a reference metric}

We have introduced the spin-2 field $h\mn$ as the perturbation about flat spacetime. When considering the theory of a field of given spin it is only natural to work with Minkowski as our spacetime metric, since the notion of spin follows from that of Poincar\'e invariance. Now when extending the theory non-linearly, we may also extend the theory about different reference metric. When dealing with a reference metric different than Minkowski, one loses the interpretation of the field as massive spin-2, but one can still get a consistent theory. One could also wonder whether it is possible to write a theory of massive gravity without the use of a reference metric at all. This interesting question was investigated in~\cite{Boulware:1973my}, where it shown that the only consistent alternative is to consider a function of the metric determinant. However as shown in~\cite{Boulware:1973my}, the consistent function of the determinant is the cosmological constant and does not provide a mass for the graviton.

\subsubsection*{Non-linear \stu}

Full diffeomorphism invariance (or covariance) indicates that the theory should be built out of scalar objects constructed out of the metric $g\mn$ and other tensors. However as explained previously a theory of massive gravity requires the notion of a reference metric\epubtkFootnote{Strictly speaking, the notion of spin is only meaningful as a representation of the Lorentz group, thus the theory of massive spin-2 field is only meaningful when Lorentz invariance is preserved, \ie when the reference metric is Minkowski. While the notion of spin can be extended to other maximally symmetric spacetimes such as AdS and dS, it loses its meaning for non-maximally symmetric reference metrics $f\mn$.} $f\mn$ (which may be Minkowski $f\mn=\eta\mn$) and
at the linearized level, the mass for gravity was not built out of the full metric $g\mn$, but rather out of the fluctuation $h\mn$ about this reference metric which does not transform as a tensor under general coordinate transformations. As a result the mass term breaks covariance.

This result is already transparent at the linear level where the mass term~\eqref{general mass term} breaks linearized diffeomorphism invariance. Nevertheless, that gauge symmetry can always be `formally' restored using the \stu trick which amounts to replacing the reference metric (so far we have been working with the flat Minkowski metric as the reference), to
\ba
\eta\mn \longrightarrow (\eta\mn -\frac2\mpl\p_{(\mu} \chi_{\nu)})\,,
\ea
and transforming $\chi_\mu$ under linearized diffeomorphism in such a way that the combination $h\mn -2\p_{(\mu}\chi_{\nu)}$ remains invariant.
Now that the symmetry is non-linearly realized and replaced by general covariance, this \stu trick should also be promoted to a fully covariant realization.

Following the same \stu trick non-linearly, one can `formally restore' covariance by including four \stu fields $\phi^a$ ($a=0,1,2,3$) and promoting the reference metric $f\mn$, which may of may not be Minkowski,  to a tensor $\tilde f\mn$~\cite{Siegel:1993sk,ArkaniHamed:2002sp},
\ba
f\mn \longrightarrow \tilde f\mn =\p_\mu \phi^a \p_\nu \phi^b f_{ab}
\label{stuH}
\ea
As we can see from this last expression, $\tilde f\mn$ transforms as a tensor under coordinate transformations as long as each of the four fields $\phi^a$ transform as scalars. We may now construct the theory of massive gravity as a scalar Lagrangian of the tensors $\tilde f\mn$ and $g\mn$. In \emph{Unitary gauge}, where the \stu fields are $\phi^a=x^a$, we simply recover $\tilde f\mn = f\mn$.

This \stu trick for massive gravity dates already from Green and Thorn~\cite{Green:1991pa} and from Siegel~\cite{Siegel:1993sk}, introduced then within the context of Open String Theory.
In the same way as the massless graviton naturally emerges in the closed string sector, open strings also have spin-2 excitations but whose lowest energy state is massive at tree level (they only become massless once quantum corrections are considered). Thus at the classical level, open strings contain a description of massive excitations of a spin-2
field, where gauge invariance is restored thanks to same \stu fields as introduced in this section. In open string theory, these \stu fields naturally arise from the ghost coordinates.
When constructing the non-linear theory of massive gravity from extra dimension, we shall see that in that context the \stu fields naturally arise at the shift from the extra dimension.

For later convenience, it will be useful to construct the following tensor quantity,
\ba
\label{Xdef}
\mathbb{X}\mupn=g^{\mu \alpha}\tilde f_{\alpha \nu}=\p^\mu \phi^a \p_\nu \phi^b f_{ab}\,,
\ea
in unitary gauge, $\mathbb{X}=g^{-1}f$.

\subsubsection*{Alternative \stu trick}

An alternative way to St\"uckelberize the reference metric $f\mn$ is to express it as
\ba
\label{Y}
g^{ac}f_{cb}\to \mathbb{Y}^a_{\, b}= g^{\mu\nu}\p_\mu \phi^a \p_\nu \phi^c f_{cb}\,.
\ea
As nicely explained in Ref.~\cite{Alberte:2013sma}, both matrices $X\mupn$ and $Y^a_{\, b}$ have the same eigenvalues, so one can choose either one of them in the definition of the massive gravity Lagrangian without any distinction.
The formulation in terms of $Y$ rather than $X$ was originally used in Ref.~\cite{Chamseddine:2010ub}, although unsuccessfully as the potential proposed there exhibits the BD ghost instability, (see for instance Ref.~\cite{Berezhiani:2010xy}).

\subsubsection*{Helicity decomposition}

If we now focus on the flat reference metric, $f\mn=\eta\mn$, we may
further split the \stu fields as $\phi^a=x^a-\frac{1}{\mpl}\chi^a$ and identify the index $a$ with a Lorentz index\epubtkFootnote{This procedure can of course be used for any reference metric, but it fails in identifying the proper physical degrees of freedom when dealing with a general reference metric. See Refs.~\cite{deRham:2011rn,deRham:2012kf} as well as Section~\ref{sec:beyondDL} for further discussions on that point.}, we obtain the non-linear generalization of the \stu trick used in Section~\ref{sec:FP}
\ba
\label{eta stuck}
\eta\mn \longrightarrow \tilde f\mn &=&\eta\mn - \frac2\mpl\p_{(\mu} \chi_{\nu)}+\frac1{\mpl^2}\p_\mu \chi^a \p_\nu \chi^b \eta_{ab}\\
&=&\eta\mn -\frac{2}{\mpl m }\p_{(\mu}A_{\nu)}-\frac{2}{\mpl m^2}\Pi\mn\\
&+&\frac{1}{\mpl^2 m^2}\p_\mu A^\alpha\p_\nu\A_\alpha+\frac{2}{\mpl^2 m^3}\p_\mu A^\alpha \Pi_{\nu\alpha}+ \frac{1}{\mpl^2 m^4}\Pi^2\mn\,,\nn
\ea
where in the second equality we have used the split performed in~\eqref{chi A pi} of $\chi^a$ in terms of the helicity-0 and -1 modes and all indices are raised and lowered with respect to $\eta\mn$.

In other words, the fluctuations about flat spacetime are promoted to the tensor $H\mn$
\ba
\label{Hcov}
h\mn=\mpl\(g\mn-\eta\mn\) \hspace{5pt}\longrightarrow\hspace{5pt}
H\mn = \mpl\(g\mn - \tilde f\mn\)
\ea
with
\ba
H\mn
&=&h\mn +2 \p_{(\mu}\chi_{\nu)}-\frac 1\mpl\eta_{ab}\p_\mu \chi^a\p_\nu \chi^b \\
&=& h\mn +\frac{2}{m} \p_{(\mu}A_{\nu)}+
\frac{2}{m^2}\Pi\mn\\
&-&\frac{1}{\mpl m^2}\p_\mu A^\alpha\p_\nu\A_\alpha-\frac{2}{\mpl m^3}\p_\mu A^\alpha \Pi_{\nu\alpha}- \frac{1}{\mpl m^4}\Pi^2\mn\nn \,.
\ea
We recognize $h\mn$ as being the helicity-2 part of the graviton, $A_\mu$ the helicity-1 part and $\pi$ is the helicity-0 . The fact that these quantities continue to correctly identify the physical degrees of freedom non-linearly in the limit $\mpl \to \infty$ is non-trivial and has been derived in~\cite{deRham:2011qq}.

\subsubsection*{Non-linear Fierz--Pauli}

The most straightforward non-linear extension of the Fierz--Pauli mass term is as follows
\ba
\label{nlFP1}
\L^{\rm (nl1)}_{\rm FP}=-m^2\mpl^2\sqrt{-g} \([(\mathbb{I}-\mathbb{X})^2]-[\mathbb{I}-\mathbb{X}]^2\)\,,
\ea
this mass term is then invariant under non-linear coordinate transformations. This non-linear formulation was used for instance in~\cite{ArkaniHamed:2002sp}. Alternatively, one may also generalize the Fierz--Pauli mass non-linearly as follows~\cite{Boulware:1973my}
\ba
\label{nlFP2}
\L^{\rm (nl2)}_{\rm FP}=-m^2\mpl^2\sqrt{-g}\sqrt{\det \X} \([(\mathbb{I}-\mathbb{X}^{-1})^2]-[\mathbb{I}-\mathbb{X}^{-1}]^2\)\,.
\ea
A prior the linear Fierz--Pauli action for massive gravity can be extended non-linearly in an arbitrary number of ways. However, as we shall see below, most of these generalizations generate a ghost non-linearly, known as the Boulware--Deser (BD) ghost. In Section~\ref{sec:ghost-free MG} we shall see that the extension of the Fierz--Pauli to a non-linear theory free of the BD ghost is unique (up to two constant parameters).

\subsection{Boulware-Deser ghost}
\label{sec:BDghost}

The easiest way to see the appearance of a ghost at the non-linear is to follow the \stu trick non-linearly and observe the appearance of an Ostrogradsky instability~\cite{Creminelli:2005qk,Deffayet:2005ys}, although the original formulation was performed in Unitary gauge in~\cite{Boulware:1973my} in the ADM language (Arnowitt, Deser and Misner, see Ref.~\cite{Arnowitt:1962hi}). In this section we shall focus on the flat reference metric, $f\mn=\eta\mn$.

Focusing solely on the helicity-0 mode $\pi$ to start with, the tensor $\mathbb{X}\mupn$ defined in~\eqref{Xdef} is expressed as
\ba
\mathbb{X}\mupn=\delta\mupn-\frac{2}{\mpl m^2}\Pi\mupn+\frac{1}{\mpl^2 m^4}\Pi^\mu_{\, \alpha}\Pi^{\alpha}_{\, \nu}\,,
\ea
where at this level all indices are raised and lowered with respect to the flat reference metric $\eta\mn$.
Then the Fierz--Pauli mass term~\eqref{nlFP1} reads
\ba
\label{FP1}
\L^{\rm (nl1)}_{{\rm FP}, \, \pi}=-\frac{4}{m^2}\([\Pi^2]-[\Pi]^2\)+\frac{4}{\mpl m^4}\([\Pi^3]-[\Pi][\Pi^2]\)+\frac{1}{\mpl^2 m^6}\([\Pi^4]-[\Pi^2]^2\)\,.
\ea
Upon integration by parts, we notice that the quadratic term in~\eqref{FP1} is a total derivative, which is another way to see the special structure of the Fierz--Pauli mass term. Unfortunately this special fact does not propagate to higher order and the cubic and quartic interactions are genuine higher order operators which lead to equations of motion with quartic and cubic derivatives. In other words these higher order operators $\([\Pi^3]-[\Pi][\Pi^2]\)$ and $\([\Pi^4]-[\Pi^2]^2\)$ propagate an additional degree of freedom which by Ostrogradsky's theorem, always enters as a ghost. While at the linear level, these operators might be irrelevant, their existence implies that one can always find an appropriate background configuration $\pi=\pi_0 +\delta \pi$, such that the ghost is manifest
\ba
\L^{\rm (nl1)}_{{\rm FP}, \, \pi}=\frac{4}{\mpl m^4}Z^{\mu\nu\alpha \beta} \p_\mu\p_\nu \delta \pi \p_\alpha\p_\beta \delta \pi\,,
\ea
with $Z^{\mu\nu\alpha\beta}=3\p^\mu\p^\alpha \pi_0 \eta^{\nu \beta}- \Box \pi_0 \eta^{\mu\alpha}\eta^{\nu \beta}-2 \p^\mu \p^\nu \pi_0 \eta^{\alpha \beta}+\cdots$.
This implies that non-linearly (or around a non-trivial background), the Fierz--Pauli mass term propagates an additional degree of freedom which is a ghost, namely the BD ghost. The mass of this ghost depends on the background configuration $\pi_0$,
\ba
m^2_{\rm ghost}\sim \frac{\mpl m^4}{\p^2 \pi_0}\,.
\ea
As we shall see below, the resolution of the vDVZ discontinuity lies in the Vainshtein mechanism for which the field takes a large vacuum expectation value, $\p^2\pi_0\gg \mpl m^2$, which in the present context would lead to a ghost with an extremely low mass, $m^2_{\rm ghost}\lesssim m^2$.

Choosing another non-linear extension for the Fierz--Pauli mass term as in~\eqref{nlFP2} does not seem to help much,
\ba
\L^{\rm (nl2)}_{{\rm FP}, \, \pi}&=&-\frac{4}{m^2}\([\Pi^2]-[\Pi]^2\)-\frac{4}{\mpl m^4}\([\Pi]^3-4[\Pi][\Pi^2]+3[\Pi^3]\)+\cdots\nn \\
&\to& \frac{4}{\mpl m^4}\([\Pi][\Pi^2]-[\Pi^3]\)+\cdots\label{FP2}
\ea
where we have integrated by parts on the second line, and we recover exactly the same type of higher derivatives already at the cubic level, so the BD ghost is also present in~\eqref{nlFP2}.

Alternatively the mass term was also generalized to include curvature invariants as in Ref.~\cite{Berkhahn:2010hc}. This theory was shown to be ghost-free at the linear level on FLRW but not yet non-linearly.

\subsubsection*{Function of the Fierz--Pauli mass term}

As an extension of the Fierz--Pauli mass term, one could instead write a more general function of it, as considered in Ref.~\cite{Boulware:1973my}
\ba
\L_{F({\rm FP})}=-m^2\sqrt{-g} F\(g\mnu g^{\alpha\beta} (H_{\mu\alpha}H_{\nu\beta}-H\mn H_{\alpha\beta})\)\,,
\ea
however one can easily see, if a mass term is actually present, \ie $F'\ne 0$, there is no analytic choice of the function $F$ which would circumvent the non-linear propagation of the BD ghost. Expanding $F$ into a Taylor expansion, we see for instance that the only choice to prevent the cubic higher-derivative interactions in $\pi$, $[\Pi^3]-[\Pi][\Pi^2]$ is $F'(0)=0$, which removes the mass term as the same time. If $F(0)\ne 0$ but $F'(0)=0$, the theory is massless about the specific reference metric, but infinitely strongly coupled about other backgrounds.

Instead to prevent the presence of the BD ghost fully non-linearly (or equivalently about any background), one should construct the mass term (or rather potential term) in such a way, that all the higher derivative operators involving the helicity-0 mode $(\p^2 \pi)^n$ are  total derivatives. This is precisely what is achieved in the ``ghost-free'' model of massive gravity presented in Part~\ref{sec:ghost-free MG}. In the next Part~\ref{part:extradim} we shall use higher dimensional GR to get some insight and intuition on how to construct a consistent theory of massive gravity.


\newpage

\part{Massive Gravity from Extra Dimensions}
\label{part:extradim}

\section{Higher-Dimensional Scenarios}
\label{sec:HigherDim Scenarios}

As seen in the previous section, the `most natural' non-linear extension of the Fierz--Pauli mass term bears a ghost. Constructing consistent theories of massive gravity has actually been a challenging task for years, and higher-dimensional scenario can provide excellent frameworks for explicit realizations of massive gravity. The main motivation behind relying on higher dimensional gravity is twofold:
\begin{itemize}
\item The five-dimensional theory is explicitly covariant.
\item A massless spin-2 field in five dimensions has five degrees of freedom which corresponds to the correct number of dofs for a massive spin-2 field in four dimensions without the pathological BD ghost.
\end{itemize}
While string theory and other higher dimensional theories give rise naturally to massive gravitons, they usually include a massless  zero-mode. Furthermore in the simplest models, as soon as the first massive mode is relevant so is an infinite tower of massive (Kaluza--Klein) modes and one is never in a regime where a single massive graviton dominates, or at least this was the situation until the Dvali--Gabadadze--Porrati model (DGP)~\cite{Dvali:2000hr,Dvali:2000rv,Dvali:2000xg}, provided the first explicit model of (soft) massive gravity, based on a higher-dimensional braneworld model.

In the DGP model the graviton has a soft mass in the sense that its propagator does not have a  simple pole at fixed value $m$, but rather admits a resonance.
Considering the K\"all\'en--Lehmann spectral representation~\cite{Kallen:1952zz,Lehmann:1954xi}, the spectral density function $\rho(\mu^2)$ in DGP is of the form
\ba
\rho_{\rm DGP}(\mu^2)\sim \frac{m_0}{\pi \mu}\frac{1}{\mu^2+m_0^2}\,,
\ea
and so DGP corresponds to a  theory of massive gravity with a resonance with width $\Delta m \sim m_0$ about $m=0$.

In a Kaluza--Klein decomposition of a flat extra dimension  we have on the other hand an infinite tower of massive modes with spectral density function
\ba
\rho_{\rm KK}(\mu^2)\sim\sum_{n=0}^\infty \delta(\mu^2-(n m_0)^2)\,.
\ea
We shall see in the section on deconstruction \ref{sec:deconstruction} how one can truncate this infinite tower by performing a discretization in real space rather than in momentum space \`a la Kaluza--Klein, so as to obtain a theory of a single massive graviton
\ba
\rho_{\rm MG}(\mu^2)\sim \delta(\mu^2-m_0^2)\,,
\ea
or a theory of multi-gravity (with $N$-interacting gravitons),
\ba
\rho_{\rm multi-gravity}(\mu^2)\sim \sum_{n=0}^N \delta(\mu^2-(n m_0)^2)\,.
\ea
In this language bi-gravity is the special case of multi-gravity where $N=2$. These different spectral representations, together with the cascading gravity extension of DGP  are represented in Figure~\ref{Fig:Spectral}.

\epubtkImage{}{%
\begin{figure}[htb]
  \centerline{\includegraphics[width=\textwidth]{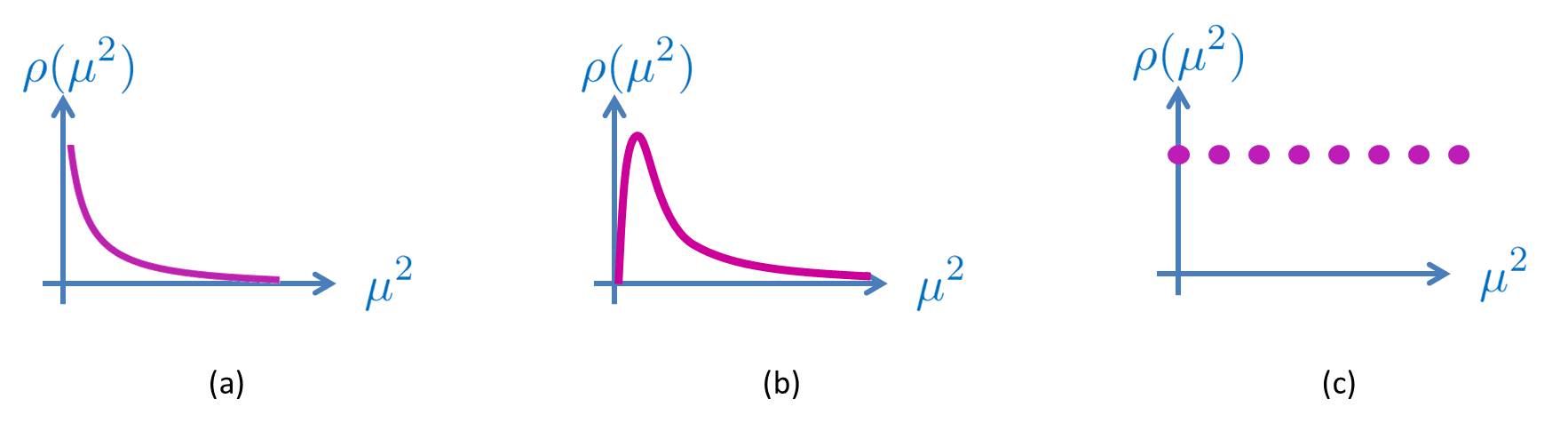}}
  \caption{Spectral representation of different models. (a) DGP (b)
    higher-dimensional cascading gravity and (c)
    multi-gravity. Bi-gravity is the special case of multi-gravity
    with one massless  mode and one massive mode. Massive gravity is
    the special case where only one massive mode couples to the rest
    of the standard model and the other modes decouple. (a) and (b)
    are models of soft massive gravity where the graviton mass can be
    thought of as a resonance.}
  \label{Fig:Spectral}
\end{figure}}

Recently another higher dimensional embedding of bi-gravity was proposed in Ref.~\cite{Yamashita:2014cra}. Rather than performing a discretization of the extra dimension, the idea behind this model is to consider a two-brane DGP model, where the radion or separation between these branes is stabilized via a Goldberger-Wise stabilization mechanism \cite{Goldberger:1999uk} where the brane and the bulk include a specific potential for the radion. At low energy the mass spectrum can be truncated to a massless mode and a massive mode, reproducing a bi-gravity theory. However the stabilization mechanism involves a relatively low scale and the correspondence breaks down above it. Nevertheless this provides a first proof of principle for how to embed such a model in a higher-dimensional picture without discretization and could be useful to tackle some of the open questions of massive gravity.

In what follows we review how five-dimensional massive gravity is a useful starting point in order to generate consistent four-dimensional theories of massive gravity, either for soft-massive gravity \`a la DGP and its extensions, or for hard massive gravity following a deconstruction framework.

The DGP model has played the role of a precursor for many developments in modified and massive gravity and it is beyond the scope of this review to summarize all of them. In this review we briefly summarize the DGP model and some key aspects of its phenomenology, and refer the reader to other reviews (see for instance~\cite{Gabadadze:2003ii,Maartens:2010ar,Gabadadze:2007dv}) for more details on the subject.

In this section,  $A,B,C\cdots=0,\ldots,4$ represent five-dimensional spacetime indices and $\mu, \nu,\alpha \cdots=0,\ldots,3$ label four-dimensional spacetime indices. $y=x^4$ represents the fifth additional dimension, $\{x^A\}=\{x^\mu, y\}$. The five-dimensional metric is given by $\uf g_{AB}(x,y)$ while the four-dimensional metric is given by $g\mn(x)$. The five-dimensional scalar curvature is $\uf R[G]$ while $R=R[g]$ is the four-dimensional scalar-curvature. We use the same notation for the Einstein tensor where $\uf G_{AB}$ is the five-dimensional one and $G\mn$ represents the four-dimensional one built out of $g\mn$.

When working in the Einstein--Cartan formalism of gravity, $\mathbbmtt{a,b,c}, \cdots$ label five-dimensional Lorentz indices and $a,b,c \cdots$ label the four-dimensional ones.

\section{The Dvali--Gabadadze--Porrati Model}
\label{sec:DGP}

The idea behind the DGP model~\cite{Dvali:2000rv,Dvali:2000hr,Dvali:2000xg} is to start with a four-dimensional braneworld in an infinite size-extra dimension.  A priori gravity would then be fully five-dimensional, with respective Planck scale $M_5$, but the matter fields localized on the brane could lead to an induced curvature term on the brane with respective Planck scale $\mpl$. See~\cite{Antoniadis:2002tr} for a potential embedding of this model within string theory.

At small distances the induced curvature dominates and gravity behaves as in four dimensions, while at large distances the leakage of gravity within the extra dimension weakens the force of gravity. The DGP model is thus a model of modified gravity in the infrared, and as we shall see, the graviton effectively acquires a soft mass, or resonance.

\subsection{Gravity induced on a brane}

We start with the five-dimensional action for the DGP model~\cite{Dvali:2000rv,Dvali:2000hr,Dvali:2000xg} with a brane localized at $y=0$,
\ba
\label{actionDGP}
S=\int \d^4 x \d y \( \frac{M_5^3}{4} \sqrt{-\uf g} \ \uf R
+\delta(y) \left[ \sqrt{-g}\frac{\mpl^2}{2}  R[g]+\mathcal{L}_m(g,\psi_i)\right]\)\,,
\ea
where $\psi_i$ represent matter field species confined to the brane with stress-energy tensor $T\mn$. This brane is considered to be an orbifold brane enjoying a $\mathbb{Z}_2$-orbifold symmetry (so that the physics at $y<0$ is the mirror copy of that at $y>0$.) We choose the convention where we consider $-\infty<y<\infty$, reason why we have a factor or $M_5^3/4$ rather than $M_5^3/2$ if we had only consider one side of the brane, for instance $y\ge 0$.

The five-dimensional Einstein equation of motion are then given by
\ba
\label{5dDGP}
M_5^3 \ \uf G_{AB}= 2 \delta(y) \uf\, T_{AB}
\ea
with
\ba
\label{5dDGPT}
\uf T_{AB}=\(- \mpl^2 G\mn + T\mn\) \delta^\mu_A \delta^\nu_B\,.
\ea
The Israel matching condition on the brane~\cite{Israel:1966rt} can be obtained by integrating this equation over $\int_{-\ep}^\ep \d y$ and taking the limit $\ep \to 0$, so that the jump in the extrinsic curvature on across the brane is related to the Einstein tensor and stress-energy tensor of the matter field confined on the brane.

\subsubsection{Perturbations about flat spacetime}
\label{sec:DGPpert}

In DGP the four-dimensional graviton is effectively massive. To see this explicitly, we look at perturbations about flat spacetime
\ba
\d s_5^2=\(\eta_{AB}+ h_{AB}(x,y)\)\d x^A \d x^B\,.
\ea
Since at this level we are dealing with five-dimensional GR, we are free to set the five-dimensional gauge of our choice and choose five-dimensional de Donder gauge (a discussion about the brane-bending mode will follow)
\ba
\label{deDonder DGP}
\p_A h^A_{\, B} =\frac 12 \p_B h^A_{\, A}\,.
\ea

In this gauge the five-dimensional Einstein tensor is simply
\ba
\uf G_{AB} = -\frac 12 \Box_5 \(h_{AB} -\frac 12 h^C_{\, C}\eta_{AB}\)\,,
\ea
where $\Box_5=\Box+\p_y^2$ is the five-dimensional d'Alembertian and $\Box$ is the four-dimensional one.

Since there is no source along the $\mu y$ or $yy$ directions ($\uf T_{\mu y}=0=\uf T_{yy}$), we can immediately infer that
\ba
\Box_5 h_{\mu y}=0 \qquad&\Rightarrow& \qquad h_{\mu y}=0\\
\Box_5 \(h_{y y}-h^\mu_{\, \mu}\)=0 \qquad&\Rightarrow& \qquad h_{y y}=h^\mu_{\, \mu}
\,, \label{hyy}
\ea
up to an homogeneous mode which in this setup we set to zero. This does not properly account for the brane-bending mode but for the sake of this analysis it will give the correct expression for the metric fluctuation $h\mn$. We will see in Section~\ref{sec:DGP decoupling} how to keep track of the brane-bending mode which is partly encoded in $h_{yy}$.

Using these relations in the five-dimensional de Donder gauge we deduce the relation for the purely four-dimensional part of the metric perturbation,
\ba
\p_\mu h^\mu_{\, \nu}=\p_\nu h^\mu_{\, \mu}\,.
\ea
Using these relations in the projected Einstein equation we get
\ba
\frac12 M_5^3\left[\Box+\p_y^2\right]\(h\mn -h \eta\mn\)
=-\delta(y)\(2 T\mn +\mpl^2 \(\Box h\mn -\p_\mu \p_\nu h \)\)\,,
\ea
where $h\equiv h^\alpha_{\, \alpha}=\eta^{\mu\nu}h\mn$ is the four-dimensional trace of the perturbations.

Solving this equation with the requirement that $h\mn \to 0$ as $y\to  \pm \infty$,  we infer the following profile for the perturbations along the extra dimension
\ba
h\mn(x,y) = e^{- |y|\, \sqrt{- \Box }}h\mn(x)\,,
\ea
where the $\Box$ should really be thought in Fourier space, and $h\mn(x)$ is set from the boundary conditions on the brane. Integrating the Einstein equation across the brane, from $-\ep$ to $+\ep$, we get
\ba
\frac12\lim_{\ep\to 0}M_5^3 \left[\p_y h\mn(x,y)-h(x,y)\eta\mn \right]^\ep_{-\ep}
+\mpl^2\(\Box h\mn(x,0)-\p_\mu \p_\nu h(x,0)\)\nn
\\ =-2 T\mn(x)\,,
\ea
yielding the modified linearized Einstein equation on the brane
\ba
\mpl^2 \left[(\Box h\mn-\p_\mu \p_\nu h)- m_0 \sqrt{-\Box}\(h\mn-h \eta\mn\)\right]=-2 T\mn\,,
\ea
where all the metric perturbations are the ones localized at $y=0$ and the constant mass scale $m_0$ is given by
\ba
m_0=\frac{M_5^3}{\mpl^2}\,.
\ea
Interestingly we see the special Fierz--Pauli combination $h\mn-h \eta\mn$ appearing naturally from the five-dimensional nature of the theory. At this level this corresponds to a linearized theory of massive gravity with a scale-dependent effective mass $m^2(\Box)=m_0 \sqrt{-\Box}$, which can be thought in Fourier space, $m^2(k)=m_0 k$. We could now follow the same procedure as derived in Section~\ref{sec:vDVZ} and obtain the expression for the sourced metric fluctuation on the brane
\ba
h\mn=-\frac{2}{\mpl^2}\frac{1}{\Box - m_0 \sqrt{-\Box}}\(T\mn -\frac 13 T \eta\mn +\frac{1}{3 m \sqrt{-\Box}}\p_\mu\p_\nu T\)\,,
\ea
where $T=\eta^{\mu\nu}T\mn$ is the trace of the four-dimensional stress-energy tensor localized on the brane.
This yields the following  gravitational exchange amplitude between two conserved sources $T\mn$ and $T'\mn$,
\ba
\mathcal{A}^{\rm DGP}_{TT'}=\int \d^4x\ h\mn T'{}\mnu = \int \d^4 x \ T'{}\mnu \frac{f^{\rm massive}_{\mu\nu\alpha\beta}}{\Box-m_0 \sqrt{-\Box}}\ T^{\alpha\beta}\,,
\ea
where the polarization tensor $f^{\rm massive}_{\mu\nu\alpha\beta}$
 the is the same as that given for Fierz--Pauli in~\eqref{polarization} in terms of $m_0$. In particular the polarization tensor includes the standard factor of $-1/3 T\eta\mn$ as opposed to $-1/2 T\eta\mn$ as would be the case in GR. This is again the manifestation of the vDVZ discontinuity which is cured by the Vainshtein mechanism as for Fierz--Pauli massive gravity. See~\cite{Deffayet:2001uk} for the explicit realization  of the Vainshtein mechanism in DGP which is where it was first shown to work explicitly.

\subsubsection{Spectral representation}

In Fourier space the propagator for the graviton in DGP is given by
\ba
\tilde G^{\rm massive}_{\mu\nu\alpha\beta}(k)=f^{\rm massive}_{\mu\nu\alpha\beta}(k,m_0)\ \tilde{\mathcal{G}}(k)\,,
\ea
with the massive polarization tensor $f^{\rm massive}$ defined in~\eqref{polarization2}
\ba
\tilde{\mathcal{G}}(k)=\frac{1}{k^2+m_0 k}\,,
\ea
which can be written in the K\"all\'en--Lehmann spectral representation as a sum of free propagators with mass $\mu$,
\ba
\label{propagator KL spectral}
\tilde{\mathcal{G}}(k)=\int_0^\infty \frac{\rho(\mu^2)}{k^2+\mu^2}\d \mu^2\,,
\ea
with the spectral density $\rho(\mu^2)$
\ba
\rho(\mu^2)=\frac{1}{\pi}\frac{m_0}{\mu}\frac{1}{\mu^2+ m_0^2}\,,
\ea
which is represented in Figure~\ref{Fig:Spectral}.
As already emphasized, the graviton in DGP cannot be thought of a single massive mode, but rather as a resonance picked about $\mu=0$.

We see that the spectral density is positive for any $\mu^2>0$, confirming the fact that about the normal (flat) branch of DGP there is no ghost.

Notice as well that in the massless limit $m_0\to 0$, we see appearing a representation of the Dirac delta function,
\ba
\lim_{m \to 0}\frac{1}{\pi}\frac{m_0}{\mu}\frac{1}{\mu^2+ m_0^2} = \delta(\mu^2)\,,
\ea
and so the massless mode is singled out in the massless limit of DGP (with the different tensor structure given by $f^{\rm massive}_{\mu\nu\alpha\beta}\ne f^{(0)}_{\mu\nu\alpha\beta}$ which is the origin of the vDVZ discontinuity see Section~\ref{sec:vDVZ}.)

\subsection{Brane-bending mode}
\label{sec:DGP decoupling}

\subsubsection*{Five-dimensional gauge-fixing}

In Section~\ref{sec:DGPpert} we have remained vague about the gauge-fixing and the implications for the brane position. The brane-bending mode is actually important to keep track off in DGP and we shall do that properly in what follows by keeping all the modes.

We work in the five-dimensional ADM split with the lapse $N=1/\sqrt{g^{yy}}=1+\frac 12 h_{yy}$, the shift $N_\mu=g_{\mu y}$ and the four-dimensional part of the metric, $g\mn(x,y)=\eta\mn+h\mn(x,y)$. The five-dimensional Einstein--Hilbert term is then expressed as
\ba
\label{DGP 5d ADM}
\L_{\rm R}^{(5)}=\frac{M_5^3}{4}\sqrt{-g}N\(R[g]+[K]^2-[K^2]\)\,,
\ea
where square brackets correspond to the trace of a tensor with respect to the four-dimensional metric $g\mn$ and $K\mn$ is the extrinsic curvature
\ba
K\mn=\frac{1}{2N}\(\p_y g\mn-D_\mu N_\nu-D_\nu N_\mu\)\,,
\ea
and $D_\mu$ is the covariant derivative with respect to $g\mn$.

First notice that the five-dimensional de Donder gauge choice~\eqref{deDonder DGP} can be made using the five-dimensional gauge fixing term
\ba
\L^{(5)}_{\rm Gauge-Fixing}&=&-\frac{M_5^3}{8}\(\p_A h^A_{\, B}-\frac 12 \p_B h^
A_{\, A}\)^2\\
&=& -\frac{M_5^3}{8}\Bigg[\(\p_\mu h\mupn-\frac 12 \p_\nu h+\p_y N_\nu-\frac 12 \p_\nu h_{yy}\)^2\\
&&\hspace{55pt}+\(\p_\mu N^\mu+\frac 12 \p_y h_{yy}-\frac 12 \p_y h\)^2\Bigg]\nn\,,
\ea
where we keep the same notation as previously, $h=\eta^{\mu\nu}h\mn$ is the four-dimensional trace.

After fixing the de Donder gauge~\eqref{deDonder DGP}, we can make the addition gauge transformation $x^A \to x^A + \xi^A$, and remain in de Donder gauge provided $\xi^A$ satisfies linearly $\Box_5 \xi^A=0$. This residual gauge freedom can be used to further fix the gauge on the brane (see~\cite{Luty:2003vm} for more details, we only summarize their derivation here).

\subsubsection*{Four-dimensional Gauge-fixing}

Keeping the brane at the fixed position $y=0$ imposes $\xi^y=0$ since we need $\xi^y(y=0)=0$ and $\xi$ should be bounded as $y\to \infty$ (the situation is slightly different in the self-accelerating branch and this mode can lead to a ghost, see Section~\ref{sec:DGP Selfacceleration} as well as~\cite{Koyama:2007za,Charmousis:2006pn}).

Using the bulk profile $h_{AB}(x,y)=e^{-\sqrt{-\Box} |y|}h_{AB}(x)$ and integrating over the extra dimension, we obtain the contribution from the bulk on the brane (including the contribution from the gauge-fixing term)
in terms of the gauge invariant quantity
\ba
\tilde h\mn=h\mn+\frac{2}{\sqrt{-\Box}}\p_{(\mu}N_{\nu)}=-\frac{2}{\sqrt{-\Box}}K\mn
\ea
\ba
\label{DGP bulk integrated}
S_{\rm bulk}^{\rm integrated}=\frac{M_5^3}{4}\int\d^4x \Bigg[-\frac 12 \tilde h^{\mu\nu}\sqrt{-\Box}\(\tilde h\mn-\frac 12 \tilde h\eta\mn\)
+\frac 12 h_{yy} \sqrt{-\Box} \(\tilde h-\frac 12 h_{yy}\)
\Bigg]\,.\nn
\ea
Notice again a factor of 2 difference from~\cite{Luty:2003vm} which arises from the fact that we integrate from $y=-\infty$ to $y=+\infty$ imposing a $\mathbb{Z}_2$-mirror symmetry at $y=0$, rather than considering only one side of the brane as in~\cite{Luty:2003vm}. Both conventions are perfectly reasonable.

The integrated bulk action~\eqref{DGP bulk integrated} is invariant under the residual linearized gauge symmetry
\ba
h\mn &\to& h\mn +2\p_{(\mu}\xi_{\nu)}\\
N_\mu &\to& N_\mu -\sqrt{-\Box} \xi_\nu\\
h_{yy} &\to& h_{yy}
\ea
which keeps both $\tilde h\mn$ and $h_{yy}$ invariant. The residual gauge symmetry can be used to set the gauge on the brane, and at this level from~\eqref{DGP bulk integrated} we can see that the most convenient gauge fixing term is~\cite{Luty:2003vm}
\ba
\L^{(4)}_{\rm Residual\ Gauge-Fixing}=-\frac{\mpl^2}{4}\(\p_\mu h\mupn-\frac 12 \p_\nu h+m_0 N_\nu\)^2\,,
\ea
with again $m_0=M_5^3/\mpl^2$, so that the induced Lagrangian on the brane (including the contribution from the residual gauge fixing term) is
\ba
\label{DGP boundary action}
S_{\rm boundary}=\frac{\mpl^2}{4}\int \d^4x \Bigg[
\frac12 h^{\mu\nu}\Box (h\mn-\frac 12 h \eta\mn)-2m_0 N_\mu \(\p_\alpha h^{\alpha \mu}-\frac 12 \p^\mu h\)
-m_0^2 N_\mu N^\mu
\Bigg]\,.
\ea
Combining the five-dimensional action from the bulk~\eqref{DGP bulk integrated} with that on the boundary
\eqref{DGP boundary action} we end up with the linearized action on the four-dimensional DGP brane~\cite{Luty:2003vm}
\ba
S_{\rm DGP}^{\rm (lin)}=\frac{\mpl^2}{4}\int \d^4 x\Bigg[
\frac{1}{2}h^{\mu\nu}\left[\Box-m_0 \sqrt{-\Box}\right](h\mn-\frac 12 h \eta\mn)
-m_0 N^\mu \p_\mu h_{yy} \\
-m_0 N^\mu \left[\sqrt{-\Box} + m_0 \right]N_\mu  
-\frac {m_0}4 h_{yy} \sqrt{-\Box }(h_{yy}-2h)
\Bigg]\nn\,.
\ea
As shown earlier we recover the theory of a massive graviton in four dimensions, with a soft mass $m^2(\Box)=m_0 \sqrt{-\Box}$. This analysis has allowed us to keep track of the physical origin of all the modes including the brane-bending mode which is especially relevant when deriving the decoupling limit as we shall see below.

The kinetic mixing between these different modes can be diagonalized by performing the change of variables~\cite{Luty:2003vm}
\ba
\label{DGP h'}
h\mn=\frac{1}{\mpl}\( h'\mn + \pi \eta\mn\)\\
\label{DGP N'}
N_\mu=\frac{1}{\mpl\sqrt{m_0}}N_\mu'+\frac{1}{\mpl m_0}\p_\mu \pi\\
\label{DGP hyy}
h_{yy}=- \frac{2\sqrt{-\Box}}{m_0\mpl}\pi\,,
\ea
so we see that the mode $\pi$ is directly related to $h_{yy}$. In the case of Section~\ref{sec:DGPpert}, we had set $h_{yy}=0$ and the field $\pi$ is then related to the brane bending mode. In either case we see that the extrinsic curvature $K\mn$ carries part of this mode.

Omitting the mass terms and other relevant operators, the action is diagonalized in terms of the different graviton modes at the linearized level $h'\mn$ (which encodes the helicity-2 mode), $N_\mu'$ (which is part of the helicity-1 mode) and $\pi$ (helicity-0 mode),
\ba
S_{\rm DGP}^{\rm (lin)}=\frac{1}{4}\int \d^4 x\Bigg[
\frac{1}{2}h'{}^{\mu\nu}\Box(h'\mn-\frac 12 h' \eta\mn)- N'{}^\mu \sqrt{-\Box} N'_\mu+3 \pi \Box \pi
\Bigg]\,.
\ea

\subsubsection*{Decoupling limit}
\label{sec:DGP DL}

We will be discussing the meaning of `decoupling limits' in more depth in the context of multi-gravity and ghost-free massive gravity in Section~\ref{sec:DL section}. The main idea behind the decoupling limit is to separate the physics of the different modes. Here we are interested in following the interactions of the helicity-0 mode without the complications from the standard helicity-2 interactions that already arise in GR. For this purpose we can take the limit $\mpl \to \infty$ while simultaneously sending $m_0=M_5^3/\mpl^2\to 0$ while keeping the scale $\Lambda=(m_0^2\mpl)^{1/3}$ fixed. This is the scale at which the first interactions arise in DGP.

In DGP the decoupling limit should be taken by considering the full five-dimensional theory, as was performed in~\cite{Luty:2003vm}. The four-dimensional Einstein--Hilbert term does not give to any operators before the Planck scale, so in order to look for the irrelevant operator that come at the lowest possible scale, it is sufficient to focus on the boundary term from the five-dimensional action. It includes operators of the form
\ba
\L_{\rm boundary}^{(5)}\supset m_0 \mpl^2 \p \(\frac{h\mn'}{\mpl}\)^n \(\frac{N_\mu'}{\sqrt{m_0} \mpl}\)^k \(\frac{\p \pi}{m_0 \mpl}\)^\ell\,,
\ea
with integer powers $n,k,\ell \ge 0$ and $n+k+\ell \ge 3$ since we are dealing with interactions. The scale at which such an operator arises is
\ba
\Lambda_{n,k, \ell}= \(\mpl^{n+k+\ell-2}m_0^{k/2+\ell-1}\)^{1/(n+3k/2+2\ell-3)}
\ea
and it is easy to see that the lowest possible scale is $\Lc=(\mpl m_0^2)^{1/3}$ which arises for $n=0, k=0$ and $\ell=3$, it is thus a cubic interaction in the helicity-0 mode $\pi$ which involves four derivatives. Since it is only a cubic interaction, we can scan all the possible ways $\pi$ enters at the cubic level in the five-dimensional Einstein--Hilbert action. The relevant piece are the ones from the extrinsic curvature in~\eqref{DGP 5d ADM}, and in particular the combination $N ([K]^2-[K^2])$, with
\ba
N&=&1+\frac12 e^{-\sqrt{-\Box} y}h_{yy}\\
K\mn&=&-\frac 12 (1-\frac 12e^{-\sqrt{-\Box}y}h_{yy})(\p_\mu N_\nu +\p_\nu N_\mu)\,.
\ea
Integrating $m_0 \mpl^2 N ([K]^2-[K^2])$ along the extra dimension, we obtain the cubic contribution in $\pi$ on the brane (using the relations~\eqref{DGP N'} and~\eqref{DGP hyy})
\ba
\L_{\Lc}=\frac{1}{2\Lc^3}(\p \pi)^2 \Box\pi\,.
\ea
So the decoupling limit of DGP arises at the scale $\Lc$ and reduces to a cubic Galileon for the helicity-0 mode with no interactions for the helicity-2 and -1 modes,
\ba
\L_{\rm DL \ DGP}&=&\frac18 h'{}^{\mu\nu}\Box \(h'\mn-\frac12 h' \eta\mn\)
-\frac 14 N'{}^\mu \sqrt{-\Box} N'_\mu\\
&& \nn +\frac32 \pi \Box \pi + \frac{1}{2\Lc^3}(\p \pi)^2 \Box\pi\,.
\ea

\subsection{Phenomenology of DGP}

The phenomenology of DGP is extremely rich and has led to many developments. In what follows we review one of the most important implications of the DGP for cosmology which the existence of self-accelerating solutions. The cosmology and phenomenology of DGP was first derived in~\cite{Deffayet:2000uy,Deffayet:2001pu} (see also~\cite{Lue:2002sw,Lue:2002fe,Lue:2004rj,Lue:2005ya}).

%

\subsubsection{Friedmann equation in de Sitter}
\label{sec:DGP Friedmann}

To get some intuition on how cosmology gets modified in DGP, we first look at de Sitter-like solutions and then infer the full Friedmann equation in a FLRW-geometry. We thus start with five-dimensional Minkowski in de Sitter slicing (this can be easily generalized to FLRW-slicing),
\ba
\label{DGP dS slicing}
\d s_5^2= b^2(y)\(\d y^2 +\gamma^{\rm (dS)}\mn \d x^\mu \d x^\nu \)\,,
\ea
where $\gamma^{\rm (dS)}\mn$ is the four-dimensional de Sitter metric with constant Hubble parameter $H$,
$\gamma^{\rm (dS)}\mn\d x^\mu \d x^\nu = - \d t^2+a^2(t)\d x^2$, and the scale factor is given by $a(t)=\exp (Ht)$. The metric~\eqref{DGP dS slicing} is indeed Minkowski in de Sitter slicing if the warp factor $b(y)$ is given by
\ba
b(y)=e^{\e H |y|}\,, \qquad{\rm with }\qquad\e =\pm 1\,,
\ea
and the mod $y$ has be imposed by the $\mathbb{Z}_2$-orbifold symmetry. As we shall see the branch $\e=+1$ corresponds to the self-accelerating branch of DGP and $\e =-1$ is the stable, normal branch of DGP.

We can now derive the Friedmann equation on the brane by integrating over the $00$-component of the Einstein equation~\eqref{5dDGP} with the source~\eqref{5dDGPT} and consider some energy density $T_{00}=\rho$. The four-dimensional Einstein tensor gives the standard contribution $G_{00}=3H^2$ on the brane and so we obtain the modified Friedmann equation
\ba
\frac{M_5^3}{2}\left[\lim_{\ep \to 0}\int_{-\ep}^\ep \uf G_{00} \d y\right]+3\mpl^2 H^2= \rho\,,
\ea
with $\uf G_{00}=3(H^2-b''(y)/b(y))$, so
\ba
\lim_{\ep \to 0}\int_{-\ep}^\ep \uf G_{00} \d y= - 6 \e H\,,
\ea
leading to the modified Friedmann equation,
\ba
\label{DGP Friedmann}
H^2-\e m_0 H = \frac{1}{3\mpl^2}\rho\,,
\ea
where the five-dimensional nature of the theory is encoded in the new term $-\e m_0 H$ (this new contribution can be seen to arise from the helicity-0 mode of the graviton and could have been derived using the decoupling limit of DGP.)

For reasons which will become clear in what follows, the choice $\e=-1$ corresponds to the \emph{stable} branch of DGP while the other choice $\e=+1$ corresponds to the \emph{self-accelerating} branch of DGP. As is already clear from the higher-dimensional perspective, when $\e=+1$, the warp factor grows in the bulk (unless we think of the junction conditions the other way around), which is already signaling towards a pathology for that branch of solution.

\subsubsection{General Friedmann equation}

This modified Friedmann equation has been derived assuming a constant $H$, which is only consistent if the energy density is constant (\ie a cosmological constant). We can now derive the generalization of this Friedmann equation for non-constant $H$. This amounts to account for $\dot H$ and other derivative corrections which might have been omitted in deriving this equation by assuming that $H$ was constant. But the Friedmann equation corresponds to the Hamiltonian constraint equation and higher derivatives (\eg $\dot H \supset \ddot a$ and higher derivatives of $H$) would imply that this equation is no longer a constraint and this loss of constraint would imply that the theory admits a new degree of freedom about generic backgrounds namely the BD ghost (see the discussion of Section~\ref{sec:Abscence of ghost proof}).

However in DGP we know that the BD ghost is absent (this is ensured by the five-dimensional nature of the theory, in five dimensions we start with five dofs, and there is thus no sixth BD mode). So the Friedmann equation cannot include any derivatives of $H$, and the Friedmann equation obtained assuming a constant $H$ is actually exact in FLRW even if $H$ is not constant. So the constraint~\eqref{DGP Friedmann} is the exact Friedmann equation in DGP for any energy density $\rho$ on the brane.

The same trick can be used for massive gravity and bi-gravity and the Friedmann equations~\eqref{Friedmann Hg}, \eqref{Friedmann Hf} and \eqref{Friedmann bigravity} are indeed free of any derivatives of the Hubble parameter.

\subsubsection{Observational viability of DGP}

Independently of the ghost issue in the self-accelerating branch of the model, there has been a vast amount of investigation on the observational viability of both the self-accelerating branch and the normal (stable) branch of DGP. First because many of these observations can apply equally well to the stable branch of DGP (modulo a minus sign in some of the cases), and second and foremost because DGP represents an excellent archetype in which ideas of modified gravity can be tested.

Observational tests of DGP fall into the following two main categories:
\begin{itemize}
\item \textbf{Tests of the Friedmann equation.} This test was performed mainly using Supernovae, but also using Baryonic Acoustic Oscillations and the CMB so as to fix the background history of the Universe~\cite{Deffayet:2001xs,Dvali:2003rk,Fairbairn:2005ue,Guo:2006ce,Maartens:2006yt,Apostolopoulos:2006si,Movahed:2007ie,Wan:2007pm,He:2007zf,Lombriser:2009xg,Tsujikawa:2010sc}.
   Current observations seem to slightly disfavor the additional $H$ term in the Friedmann equation of DGP, even in the normal branch where the late-time acceleration of the Universe is due to a cosmological constant as in $\Lambda$CDM.
    These put bounds on the graviton mass in DGP to the order of $m_0 \lesssim 10^{-1} H_0$, where $H_0$ is the Hubble parameter today (see Ref.~\cite{Xu:2013ega} for the latest bounds at the time of writing, including data from Planck). Effectively this means that in order for DGP to be consistent with observations, the graviton mass can have no effect on the late-time acceleration of the Universe.

\item \textbf{Tests of an extra fifth force}, either within the solar system, or during structure formation (see for instance~\cite{Koyama:2005kd,Gong:2008fh,Song:2006jk,Song:2007wd,Fang:2008kc,Wei:2008ig} Refs.~\cite{Stabenau:2006td,Khoury:2009tk,Schmidt:2009sg} for N-body simulations as well as Ref.~\cite{Amendola:2007rr,Schmidt:2008hc} using weak lensing).

    Evading fifth force experiments will be discussed in more detail within the context of the Vainshtein mechanism in Section~\ref{sec:Vainshtein} and thereafter, and we save the discussion to that section. See Refs.~\cite{Lue:2002sw,Lue:2002fe,Lue:2004rj,Lue:2005ya,Scoccimarro:2009eu} for a five-dimensional study dedicated to DGP.  The study of cosmological perturbations within the context of DGP was also performed in depth for instance in~\cite{Koyama:2007ih,Cardoso:2007xc}.
\end{itemize}

\subsection{Self-acceleration branch}
\label{sec:DGP Selfacceleration}

The cosmology of DGP has led to a major conceptual breakthrough, namely the realization that the Universe could be `self-accelerating'. This occurs when choosing the $\e=+1$ branch of DGP, the Friedmann equation in the vacuum reduces to~\cite{Deffayet:2000uy,Deffayet:2001pu}
\ba
H^2-m_0 H=0\,,
\ea
which admits a non-trivial solution $H=m_0$ in the absence of any cosmological constant nor vacuum energy. In itself this would not solve the old cosmological constant problem as the vacuum energy ought to be set to zero on its own, but it can lead to a model of `dark gravity' where the amount of acceleration is governed by the scale $m_0$ which is stable against quantum corrections.

This realization has opened a new field of study in its own right. It is beyond the scope of this review on massive gravity to summarize all the interesting developments that arose in the past decade and we simply focus on a few elements namely the presence of a ghost in this self-accelerating branch as well as a few cosmological observations.

\subsubsection*{ghost}

The existence of a ghost on the self-accelerating branch of DGP was first pointed out in the decoupling limit~\cite{Luty:2003vm,Nicolis:2004qq}, where the helicity-0 mode of the graviton is shown to enter with the wrong sign kinetic in this branch of solutions. We emphasize that the issue of the ghost in the self-accelerating branch of DGP is completely unrelated to the sixth BD ghost on some theories of massive gravity. In DGP there are five dofs one of which is a ghost.
The analysis was then generalized in the fully fledged five-dimensional theory by K.~Koyama in
\cite{Koyama:2005tx} (see also~\cite{Gorbunov:2005zk,Koyama:2007za} and~\cite{Charmousis:2006pn}).

When perturbing about Minkowski, it was shown that the graviton has an effective mass $m^2=m_0\sqrt{-\Box}$. When perturbing on top of the self-accelerating solution a similar analysis can be performed and one can show that in the vacuum the graviton has an effective mass at precisely the Higuchi-bound, $m^2_{\rm eff}=2 H^2$ (see Ref.~\cite{Higuchi:1986py}). When matter or a cosmological constant is included on the brane, the graviton mass shifts either inside the forbidden Higuchi-region $0<m^2_{\rm eff}<2H^2$, or outside $m^2_{\rm eff}>2H^2$. We summarize the three case scenario following~\cite{Koyama:2005tx,Charmousis:2006pn}
\begin{itemize}
\item In~\cite{Higuchi:1986py} it was shown that when the effective mass is within the forbidden Higuchi-region, the helicity-0 mode of graviton has the wrong sign kinetic term and is a ghost.
\item Outside this forbidden region, when $m^2_{\rm eff}>2 H^2$, the zero-mode of the graviton is healthy but
there exists a new normalizable brane-bending mode in the self-accelerating branch\epubtkFootnote{In the normal branch of DGP, this brane-bending mode turns out not to be normalizable. The normalizable brane-bending mode which is instead present in the normal branch fully decouples and plays no role.} which is a genuine degree of freedom. For $m^2_{\rm eff}>2 H^2$ the brane-bending mode was shown to be a ghost.
\item Finally at the critical mass $m^2_{\rm eff}=2 H^2$ (which happens when no matter nor cosmological constant is present on the brane), the brane-bending mode takes the role of the helicity-0 mode of the graviton, so that the theory graviton still has five degrees of freedom, and this mode was shown to be a ghost as well.
\end{itemize}
In summary, independently of the matter content of the brane, so long as the graviton is massive $m^2_{\rm eff}>0$, the self-accelerating branch of DGP exhibits a ghost. See also~\cite{Dvali:2006if} for an exact non-perturbative argument studying domain walls in DGP. In the self-accelerating branch of DGP domain walls bear a negative gravitational mass. This non-perturbative solution can also be used as an argument for the instability of that branch.

\subsubsection*{Evading the ghost?}

Different ways to remove the ghosts were discussed for instance in~\cite{Izumi:2006ca} where a second brane was included. In this scenario it was then shown that the graviton could be made stable but at the cost of including a new spin-0 mode (that appears as the mode describing the distance between the branes).

Alternatively it was pointed out in~\cite{Gabadadze:2006xm} that if the sign of the extrinsic curvature was flipped sign, the self-accelerating solution on the brane would be stable.

Finally, a stable self-acceleration was also shown to occur in the massless case $m^2_{\rm eff}=0$ by relying on Gauss--Bonnet terms in the bulk and a self-source AdS$_5$ solution~\cite{deRham:2006pe}. The five-dimensional theory is then similar as that of DGP~\eqref{actionDGP} but with the addition of a five-dimensional Gauss--Bonnet term $\mathcal{R}^2_{\rm GB}$ in the bulk and the wrong sign five-dimensional Einstein--Hilbert term,
\ba
S= \int \d^5 x\Bigg[\sqrt{- \uf g}\(-\frac{M_5^3}{4}\uf R[\uf g]-\frac{M_5^3\ell^2}{4}\uf \mathcal{R}^2_{\rm GB}[\uf g]\)\\
+\delta(y) \left[ \sqrt{-g}\frac{\mpl^2}{2} R+\mathcal{L}_m(g,\psi_i)\right]\Bigg]\,.\nn
\ea
The idea is not so dissimilar as in new massive gravity (see Section~\ref{sec:NMG}), where here the wrong sign kinetic term in five-dimensions is balanced by the Gauss--Bonnet term in such a way that the graviton has the correct sign kinetic term on the self-sourced AdS$_5$ solution. The length scale $\ell$ is related to this AdS length scale, and the self-accelerating branch admits a stable (ghost-free) de Sitter solution with $H\sim \ell^{-1}$.

We do not discuss this model any further in what follows since the graviton admits a zero (massless) mode. It is feasible that this model can be understood as a bi-gravity theory where the massive mode is a resonance. It would also be interested to see how this model fits in with the Galileon theories~\cite{Nicolis:2008in} which admit stable self-accelerating solution.

In what follows we go back to the standard DGP model be it the self-accelerating branch ($\e=1$) or the normal branch ($\e=-1$).

\subsection{Degravitation}
\label{sec:degravitation}

One of the main motivations behind modifying gravity in the infrared is to tackle the Old Cosmological constant problem. The idea behind `degravitation'~\cite{Dvali:2002fz,Dvali:2002pe,ArkaniHamed:2002fu,Dvali:2007kt} is if gravity is modified in the IR, then a cosmological constant (or the vacuum energy) could have a smaller impact on the geometry. In these models, we would live with a large vacuum energy (be it at the TeV scale or at the Planck scale) but only observe a small amount of late-acceleration due to the modification of gravity. In order for a theory of modified gravity to potentially tackle the Old Cosmological Constant Problem via degravitation it needs to have the two following properties:
\begin{enumerate}
\item First gravity must be weaker in the infrared and effectively massive~\cite{Dvali:2007kt} so that the effect of IR sources can be degravitated.
\item Second there must exist some (nearly) static attractor solutions towards which the system can evolve at late-time for arbitrary value of the vacuum energy or cosmological constant.
\end{enumerate}

\subsubsection*{Flat solution with a cosmological constant}

The first requirement is present in DGP, but as was shown in~\cite{Dvali:2007kt} in DGP gravity is not `sufficiently weak' in the IR to allow degravitation solutions. Nevertheless it was shown in~\cite{Deffayet:2001aw} that the normal branch of DGP satisfies the second requirement for any negative value of the cosmological constant. In these solutions the five-dimensional spacetime is not Lorentz invariant, but in a way which would not (at this background level) be observed when confined on the four-dimensional brane.

For positive values of the cosmological constant, DGP does not admit a (nearly) static solution. This can be understood at the level of the decoupling limit using the arguments of~\cite{Dvali:2007kt} and generalized for other mass operators.

Inspired by the form of the graviton in DGP, $m^2(\Box)=m_0 \sqrt{-\Box}$, we can generalize the form of the graviton mass to
\ba
\label{m alpha}
m^2(\Box)=m_0^2\(\frac{-\Box}{m_0^2}\)^\alpha\,,
\ea
with $\alpha$ a positive dimensionless constant. $\alpha=1$ corresponds to a modification of the kinetic term. As shown in~\cite{deRham:2013tfa}, any such modification leads to ghosts, so we do not consider this case here. $\alpha>1$ corresponds to a UV modification of gravity, and so we focus on $\alpha<1$.

In the decoupling limit the helicity-2 decouples from the helicity-0 mode which behaves (symbolically) as follows~\cite{Dvali:2007kt}
\ba
\label{alpha DL}
3 \Box \pi- \frac{1}{\mpl m_0^{4(1-\alpha)}}\Box \(\Box^{1-\alpha}\pi\)^2+\cdots =-\frac{1}{\mpl}T\,,
\ea
where $T$ is the trace of the stress-energy tensor of external matter fields. At the linearized level, matter couples to the metric $g\mn=\eta\mn+\frac{1}{\mpl}(h'\mn+\pi\eta\mn)$. We now check under which conditions we can still recover a nearly static metric in the presence of a cosmological constant $T\mn=-\Lambda_{\rm CC}g\mn$. In the linearized limit of GR this leads to the profile for the helicity-2 mode (which in that case corresponds to a linearized de Sitter solution)
\ba
h'\mn=-\frac{\Lambda_{\rm CC}}{6\mpl}\eta_{\rho \sigma}x^\rho x^\sigma \eta\mn\,.
\ea
One way we can obtain a static solution in this extended theory of massive gravity at the linear level is by ensuring that the solution for $\pi$ cancels out that of $h'\mn$ so that the metric $g\mn$ remains flat. $\pi=+\frac{\Lambda_{\rm CC}}{6\mpl}\eta\mn x^\mu x^\nu$ is actually the solution of~\eqref{alpha DL} when only the term $3 \Box \pi$ contributes and all the other operators vanish for $\pi \propto x_\mu x^\mu$. This is the case if $\alpha<1/2$ as shown in~\cite{Dvali:2007kt}. This explains why in the case of DGP which corresponds to border line scenario $\alpha=1/2$, one can never fully degravitate a cosmological constant.

\subsubsection*{Extensions}

This realization has motivated the search for theories of massive gravity with $0\le \alpha<1/2$, and especially the extension of DGP to higher dimensions where the parameter $\alpha$ can get as close to zero as required. This is the main motivation behind higher dimensional DGP~\cite{Kolanovic:2003am,Gabadadze:2003ck} and cascading gravity~\cite{deRham:2007xp,deRham:2007rw,deRham:2008qx,deRham:2009wb} as we review in what follows.
(In~\cite{Porrati:2004yi} it was also shown how a regularized version of higher dimensional DGP could be free of the strong coupling and ghost issues).

Note that $\alpha \equiv 0$ corresponds to a hard mass gravity. Within the context of DGP, such a model with an `auxiliary' extra dimension was proposed in~\cite{Gabadadze:2009ja,deRham:2009rm} where we consider a finite-size large extra dimension which breaks five-dimensional Lorentz invariance. The five-dimensional action is motivated by the five-dimensional gravity with scalar curvature in the ADM decomposition $\uf R=R[g]+[K]^2-[K^2]$, but discarding the contribution from the four-dimensional curvature $R[g]$. Similarly as in DGP, the four-dimensional curvature still appears induced on the brane
\ba
S=\frac{\mpl^2}{2}\int_0^{\ell} \d y \int d^4x \sqrt{-g}\(m_0 \([K]^2-[K^2]\)+\delta(y) R[g]\)\,,
\ea
where $\ell$ is the size of the auxiliary extra dimension and $g\mn$ is a four-dimensional metric and we set the lapse to one (this shift can be kept and will contribute to the four-dimensional \stu field which restores four-dimensional invariance, but at this level it is easier to work in the gauge where the shift is set to zero and reintroduce the \stu fields directly in four dimensions). Imposing the Dirichlet conditions $g\mn(x, y=0)=f\mn$, we are left with a theory of massive gravity at $y=0$, with reference metric $f\mn$ and hard mass $m_0$. Here again the special structure $\([K]^2-[K^2]\)$ inherited (or rather inspired) from five-dimensional gravity ensures the Fierz--Pauli structure and the absence of ghost at the linearized level. Up to cubic order in perturbations it was shown in~\cite{deRham:2010gu} that the theory is free of ghost and its decoupling limit is that of a Galileon.

Furthermore it was shown in~\cite{deRham:2009rm} that it satisfies both requirements presented above to potentially help degravitating a cosmological constant. Unfortunately at higher orders this model is plagued with the BD ghost~\cite{Hassan:2011zr} unless the boundary conditions are chosen appropriately~\cite{Berezhiani:2011nc}. For this reason we will not review this model any further in what follows and focus instead on the ghost-free theory of massive gravity derived in~\cite{deRham:2010iK,deRham:2010kj}

\subsubsection{Cascading gravity}

\subsubsection*{Deficit angle}

It is well known that a tension on a cosmic string does not cause the cosmic strong to inflate but rather creates a deficit angle in the two spatial dimensions orthogonal to the string. Similarly if we consider a four-dimensional brane embedded in six-dimensional gravity, then a tension on the brane leads to the following flat geometry
\ba
\d s_6^2=\eta\mn \d x^\mu \d x^\nu + \d r^2 +r^{2}\(1-\frac{\Delta\theta}{2\pi}\)\d \theta^2\,,
\ea
where the two extra dimensions are expressed in polar coordinates $\{r, \theta\}$ and $\Delta \theta$ is a constant which parameterize the deficit angle in this canonical geometry. This deficit angle is related to the tension on the brane $\Lambda_{\rm CC}$ and the six-dimensional Planck scale (assuming six-dimensional gravity)
\ba
\Delta \theta= 2 \pi \frac{\Lambda_{\rm CC}}{M_6^4}\,.
\ea
For a positive tension $\Lambda_{\rm CC}>0$, this creates a positive deficit angle and since $\Delta \theta$ cannot be smaller than $2\pi$, the maximal tension on the brane is $M_6^4$. For a negative tension on the other hand, there is no such bound as it creates a surplus of angle, see Figure~\ref{Fig:Deficit}.

\epubtkImage{}{%
\begin{figure}[htb]
  \centerline{\includegraphics[width=\textwidth]{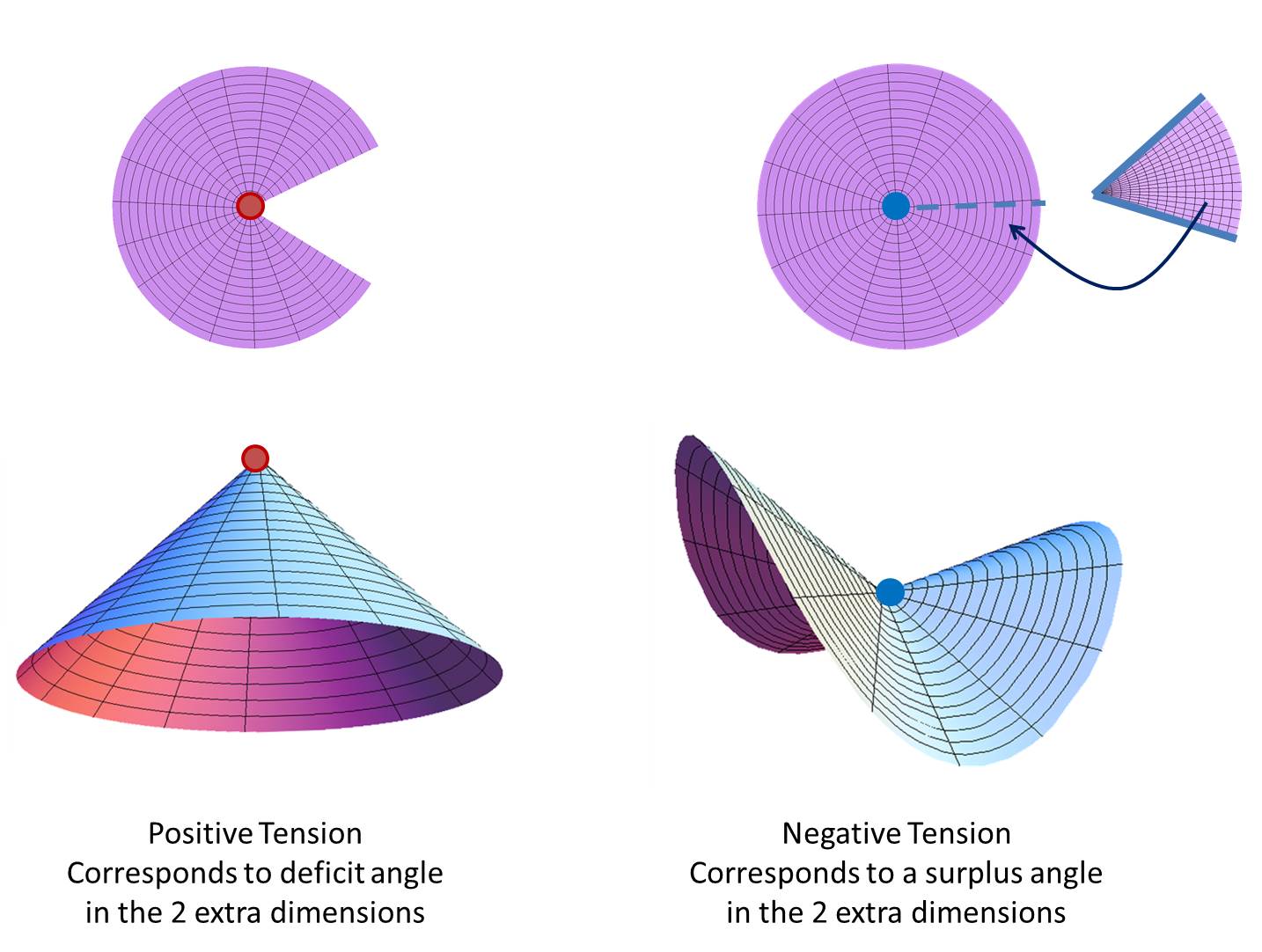}}
  \caption{Codimension-2 brane  with positive (resp.\ negative) tension brane leading to a positive (resp. negative) deficit angle in the two extra dimensions.}
  \label{Fig:Deficit}
\end{figure}}

This interesting feature has lead to many potential ways to tackle the cosmological constant by considering our Universe to live in a $3+1$-dimensional brane embedded in two or more large extra dimensions. (See Refs.~\cite{Aghababaie:2003wz,Aghababaie:2003ar,Navarro:2003bf,Nilles:2003km,Burgess:2004kd,Vinet:2004bk,Tolley:2006ht,Tolley:2007et,Burgess:2004dh,Burgess:2006ds,Garriga:2004tq,Kehagias:2004fb,Vinet:2005dg,Burgess:2004ib,Peloso:2006cq,Navarro:2004di,Lee:2003wg,Burgess:2004yq,Tolley:2005nu,deRham:2005ci} for the Supersymmetric Large Extra Dimension scenario as an alternative way to tackle the cosmological constant problem).
Extending the DGP to more than one extra dimension could thus provide a natural way to tackle the cosmological constant problem.

\subsubsection*{Spectral representation}

Furthermore in $n$-extra dimensions the gravitational potential is diluted as $V(r)\sim r^{-1-n}$. If the propagator has a K\"all\'en--Lehmann spectral representation with spectral density $\rho(\mu^2)$, the Newtonian potential has the following spectral representation
\ba
V(r)=\int_0^\infty\frac{\rho(\mu^2)e^{- \mu r}}{r}\d \mu^2\,.
\ea
In a higher-dimensional DGP scenario, the gravitation potential behaves higher dimensional at large distance, $V(r)\sim r^{-(1+n)}$ which implies $\rho(\mu^2)\sim \mu^{n-2} $ in the IR  as depicted in Figure~\ref{Fig:Spectral}.

Working back in terms of the spectral representation of the propagator as given in~\eqref{propagator KL spectral}, this means that the propagator goes to  $1/k$ in the IR as $\mu \to 0$ when $n=1$ (as we know from DGP), while it goes to a constant for $n>1$. So for more than one extra dimension, the theory tends towards that of a hard mass graviton in the far IR, which corresponds to $\alpha \to 0$ in the parametrization of~\eqref{m alpha}. Following the arguments of~\cite{Dvali:2007kt} such a theory should thus be a good candidate to tackle the cosmological constant problem.

\subsubsection*{A brane on a brane}

Both the spectral representation and the fact that codimension-two (and higher) branes can accommodate for a cosmological constant while remaining flat has made the field of higher-codimension branes particularly interesting.

 However as shown in~\cite{Gabadadze:2003ck} and~\cite{deRham:2007xp,deRham:2007rw,deRham:2008qx,deRham:2009wb}, the straightforward extension of DGP to two large extra dimensions leads to ghost issues (sixth mode with the wrong sign kinetic term, see also~\cite{Hassan:2010ys,Berkhahn:2012wg}) as well as divergences problems (see Refs.~\cite{Goldberger:2001tn,deRham:2007pz,deRham:2007dg,Papantonopoulos:2006dv,Papantonopoulos:2007fk,Kobayashi:2008bm,Burgess:2008yx}).

To avoid these issues, one can consider simply applying the DGP procedure step by step and consider a $4+1$-dimensional DGP brane embedded in six dimension. Our Universe would then be on a and $3+1$-dimensional DGP brane embedded in the $4+1$ one, (note we only consider one side of the brane here which explains the factor of 2 difference compared with~\eqref{actionDGP})
\ba
S=\frac{M_6^4}{2}\int \d^6x \sqrt{-g_6} \, ^{(6)}\hspace{-2pt}R
+\frac{M_5^3}{2}\int \d^5x \sqrt{-g_5} \, ^{(5)}\hspace{-2pt}R\\
+\frac{\mpl^2}{2}\int \d^4x \sqrt{-g_4} \, ^{(4)}\hspace{-2pt}R
+\int \d^4x \L_{\rm matter}(g_4, \psi)\,. \nn
\ea
This model has two cross-over scales: $m_5=M_5^3/\mpl^2$ which characterizes the scale at which one crosses from the four-dimensional to the five-dimensional regime, and $m_6=M_6^4/M_5^3$ yielding the crossing from a five-dimensional to a six-dimensional behavior. Of course we could also have a simultaneous crossing if $m_5=m_6$. In what follows we focus on the case where $\mpl> M_5 > M_6$.

Performing the same linearized analysis as in Section~\ref{sec:DGPpert} we can see that the four-dimensional theory of gravity is effectively massive with the soft mass in Fourier space
\ba
m^2(k)= \frac{\pi m_5}{4}\frac{\sqrt{m_6^2-k^2}}{{\rm arcth}\sqrt{\frac{m_6-k}{m_6+k}}}\,.
\ea
We see that the $4+1$-dimensional brane plays the role of a regulator (a divergence occurs in the limit $m_5\to 0$).

In this six-dimensional model, there are effectively two new scalar degrees of freedom (arising from the extra dimensions). We can ensure that both of them have the correct sign kinetic term by
\begin{itemize}
  \setlength{\itemsep}{0pt}
   \setlength{\parskip}{0pt}
   \setlength{\parsep}{0pt}
\item Either smoothing out the brane~\cite{Gabadadze:2003ck,deRham:2007rw} (this means that one should really consider a six-dimensional curvature on both the smoothed $4+1$ and on the $3+1$-dimensional branes, which is something one would naturally expect\epubtkFootnote{Note that in DGP, one could also consider a smooth brane first and the results would remain unchanged.}).
 \item Or by including some tension on the $3+1$ brane (which is also something natural since the setup is designed to degravitate a large cosmological constant on that brane). This was shown to be  ghost free in the decoupling limit in~\cite{deRham:2007xp} and in the full theory in~\cite{deRham:2010rw}.
\end{itemize}

 As already mentioned in two large extra dimensional models there is to be a maximal value of the cosmological constant that can be considered which is related to the six-dimensional Planck scale. Since that scale is in turn related to the effective mass of the graviton and since observations set that scale to be relatively small, the model can only take care of a relatively small cosmological constant. Nevertheless it still provides a proof of principle on how to evade Weinberg's no-go theorem~\cite{Weinberg:1988cp}.

The extension of cascading gravity to more than two extra dimensions was considered in~\cite{deRham:2009wb}. It was shown in that case how the $3+1$ brane remains flat for arbitrary values of the cosmological constant on that brane (within the regime of validity of the weak-field approximation). See Figure~\ref{Fig:7dCascading} for a picture on how the scalar potential adapts itself along the extra dimensions to accommodate for a cosmological constant on the brane.

\epubtkImage{}{%
\begin{figure}[htb]
  \centerline{
    \includegraphics[width=8cm]{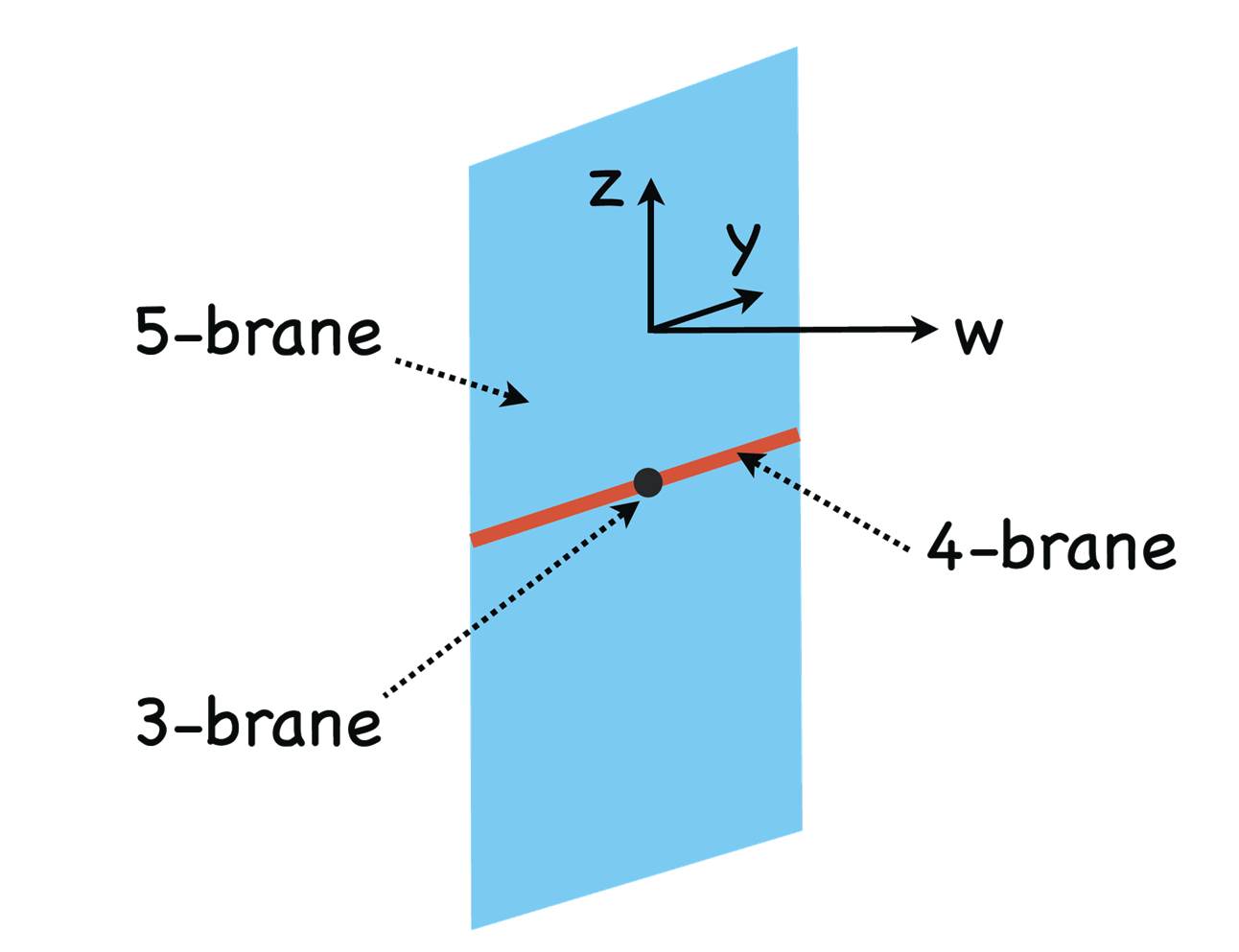}\qquad
    \includegraphics[width=6cm]{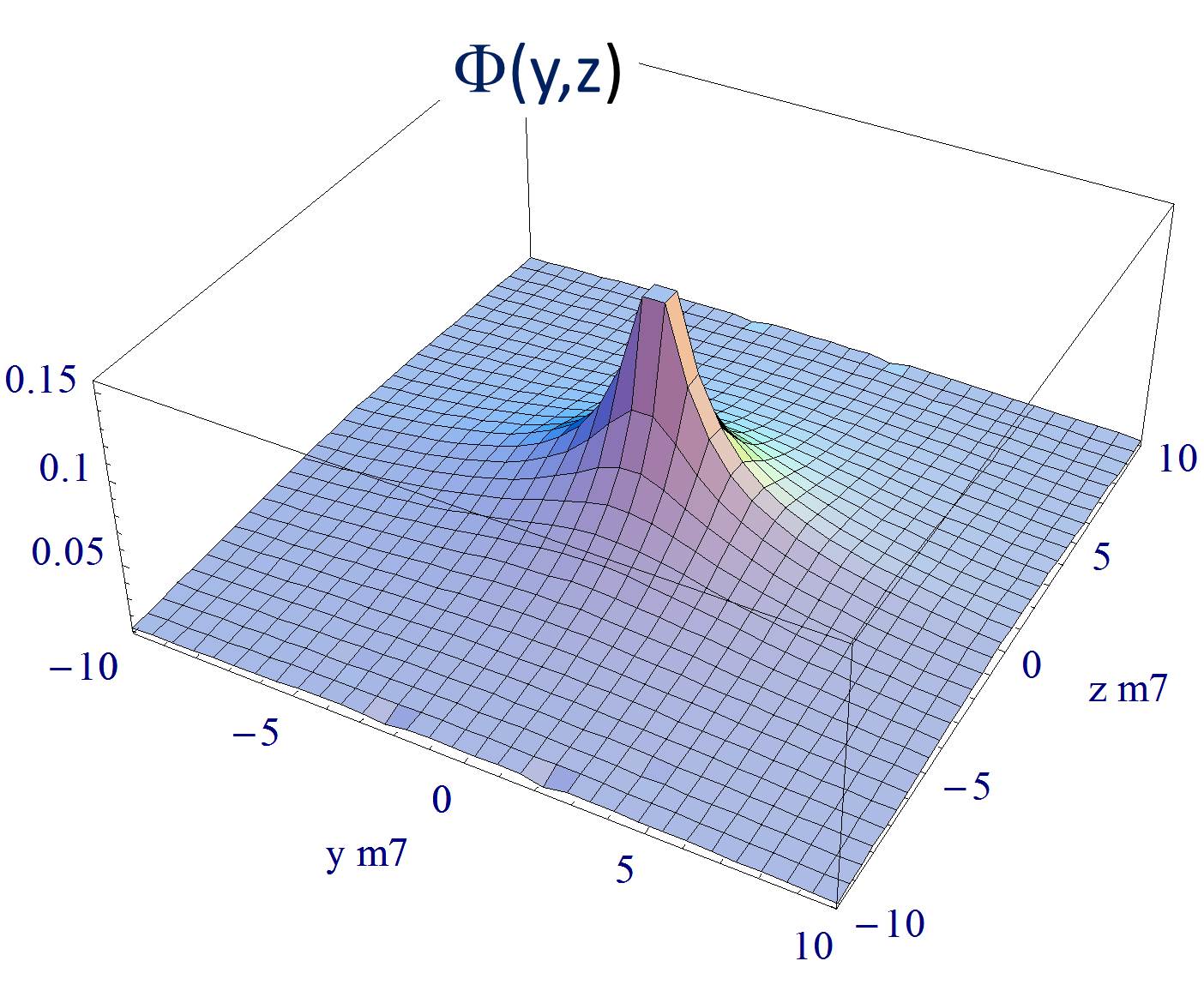}
  }
  \caption{Seven-dimensional cascading scenario and solution for one
    the metric potential $\Phi$ on the $(5+1)$-dimensional brane in a
    7-dimensional Cascading gravity scenario with tension on the
    $(3+1)$-dimensional brane located at $y=z=0$,  in the case where
    $M_6^4/M_5^3 = M_7^5/M_6^4=m_7$. $y$ and $z$ represent the two
    extra dimensions on the $(5+1)$-dimensional
    brane. From~\cite{deRham:2009wb}.}
  \label{Fig:7dCascading}
\end{figure}}



\section{Deconstruction}
\label{sec:deconstruction}

As for DGP and its extensions,
to get some insight on how to construct a four-dimensional theory of single massive graviton, we can start with five-dimensional General Relativity. This time we consider the extra dimension to be compactified and of finite size $R$, with periodic boundary conditions. It is then natural to perform a Kaluza--Klein decomposition and to obtain a tower of Kaluza--Klein graviton mode in four dimensions. The zero mode is then massless and the higher modes are all massive with mass separation $m=1/R$. Since the graviton mass is constant in this formalism we omit the subscript $_0$ in the rest of this review.

Rather than starting directly with a Kaluza--Klein decomposition (discretization in Fourier space), we perform instead a discretization in real space, known as ``deconstruction'' of five-dimensional gravity~\cite{ArkaniHamed:2001ca,ArkaniHamed:2001nc,Deffayet:2005yn,Deffayet:2003zk,ArkaniHamed:2003vb,Schwartz:2003vj,Kiritsis:2008at}. The deconstruction framework helps making the connection with massive gravity more explicit. However we can also obtain multi-gravity out of it which is then completely equivalent to the Kaluza--Klein decomposition (after a non-linear field redefinition).

The idea behind deconstruction is simply to `replace' the continuous fifth dimension $y$ by a series of $N$ sites $y_j$ separated by a distance $\ell=R/N$. So that the five-dimensional metric is replaced by a set of $N$ interacting metrics depending only on $x$.

In what follows we review the procedure derived in~\cite{deRham:2013awa} to recover four-dimensional ghost-free massive gravity as well as bi- and multi-gravity out of five-dimensional GR. The procedure works in any dimensions and we only focus to deconstructing five-dimensional GR for sake of concreteness.

\subsection{Formalism}

\subsubsection{Metric versus Einstein--Cartan formulation of GR}

Before going further, let us first describe five-dimensional general relativity in its Einstein--Cartan formulation, where we introduce a set of vielbein $e^\mathbbmtt{a}_A$, so that the relation between the metric and the vielbein is simply,
\ba
g_{AB}(x,y)=e^\mathbbmtt{a}_A(x,y)e^\mathbbmtt{b}_B(x,y)\eta_{\mathbbmtt{ab}}\,,
\ea
where as mentioned previously the capital latin letters label five-dimensional spacetime indices while  letters $\mathbbmtt{a,b,c}\cdots$ label five-dimensional Lorentz indices.

Under the torsionless condition,  $\d e + \omega \wedge e = 0$, the antisymmetric spin connection  $\omega$, is uniquely determined in terms of the vielbeins
\ba
\label{connection vielbein 5d}
\omega^{\mathbbmtt{ab}}_A = \frac{1}{2}e_A^\mathbbmtt{c}(O^{\mathbbmtt{ab}}_{\ \ \ \ \mathbbmtt{c}}-O_\mathbbmtt{c}^{\ \ \mathbbmtt{ab}}-O^{\mathbbmtt{b}\ \ \mathbbmtt{a}}_{\ \ \mathbbmtt{c}})\,,
\ea
with $O^{\mathbbmtt{ab}}_{\ \ \ \ \mathbbmtt{c}}= 2 e^{\mathbbmtt{a}A} e^{\mathbbmtt{b}B}\, \partial_{[A}e_{B] \mathbbmtt{c}}$. In the Einstein--Cartan formulation of GR, we introduce a 2-form Riemann curvature,
\ba
\mathcal{R}^{\mathbbmtt{ab}}=\d \omega^{\mathbbmtt{ab}}+\omega^{\mathbbmtt{a}}_{\ \ \mathbbmtt{c}}\wedge\omega^{\mathbbmtt{cb}}\,,
\ea
and up to boundary terms, the Einstein--Hilbert action is then given in the respective metric and the vielbein languages by (here in five dimensions for definiteness),
\ba
\label{S5d_1}
S_{\rm EH}^{(5)}&=&\frac{M_5^3}{2}\int \d^4x\d y \sqrt{-g}\, R^{(5)}[g]\\
&=&\frac{M_5^3}{2\times 3!}\int\ep_{\mathbbmtt{abcde}}  \, \R^{\mathbbmtt{ab}}\wed e^\mathbbmtt{c}\wed e^\mathbbmtt{d} \wed e^\mathbbmtt{e}\,,
\label{S5d_2}
\ea
where $R^{(5)}[g]$ is the scalar curvature built out of the five-dimensional metric $g\ab$ and $M_5$ is the five-dimensional Planck scale.

The counting of the degrees of freedom in both languages is of course equivalent and goes as follows: In $d$-spacetime dimensions,  the metric has $d(d+1)/2$ independent components. Covariance removes $2d$ of them\epubtkFootnote{The local gauge invariance associated with Covariance leads to $d$ first class constraints which remove $2d$ degrees of freedom, albeit in phase space. For global symmetries such as Lorentz invariance, there is no first-class constraints associated with them, and that global symmetry only removes $d(d-1)/2$ degrees of freedom. Technically the counting should be performed in phase space, but the results remains the same. See Section~\ref{sec:ADMHamiltonian} for a more detailed review  on the counting of degrees of freedom.}, which leads to $\mathcal{N}_d=d(d-3)/2$ independent degrees of freedom. In recover we recover the usual $\mathcal{N}_4=2$ independent polarizations for Gravitational waves. In five-dimensions, this leads to $\mathcal{N}_5=5$ degrees of freedom which is the same number  of degrees of freedom as a massive spin-2 field in four dimensions. This is as expect from the Kaluza--Klein philosophy (massless bosons in $d+1$ dimensions have the same number of degrees of freedom as massive bosons in $d$ dimensions -- this counting does not directly apply to fermions).

In the Einstein--Cartan formulation, the counting goes as follows: The vielbein has $d^2$ independent components. Covariance removes $2d$ of them, and the additional Global Lorentz invariance removes an additional $d(d-1)/2$, leading once again to a total of  $\mathcal{N}_d=d(d-3)/2$ independent degrees of freedom.

In GR one usually considers the metric and the vielbein formulation as being fully equivalent. However this perspective is true only in the bosonic sector. The limitations of the metric formulation becomes manifest when  coupling gravity to fermions. For such couplings one requires the vielbein formulation of GR. For instance, in four spacetime dimensions, the covariant action for a Dirac Fermion $\psi$ at the quadratic order is given by
(see Ref.~\cite{MacDowell:1977jt}),
\ba
\label{SDirac}
S_{\rm Dirac}=\int \frac{1}{3!}\ep_{abcd}\ e^a\wedge e^b \wedge e^c\ \left[
\frac{i}{2}\bar \psi \gamma^d\, \overleftrightarrow{D}\, \psi-\frac{m}{4}e^d \bar \psi \psi
\right]\,,
\ea
where the $\gamma^a$'s  are the Dirac matrices and $D$ represents the covariant derivative, $D \psi = d \psi -\frac 18 \omega^{ab}[\gamma_a,\gamma_b]\psi$.

In the bosonic sector, one can convert the covariant action of bosonic fields (\eg of scalar, vector fields, etc\ldots) between the vielbein and the metric language without much confusion, however this is not possible for the covariant Dirac action,  or other half-spin fields. For these types of matter fields, the Einstein--Cartan Formulation of GR is more fundamental than its metric formulation. In doubt, one should always start with the vielbein formulation. This is especially important in the case of deconstruction when a discretization in the metric language is not equivalent to a discretization in the vielbein variables. The same holds for Kaluza--Klein decomposition, a point which might have been under-appreciated in the past.

\subsubsection{Gauge-fixing}

The discretization process breaks covariance and so before staring this procedure it is wise to fix the gauge (failure to do so leads to spurious degrees of freedom which then become ghost in the four-dimensional description). We thus start in five spacetime dimensions by setting the gauge
\ba
\label{5dGauge}
G_{AB}(x,y)\d x^A \d x^B= \d y^2 + g\mn(x,y)\d x^\mu \d x^\nu\,,
\ea
meaning that the lapse is set to unity and the shift to zero. Notice that one could in principle only set the lapse to unity and keep the shift present throughout the discretization. From a four-dimensional point of view, the shift will then `morally' play the role of the \stu fields, however they do so only after a cumbersome field redefinition. So for sake of clarity and simplicity, in what follows we first gauge-fix the shift and then once the four-dimensional theory is obtained to restore gauge invariance by use of the \stu trick presented previously.

In vielbein language,
we fix the five-dimensional coordinate system and use four Lorentz transformations to set
\ba
\label{VielbeinGauge1}
e^\mathbbmtt{a}=\(\begin{array}{c}e^a_\mu \d x^\mu \\ \d y \end{array}\)\,,
\ea
and use the remaining six Lorentz transformations to set
\ba
\label{GaugeViel}
\omega^{ab}_y= e^{\mu [ a}\p_y e^{b]}_\mu =0\,.
\ea

In this gauge, the five-dimensional Einstein--Hilbert term (\ref{S5d_1}, \ref{S5d_2}) is given by
\ba
\label{S5d_3}
S_{\rm EH}^{(5)}&=&\frac{M_5^3}{2}\int \d^4x\d y \sqrt{-g}\, \(R[g]+[K]^2-[K^2]\)\\
&=&\frac{M_5^3}{4}\int\Big(\ep_{abcd}  \, R^{ab}\wed e^c\wed e^d -K^a\wed K^b\wed e^c \wed e^d\\
&&+2 K^a\wed \p_y e^b\wed e^c \wed e^d\Big)\wed \d y\,,\nn
\label{S5d_4}
\ea
where $R[g]$, is the four-dimensional curvature built out of the four-dimensional metric $g\mn$, $R^{ab}$ is the 2-form curvature built out of the four-dimensional vielbein $e^a_\mu$ and its associated connection $\omega^{ab}=\omega^{ab}_\mu \d x^\mu$, $R^{ab}=\d \omega^{ab}+\omega^a_{\ c}\wed \omega^{cb}$, and $K\mupn=g^{\mu\alpha} K_{\alpha \nu}$ is the extrinsic curvature,
\ba
K\mn&=&\frac12 \p_y g\mn=e^a_{(\mu}\p_y e^b_{\nu)}\, \eta_{ab}\\
K^a_\mu&=& e^{\nu a}K\mn\,.
\ea

\subsubsection{Discretization in the vielbein}

One could in principle go ahead and perform the discretization directly at the level of the metric but first this would not lead to a consistent truncated theory of massive gravity\epubtkFootnote{Discretizing at the level of the metric leads to a mass term similar to~\eqref{nlFP1} which as we have seen contains a BD ghost. }. As explained previously, the vielbein is more fundamental than the metric itself, and in what follows we discretize the theory keeping the vielbein as the fundamental object.
\ba
\label{disc1}
y &\hookrightarrow& y_j\\
\label{disc2}
e^a_\mu(x,y) &\hookrightarrow& e_{j}{}^a_\mu(x)=e^a_\mu(x,y_j) \\
\label{disc3}
\p_y e^a_\mu(x,y) &\hookrightarrow& m_N\(e_{j+1}{}^a_\mu-e_j{}^a_\mu\)\,.
\ea
The gauge choice~\eqref{GaugeViel} then implies
\ba
\label{SymmetricVielbein}
\omega^{ab}_y=e^{\mu [ a}\p_y e^{b]}_\mu =0&\hookrightarrow&e_{j+1}{}^{\mu [a} e_j{}^{b]}_\mu=0\,,
\ea
where the arrow $\hookrightarrow$ represents the deconstruction of five-dimensional gravity.  We have also introduced the `truncation scale', $m_N=N m =\ell^{-1}=N R^{-1}$, \ie the scale of the highest mode in the discretized theory.
After discretization, we see the Deser--van Nieuwenhuizen~\cite{Deser:1974cy} condition appearing in Eq.~\eqref{SymmetricVielbein},
which corresponds to the symmetric vielbein condition. This is a sufficient condition to allow for a formulation back into the metric language~\cite{Nibbelink:2006sz,Hinterbichler:2012cn,Deffayet:2012zc}. Note however that as mentioned in~\cite{deRham:2013awa}, we have not assumed that this symmetric vielbein condition was true, we simply derived it from the discretization procedure in the five-dimensional gauge choice $\omega^{ab}_{y}=0$.

In terms of the extrinsic curvature, this implies
\ba
\label{disc3}
K^a_\mu&\hookrightarrow&m_N\(e_{j+1}{}^a_\mu-e_{j}{}^a_\mu\)\,.
\ea
This can be written back in the metric language as follows
\ba
\label{disc4}
g\mn(x,y)&\hookrightarrow&g_{j, \mu\nu}(x)=g\mn(x,y_j)\\
K\mupn&\hookrightarrow&-m_N \K\mupn[g_{j}, g_{j+1}]\equiv-m_N\(\delta\mupn -\(\sqrt{g_{j}^{-1}\, g_{j+1}}\)\mupn\)\,,
\label{disc5}
\ea
where the square root in the extrinsic curvature appears after converting back to the metric language. The square root exists as long as the metrics $g_{j}$ and $g_{j+1}$ have the same signature and $g_{j}^{-1}\, g_{j+1}$ has positive eigenvalues so if both metrics were diagonal the `time' direction associated with each metric would be the same, which is a meaningful requirement.

From the metric language, we thus see that the discretization procedure amounts to converting the extrinsic curvature to an interaction between neighboring sites through the building block $\K\mupn[g_j,g_{j+1}]$.

\subsection{Ghost-free massive gravity}
\label{sec:MG}

\subsubsection{Simplest discretization}

In this subsection we focus on deriving a consistent theory of massive gravity from the discretization procedure (\ref{disc4}, \ref{disc5}). For this we consider a discretization with only two sites $j=1,2$
and will only be considered in the four-dimensional action induced on one site (say site 1), rather than the sum of both sites. This picture is analogous in spirit to a braneworld picture where we induce the action at one point along the extra dimension. This picture gives the theory of a unique dynamical metric, expressed in terms of a reference metric which corresponds to the fixed metric on the other site. We emphasize that this picture corresponds to a \emph{trick} to build a consistent theory of massive gravity, and would otherwise be more artificial than its multi-gravity extension. However as we shall see later, massive gravity can be seen as a perfectly consistent limit of multi (or bi-)gravity where the massless spin-2 field (and other fields in the multi-case) decouple and is thus perfectly acceptable.

To simplify the notation for this two-site case, we write the vielbein on both sites as $e_1=e$, $e_2=f$, and  similarly for the metrics $g_{1, \mu\nu}=g\mn$ and $g_{2, \mu\nu}=f\mn$. Out of the five-dimensional action for GR, we obtain the theory of massive gravity in four dimensions, (on site 1),
\ba
S^{(5)}_{\rm EH}&\hookrightarrow&S^{(4)}_{\rm mGR}\,,
\ea
with
\ba
\label{S4d_1}
S^{(4)}_{\rm mGR}&=&\frac{\mpl^2}{2}\int \d^4x \sqrt{-g}\, \(R[g]+m^2\([\K]^2-[\K^2]\)\)\\
&=&\frac{\mpl^2}{4}\int\,\ep_{abcd}  \( R^{ab}\wed e^c\wed e^d
+ m^2\A^{abcd}(e,f)\)\,,
\label{S4d_2}
\ea
with the mass term in the vielbein language
\ba
\label{Aabcd1}
\A^{abcd}(e,f)=(f^{a}-e^{a})\wed(f^{a}-e^{a})\wed e^{c} \wed e^{d }\,,
\ea
or the mass term building block in the metric language,
\ba
\label{eqK1}
\K\mupn=\delta\mupn -\(\sqrt{g^{-1}f}\)\mupn\,.
\ea
and we introduced the four-dimensional Planck scale, $\mpl^2=M_5^3\int \d y$, where in this case we limit the integral about one site.

The theory of massive gravity~\eqref{S4d_1}, or equivalently~\eqref{S4d_2} is one special example  of a ghost-free theory of massive gravity (\ie for which the BD ghost is absent). In terms of the `St\"uckelbergized" tensor $\mathbb{X}$ introduced in Eq.~\eqref{Xdef}, we see that
\ba
\K\mupn=\delta\mupn -\(\sqrt{\mathbb{X}}\)\mupn\,,
\ea
or in other words,
\ba
\mathbb{X}\mupn=\delta\mupn-2 \K\mupn+\K^\mu_\alpha \K^\alpha_\nu\,,
\ea
and the mass term can be written as
\ba
\label{Lmass K2}
\L_{\rm mass}&=&-\frac{m^2\mpl^2}{2}\sqrt{-g}\([\K^2]-[\K]^2\)\\
&=&-\frac{m^2\mpl^2}{2}\sqrt{-g} \([(\mathbb{I}-\sqrt{\mathbb{X}})^2]-[\mathbb{I}-\sqrt{\mathbb{X}}]^2\)\,.
\ea
This also a generalization of the Fierz--Pauli mass term, albeit more complicated on first sight than the ones considered in~\eqref{nlFP1} or~\eqref{nlFP2}, but as we shall see, a generalization of the Fierz--Pauli mass term which remains free of the BD ghost.

\subsubsection{Generalized mass term}

This mass term is not the unique ghost-free generalization of Fierz--Pauli gravity and by considering more general discretization procedures we can generate the entire 2-parameter family of acceptable potentials for gravity.
Rather than considering the straight-forward discretization $e(x,y)\hookrightarrow e_j(x)$, we could consider the average value on one site, pondered with arbitrary weight $r$,
\ba
e(x,y)\hookrightarrow r e_j+(1-r)e_{j+1}\,.
\ea
The mass term at one site is then generalized to
\ba
\label{Ka}
K^a\wedge K^b \wedge e^c \wedge e^d \hookrightarrow  m^2 \A^{abcd}_{r,s}(e_j,e_{j+1})\,,
\ea
and the most general action for massive gravity with reference vielbein $f$ is thus\epubtkFootnote{This special fully non-linear and Lorentz invariant theory of massive gravity, which has been proven to be free of the BD ghost has since then be dubbed `dRGT' theory. To avoid any confusion, we thus also call this ghost-free theory of massive gravity, the dRGT theory.}
\ba\label{dRGT_vielbein}
S_{\rm mGR}=\frac{\mpl^2}{4}\int\,\ep_{abcd}  \( R^{ab}\wed e^c\wed e^d
+ m^2\A^{abcd}_{r,s}(e,f)\)\,,
\ea
with
\ba
\label{Aabcd}
\A^{abcd}_{r,s}(e,f)=(f^a-e^a)\wedge(f^b-e^b)\wedge((1-r) e^c+r f^c)
\wedge ((1-s) e^d+s f^d)\,,\nn
\ea
for any $ r, s\in \mathbb{R}$.

In particular for the two-site case, this generates the two-parameter family of mass terms
\ba
\label{def_A}
\A_{r,s}^{abcd}(e,f)&=&c_0\, e^a\wedge e^b \wedge e^c \wedge e^d+c_1\, e^a\wedge e^b \wedge e^c \wedge f^d\\
&+&c_2\, e^a\wedge e^b \wedge f^c \wedge f^d+c_3\, e^a\wedge f^b \wedge f^c \wedge f^d
+c_4\, f^a\wedge f^b \wedge f^c \wedge f^d\nn\\
&\equiv&\A_{1-r,1-s}(f,e)\,,
\ea
with $c_0=(1-s)(1-r)$, $c_1=(-2+3s+3r-4 rs)$, $c_2=(1-3s-3r+6 rs)$, $c_3=(r+s-4rs)$ and $c_4=r s$. This corresponds to the most general potential which, by construction, includes no cosmological constant nor tadpole.

We see that in the vielbein language, the expression for the mass term is extremely natural and simple. In fact this form was guessed at already for special cases in Ref.~\cite{Nibbelink:2006sz} and even earlier in~\cite{Zumino:1970tu}. However the crucial analysis on the absence of ghosts and the reason for these terms was incorrect in both of these presentations. Subsequently after the development of the consistent metric formulation, the generic form of the mass terms was given in Refs.~\cite{Chamseddine:2011mu}\epubtkFootnote{The analysis performed in Ref.~\cite{Chamseddine:2011mu} was unfortunately erroneous, and the conclusions of that paper are thus incorrect.} and \cite{Hinterbichler:2012cn}.


In the metric Language, this corresponds to the following Lagrangian for ghost-free  massive gravity~\cite{deRham:2010kj},
\ba\label{dRGT_metric}
\L_{\rm mGR}=\frac{\mpl^2}{2}\int \d^4 x\sqrt{-g}\(R+\frac{m^2}{2}\(\L_2[\K]+\alpha_3 \L_3[\K]+\alpha_4 \L_4[\K]\)\)\,,
\ea
where the two parameters $\alpha_{3,4}$ are related to the two discretization parameters $r,s$ as
\ba
\label{alpha <-> r,s}
\alpha_3=r+s, \hspace{10pt}{\rm and}\hspace{10pt}
\alpha_4=rs \,,
\ea
and for any tensor $Q$, we define the scalar $\L_n$ symbolically as
\ba
\label{genericLn}
\L_n[Q]=\ep \ep Q^n\,,
\ea
for any $n=0,\cdots, d$, where $d$  is the number of spacetime dimensions.
$\ep$ is the Levi-Cevita antisymmetric symbol, so for instance in four dimensions, $\L_2[Q]=\ep^{\mu\nu\alpha\beta}\ep_{\mu'\nu'\alpha\beta} Q^{\mu'}_\mu Q^{\nu'}_\nu=2!([Q]^2-[Q^2])$, so we recover the mass term expressed in~\eqref{Lmass K2}.
Their explicit form is given in what follows in the relations (\ref{L2}-\ref{L4}) or (\ref{L2 X}-\ref{L4 X}).

This procedure is easily generalizable to any number of dimensions, and massive gravity in $d$ dimensions has $(d-2)$-free parameters which are related to the $(d-2)$ discretization parameters.

\subsection{Multi-gravity}

In the previous section, we showed how to obtain massive gravity from considering the five-dimensional Einstein--Hilbert action on one site\epubtkFootnote{In the previous section we obtained directly a theory of massive gravity, this should be seen as a \emph{trick} to obtain a consistent theory of massive gravity. However we shall see that we can take a decoupling limit of bi- (or even multi-)gravity so as to recover massive gravity and a \emph{decoupled} massless spin-2 field. In this sense massive gravity is a perfectly consistent limit of bi-gravity.}. Instead in this section, we integrate over the whole of the extra dimension,  which corresponds to summing over all the sites after discretization. Following the procedure of~\cite{deRham:2013awa}, we consider $N=2M+1$ sites to start which leads to multi-gravity~\cite{Hinterbichler:2012cn}, and then focus on the two-site case leading to bi-gravity~\cite{Hassan:2011zd}.

Starting with the five-dimensional action~\eqref{S5d_4} and applying the discretization procedure~\eqref{Ka} with $\A^{abcd}_{r,s}$ given in~\eqref{Aabcd}, we get
\ba
\label{Multigravity_1}
S_{N\, {\rm mGR}}&=&\frac{M_4^2}{4}\sum_{j=1}^N\int\,\ep_{abcd}  \( R^{ab}[e_j]\wed e_j^c\wed e_j^d
+ m_N^2\A_{r_j, s_j}^{abcd}(e_j,e_{j+1})\)\\
&=&\frac{M_4^2}{2}\sum_{j=1}^N\int \d^4 x \sqrt{-g_j}\(R[g_j]+\frac{m_N^2}{2}\sum_{n=0}^4\alpha^{(j)}_n \L_n(\K_{j, j+1})
\)\,, \nn
\label{Multigravity_2}
\ea
with $M_4^2=M_5^3 R =M_5^3/m$, $\alpha_2^{(j)}=-1/2$, and in this deconstruction framework we obtain no Cosmological constant nor tadpole, $\alpha_0^{(j)}=\alpha_1^{(j)}=0$ at any site $j$, (but we keep them for generality). In the mass Lagrangian, we use the shorthand notation $\K_{j, j+1}$ for the tensor $\K\mupn [g_j, g_{j+1}]$.
This is a special case of multi-gravity presented in~\cite{Hinterbichler:2012cn} (see also~\cite{Noller:2013yja} for other `topologies' in the way the multiple gravitons interact), where each metric only interacts with two other metrics, \ie with its closest neighbors, leading to $2N$-free parameters.  For any fixed $j$, one has $\alpha_3^{(j)}=(r_j+s_j)$, and $\alpha_4^{(j)}=r_j s_j$.

To see the mass spectrum of this multi-gravity theory, we perform a Fourier decomposition, which is what one would obtain (after a field redefinition) by performing a KK decomposition rather than a real space discretization. KK decomposition and deconstruction are thus perfectly equivalent (after a non-linear - but benign\epubtkFootnote{The field redefinition is local so no new degrees of freedom or other surprises hide in that field redefinition.} - field  redefinition). We define the discrete Fourier transform of the vielbein variables,
\ba
\tilde{e}^a_{\mu, n}=\frac{1}{\sqrt{N}}\sum_{j=1}^N e^a_{\mu, j} e^{i \frac{2\pi }{N}j}\,,
\ea
with the inverse map,
\ba
e^a_{\mu, j}=\frac{1}{\sqrt{N}}\sum_{n=-M}^M \tilde e^a_{\mu, n} e^{-i \frac{2\pi }{N}n}\,.
\ea
In terms of the Fourier transform variables, the multi-gravity action then reads at the linear level
\ba
\label{Lmulti_Fourier}
\L=\sum_{n=-M}^{M}\Big[(\p \tilde h_n)(\p \tilde h_{-n})+m_n^2\tilde h_n \tilde h_{-n} \Big]+\mathcal{L}_{\rm int}
\ea
with $\mpl^{-1}\tilde h_{\mu\nu, n}=\tilde{e}_{\mu, n}^a \tilde{e}_{\nu, n}^b\eta_{ab}-\eta\mn$ and $\mpl$ represents the four-dimensional Planck scale, $\mpl=M_4/\sqrt{N}$. The reality condition on the vielbein imposes $\tilde{e}_n=\tilde{e}_{-n}^*$ and similarly for $\tilde h_n$. The mass spectrum is then
\ba
m_n=m_N \sin \(\frac{n}{N}\) \approx n m \hspace{10pt}{\rm for}\hspace{10pt} n\ll N.
\ea

The counting of the degrees of freedom in multi-gravity goes as follows: the theory contains $2M$ massive spin-2 fields with five degrees of freedom each and one massless spin-2 field with two degrees of freedom, corresponding to a total of $10M+2$ degrees of freedom. In the continuum limit, we also need to account for the zero mode of the lapse and the shift which have been gauged fixed in five dimensions (see Ref.~\cite{Schwartz:2003vj} for a nice discussion of this point). This leads to three additional degrees of freedom, summing up to a total of $5N$ degrees of freedom of the four coordinates $x^a$.

\subsection{Bi-gravity}
\label{sec:bi-gravity}

Let us end this section with the special case of bi-gravity. Bi-gravity can also be derived from the deconstruction paradigm, just as massive gravity and multi-gravity, but the idea has been investigated for many years (see for instance~\cite{Rosen:1978mb,Israelit:1986ez}). Like massive gravity, bi-gravity was for a long time thought to host a BD ghost parasite, but a ghost-free realization was recently proposed by Hassan and Rosen~\cite{Hassan:2011zd} and bi-gravity is thus experiencing a revived amount of interested. This extensions is nothing other than the ghost-free massive gravity Lagrangian for a dynamical reference metric with the addition of an Einstein--Hilbert term for the now dynamical reference metric.

\subsubsection*{Bi-gravity from deconstruction}

Let us consider a two-site discretization with periodic boundary conditions,  $j=1,2,3$ with quantities at the site $j=3$ being identified with that at the site $j=1$. Similarly as in Section~\ref{sec:MG} we denote by $g\mn=e^a_{\, \mu} e^b_{\, \nu} \eta_{ab}$ and by  $f\mn=$\texthtbardotlessj$ ^a_{\, \mu}\ $\texthtbardotlessj$^b_{\, \nu} \eta_{ab}$ the metrics and vielbeins at the respective locations $y_1$ and $y_2$.

Then applying the discretization procedure highlighted in Eqns.~(\ref{disc1}, \ref{disc2}, \ref{disc3}, \ref{disc4} and \ref{disc5}) and summing over the extra dimension, we obtain the bi-gravity action
\ba
\label{bigravity}
S_{\rm bi-gravity}&=&\frac{\mpl^2}{2} \int \d^4x \sqrt{-g} R[g]+ \frac{M_f^2}{2} \int \d^4x \sqrt{-f} R[f]\\
&+& \frac{\mpl^2 m^2}{4} \int \d^4 x \sqrt{-g} \sum_{n=0}^4\alpha_n \L_n[\K[g,f]]\,,\nn
\ea
where $\K[g,f]$ is given in~\eqref{eqK1} and we use the notation $M_g=\mpl$.
We can equivalently well write the mass terms  in terms of $\K[f,g]$ rather than $\K[g,f]$ as performed in~\eqref{opposite Argument}.

Notice that the most naive discretization procedure would lead to $M_g =\mpl =M_f$, but these can be generalized either `by hand' by changing the weight of each site during the discretization, or by considering a non-trivial configuration along the extra dimension (for instance warping along the extra dimension\epubtkFootnote{See Refs.~\cite{Deffayet:2004ws,Randall:2005me,Gallicchio:2005mh,Kiritsis:2008at} for additional work on deconstruction in five-dimensional AdS, and how this tackles the strong coupling issue.}), or most simply by performing a conformal rescaling of the metric at each site.

Here $\L_0[\K[g,f]]$ corresponds to a cosmological constant for the metric $g\mn$ and the special combination $\sum_{n=0}^4 (-1)^nC_4^n\L_n[\K[g,f]]$, where the $C^m_n$ are the binomial coefficients  is the cosmological constant for the metric $f\mn$, so only $\L_{2,3,4}$ correspond to genuine interactions between the two metrics.

In the deconstruction framework, we naturally obtain $\alpha_2=1$ and no tadpole nor cosmological constant for either metrics.

\subsubsection*{Mass eigenstates}

In this formulation of bi-gravity, both metrics $g$ and $f$ carry a superposition of the massless and the massive spin-2 field. As already emphasize the notion of mass (and of spin) only makes sense for a field living in Minkowski, and so to analyze the mass spectrum, we expand both metrics about flat spacetime,
\ba
g\mn&=& \eta\mn + \frac{1}{\mpl} \delta g\mn \\
f\mn&=& \eta\mn + \frac{1}{M_f} \delta f\mn\,.
\ea
Then to quadratic order in $h$, the action for bi-gravity reads (for $\alpha_0=\alpha_1=0$ and $\alpha_2=-1/2$),
\ba
S^{(2)}_{\rm bi-gravity}=\int \d^4 x\Big[
-\frac 14 \delta g^{\mu\nu}\Ein^{\alpha\beta}\mn \delta g\ab -\frac 14\delta f^{\mu\nu}\Ein^{\alpha\beta}\mn \delta f\ab
-\frac 18 m_{\rm eff}^2 \(h\mn^2-h^2\)
\Big]\,,
\ea
where all indices are raised and lowered with respect to the flat Minkowski metric and the Lichnerowicz operator $\Ein^{\alpha\beta}\mn$ was defined in~\eqref{Lichnerowicz operator}.
We see appearing the Fierz--Pauli mass term combination $h\mn^2-h^2$ introduced in~\eqref{FPmass} for the massive field   with the effective mass $M_{\rm eff}$ defined as~\cite{Hassan:2011zd}
\ba
M_{\rm eff}^2&=&\(\mpl^{-2}+M_f^{-2}\)^{-1}\\
m^2_{\rm eff}&=& m^2 \frac{\mpl^2}{M_{\rm eff}^2}\,.
\ea
The massive field $h$ is given by
\ba
h\mn= M_{\rm eff}\(\frac{1}{\mpl}\delta g\mn-\frac{1}{M_f}\delta f\mn\)=M_{\rm eff} \(g\mn-f\mn\)\,,
\ea
while the other combination represents the massless field $\ell\mn$,
\ba
\ell\mn= M_{\rm eff}\(\frac{1}{M_f}\delta g\mn+\frac{1}{\mpl}\delta f\mn\)\,,
\ea
so that in terms of the light and heavy spin-2 fields (or more precisely in terms of the two mass eigenstates $h$ and $\ell$), the quadratic action for bi-gravity reproduces that of a massless spin-2 field $\ell$ and a Fierz--Pauli massive spin-2 field $h$ with mass $m_{\rm eff}$,
\ba
S^{(2)}_{\rm bi-gravity}=\int \d^4 x\Big[
\hspace{-15pt}&& -\frac 14h^{\mu\nu}\Big[\Ein^{\alpha\beta}\mn +\frac 12 m^2_{\rm eff}\(\delta^\alpha_\mu \delta^\beta_\nu-\eta\abu\eta\mn\)\Big]h\ab \\
&& -\frac 14  \ell^{\mu\nu}\Ein^{\alpha\beta}\mn \ell\ab
\Big]\nn
\,.
\ea
As explained in~\cite{Hassan:2011zd}, in the case where there is a large Hierarchy between the two Planck scales $\mpl$ and $M_f$, the massive particles is always the one that enters at the lower Planck mass and the massless one the one that has a large Planck scale. For instance if $M_f \gg \mpl$, the massless particle is mainly given by $\delta f\mn $ and the massive one mainly by $\delta g\mn$. This means that in the limit $M_f \to \infty$ while keeping $\mpl$ fixed, we recover the theory of a massive gravity and a fully decoupled massless graviton as will be explained in Section~\ref{sec:MG DL of bigravity}.

\subsection{Coupling to matter}
\label{sec:Matter}

So far we have only focus on an empty five-dimensional bulk with no matter. It is natural though to consider matter fields living in five dimensions, $\chi(x,y)$ with Lagrangian (in the gauge choice~\eqref{5dGauge})
\ba
\L_{\rm matter}=\sqrt{-g}\(-\frac12 (\p_\mu \chi)^2-\frac 12 (\p_y\chi)^2-V(\chi)\)\,,
\ea
in addition to arbitrary potentials (we focus on the case of a scalar field for simplicity, but the same philosophy can be applied to higher-spin species be it bosons or fermions). Then applying the same discretization scheme used for gravity, every matter field then comes in $N$ copies
\ba
\chi(x,y)\hookrightarrow \chi^{(j)}(x)=\chi(x,y_j)\,,
\ea
for $j=1,\cdots, N$ and each field $\chi^{(j)}$ is coupled to the associated vielbein $e^{(j)}$ or metric $g^{(j)}\mn=e^{(j)\, a}_\mu e^{(j)\, b}_\nu\eta_{ab}$ at the same site. In the discretization procedure, the gradient along the extra dimension  yields a mixing (interaction) between fields located on  neighbouring sites,
\ba
\int \d y \, (\p_y\chi)^2 \hookrightarrow  R \sum_{j=1}^N\, m^2 (\chi^{(j+1)}(x)- \chi^{(j)}(x))^2\,,
\ea
(assuming again periodic boundary conditions, $\chi^{(N+1)}=\chi^{(1)}$). The discretization procedure could be  also performed using a more complicated definition of the derivative along $y$ involving more than two sites, which leads to further interactions between the different fields.

In the two-sight derivative formulation,  the action for matter is then
\ba
S_{\rm matter} \hookrightarrow \frac{1}{m}\int \d^4 x \sum_j  \sqrt{-g^{(j)}} \Big( - \frac{1}{2} g^{(j)\,\mu \nu }\partial_{\mu} \chi^{(j)} \partial_{\nu} \chi^{(j)}  \\
- \frac{1}{2} m^2 (\chi^{(j+1)}-\chi^{(j)})^2 -V(\chi^{(j)})\Big)\,.\nn
\ea
The coupling to gauge fields or fermions can be derived in the same way, and the vielbein formalism makes it natural to extend the action~\eqref{SDirac} to five dimensions and applying the discretization procedure. Interestingly in the case of fermions, the fields $\psi^{(j)}$ and $\psi^{(j+1)}$ would not directly couple to one another, but they would couple to both the vielbein $e^{(j)}$ at the same site and the one $e^{(j-1)}$ on the neighboring site.

Notice however that the current full proofs for the absence of the BD ghost do not include such couplings between matter fields living on different metrics (or vielbeins), nor matter fields coupling directly to more than one metric (vielbein).

\subsection{No new kinetic interactions}
\label{sec:NoNewKin}

In GR, diffeomorphism invariance uniquely fixes the kinetic term to be the
 Einstein--Hilbert one
\ba
\L_{\rm EH}=\sqrt{-g}R\,,
\label{EH}
\ea
(see for instance Refs.~\cite{Gupta:1954zz,Weinberg:1965rz,Deser:1969wk,Feynman:1996kb,Boulware:1974sr} for the uniqueness of GR for the theory of  a massless spin-2 field).

In more than four dimensions, the GR action can be supplemented by additional Lovelock invariants~\cite{Lovelock:1971yv} which respect diffeomorphism invariance and are expressed in terms of higher powers of the Riemann curvature but lead to second order equations of motion. In four dimensions there is only one non-trivial additional Lovelock invariant corresponding the Gauss--Bonnet term but it is topological and thus does not affect the theory, unless other degrees of freedom such as a scalar field is included.

So when dealing with the theory of a single massless spin-2 field in four dimensions the only allowed kinetic term is the well-known Einstein--Hilbert one. Now when it comes to the theory of a massive spin-2 field, diffeomorphism invariance is broken and so in addition to the allowed potential terms described in (\ref{L0}\,--\,\ref{L4}), one could consider other kinetic terms which break diffeomorphism.

This possibility was explored in Refs.~\cite{Folkerts:2011ev,Hinterbichler:2013eza,Folkerts:2013mra} where it was shown that in four dimensions, the following derivative interaction $\L^{\rm (der)}_{3}$  is ghost-free at leading order (\ie there is no higher derivatives for  the \stu fields when introducing the \stu fields associated with \emph{linear} diffeomorphism),
\ba
\label{Lk3}
\L^{\rm (der)}_{3}= \ep^{\mu \nu \rho \sigma} \ep^{\mu' \nu' \rho' \sigma'} h_{\sigma\sigma'}\p_\rho h_{\mu\mu'} \p_{\rho'}h_{\nu \nu'}\,.
\ea
So this new derivative interaction would be allowed for a theory of a massive spin-2 field which \emph{does not couple to matter}. Note that this interaction can only be considered if the spin-2 field is massive in the first place, so this interaction can only be present if the Fierz--Pauli mass term~\eqref{FPmass} is already present in the theory.

Now let us turn to a theory of gravity. In that case we have seen that the coupling to matter forces \emph{linear} diffeomorphisms to be extended to fully \emph{non-linear} diffeomorphism.  So to be viable in a theory of massive gravity, the derivative interaction~\eqref{Lk3} should enjoy a ghost-free non-linear completion (the absence of ghost non-linearly can be checked for instance by restoring \emph{non-linear} diffeomorphism using the \emph{non-linear} \stu decomposition~\eqref{Hcov} in terms of the helicity-1 and -0 modes given in~\eqref{chi A pi}, or by performing an ADM analysis as will be performed for the mass term in Section~\ref{sec:Abscence of ghost proof}.) It is easy to check that by itself $\L^{\rm (der)}_{3}$ has a ghost at quartic order and so other non-linear interactions should be included for this term to have any chance of being ghost-free.

Within the deconstruction paradigm, the non-linear completion of $\L^{\rm (der)}_{3}$ could have a natural interpretation as arising from the five-dimensional Gauss--Bonnet term after discretization. Exploring the avenue would indeed lead to a new kinetic interaction of the form $\sqrt{-g}\K\mn \K\ab {} ^{*} \hspace{-3pt} R^{\mu\nu\alpha\beta}$, where $^{*} \hspace{-3pt} R$ is the dual Riemann tensor~\cite{Kimura:2013ika,deRham:2013tfa}. However a simple ADM analysis shows that such a term propagates more than five degrees of freedom and thus has an Ostrogradsky ghost (similarly as the BD ghost). As a result this new kinetic interaction~\eqref{Lk3} does not have a natural realization from a five-dimensional point of view (at least in its metric formulation, see Ref.~\cite{deRham:2013tfa} for more details.)

We can push the analysis even further and show that no matter what the higher order interactions are, as soon as $\L^{\rm (der)}_{3}$ is present it will \emph{always} lead to a ghost and so such an interaction is never acceptable~\cite{deRham:2013tfa}.

\emph{As a result, the Einstein--Hilbert kinetic term is the only allowed kinetic term in Lorentz-invariant (massive) gravity.}

This result shows how special and unique the Einstein--Hilbert term is. Even without imposing diffeomorphism invariance, the stability of the theory fixes the kinetic term to be nothing else than the Einstein--Hilbert term and thus forces diffeomorphism invariance at the level of the kinetic term. Even without requiring coordinate transformation invariance, the Riemann curvature remains the building block of the kinetic structure of the theory, just as in GR.

Before summarizing the derivation of massive gravity from higher dimensional deconstruction\,/\,Kaluza--Klein decomposition, we briefly comment on other `apparent' modifications of the kinetic structure like in $f(R)$ -- gravity (see for instance Refs.~\cite{Cai:2013lqa,Kluson:2013yaa,Bamba:2013fha} for $f(R)$ massive gravity and their implications to cosmology).

Such kinetic terms \`a la $f(R)$  are also possible without a mass term for the graviton. In that case diffeomorphism invariance allows us to perform a change of frame. In the Einstein-frame $f(R)$ gravity is seen to correspond to a theory of gravity with a scalar field, and the same result will hold in $f(R)$ massive gravity (in that case the scalar field couples non-trivially to the \stu fields). As a result $f(R)$ is not
a genuine modification of the kinetic term but rather a standard Einstein--Hilbert term and the addition of a new scalar degree of freedom which not a degree of freedom of the graviton but rather an independent scalar degree of freedom which couples non-minimally to matter (see Ref.~\cite{DeFelice:2010aj} for a review on $f(R)$-gravity.)


\newpage

\part{Ghost-free Massive Gravity}
\label{sec:ghost-free MG}

\section{Massive, Bi- and Multi-Gravity Formulation: A Summary}
\label{sec:dRGTsummary}

The previous `deconstruction' framework gave a intuitive argument for the emergence of a potential of the form~\eqref{dRGT_metric} (or~\eqref{dRGT_vielbein} in the vielbein language) and its bi- and multi-metric generalizations. In deconstruction or Kaluza--Klein decomposition a certain type of interaction arises naturally and we have seen that the whole spectrum of allowed potentials (or interactions) could be generated by extending the deconstruction procedure to a more general notion of derivative or by involving the mixing of more sites in the definition of the derivative along the extra dimensions. We here summarize the  most general formulation for the theories of massive gravity about a generic reference metric, bi-gravity and multi-gravity and provide a dictionary between the different languages used in the  literature.

The general action for ghost-free (or dRGT) massive gravity in the vielbein language is
\ba\label{dRGT_vielbein}
\boxed{
S_{\rm mGR}=\frac{\mpl^2}{4}\int\,  \Big( \ep_{abcd} R^{ab}\wed e^c\wed e^d
+ m^2 \L^{\rm (mass)}(e,f)\Big)\,,
}
\ea
with
\ba\label{dRGT_vielbein_mass}
\boxed{
\begin{array}{rcl}
 \L^{\rm (mass)}(e,f)&=&\ep_{abcd} \Big[c_0\, e^a\wedge e^b \wedge e^c \wedge e^d+c_1\, e^a\wedge e^b \wedge e^c \wedge f^d\\[5pt]
& &
+c_2\, e^a\wedge e^b \wedge f^c \wedge f^d
+c_3\, e^a\wedge f^b \wedge f^c \wedge f^d\\[5pt]
& &
+c_4\, f^a\wedge f^b \wedge f^c \wedge f^d\Big]\,,
\end{array}
}
\ea
or in the metric language,
\ba\label{dRGT_metric}
\boxed{
S_{\rm mGR}=\frac{\mpl^2}{2}\int \d^4 x\sqrt{-g}\(R+\frac{m^2}{2}\sum_{n=0}^4 \alpha_n \L_n[\K[g,f]]\)\,.
}
\ea
In what follows we will use the notation for the overall potential of massive gravity
\ba
\label{U}
\mathcal{U}=-\frac{\mpl^2}{4}\sqrt{-g}\sum_{n=0}^4\alpha_n \L_n[\K[g,f]]=- \L^{\rm (mass)}(e,f) \,,
\ea
so that
\ba
\label{LmGR}
\L_{\rm mGR}=\mpl^2\L_{\rm GR}[g]-m^2\ \mathcal{U}[g,f]\,,
\ea
where $\L_{\rm GR}[g]$ is the standard GR Einstein--Hilbert Lagrangian for the dynamical metric $g\mn$ and $f\mn$ is the reference metric
and for bi-gravity,
\ba
\label{bi-gravity}
\L_{\rm bi-gravity}=\mpl^2 \L_{\rm GR}[g]+M_f^2\L_{\rm GR}[f]-m^2\ \mathcal{U}[g,f]\,,
\ea
where both $g\mn$ and $f\mn$ are then dynamical metrics.

Both massive gravity and bi-gravity break one copy of diff invariance and so the \stu fields can be introduced in exactly the same way in both cases $\U[g,f]\to \U[g,\tilde f]$ where the St\"uckelbergized metric $\tilde f\mn$ was introduced in~\eqref{stuH} (or alternatively $\U[g,f]\to \U[\tilde g,f]$). Thus bi-gravity is by no means an alternative to introducing the \stu fields as is sometimes stated.

In these formulations,  $\L_0$  (or the term proportional to $c_0$) correspond to a cosmological constant, $\L_1$ to a tadpole, $\L_2$ to the mass term and $\L_{3,4}$ to allowed higher order interactions. The presence of the tadpole $\L_1$ would imply a non-zero vev. The presence of the potentials $\L_{3,4}$ without $\L_2$ would lead to infinitely strongly coupled degrees of freedom and would thus be pathological.  We recall that $\K[g,f]$ is given in terms of the metrics $g$ and $f$ as
\ba
\label{eqK}\boxed{
\K\mupn[g,f]=\delta\mupn -\(\sqrt{g^{-1}f}\)\mupn\,,}
\ea
and  the Lagrangians $\L_n$ are defined in as follows,
\ba
\label{L0}
\L_0[Q]&=&\ep^{\mu\nu\alpha\beta}\ep_{\mu\nu\alpha\beta}\\
\label{L1}
\L_1[Q]&=&\ep^{\mu\nu\alpha\beta}\ep_{\mu'\nu\alpha\beta}\, Q^{\mu'}_\mu\\
\label{L2}
\L_2[Q]&=&\ep^{\mu\nu\alpha\beta}\ep_{\mu'\nu'\alpha\beta}Q^{\mu'}_\nu Q^{\nu'}_\nu\\
\label{L3}
\L_3[Q]&=&\ep^{\mu\nu\alpha\beta}\ep_{\mu'\nu'\alpha'\beta}Q^{\mu'}_\nu Q^{\nu'}_\nu Q^{\alpha'}_\alpha\\
\L_4[Q]&=&\ep^{\mu\nu\alpha\beta}\ep_{\mu'\nu'\alpha'\beta'}Q^{\mu'}_\nu Q^{\nu'}_\nu Q^{\alpha'}_\alpha Q^{\beta'}_\beta\,.\label{L4}
\ea

We have introduced the constant $\L_0$ ($\L_0=4!$ and $\sqrt{-g}\L_0$ is nothing other than the cosmological constant) and the tadpole $\L_1$ for completeness.
Alternatively we may express these scalars as follows
\ba
\label{L0 X}
\L_0[Q]&=&4!\\
\label{L1 X}
\L_1[Q]&=&3!\, [Q]\\
\label{L2 X}
\L_2[Q]&=&2!([Q]^2-[Q^2])\\
\label{L3 X}
\L_3[Q]&=&([Q]^3-3[Q][Q^2]+2[Q^3])\\
\label{L4 X}
\L_4[Q]&=&([Q]^4-6[Q]^2[Q^2]+3[Q^2]^2+8[Q][Q^3]-6[Q^4])\,.
\ea
These are easily generalizable to any number of dimensions, and in $d$ dimensions we find $d$ such independent scalars.

The multi-gravity action is a generalization to multiple interacting spin-2 fields with the same form for the interactions, and bi-gravity is the special case of two metrics ($N=2$),
\ba
\label{Multigravity_1}\boxed{
S_{N}=\frac{\mpl^2}{4}\sum_{j=1}^N\int\,  \(\ep_{abcd} R^{ab}[e_j]\wed e_j^c\wed e_j^d
+ m_N^2\L^{\rm (mass)}(e_j,e_{j+1})\)\,,}
\ea
or
\ba\boxed{
S_{N}=\frac{\mpl^2}{2}\sum_{j=1}^N\int \d^4 x \sqrt{-g_j}\(R[g_j]+\frac{m_N^2}{2}\sum_{n=0}^4\alpha^{(j)}_n \L_n[\K[g_j, g_{j+1}]]
\)\,.
\label{Multigravity_2}}
\ea

\subsubsection*{Inverse argument}

We could have written this set of interactions in terms of $\K[f,g]$ rather than $\K[g,f]$,
\ba
\U&=& \frac{\mpl^2 m^2}{4} \int \d^4 x \sqrt{-g} \sum_{n=0}^4\alpha_n \L_n[\K[g,f]]\nn \\
&=& \frac{\mpl^2 m^2}{4}\int \d^4 x \sqrt{-f} \sum_{n=0}^4\tilde \alpha_n \L_n[\K[f,g]]\,,
\label{opposite Argument}
\ea
with
\ba
\(\begin{array}{c} \tilde \alpha_0\\ \tilde \alpha_1\\\tilde \alpha_2\\\tilde \alpha_3\\ \tilde \alpha_4 \end{array}\)= \(\begin{array}{ccccc}
1 & 0 & 0 & 0 & 0\\
-4 & -1 & 0 & 0 & 0\\
6 & 3 & 1 & 0 & 0\\
-4 & -3 & -2 & -1 & 0\\
1 & 1 & 1 & 1 &  1 \\
\end{array}\)
\(\begin{array}{c}
\alpha_0\\
\alpha_1\\ \alpha_2\\ \alpha_3\\  \alpha_4
\end{array}\)\,.
\ea
Interestingly, the absence of tadpole and cosmological constant for say the metric $g$ implies  $\alpha_0=\alpha_1=0$ which in turn implies the absence of tadpole and cosmological constant for the other metric $f$, $\tilde \alpha_0=\tilde \alpha_1=0$, and thus $\tilde \alpha_2=\alpha_2=1$.

\subsubsection*{Alternative variables}

Alternatively another fully equivalent convention has also been used in the literature~\cite{Hassan:2011vm} in terms of $\X\mupn=g^{\mu \alpha}f_{\alpha \nu}$ defined in~\eqref{Xdef},
\ba
\label{mGR2}
\mathcal{U}=-\frac{\mpl^2}{4}\sqrt{-g} \sum_{n=0}^4\frac{\beta_n}{n!} \L_n[\sqrt{\X}]\,,
\ea
which is equivalent to~\eqref{U} with  $\L_0=4!$ and
\ba
\label{beta<->alpha}
\(\begin{array}{c}
\beta_0 \\ \beta_1 \\ \beta_2 \\ \beta_3\\ \beta_4
\end{array}\)=
\(\begin{array}{ccccc}
1 & 1 & 1& 1 & 1\\
0 & -1 & -2 & -3 & -4 \\
0 & 0 & 2 & 6 & 12\\
0 & 0 & 0 & -6 & -24 \\
0 & 0 & 0 & 0 & 24
\end{array}\)
\(\begin{array}{c}
\alpha_0 \\ \alpha_1 \\ \alpha_2 \\ \alpha_3\\ \alpha_4
\end{array}\)\,,
\ea
or the inverse relation,
\ba
\(\begin{array}{c}
\alpha_0 \\ \alpha_1 \\ \alpha_2 \\ \alpha_3\\ \alpha_4
\end{array}\)=
\frac 1 {24}\(\begin{array}{ccccc}
24 & 24 & 12& 4 & 1\\
0 & -24 & -24 & -12 & -4 \\
0 & 0 & 12 & 12 & 6 \\
0 & 0 & 0 & -4 & -4 \\
0 & 0 & 0 & 0 & 1
\end{array}\)
\(\begin{array}{c}
\beta_0 \\ \beta_1 \\ \beta_2 \\ \beta_3\\ \beta_4
\end{array}\)\,,
\ea
so that in order to avoid a tadpole and  a cosmological constant we need to set for instance $\beta_4=-(24 \beta_0+24 \beta_1+12 \beta_2+4 \beta_3)$ and $\beta_3=-6(4\beta_0+3 \beta_1+\beta_2)$.

\subsubsection*{Expansion about the reference metric}

In the vielbein language the mass term is extremely simple, as can be seen in  Eqn.~\eqref{dRGT_vielbein} with $\A$ defined in~\eqref{A m to 0}. Back to the metric language, this means that the mass term takes a remarkably simple form when writing the dynamical metric $g\mn$ in terms of the reference metric $f\mn$ and a difference $\tilde h\mn=2h\mn+h^2\mn$ as
\ba
g\mn=f\mn+2h\mn+h_{\mu\alpha}h_{\nu\beta}f^{\alpha \beta}\,,
\ea
where $f^{\alpha\beta}=(f^{-1})^{\alpha\beta}$. The mass terms is then expressed as
\ba
\mathcal{U}=-\frac{\mpl^2}{4}\sqrt{-f}\sum_{n=0}^4 \kappa_n {\L}_n[f^{\mu\alpha}h_{\alpha\nu}]\,,
\label{LmassVielbeininspired}
\ea
where the $\L_n$ have the same expression as the $\L_n$ in (\ref{L0}-\ref{L4}) so $\tilde \L_n$ is genuinely $n^{\rm th}$ order in $h\mn$. The expression~\eqref{LmassVielbeininspired} is thus at most quartic order in $h\mn$ but is valid to \emph{all orders} in $h\mn$, (there is no assumption that $h\mn$ be small). In other words, the mass term~\eqref{LmassVielbeininspired}  is not an expansion in $h\mn$ truncated to a finite (quartic) order, but rather a fully equivalent way to rewrite the mass Lagrangian in terms of the variable $h\mn$ rather than $g\mn$. The relation between the coefficients $\kappa_n$ and $\alpha_n$ is given by
\ba
\label{beta<->alpha}
\(\begin{array}{c}
\kappa_0 \\ \kappa_1 \\ \kappa_2 \\ \kappa_3\\ \kappa_4
\end{array}\)=
\(\begin{array}{ccccc}
1 & 0 & 0& 0 & 0 \\
4 & 1 & 0 & 0 & 0 \\
6 & 3 & 1 & 0 & 0\\
4 & 3 & 3 & 1 & 0 \\
1 & 1 & 1 & 1 & 1
\end{array}\)
\(\begin{array}{c}
\alpha_0 \\ \alpha_1 \\ \alpha_2 \\ \alpha_3\\ \alpha_4
\end{array}\)\,.
\ea
The quadratic expansion about a background different from the reference metric was derived in Ref.~\cite{Guarato:2013gba}.

\section{Evading the BD Ghost in Massive Gravity}
\label{sec:Abscence of ghost proof}

The deconstruction framework gave an intuitive approach on how to construct a theory of massive gravity or multiple interacting `gravitons'. This lead to the ghost-free dRGT theory of massive gravity and its bi- and multi-gravity extensions in a natural way. However these developments were only possible \textit{a posteriori}.

The deconstruction  framework was proposed earlier (see Refs.~\cite{ArkaniHamed:2001ca,ArkaniHamed:2001nc,Deffayet:2003zk,ArkaniHamed:2003vb,Schwartz:2003vj,Deffayet:2003zk,Deffayet:2005yn}) directly in the metric language and despite starting from a perfectly healthy five-dimensional  theory of GR, the discretization in the metric language leads to the standard BD issue (this also holds in a KK decomposition when truncating the KK tower at some finite energy scale). Knowing that massive gravity (or multi-gravity) can be naturally derived from a healthy five-dimensional theory  of GR is thus not a sufficient argument for the absence of the BD ghost, and a great amount of effort was devoted to that proof, which is known by now a multitude of different forms and languages.

Within this review one cannot make justice to all the independent proofs that have been formulated by now in the literature. We thus focus on a few of them - the Hamiltonian analysis in the ADM language - as well as the analysis in the \stu language. One of the proofs in the vielbein formalism will be used in the multi-gravity case, and thus we do not emphasize that proof in the context of massive gravity, although it is perfectly applicable (and actually very elegant) in that case.   Finally after deriving the decoupling limit in Section~\ref{sec:DL MG}, we also briefly review how it can be used to prove the absence of ghost more generically.

We note that even though the original argument on how the BD ghost could be circumvented in the full nonlinear theory was presented in~\cite{deRham:2010iK} and~\cite{deRham:2010kj}, the absence of BD ghost in ``ghost-free massive gravity" or dRGT has been the subject of many discussions~\cite{Alberte:2010it,Alberte:2010qb,Kluson:2011qe,Kluson:2011aq,Chamseddine:2011mu,Kluson:2011rt,Kluson:2012zz,Chamseddine:2013lid} (see also~\cite{Kluson:2012ps,Kluson:2013cy,Kluson:2013lza,Kluson:2013aca,KLUSON:2013rha} for related discussions in bi-gravity). By now the confusion has been clarified, and see for instance~\cite{Hassan:2011hr,Hassan:2011ea,Mirbabayi:2011aa,Kluson:2012gz,Kluson:2012wf,Hassan:2012qv,Alexandrov:2013rxa,Golovnev:2014upa}
for thorough proofs addressing all the issues raised in the previous literature. (See also~\cite{Kluson:2013hoa} for the proof of the absence of ghosts in other closely related models).

\subsection{ADM formulation}
\label{sec:ADMHamiltonian}

\subsubsection{ADM formalism for GR}

Before going onto the subtleties associated with massive gravity, let us briefly summarize how the counting of the number of degrees of freedom can be performed in the ADM language using the Hamiltonian for GR. Using an ADM decomposition (where this time, we single out the time, rather than the extra dimension as was performed in Part~\ref{part:extradim}),
\ba
\label{ADM}
\d s^2=-N^2 \d t^2+\gamma_{ij}\(\d x^i + N^i \d t\)\(\d x^j + N^j \d t\)\,,
\ea
with the lapse $N$, the shift $N^i$ and the 3-dimensional space metric $\gamma_{ij}$. In this section indices are raised and lowered with respect to $\gamma_{ij}$ and dots represent derivatives with respect to $t$. In terms of these variables, the Lagrangian density for GR is
\ba
\L_{\rm GR}&=&\frac{\mpl^2}{2}\int \d t \(\sqrt{-g}R +\p_t\left[\sqrt{-g}[k]\right]\)\\
&=& \frac{\mpl^2}{2}\int \d t N \sqrt{\gamma} \(^{(3)}\! R[\gamma]+[k]^2-[k^2]\)\,,
\ea
where $^{(3)}R$ is the three-dimensional scalar curvature built out of $\gamma$ (no time derivatives in $^{(3)}R$) and $k_{ij}$ is the three-dimensional extrinsic curvature,
\ba
k_{ij}=\frac 1 {2N}\( \dot{\gamma}_{ij}-\nabla_{(i} N_{j)}\)\,.
\ea
The GR action can thus be expressed in a way which has no double or higher time derivatives and only first time-derivatives squared of $\gamma_{ij}$. This means that neither the shift nor the lapse are truly dynamical and they do not have any associated conjugate momenta.
The conjugate momentum associated with $\gamma$ is,
\ba
p^{ij}=\frac{\p \sqrt{-g}R}{\p \dot{\gamma}_{ij}}\,.
\ea
We can now construct the Hamiltonian density for GR in terms of the 12 phase space variables ($\gamma_{ij}$ and $p^{ij}$ carry 6 component each),
\ba
\label{H GR}
\mathcal{H}_{\rm GR}=N \mathcal{R}_0(\gamma, p)+N^i \mathcal{R}_i(\gamma, p)\,.
\ea
So we see that in GR, both the shift and the lapse play the role of Lagrange multipliers. Thus they propagate a first-class constraint each which removes 2 phase space degrees of freedom per constraint. The counting of the number of degrees of freedom in phase space thus goes as follows:
\ba
(2\times 6)-2\ \text{lapse constraints} -2\times 3\ \text{shift constraints}=4=2\times 2\,,
\ea
corresponding to a total of 4 degrees of freedom in phase space, or 2 independent degrees of freedom in field space. This is the very well-known and established result that in four dimensions GR propagates 2 physical degrees of freedom, or gravitational waves have two polarizations.

This result is fully generalizable to any number of dimensions, and in $d$ spacetime dimensions, gravitational waves carry $d(d-3)/2$ polarizations. We now move to the case of massive gravity.

\subsubsection{ADM counting in massive gravity}

We now amend the GR Lagrangian with a potential $\mathcal{U}$. As already explained, this can only be performed by breaking covariance (with the exception of a cosmological constant). This potential could be a priori an arbitrary function of the metric, but contains no derivatives and so does not affect the definition of the conjugate momenta $p^{ij}$ This translates directly into a potential at the level of the Hamiltonian density,
\ba
\mathcal{H}=N \mathcal{R}_0(\gamma, p)+N^i \mathcal{R}_i(\gamma, p)+m^2\mathcal{U}\(\gamma_{ij}, N^i, N\)\,,
\ea
where the overall potential for ghost-free massive gravity is given in~\eqref{U}.

If $\mathcal{U}$ depends non-linearly on the shift or the lapse then these are no longer directly Lagrange multipliers (if they are non-linear, they still appear at the level of the equations of motion, and so they do not propagate a constraint for the metric but rather for themselves). As a result for an arbitrary potential one is left with $(2\times 6)$ degrees of freedom in the three-dimensional metric and its momentum conjugate and no constraint is present  to reduce the phase space. This leads to $6$ degrees of freedom in field space: the two usual transverse polarizations for the graviton (as we have in GR), in addition to two `vector' polarizations and two `scalar' polarizations.

These 6 polarizations correspond to the five healthy massive spin-2 field degrees of freedom in addition to the sixth BD ghost, as explained in Section~\ref{sec:BDghost} (see also Section~\ref{sec:stulanguage}).

This counting is also generalizable to an arbitrary number of dimensions, in $d$ spacetime dimensions, a massive spin-2 field should propagate the same number of degrees of freedom as a massless spin-2 field in $d+1$ dimensions, that is $(d+1)(d-2)/2$ polarizations. However an arbitrary potential would allow for $d(d-1)/2$ independent degrees of freedom, which is 1 too many excitations, always corresponding to one BD ghost degree of freedom in an arbitrary number of dimensions.

The only way this counting can be wrong is if the constraints for the shift and the lapse cannot be inverted for the shift and the lapse themselves, and thus at least one of the equations of motion from the shift or the lapse imposes a constraint on the three-dimensional metric $\gamma_{ij}$. This loophole was first presented in~\cite{deRham:2010gu} and an example was provided in~\cite{deRham:2010iK}. It was then used in~\cite{deRham:2010kj} to explain how the `no-go' on the presence of a ghost in massive gravity could be circumvented. Finally, this argument was then carried through fully non-linearly in~\cite{Hassan:2011hr} (see also~\cite{Kluson:2011aq} for the analysis in $1+1$ dimensions as presented in~\cite{deRham:2010kj}).

\subsubsection{Eliminating the BD ghost}

\subsubsection*{Linear Fierz--Pauli massive gravity}

Fierz--Pauli massive gravity is special in that at the linear level (quadratic in the Hamiltonian), the lapse remains linear, so it still acts as a Lagrange multiplier generating a primary second-class constraint. Defining the metric as $h\mn=\mpl(g\mn-\eta\mn)$, (where for simplicity and definiteness we take Minkowski as the reference metric $f\mn=\eta\mn$, although most of what follows can be easily generalizable to an arbitrary reference metric $f\mn$).
Expanding the lapse as $N=1+\delta N$, we have $h_{00}=\delta N +\gamma_{ij}N^i N^j$ and $h_{0i}=\gamma_{ij}N^j$. In the ADM decomposition, the Fierz--Pauli mass term is then (see Eq.~\eqref{FPmassunitary})
\ba
\mathcal{U}^{(2)}=-m^{-2}\L_{\rm FP\, mass}&=& \frac 18  \(h\mn^2- h^2\)\nn \\
&=&\frac 18 \(h_{ij}^2-(h^i_i)^2 -2\( N_i^2  - \delta N h^i_i\)\)\,,
\ea
and is \emph{linear} in the lapse. This is sufficient to deduce that it will keep imposing a constraint on the three-dimensional phase space variables $\{\gamma_{ij}, p^{ij}\}$ and remove at least half of the unwanted BD ghost. The shift on the other hand is non-linear already in the Fierz--Pauli theory, so their equations of motion impose a relation for themselves rather than a constraint for the three-dimensional metric. As a result the Fierz--Pauli theory (at that order) propagates three additional degrees of freedom than GR, which are the usual five degrees of freedom of a massive spin-2 field. Non-linearly however the Fierz--Pauli mass term involve a non-linear term in the lapse in such a way that the constraint associated with it disappears and Fierz--Pauli massive gravity has a ghost at the non-linear level, as pointed out in~\cite{Boulware:1973my}. This is in complete agreement with the discussion in Section~\ref{sec:BDghost}, and is a complementary way to see the issue.

In Ref.~\cite{Creminelli:2005qk}, the most general potential was considered up to quartic order in the $h\mn$, and it was shown that there is no choice of such potential (apart from a pure cosmological constant) which would prevent the lapse from entering non-linearly. While this result is definitely correct, it does not however imply the absence of a constraint generated by the set of shift and lapse $N^\mu =\{N, N^i\}$. Indeed there is no reason to believe that the lapse should necessarily be the quantity to generates  the constraint necessary to remove the BD ghost. Rather it can be any combination of the lapse and the shift.

\subsubsection*{Example on how to evade the BD ghost non-linearly}

As an instructive example presented in~\cite{deRham:2010iK}, consider the following Hamiltonian,
\ba
\mathcal{H}= N \tilde{\mathcal{C}}_0(\gamma, p)+N^i \tilde{\mathcal{C}}_i(\gamma, p)+m^2\U \,,
\label{Htoy_1}
\ea
with the following example for the potential
\ba
\label{Utoy1}
\U=V(\gamma,p) \frac{\gamma_{ij}N^iN^j}{2N}\,.
\ea
In this example neither the lapse nor the shift enter linearly, and one might worry on the loss of the constraint to project out the BD ghost. However upon solving for the shift and substituting back into the Hamiltonian (this is possible since the lapse is not dynamical), we get
\ba
\mathcal{H}= N \(\tilde{\mathcal{C}}_0(\gamma, p)-\frac{\gamma^{ij}\tilde{\mathcal{C}}_i \tilde{\mathcal{C}}_j}{2m^2 V(\gamma,p)}\)\,,
\label{Htoy_2}
\ea
and the lapse now appears as a Lagrange multiplier generating a constraint, even though it was not linear in~\eqref{Htoy_1}.  This could have been seen more easily, without the need to explicitly integrating out the shift by computing the Hessian
\ba
\label{HEssianL}
L\mn=\frac{\p^2 \mathcal{H}}{\p N^\mu \p N^\nu}= m^2 \frac{\p^2 \mathcal{U}}{\p N^\mu \p N^\nu}\,.
\ea
In the example~\eqref{Htoy_1}, one has
\ba
L\mn=\frac{m^2 V(\gamma,p)}{N^3} \(\begin{array}{cc}
N_i^2 & -N \, N_i\\
- N\, N_j & N^2 \,\gamma_{ij}
\end{array}\)\qquad\Longrightarrow \qquad
\det\(L\mn\)=0 \,.
\ea
The Hessian cannot be inverted which means that the equations of motion cannot be solved for all the shift and the lapse. Instead one of these ought to be solved for the three-dimensional phase space variables which corresponds to the primary second-class constraint. Note that this constraint is not associated with a symmetry in this case and while the Hamiltonian is then pure constraint in this toy example, it will not be in general.

Finally one could also have deduce the existence of a constraint by performing the \emph{linear} change of variable
\ba
\label{ShiftChange}
N_i \to n_i = \frac{N_i}{N}\,,
\ea
in terms of which the Hamiltonian is then explicitly linear in the lapse,
\ba
\mathcal{H}= N \(\tilde{\mathcal{C}}_0(\gamma, p)+ n^i \tilde{\mathcal{C}}_i(\gamma, p)+
m^2 V(\gamma,p) \frac{\gamma_{ij}n^in^j}{2}\)\,,
\label{Htoy_3}
\ea
and generates a constraint that can be read for $\{n_i, \gamma_{ij}, p^{ij}\}$.

\subsubsection*{Condition to evade the ghost}

To summarize, the condition to eliminate (at least half of) the BD ghost is that the det of the Hessian~\eqref{HEssianL} $L\mn$ vanishes as explained in~\cite{deRham:2010kj}. This was shown to be the case in the ghost-free theory of massive gravity~\eqref{dRGT_metric} (\eqref{dRGT_vielbein}) exactly in some cases and up to quartic order, and then fully non-linearly in~\cite{Hassan:2011hr}. We summarize the derivation in the general case in what follows.

Ultimately, this means that in massive gravity we should be able to find a new shift $n^i$ related to the original one as follows $N^i=f_0(\gamma,  n)+N f_1(\gamma,  n)$, such that the Hamiltonian takes the following factorizable form
\ba
\label{Ham factorizable}
\mathcal{H}=\(\mathcal{A}_1(\gamma, p)+N \mathcal{C}_1(\gamma,p)\)\mathcal{F}(\gamma,p,  n)+ \(\mathcal{A}_2(\gamma, p)+N \mathcal{C}_2(\gamma,p)\)\,.
\ea
In this form, the equation of motion for the shift is manifestly independently of the lapse and integrating over the shift $n^i$ manifestly keeps the Hamiltonian linear in the lapse and has the constraint $\mathcal{C}_1(\gamma,p)\mathcal{F}(\gamma,p, n^i(\gamma))+\mathcal{C}_2(\gamma,p)=0$. However such a field redefinition has not (yet) been found. Instead the new shift $n^i$ found below does the next best thing (which is entirely sufficient) of \textit{a.} Keeping the Hamiltonian linear in the lapse and \textit{b.} Keeping its own equation of motion independent of the lapse, which is sufficient to infer the presence of a primary constraint.

\subsubsection*{Primary constraint}

We now proceed by deriving the primary first-class constraint present in ghost-free (dRGT) massive gravity. The proof works equally well for any reference at no extra cost, and so we consider a general reference metric $f\mn$ in its own ADM decomposition, while keep the dynamical metric $g\mn$ in its original ADM form (since we work in unitary gauge, we may not simplify the metric further),
\ba
g\mn\d x^\mu \d x^\nu&=&-N^2 \d t^2+\gamma_{ij}\(\d x^i + N^i \d t\)\(\d x^j + N^j \d t\)\\
f\mn\d x^\mu \d x^\nu&=&-\Nf^2\d t^2+\bar{f}_{ij}\(\d x^i + \Nf^i \d t\)\(\d x^j + \Nf^j \d t\)\,,
\ea
and denote again by $p^{ij}$ the conjugate momentum  associated with $\gamma_{ij}$. $\bar{f}_{ij}$ is not dynamical in massive gravity so there is no conjugate momenta associated with it. The bars on the reference metric are there to denote that these quantities are parameters of the theory and not dynamical variables, although the proof for a dynamical reference metric and multi-gravity works equally well, this is performed in Section~\ref{sec:multigravity}.

Proceeding similarly as in the previous example, we perform a change of variables  similar as in~\eqref{ShiftChange} (only more complicated, but which remains linear in the lapse when expressing $N^i$ in terms of $n^i$)~\cite{Hassan:2011hr}
\ba
N^i\to n^i\qquad\text{defined as}\qquad
N^i-\Nf^i=\(\Nf\delta^i_j+N D^i_j\)n^j\,,
\ea
where the matrix $D^i_j$ satisfies the following relation
\ba
\label{defD}
D^i_k D^k_j=(P^{-1})^i_k\gamma^{k\ell}\bar{f}_{\ell j}\,,
\ea
with
\ba
P^i_j=\delta^i_j+(n^i \bar{f}_{j \ell} n^\ell-n^k \bar{f}_{k \ell} n^\ell\delta^i_j)\,.
\ea
In what follows we use the definition
\ba
\tilde D^i_{\ j}=\kappa D^i_{\ j}\,,
\ea
with
\ba
\kappa=\sqrt{1-n^i n^j \bar{f}_{ij}}\,.
\ea

The field redefinition naturally involves a square root through the expression of the matrix $D$ in~\eqref{defD}, which should come as no surprise from the square root structure of the potential term. For the potential to be writable in the metric language, the square root in the definition of the tensor $\K\mupn$ should exist, which in turns imply that the square root in the definition of $D^i_j$ in~\eqref{defD} must also exist. While complicated, the important point to notice is that this field redefinition remains linear in the lapse (and so does not spoil the standard constraints of GR).

The Hamiltonian for massive gravity is then
\ba
\mathcal{H}_{\rm mGR}&=&\mathcal{H}_{\rm GR}+m^2 \mathcal{U}\nn \\
\label{H mGR}
&=&N\, \mathcal{R}_0(\gamma, p)+\(\Nf^i+\(\Nf\delta^i_j+N D^i_j\)n^j\)\, \mathcal{R}_i(\gamma, p)\\
&&\nn +m^2 \, \mathcal{U}(\gamma, N^i(n), N)\,,
\ea
where $\mathcal{U}$ includes the new contributions from the mass term. $\mathcal{U}(\gamma, N^i, N)$ is neither linear in the lapse $N$, nor in the shift $N^i$. There is actually \emph{no} choice of potential $\mathcal{U}$ which would keep it linear in the lapse beyond cubic order~\cite{Creminelli:2005qk}. However as we shall see, when expressed in terms of the redefined shift $n^i$, the non-linearities in the shift absorb all the original non-linearities in the  lapse and $\mathcal{U}(\gamma, n^i, N)$. In itself this is not sufficient to prove the presence of a Constraint, as the integration over the shift $n^i$ could in turn lead to higher order lapse  in the Hamiltonian,
\ba
\label{U ADM}
\mathcal{U}(\gamma, N^i(n^j), N)= N\, \mathcal{U}_0(\gamma,n^j)+\Nf\,  \mathcal{U}_1(\gamma,n^j)\,,
\ea
with
\ba
\mathcal{U}_0 &=& -\frac{\mpl^2}{4}\sqrt{\gamma}\sum_{n=0}^3\frac{(4-n)\beta_n}{n!}   \L_n[\tilde D^i_{\, j}]\\
\mathcal{U}_1 &=& -\frac{\mpl^2}{4} \sqrt{\gamma}
\Big(3! \beta_1 \kappa+2 \beta_2 D^i_{\, j}P^j_{\, i}\\
&+&\beta_3 \kappa\left[2 D^{[k}_{\  k}n^{i]} \bar{f}_{ij}  D^j_{\, \ell} n^\ell + D^{[i}_{\ i}D^{j]}_{\ j}\right]\Big)-\frac{\mpl^2}{4}\beta_4\sqrt{\bar f}\,,\nn
\ea
where the $\beta$'s are expressed in terms of the $\alpha$'s as in~\eqref{beta<->alpha}. For the purpose of this analysis it is easier to work with that notation.

The structure of the potential is so that the equations of motion with respect to the shift are independent of the lapse $N$ and impose the following relations in terms of $\bar n_i=n^j \bar f_{ij}$,
\ba
\label{shift constraint}
m^2\sqrt{\gamma}\left[3! \beta_1 \bar n_i + 4 \beta_2 \tilde D^j_{\, [j}\bar n_{i]}
+\beta_3 \tilde D^{[j}_{\ j}\(\tilde D^{k]}_{\ k}\bar n_i-2 \tilde D^{k]}_{\ i}\bar n_k\)\right]= \kappa \mathcal{R}_i(\gamma, p)\,,
\ea
which entirely fixes the three shifts $n^i$ in terms of $\gamma_{ij}$ and $p^{ij}$ as well as the reference metric $\bar f_{ij}$ (note that $\Nf^i$ entirely disappears from these equations of motion).

The two requirements defined previously are thus satisfied: \emph{a.} The Hamiltonian is linear in the lapse and \emph{b.} the equations of motion with respect to the shift $n^i$ are independent of the lapse, which is sufficient to infer the presence of a primary constraint. This primary constraint is derived by varying with respect to the lapse and evaluating the shift on the constraint surface~\eqref{shift constraint},
\ba
\label{Hamiltonian constraint}
\mathcal{C}_0=\mathcal{R}_0(\gamma,p)+D^i_{\, j}n^j\mathcal{R}_i(\gamma,p)+m^2 \mathcal{U}_0(\gamma, n(\gamma,p))\approx 0\,,
\ea
where the symbol $``\approx"$ means on the constraint surface.
The existence of this primary constraint is sufficient to infer the absence of BD ghost. If we were dealing with a generic system (which could allow for some spontaneous parity violation),  it could still be in principle that there are no  secondary constraints associated with $\mathcal{C}_0=0$ and the theory  propagates 5.5 physical degrees of freedom (11 dofs in phase space). However physically this never happens in the theory of gravity we are dealing with, and as we shall see below, there ghost-free massive gravity has a  secondary constraint which was explicitly found in~\cite{Hassan:2011ea}.

\subsubsection*{Secondary constraint}

Let us imagine we  start with initial conditions which satisfy the constraints of the system, in particular the modified Hamiltonian constraint~\eqref{Hamiltonian constraint}. As the system evolves the constraint~\eqref{Hamiltonian constraint} needs to remain satisfied. This means that the modified Hamiltonian constraint ought to be independent of time, or in other words it should commute with the Hamiltonian. This requirement generates a secondary constraint,
\ba
\mathcal{C}_{2}\equiv \frac{\d}{\d t}\mathcal{C}_0=\{\mathcal{C}_0,H_{\rm mGR}\}\approx
\{\mathcal{C}_0,H_1\}\approx  0\,,
\ea
with $H_{{\rm mGR}, 1}=\int \d^3 x \mathcal{H}_{{\rm mGR}, 1}$ and
\ba
\mathcal{H}_1=\(\Nf^i+\Nf n^i(\gamma,p)\)\mathcal{R}_i+m^2 \Nf \mathcal{U}_1(\gamma,n(\gamma,p))\,.
\ea
Finding the precise form of this secondary constraint requires a very careful analysis of the Poisson bracket algebra of this system. This formidable task lead to some confusions at first (see Refs.~\cite{Kluson:2011qe}) but was then successfully derived in~\cite{Hassan:2011ea} (see also~\cite{Golovnev:2011aa,Golovnev:2014upa} and~\cite{Kluson:2012wf}). Deriving the whole  set of Poisson brackets is beyond the scope of this review and we simply give the expression for the secondary constraint,
\ba
\label{secondary constraint}
\mathcal{C}_2&\equiv&\mathcal{C}_0 \nabla_i\(\Nf^i+\Nf n^i\)+m^2 \Nf \(\gamma_{ij}p^\ell_\ell-2 p_{ij}\)\mathcal{U}_1^{ij}\\ && +2m^2 \Nf \sqrt{\gamma}\nabla_i\mathcal{U}_{1\, j}^i D^j_{\ k}n^k
+\(\mathcal{R}_jD^i_{\ k}n^k-\sqrt{\gamma}\bar{\mathcal{B}}_j^{\ i}\)\nabla_i\(\Nf^j+\Nf n^j\)\nn\\
&&+\(\nabla_i\mathcal{R}_0+\nabla_i\mathcal{R}_jD^j_{\ k}n^k\)\(\Nf^i+\Nf n^i\)
\nn\,,
\ea
where unless specified otherwise, all  indices are raised and lowered with respect to the dynamical metric $\gamma_{ij}$, and the covariant derivatives are also taken with respect to the same metric. We also define
\ba
\mathcal{U}_1^{ij}&=&\frac{1}{\sqrt{\gamma}}\frac{\p \mathcal{U}_1}{\p \gamma_{ij}}\\
\bar{\mathcal{B}}_{ij}&=&-\frac{\mpl^2}{4}\Big[(\tilde D^{-1})^k_{\ j} \bar f_{ik}\(3\beta_1 \L_0[ \tilde D]+2\beta_2\L_1[ \tilde D]+\frac{\beta_3}{2} \L_2[\tilde D]\)\\
&&-\beta_2 \bar f_{ij}+2 \beta_3 \bar f_{i[k} \tilde D^k_{\, j]}\Big]\,.\nn
\ea

The important point to notice is that the secondary constraint~\eqref{secondary constraint} only depends
on the phase space variables $\gamma_{ij}, p^{ij}$ and not on the lapse $N$. Thus it constraints the phase space variables rather than the lapse and provides a genuine secondary constraint in addition to the primary one~\eqref{Hamiltonian constraint} (indeed one can check  that $\mathcal{C}_2|_{\mathcal{C}_0=0}\ne 0$.).

Finally we should also check that this secondary constraint is also maintained in time. This was performed~\cite{Hassan:2011ea}, by inspecting the condition
\ba
\label{tertiary constraint}
\frac{\d}{\d t}\mathcal{C}_2=\{\mathcal{C}_2,H_{\rm mGR}\}\approx 0\,.
\ea
This condition should be satisfied without further constraining the phase space  variables, which would otherwise imply that fewer than five degrees of freedom are propagating. Since five fully fledged dofs are propagating at the linearized level, the same must happen non-linearly\epubtkFootnote{Some dofs may `accidentally' disappear about some special backgrounds, but dofs cannot disappear non-linearly if they were present at the linearized level.}. Rather than a constraint on $\{\gamma_{ij}, p^{ij}\}$,~\eqref{tertiary constraint} must be solved for the lapse. This is only possible if both the two following conditions are satisfied
\ba
\{\mathcal{C}_2(x), \mathcal{H}_1(y)\}\napprox 0 \hspace{10pt}{\rm and}\hspace{10pt}
\{\mathcal{C}_2(x), \mathcal{C}_0(y)\}\napprox 0\,.
\ea
As shown in~\cite{Hassan:2011ea}, since these conditions do not vanish at the linear level (the constraints reduce to the Fierz--Pauli ones in that case), we can deduce that they cannot vanish non-linearly and thus the condition~\eqref{tertiary constraint} fixes the expression for the lapse rather than constraining further the phase space dofs. Thus there is no tertiary constraint on the phase space.

To conclude we have  shown in this section that ghost-free (or dRGT) massive gravity is indeed free from the BD ghost and  the theory propagates five physical dofs about generic backgrounds. We now present the proof in other languages, but stress that the proof developed in this section is sufficient to  infer the absence of BD ghost.

\subsubsection*{Secondary constraints in bi and multi-gravity}
\label{sec:bifurvation}

In bi- or multi-gravity where all the metrics are dynamical the Hamiltonian is pure constraint (every term is linear in the one of the lapses as can be seen explicitly already from~\eqref{H mGR} and~\eqref{U ADM}).

 In this case the evolution equation of the primary constraint can always be solved for their respective Lagrange multiplier (lapses) which can always be set to zero. Setting the lapses to zero would be unphysical in a theory of gravity and  instead one should take a \emph{`bifurcation'} of the Dirac constraint analysis as explained in~\cite{Banados:2013fda}. Rather than solving for the Lagrange multipliers we can \emph{choose} to use the evolution equation of some of the primary constraints to provide additional secondary constraints instead of solving them for the lagrange multipliers.

 Choosing this bifurcation leads to statements which are then continuous with the massive gravity case and one recovers the correct number of degrees of freedom. See Ref.~\cite{Banados:2013fda} for an enlightening discussion.

\subsection{Absence of ghost in the \stu language}
\label{sec:stulanguage}

\subsubsection{Physical degrees of freedom}

Another way to see the absence of ghost in massive gravity is to work directly in the \stu language for massive spin-2 fields introduced in Section~\ref{sec:nonlinearStuck}. If the four scalar fields $\phi^a$ were dynamical, the theory would propagate six degrees of freedom (the two usual helicity-2 which dynamics is encoded in the standard Einstein--Hilbert term, and the four \stu fields). To remove the sixth mode, corresponding to the BD ghost, one needs to check that not all four \stu fields are dynamical but only three of them. See also~\cite{Alberte:2013sma} for a theory of two \stu fields.

Stated more precisely, in the \stu language beyond the DL, if $\E^a$ is the equation of motion with respect to the field $\phi^a$,
the correct requirement for the absence of ghost is that the Hessian $\mathcal{A}_{ab}$ defined as
\ba
\mathcal{A}_{ab}=-\frac{\delta \E_a}{\delta \ddot \phi^b}=\frac{\delta^2 \L}{\delta \dot \phi^a\delta \dot \phi^b}
\ea
be not invertible, so that the dynamics of not all four \stu may be derived from it. This is the case if
\ba
\det \( \mathcal{A}_{ab}\) =0\,,
\ea
as first explained in Ref.~\cite{deRham:2011rn}. This condition was successfully shown to arise in a number of situations for the ghost-free theory of massive gravity
with potential given in~\eqref{dRGT_metric} or equivalently in~\eqref{dRGT_vielbein} in Ref.~\cite{deRham:2011rn} and then more generically in Ref.~\cite{Hassan:2012qv}.\epubtkFootnote{More recently, Alexandrov impressively performed the full analysis for bi-gravity and massive gravity in the vielbein language~\cite{Alexandrov:2013rxa} determining the full set of primary and secondary constraints, confirming again the absence of BD ghost. This resolves the potential sources of subtleties raised in
Refs.~\cite{Chamseddine:2013lid,Kluson:2013cy,Kluson:2013lza,Kluson:2013aca}. }
For illustrative purposes, we start by showing how this constraint arises in  simple two-dimensional realization of ghost-free massive gravity before deriving the more general proof.

\subsubsection{Two-dimensional case}

Consider massive gravity on a two-dimensional space-time, $\d s^2=-N^2 \d t^2 + \gamma\(\d x+N_x \d t\)^2$, with the two \stu fields $\phi^{0,1}$~\cite{deRham:2011rn}. In  this case the graviton potential can only have one independent non-trivial term,  (excluding the tadpole),
\ba
\mathcal{U}=-\frac{\mpl^2}{4}N \sqrt{\gamma} \(\L_2(\K)+1\)\,.
\ea
In light-cone coordinates,
\ba
\phi^\pm&=&\phi^0\pm \phi^1\\
\mathcal{D}_\pm&=& \frac{1}{\sqrt{\gamma}}\p_x\pm\frac 1 N\Big[\p_t-N_x \p_x\Big]\,,
\ea
the potential is thus
\ba
\label{2dLm}
\U=-\frac{\mpl^2}{4}N \sqrt{\gamma} \sqrt{(\mathcal{D}_-\phi^-)(\mathcal{D}_+\phi^+)}\,.
\ea
The Hessian of this Lagrangian with respect to the two \stu fields $\phi^\pm$ is then
\ba
\mathcal{A}_{ab}&=&\frac{\delta^2 \L_{\rm mGR}}{\delta \dot \phi^a \delta \dot \phi^b}
= - m^2 \frac{\delta^2 \U}{\delta \dot \phi^a \delta \dot \phi^b}\nn\\
&\propto&
\(\begin{array}{cc}
(\mathcal{D}_-\phi^-)^2 & - (\mathcal{D}_-\phi^-)(\mathcal{D}_+\phi^+)\\
-(\mathcal{D}_-\phi^-)(\mathcal{D}_+\phi^+)& (\mathcal{D}_+\phi^+)^2
\end{array}\)\,,
\ea
and is clearly non-invertible, which shows that not both \stu fields are dynamical. In this special case, the Hamiltonian is actually pure constraint as shown in~\cite{deRham:2011rn}, and there are no propagating degrees of freedom. This is as expected for a massive spin-two field in two dimensions.

As shown in Refs.~\cite{deRham:2010kj,deRham:2011rn} the square root can be traded for an auxiliary  non-dynamical variable $\lambda\mupn$. In this two-dimensional example,  the mass term~\eqref{2dLm} can be rewritten with the help of an auxiliary non-dynamical variable $\lambda$ as
\ba
\U=-\frac{\mpl^2}{4}N \sqrt{\gamma} \(\lambda+\frac 1{2 \lambda}(\mathcal{D}_-\phi^-)(\mathcal{D}_+\phi^+)\)\,.
\ea
A similar trick will be used in the full proof.

\subsubsection{Full proof}

The full proof in the minimal model (corresponding to $\alpha_2=1$ and $\alpha_{3}=-2/3$ and $\alpha_4=1/6$ in~\eqref{dRGT_metric} or $\beta_2=\beta_3=0$ in the alternative formulation~\eqref{mGR2}), was derived in Ref.~\cite{Hassan:2012qv}. We briefly review the essence of the argument, although the full technical derivation is beyond the scope of this review and refer the reader to Refs.~\cite{Hassan:2012qv} and~\cite{Alexandrov:2013rxa} for a fully-fledged derivation.

Using a set of auxiliary variables $\lambda^a_{\, b}$ (with $\lambda_{ab}=\lambda_{ba}$, so these auxiliary variables contain ten elements in four dimensions) as explained previously, we can rewrite the potential term in the minimal model as~\cite{Buchbinder:2012wb,Kluson:2011aq},
\ba
\U=\frac{\mpl^2}{4}\sqrt{-g}\([\lambda]+[\lambda^{-1} \cdot Y ]\)\,,
\ea
where the matrix $Y$ has been defined in~\eqref{Y} and is equivalent to $X$ used previously. Upon integration over the auxiliary variable $\lambda$ we recover the square-root structure as mentioned in Ref.~\cite{deRham:2010kj}. We now perform an ADM decomposition as in~\eqref{ADM}
which implies the ADM decomposition on the matrix $Y$,
\ba
Y^a_{\, b}= g^{\mu\nu}\p_\mu \phi^a \p_\nu \phi^c f_{cb}=-\D_t \phi^a \D_t \phi^c f_{cb}+V^a_{\, b}\,,
\ea
with
\ba
\D_t&=&\frac 1N \(\p_t-N^i\p_i\)\\
V^a_{\ b}&=&\gamma^{ij}\p_i \phi^a\p_j \phi^c f_{cb}\,.
\ea
Since the matrix $V$ uses a projection along the 3 spatial directions it is genuinely a rank-3 matrix rather than rank 4. This implies that $\det V=0$.
Notice that we consider an arbitrary reference metric $f$, as the proof does not depend on it and can be done for any $f$ at no extra cost~\cite{Hassan:2012qv}. The canonical momenta conjugate to $\phi^a$ is given by
\ba
p_a=\frac 12 \tilde \alpha (\lambda^{-1})_{ab}\D_0 \phi^b\,,
\ea
with
\ba
\tilde \alpha=2\mpl^2 m^2 \sqrt{\gamma}\,.
\ea
In terms of these conjugate momenta, the equations of motion with respect to $\lambda^{ab}$ then imposes the relation (after multiplying with the matrix\epubtkFootnote{We stress that multiplying with the matrix $\lambda$ is \emph{not} a projection, the equation~\eqref{eomSTu} contains as much information as the equation of motion with respect to $\lambda$, multiplying the with the matrix $\lambda$ on both sides simply make the rank of the equation more explicit. } $\alpha\lambda$ on both side),
\ba
\label{eomSTu}
\lambda^{ac}C_{ab}\lambda^{bd}=V^{ab}\,,
\ea
with the matrix $C_{ab}$ defined as
\ba
C_{ab}=\tilde{\alpha}^2f_{ab}+p_a p_b\,.
\ea
Since $\det V=0$, as mentioned previously, the equation of motion~\eqref{eomSTu} is only consistent if we also have $\det C=0$. This is the first constraint found in~\cite{Hassan:2012qv} which is already sufficient to remove (half) the BD ghost,
\ba
\mathcal{C}_1\equiv\frac{\det C}{\det f}= \tilde{\alpha}^2+(f^{-1})^{ab}p_a p_b = 0\,,
\ea
which is the primary constraint on a subset of physical phase space variables $\{\gamma_{ij}, p_a\}$, (by construction $\det f \ne 0$). The secondary constraint is then derived  by commuting $\C_1$ with the Hamiltonian. Following the derivation of~\cite{Hassan:2012qv}, we get on the constraint surface
\ba
\C_2&=&\frac{1}{\tilde{\alpha}^2 N}\frac{\d \C_1}{\d t}=\frac{1}{\tilde{\alpha}^2 N}\int \d y\, \{\C_1(y),H(x)\}\\
&\propto&-\gamma^{-1/2} \gamma_{ij}\pi^{ij} -2\frac{\tilde{\alpha}}{\gamma}(\lambda^{-1})_{ab}\p_i \phi^a \gamma^{ij}\nabla_j^{(f)}p^b \\
&\equiv& 0\,,\nn
\ea
where $\pi^{ij}$ is the momentum conjugate associated with $\gamma_{ij}$, and $\nabla^{(f)}$ is the covariant derivative associated with $f$.


\subsubsection{\stu method on arbitrary backgrounds}
\label{sec:ghostFreeBeyondDL}

When working about different non-Minkowski backgrounds, one can instead generalize the definition of the helicity-0 mode as was performed in~\cite{Mirbabayi:2011aa}. The essence of the argument is to perform a rotation in field space so that the fluctuations of the \stu fields about a curved background form a vector field in the new basis, and one can then employ the standard treatment for a vector field. See also~\cite{Alberte:2011ah} for another study of the \stu fields in an FLRW background.

Recently a covariant \stu analysis valid about any background was performed in Ref.~\cite{Kugo:2014hja} using the BRST formalism. Interestingly this method also allows to derive the decoupling limit of massive gravity about any background.

In what follows we review the approach derived in~\cite{Mirbabayi:2011aa} which provides yet another independent argument for the absence of ghost in all generalities. The proofs presented in the previous section work to all orders about a trivial background while in~\cite{Mirbabayi:2011aa}, the proof is performed about a generic (curved) background, and the analysis can thus stop at quadratic order in the fluctuations.  Both types of analysis are equivalent so long as the fields are analytic, which is the case if one wishes to remain within the regime of validity of the theory.

Consider a generic background metric, which in unitary gauge (\ie in the coordinate system $\{x\}$ where the \stu background fields are given by $\phi^a(x)=x^\mu \delta^a_\mu$), the background metric is given by
$g^{\rm bg}\mn =e^a_{\, \mu}(x)e^b_{\, \nu}(x)\eta_{ab}$, and the background \stu fields are given by  $\phi^a_{\rm bg}(x)= x^a-A^a_{\rm bg}(x)$.


We now add fluctuations about that background,
\ba
\phi^a&=&\phi^a_{\rm bg}-a^a=x^a-A^a\\
g\mn&=& g^{\rm bg}\mn +h\mn\,,
\ea
with $A^a=A^a_{\rm bg}+a^a$.

\subsubsection*{Flat background metric}

First note that
if we consider a flat background metric to start with,
then at zeroth order in $h$, the ghost-free potential is of the form~\cite{Mirbabayi:2011aa}, (this can also be seen from~\cite{Gabadadze:2013ria,Ondo:2013wka})
\ba
\mathcal{L}_{A}=-\frac 14 F F \(1+ \p A+\cdots\)\,,
\ea
with $F_{ab}=\p_a A_b-\p_b A_a$.
This means that for a symmetric \stu background configuration, \ie if the matrix $\p_\mu \phi^a_{\rm bg}$ is symmetric, then $F_{ab}^{\rm bg}=0$,  and at quadratic order in the fluctuation $a$, the action has a $U(1)$-symmetry. This symmetry is lost non-linearly, but is still relevant when looking at quadratic fluctuations about arbitrary backgrounds.
Now using the split about the background, $A^a=A^a_{\rm bg}+a^a$, this means that up to quadratic order in the fluctuations $a^a$, the action at zeroth order in the metric fluctuation is of the form~\cite{Mirbabayi:2011aa}
\ba
\label{L2 a}
\L_{a}^{(2)}=\bar B^{\mu\alpha\nu\beta}f\mn f\ab\,,
\ea
with $f\mn=\p_\mu a_\nu-\p_\nu a_\mu$ and $\bar B^{\mu\alpha\nu\beta}$ is a set of constant  coefficients which depends on $A^a_{\rm bg}$. This quadratic action has an accidental $U(1)$-symmetry which is responsible for projecting out one of the four dofs naively present in the four \stu fluctuations  $a^a$. Had we considered any other potential term, the $U(1)$ symmetry would have been generically lost and all four \stu fields would have been dynamical.

\subsubsection*{Non-symmetric background \stu}

If the background configuration is not symmetric, then at every point one needs to perform first an internal Lorentz transformation $\Lambda(x)$ in the \stu field space, so as to align them with the coordinate basis and recover a symmetric configuration for the background \stu fields. In this new Lorentz frame, the \stu fluctuation is $\tilde a^\mu =\Lambda^\mu_{\, \nu}(x)a_\nu$. As a result, to quadratic order in the \stu fluctuation the part of the ghost-free potential which is independent of the metric fluctuation and its curvature  goes symbolically as~\eqref{L2 a} with $f$ replaced by $f\to \tilde f +(\p \Lambda)\Lambda^{-1}\tilde a$, (with $\tilde f\mn=\p_\mu \tilde a_\nu-\p_\nu \tilde a_\mu$). Interestingly the Lorentz boost $(\p \Lambda)\Lambda^{-1}$ now plays the role of a mass term for what looks like a gauge field $\tilde a$. This mass term breaks the $U(1)$ symmetry, but there is still no kinetic term for $\tilde a_0$, very much as in a Proca theory. This part of the potential is thus manifestly ghost-free (in the sense that it provides a dynamics for only three of the four \stu fields, independently of the background).

Next we consider the mixing with  metric fluctuation $h$ while still assuming zero curvature.
At linear order in $h$, the ghost-free potential, \eqref{dRGT_metric} goes as follows
\ba
\label{stu Gback h}
\L_{Ah}^{(2)}=h^{\mu\nu}\sum_{n=1}^3 c_n X\mn^{(n)} + h F (\p A+\cdots)\,,
\ea
where the tensors $X^{(n)}\mn$ are similar to the ones  found in the decoupling limit, but now expressed in terms  of the symmetric full four \stu fields rather than just $\pi$, \ie replacing $\Pi\mn$ by $\p_\mu A_\nu+\p_\nu A_\mu$ in the respective expressions (\ref{X1}, \ref{X2} and \ref{X3}) for $X^{(1,2,3)}\mn$.
Starting with the symmetric configuration for the \stu fields, then since we are working at the quadratic level in perturbations, one of the $A_\mu$ in the $X^{(n)}\mn$ is taken to be the fluctuation $a_\mu$, while the others are taken to be the background field $A_\mu^{\rm bg}$. As a result in the first terms in $h X$ in ~\eqref{stu Gback h} $\p_0 a_0$ cannot come at the same time as $h_{00}$ or $h_{0i}$, and we can thus integrate by parts the time derivative acting on any $a_0$, leading to a harmless first time derivative on $h_{ij}$, and no time evolution for $a_0$.

As for the second type of term in~\eqref{stu Gback h}, since $F=0$ on the background field $A_\mu^{\rm bg}$, the second type of terms is forced to be proportional to $f\mn$ and cannot involve any $\p_0 a_0$ at all. As a result $a_0$ is not dynamical, which ensures that the theory is free from the BD ghost.

This part of the argument generalizes easily  for non symmetric background \stu configurations, and the same replacement $f\to \tilde f +(\p \Lambda)\Lambda^{-1}\tilde a$ still ensures that $\tilde a_0$ acquires no dynamics from~\eqref{stu Gback h}.

\subsubsection*{Background curvature}

Finally to complete the argument we consider the effect from background curvature, then $g^{\rm bg}\mn\ne \eta\mn$, with $g^{\rm bg}\mn=e^a_{\, \mu}(x)e^b_{\, \nu}(x)$. The space-time curvature is another source of `misalignment' between the coordinates and the \stu fields.
To rectify for this misalignment, we could go two ways: Either perform a local change of coordinate so as to align the background metric $g^{\rm bg}\mn$ with the flat reference metric $\eta\mn$ (\ie going to local inertial frame), or the other way around: \ie express the flat reference metric in terms of the curved background metric, $\eta_{ab} = e^\mu_{\, a}e^\nu_{\, b}g^{\rm bg}\mn$, in terms of the inverse vielbein, $e^\mu_{\, a}\equiv (e^{-1})^\mu_a$. Then the building block of ghost-free massive gravity is the matrix $\mathbb{X}$, defined previously as
\ba
\mathbb{X}\mupn=\(g^{-1}\eta\)\mupn=g^{\mu\gamma} (e^\alpha_{\, a}\p_\gamma \phi^a)
(e^\beta_{\, b}\p_\nu \phi^b)g^{\rm bg}\ab\,.
\ea
As a result the whole formalism derived previously is directly applicable with the only subtlety that the \stu fields $\phi^a$ should be replaced by their `vielbein-dependent' counterparts, \ie $\p_\mu A_\nu\to g^{\rm bg}\mn-g^{\rm bg}_{\nu \alpha}e^\alpha_{\, a}\p_\mu \phi^a$. In terms of the \stu field fluctuation $a^a$, this implies the replacement $a^a\to \bar a_\mu=g^{\rm bg}\mn e^\nu_{\ a}a^a$, and symbolically, $f\to \bar f +( \p \Sigma )\Sigma^{-1}\bar a$, with $\Sigma= g \dot e$. The situation is thus the same as when we were dealing with a non-symmetric \stu background configuration, after integration by parts (which might involve curvature harmless contributions), the potential can be written in a way which never involves any time derivative on $\bar a_0$. As a result  $\bar a_\mu$ plays the role of an effective Proca vector field which only propagates three degrees of freedom, and this about any curved background metric.
The beauty of this argument lies in the correct identification of the proper degrees of freedom when dealing with a curved background metric.


\subsection{Absence of ghost in the vielbein formulation}

Finally, we can also prove the absence of ghost for dRGT in the Vielbein formalism, either directly at the level of the Lagrangian in some special cases as shown in~\cite{Deffayet:2012nr} or in full generality in the Hamiltonian formalism, as shown in~\cite{Hinterbichler:2012cn}. The later proof also works in all generality for a multi-gravity theory and will thus be presented in more depth in what follows, but we first focus on a special case  presented in Ref.~\cite{Deffayet:2012nr}.

Let us start with massive gravity in the vielbein formalism~\eqref{dRGT_vielbein}. As was the case in Section~\ref{sec:ghost-free MG}, we work with the symmetric vielbein condition, $e^a_\mu f^b_\nu \eta_{ab}=e^a_\nu f^b_\mu \eta_{ab}$. For simplicity we specialize further to the case where $f^a_\mu =\delta^a_\mu$, so that the symmetric vielbein condition imposes $e^{a\mu}=e^{\mu a}$. Under this condition, the vielbein contains as many independent components as the metric. In $d$ spacetime dimensions, there is a priori $d(d+1)/2$ independent components in the symmetric vielbein.

Varying the action~\eqref{dRGT_vielbein} with respect to the vielbein leads to the modified Einstein equation,
\ba
G_a=t_a=-\frac{m^2}{2}\ep_{abcd}\Big(4 c_0\, e^b \wedge e^c \wedge e^d+3 c_1\, e^b \wedge e^c \wedge f^d\\
+2c_2\, e^b \wedge f^c \wedge f^d+c_3\, f^b \wedge f^c \wedge f^d\Big)\,,
\ea
with $G_a=\ep_{abcd}\omega^{bc}\wedge e^d$. From the Bianchi identity, $\mathcal{D} G_a = \d G_a -\omega^b_{\, a} G_b$, we infer the $d$ constraints
\ba
\label{Bianchi_vielbein_1}
\mathcal{D}t_a = \d t_a -\omega^b_{\ a} t_b=0\,,
\ea
leading to $d(d-1)/2$ independent components in the vielbein. This is still one too many component, unless an additional constraint is found. The idea behind the proof in Ref.~\cite{Deffayet:2012nr}, is then to use the Bianchi identities to infer an additional constraint of the form,
\ba
\label{Extra constraint}
m^a \wedge G_a = m^a \wedge t_a\,,
\ea
where $m^a$ is an appropriate one-form  which depends  on the specific coefficients of the theory. Such a  constrain is present at the linear level for Fierz--Pauli massive gravity, and it was further shown in Ref.~\cite{Deffayet:2012nr} that special choices of coefficients for the theory lead to remarkably simple analogous relations fully non-linearly. To give an example, we consider all the coefficients $c_n$ to vanish but $c_1\ne 0$. In that case the Bianchi identity~\eqref{Bianchi_vielbein_1} implies
\ba
\label{Dt 2}
\mathcal{D}t_a = 0 \qquad\Longrightarrow\qquad\omega^b_{\ c b}=0\,,
\ea
where similarly as in~\eqref{connection vielbein 5d}, the torsionless connection is given in term of the vielbein as
\ba
\label{connection vielbein 4d}
\omega^{ab}_\mu = \frac{1}{2}e_\mu^c(o^{ab}_{\ \ c}-o_c^{\ ab}-o^{b\ \ a}_{\ c})\,,
\ea
with $o^{ab}_{\ \ c}= 2 e^{a\mu} e^{b\nu}\, \partial_{[\mu}e_{\nu] c}$. The Bianchi identity~\eqref{Dt 2} then implies $e_a^{\ b}\p_{[b}e^a_{a]}=0$, so  that we obtain an extra constraint of the form~\eqref{Extra constraint} with $m^a=e^a$.
Ref.~\cite{Deffayet:2012nr} derived  similar constraints for other parameters of the theory.

\subsection{Absence of ghosts in multi-gravity}
\label{sec:multigravity}

We now turn to the proof for the absence of ghost in multi-gravity and follow the vielbein formulation of Ref.~\cite{Hinterbichler:2012cn}. In this subsection we use the notation that upper case latin indices represent $d$-dimensional Lorentz indices, $A, B, \cdots=0, \cdots, d-1$, while lower case latin indices represent the $d-1$-dimensional Lorentz indices along the space directions after ADM decomposition, $a, b, \cdots =1, \cdots, d-1$. Greek indices represent $d$-dimensional spacetime indices $\mu, \nu=0, \cdots, d-1$, while the `middle' of the latin alphabet indices $i,j\cdots$ represent pure space indices $i,j \cdots=1,\cdots,d-1$. Finally capital indices label the metric and span over $I,J,K,\cdots =1, \cdots, N$.

Let us start with $N$ \emph{non-interacting} spin-2 fields. The theory has then $N$ copies of coordinate transformation invariance (the coordinate system associated with each metric can be changed separately), as well as $N$ copies of Lorentz invariance. At this level may, for each vielbein $e_{(J)}$, $J=1,\cdots,N$ we may use part of the Lorentz freedom to work in the upper triangular form for the vielbein,
\ba
\label{VielbeinUpperDiagonal}
e_{(J)}{}^A_{\ \mu}=\(\begin{array}{cc}
N_{(J)} & N_{(J)}^i e_{(J)}{}^a_{\  i}\\
0 & e_{(J)}{}^a_{\ i}
\end{array}\)\,, \qquad
e_{(J)}{}^\mu_{\ A}=\(\begin{array}{cc}
N_{(J)}{}^{-1} & 0 \\
-N_{(J)}^i N^{-1} & e_{(J)}{}^i_{\ a}
\end{array}\)\,,
\ea
leading to the standard ADM decomposition for the metric,
\ba
g_{(J)}{}\mn \d x^\mu \d x^\nu &=& e_{(J)}{}^A_{\ \mu }e_{(J)}{}^B _{\ \nu}\eta_{AB} \d x^\mu \d x^\nu\nn\\
&=& -N_{(J)}{}^2 \d t^2 +\gamma_{(J)}{}_{ij}\(\d x^i + N_{(J)}^i \d t\)\(\d x^j + N_{(J)}{}^j \d t\)\,,
\ea
with the three-dimensional metric $\gamma_{(J)}{}_{ij}= e_{(J)}{}^a_{\ i}e_{(J)}{}^b_{\ j}\delta_{ab}$.
Starting with non-interacting fields,  we simply take $N$ copies of the GR action,
\ba
L_{N {\rm GR}}=\int \d t \sum_{J=1}^N\sqrt{-g_{(J)}}R_{(J)}\,,
\ea
and the Hamiltonian in terms of the vielbein variables then takes the form~\eqref{H GR}
\ba
\mathcal{H}_{N {\rm GR}}=\int \d^d x\sum_{J=1}^N\(\pi_{(J)}{}_a^{\ i} \dot{e}_{(J)}{}^a_{\ i}
+ N_{(J)}{} \mathcal{C}_{(J)}{}_0 +N_{(J)}^i\mathcal{C}_{(J)}{}_i-\frac 12 \lambda_{(J)}^{ab}\mathcal{P}_{(J)}{}_{ab}\)\,,
\ea
where $\pi_{(J)}{}_a^{\ i}$ is the conjugate momentum associated with the vielbein $e_{(J)}{}^a_{\ i}$ and the constraints $\mathcal{C}_{(J)}{}_{0,i}=\mathcal{C}_{0,i}(e_{(J)}, \pi_{(J)})$ are the ones mentioned previously in~\eqref{H GR} (now expressed in the vielbein variables) and are related to diffeomorphism invariance. In the vielbein language there is an addition $d(d-1)/2$ primary constraints for each vielbein field
\ba
\mathcal{P}_{(J)}{}_{ab}=e_{(J)}{}_{[a i}\ \pi_{(J)}{}^{\ i}_{b]},
\ea
related to the residual local Lorentz symmetry still present after fixing the upper triangular form for the vielbeins.

Now rather than setting part of the $N$ Lorentz frames to be on the upper diagonal form for all the $N$ vielbein~\eqref{VielbeinUpperDiagonal} we only use one Lorentz boost to set one of the vielbein in that form, say $e_{(1)}$, and `unboost' the $N-1$ other frames, so that for any of the other vielbein one has
\ba
e_{(J)}{}^A_{\ \mu}&=&\(\begin{array}{cc}
N_{(J)} \tilde \gamma_{(J)} +N^i_{(J)}e_{(J)}{}^a_{\ i} p_{(J)}{}_a& N_{(J)} p_{(J)}^a+ N_{(J)}^i e_{(J)}{}^b_{\  i}S_{(J)}{}^a_b\\
e_{(J)}{}^a_{\ i} & e_{(J)}{}^b_{\ i}S_{(J)}{}^a_b
\end{array}\)\\
S_{(J)}{}^a_b&=&\delta^a_b+\tilde \gamma_{(J)}^{-1}p_{(J)}^a p_{(J)}{}_b\\
\tilde \gamma_{(J)}&=&\sqrt{1+p_{(J)}{}_a p_{(J)}^a}
\ea
where $p_{(J)}{}_a$ is the boost that would bring that vielbein in the upper diagonal form.

We now consider arbitrary interactions between the $N$ fields of the form~\eqref{dRGT_vielbein},
\ba
L_{N\, {\rm int}} = \sum_{J_1,\cdots,J_d=1}^N \alpha_{J_1,\cdots,J_d} \ep_{a_1 \cdots a_d}\, e_{(J_1)}^{a_1}\wedge\cdots \wedge e_{(J_d)}^{a_d}\,,
\label{Interactions Vielbein}
\ea
where for concreteness we assume $d\le N$, otherwise the formalism is exactly the same (there is some redundancy in this formulation, \ie some interactions are repeated in this formulation, but this has no consequence for the argument). Since the vielbeins $e_{(J)}{}^{A}_{\ 0}$ are linear in their respective shifts and lapse $N_{(J)}, N_{(J)}^i$ and the vielbeins
$e_{(J)}{}^{A}_{\ i}$ do not depend any shift nor lapse, it is easy to see that the general set of interactions
\eqref{Interactions Vielbein} lead to a Hamiltonian which is also linear in every shift and lapse,
\ba
\mathcal{H}_{N\, {\rm int}}&=&\sum_{J=1}^N\(N_{(J)} \mathcal{C}^{\rm int}_{(J)}(e,p)+N_{(J)}^i \mathcal{C}^{\rm int}_{(J)}{}_{i}(e,p)\)\,.
\ea
Indeed the wedge structure of~\eqref{dRGT_vielbein} or~\eqref{Interactions Vielbein} ensures that there is one and only one vielbein with time-like index $e_{(J)}{}^{A}_{\ 0}$ for every term $ \ep_{a_1 \cdots a_d}\, e_{(J_1)}^{a_1}\wedge\cdots \wedge e_{(J_d)}^{a_d}$.

Notice that for the interactions, the terms $\mathcal{C}^{\rm int}_{(J)}{}_{0,i}$ can depend on \emph{all} the $N$ vielbeins $e_{(J')}$ and all the $N-1$ `boosts'  $p_{(J')}$, (as mentioned previously, part of one Lorentz frame is set so that $p_{(1)}=0$ and $e_{(1)}$ is in the upper diagonal form).
Following the procedure of~\cite{Hinterbichler:2012cn}, we can now solve for the $N-1$ remaining boosts by using $(N-1)$ of the $N$  shift equations of motion
\ba
\label{ShiftEq}
\mathcal{C}_{(J)\, i}(e, \pi) + \mathcal{C}_{(J)\, i}(e, p)=0 \qquad\forall \hspace{5pt} J=1,\cdots,N\,.
\ea
Now assuming that all $N$ vielbein are interacting\epubtkFootnote{If only $M$ vielbein of the $N$ vielbein are interacting there will be $(N-M+1)$ copies of diffeomorphism invariance and $M-1$ additional Hamiltonian constraints, leading to the correct number of dofs for $(N-M+1)$ massless spin-2 fields and $M-1$ massive spin-2 fields.}, (\ie there is no vielbein $e_{(J)}$ which does not appear at least once in the interactions~\eqref{Interactions Vielbein} which mix different vielbeins), the shift equations~\eqref{ShiftEq} will involve all the $N-1$ boosts and can be solved for them without spoiling the linearity in any of the $N$ lapses $N_{(J)}$. As a result, the $N-1$ lapses $N_{(J)}$ for $J=2,\cdots,N$ are Lagrange multiplier for $(N-1)$ first class constraints. The lapse $N_{(1)}$ for the first vielbein combines with the remaining shift $N_{(d)}^i$ to generate the one remaining copy of diffeomorphism invariance.

We now have all the ingredients to count the  number of dofs in phase space: We start with $d^2$ components in each of the $N$ vielbein $e^{\ a}_{{(J)}\ i}$ and associated conjugate momenta, that is a total of $2\times d^2\times N$ phase space variables. We then have $2\times d(d-1)/2 \times N$ constraints\epubtkFootnote{Technically, only one of them generates a first class constraint, while the $N-1$ others generate a second-class  constraint. There are therefore $(N-1)$ additional secondary constraints to be found by commuting the primary constraint with the Hamiltonian, but the presence of these constraints at the linear level ensures that they must exist at the non-linear level. There is also another subtlety in obtaining the secondary constraints associated with the fact that the Hamiltonian is pure constraint, see the discussion in Section~\ref{sec:bifurvation} for more details.} associated with the $\lambda_{(J)}^{ab}$. There is one  copy of diffeomorphism removing $2\times(d+1)$ phase space dofs (with Lagrange multiplier $N_{(1)}$ and $N_{(1)}^i$) and $(N-1)$ additional first-class constraints with Lagrange multipliers $N_{(J\ge2)}$ removing $2\times(N-1)$ dofs. As a result we end up with
\ba
&& \(2\times \frac{d(d-1)}{2} \times N\)- 2\times \frac{d(d-1)}{2}\times N- 2 \times(d+1)-2 \times (N-1)\nn \\
&&= \(d^2N -2N+d(N-2)\)\text{phase space dofs}\nn\\
&&=\frac 12 \(d^2N -2N+d(N-2)\)\text{field space dofs}\\
&&=\frac 12 \(d^2-d-2\)\text{dofs for a massless  spin-2 field}\nn\\
&&\ +\frac 12 \(d^2+d-2\)\times (N-1) \ \text{ dofs for $(N-1)$ a massive  spin-2 fields}\,,
\ea
which is the correct counting in $(d+1)$ spacetime dimensions, and the theory is thus free of any BD ghost.

\section{Decoupling Limits}
\label{sec:DL section}

\subsection{Scaling versus decoupling}

Before moving to the decoupling of massive gravity and bi-gravity, let us make a brief interlude concerning the correct identification of degrees of freedom. The \stu trick used previously to identify the correct degrees of freedom works in all generality, but care must be used when taking a ``decoupling limit'' (\ie scaling limit) as will be done in Section~\ref{sec:MG DL of bigravity}.

Imagine the following gauge field theory
\ba
\L= -\frac 12 m^2 A_\mu A^\mu\,,
\ea
\ie the Proca mass term without any kinetic Maxwell term for the gauge field. Since there are no dynamics in this theory, there is no degrees of freedom. Nevertheless, one could still proceed and use the same split $A_\mu = A_\mu+\p_\mu \chi/m$ as performed previously,
\ba
\label{A^2 with chi}
\L=-\frac 12 m^2 A_{\mu}A^{\mu}+m (\p_\mu A^\mu)\chi-\frac 12  (\p \chi)^2\,,
\ea
so as to introduce what appears to be a kinetic term for the mode $\chi$. At this level the theory is still invariant under $\chi \to \chi + m \xi$ and $A^\bot_\mu \to A^\bot_\mu -\p_\mu \xi$, and so while there appears to be a dynamical degree of freedom $\chi$, the symmetry  makes that degree of freedom unphysical, so that~\eqref{A^2 with chi} still propagates no physical degree of freedom.

Now consider the $m\to 0$ scaling limit of~\eqref{A^2 with chi} while keeping $A_\mu$ and $\chi$ finite. In that \emph{scaling} limit, the theory reduces to
\ba
\label{scaling1}
\L_{m\to 0}=-\frac 12 (\p \chi)^2\,,
\ea
\ie one degree of freedom with no symmetry which implies that the theory~\eqref{scaling1} propagates one degree of freedom. This is correct and thus means that~\eqref{scaling1} is \emph{not} a consistent \emph{decoupling limit} of~\eqref{A^2 with chi} since the number of degrees of freedom is different already at the linear level. In the rest of this review we will call a \emph{decoupling limit} a specific type of scaling limit which preserves the same number of physical propagating degrees of freedom in the linear theory. As suggested by the name, a decoupling limit is a special kind of limit in which some of the degrees of freedom of the original theory might \emph{decouple} from the rest, but the total number of degrees of freedom remains identical. For the theory~\eqref{A^2 with chi}, this means that the scaling ought to be taken not with $A_\mu$ fixed but rather with $\tilde A_\mu = A_\mu/m$ fixed. This is indeed a consistent rescaling which leads to finite contributions in the limit $m\to 0$,
\ba
\L_{m\to 0}=-\frac 12 \tilde A_{\mu}\tilde A^{\mu}+ (\p_\mu \tilde A^\mu)\chi-\frac 12  (\p \chi)^2\,,
\ea
which clearly propagates no degrees of freedom.

This procedure is true in all generality: a decoupling limit is a special scaling limit where all the fields in the original theory are scaled with the highest possible power of the scale in such a way that the decoupling limit is finite.

\emph{A decoupling limit of a theory never changes the number of physical degrees of freedom of a theory. At best it `decouples' some of them in such a way that they are inaccessible from another sector}.

Before looking at the massive gravity limit of bi-gravity and other decoupling limits of massive and bi-gravity, let us start by describing the different scaling limits that can be taken. We start with a bi-gravity theory where the two spin-2 fields have respective Planck scales $M_g$ and $M_f$ and the interactions between the two metrics arises at the scale $m$. In order to stick to the relevant points we perform the analysis in four dimensions, but the following arguments are extend trivially to arbitrary dimensions.

\begin{itemize}
\item \textbf{Non-interacting Limit:} The most natural question to ask is what happens in the limit where the interactions between the two fields are `switched off', \ie when sending the scale $m\to$, (the limit $m\to 0$ is studied more carefully in sections~\ref{sec:DL MG} and \ref{sec:DL bigravity}). In that case if the two Planck scales $M_{g,f}$ remain fixed  as $m\to 0$, we then recover two massless non-interacting spin-2 fields (carrying both 2 helicity-2 modes), in addition to a decoupled sector containing a helicity-0 mode and a helicity-1 mode. In bi-gravity matter fields couple only to one metric, and this remains the case in the limit $m\to 0$, so that the two massless spin-2 fields live in two fully decoupled sectors even when matter in included.
\item \textbf{Massive Gravity:}
Alternatively, we may look at the limit where one of the spin-2 fields (say $f\mn$) decouples. This can be studied by sending its respective Planck scale to infinity. The resulting limit corresponds to a massive spin-2 field (carrying five dofs) and a \emph{decoupled} massless spin-2 field carrying 2 dofs. This is nothing other than the massive gravity limit of bi-gravity (which includes a fully decoupled massless sector).

If one considers matter coupling to the metric $f\mn$ which scales in such a way that a non-trivial solution for $f\mn$ survives in the $M_f\to \infty$ limit $f\mn \to \bar f\mn$, we then obtain a massive gravity sector on an arbitrary non-dynamical reference metric $\bar f\mn$. The dynamics of the massless spin-2 field fully decouples from that massive sector.
\item \textbf{Other Decoupling Limits} Finally, one can look at combinations of the previous limits, and the resulting theory depends on how fast $M_f, \mpl \to \infty$ compared to how fast $m\to0$. For instance if one takes the limit $M_f, \mpl \to \infty $ and $m \to 0$, while keeping both $m_g/M_f$ and $\Lambda_3^3=M_g m^2$ fixed, then we obtain what is called the $\Lambda_3$-decoupling limit of bi-gravity (derived in Section~\ref{sec:DL bigravity}), where the dynamics of the two helicity-2 modes (which are both massless in that limit), and that of the helicity-1 and -0 modes can followed without keeping track of the standard non-linearities of GR.\\
    If on top of this $\Lambda_3$-decoupling limit one further takes  $M_f\to \infty$, then one of the massless spin-2 fields fully decoupled (no communication between that field and the helicity-1 and -0 modes). If on the other hand we take the additional limit $m \to 0$ on top of the $\Lambda_3$-decoupling limit, then the helicity-0 and -1 modes fully decouple from both helicity-2 modes.
\end{itemize}

In all of these decoupling limits, the number of dofs remains the same as in the original theory, some fields are simply decoupled from the rest of the standard gravitational sector. These prevents any communication between these decoupled fields and the gravitational sector, and so from the gravitational sector view point it appears as if these decoupled fields did not exist.

It is worth stressing that all of these limits are perfectly sensible and lead to sensible theories, (from a theoretical view point). This is important since if one of these scaling limits lead to a pathological theory, it would have severe consequences for the parent bi-gravity theory itself.

Similar decoupling limit could be taken in multi-gravity and out of $N$ interacting spin-2 fields, we could obtain for instance $N$ decoupled massless spin-2 fields and $3(N-1)$ decoupled dofs in the helicity-0 and -1 modes.

In what follows we focus on massive gravity limit of bi-gravity when $M_f\to \infty$.

\subsection{Massive gravity as a decoupling limit of bi-gravity}
\label{sec:MG DL of bigravity}

\subsubsection{Minkowski reference metric}

In the following two sections we review the decoupling arguments given previously in the literature, (see for instance~\cite{deRham:2012kf}).
We start with the theory of bi-gravity presented in Section~\ref{sec:bi-gravity} with the action~\eqref{bigravity}
\ba
\label{Bigravity_Lagrangian}
\mathcal{L}_{\rm bi-gravity}&=&\frac{M_g^2}{2}\sqrt{-g}R[g]+\frac{M_f^2}{2}\sqrt{-f}R[f]
+\frac{1}{4}m^2 \mpl^2\sqrt{-g}\, \L_m(g,f) \nn \\
&&+\sqrt{-g}\L_{g}^{\rm (matter)}(g\mn, \psi_g)
+\sqrt{-f}\L_{f}^{\rm (matter)}(f\mn, \psi_f)\,,
\ea
with $\L_m(g,f)=\sum_{n=0}^4\alpha_n \L_n[\K(g,f)]$ as defined in~\eqref{dRGT_metric} and where
$\mathcal{K}^{\mu}_{\nu}=\delta^{\mu}_{\nu}-\sqrt{g^{\mu\alpha}f_{\alpha\nu}}$. We also allow for the coupling to matter with different species  $\psi_{g,f}$ living on each metrics.

We now consider matter fields $\psi_{f}$ such that $f\mn=\eta\mn$ is a solution to the equations of motion (so for instance there is no overall cosmological constant living on the  metric $f\mn$).
In that case we can write that metric $f\mn$ as
\ba
f\mn&=&\eta\mn + \frac{1}{M_f}\chi\mn\,,
\ea
We may now take the limit $M_f \to \infty$, while keeping the scales $M_g$ and $m$ and all the fields $\chi, g, \psi_{f,g}$ fixed. We then recover massive gravity plus a completely decoupled massless spin-2 field $\chi\mn$, and a fully decoupled matter sector $\psi_f$ living on flat space
\ba
\mathcal{L}_{\rm bi-gravity}\xrightarrow{M_f\to \infty}
\L_{\rm MG}(g,\eta)+\sqrt{-g}\L_{g}^{\rm (matter)}(g\mn, \psi_g)\\
\nn+\frac 12 \chi^{\mu\nu}\Ein^{\alpha\beta}\mn \chi_{\alpha\beta}+ \L_{f}^{\rm (matter)}(\eta\mn, \psi_f)\,,
\ea
with the massive gravity Lagrangian $\L_{\rm MG}$ is expressed in~\eqref{dRGT_metric}. That massive gravity Lagrangian remains fully non-linear in this limit and is expressed in terms of the full metric $g\mn$ and the reference metric $\eta\mn$. While the metric $f\mn$ is `frozen' in this limit, we emphasize however that the  massless spin-2 field $\chi\mn$ is itself \emph{not} frozen -- its dynamics is captured through the kinetic term $\chi^{\mu\nu}\Ein^{\alpha\beta}\mn \chi_{\alpha\beta}$, but that spin-2 field decouple from its own matter sector $\psi_f$, (although this can be accommodated for by scaling the matter fields $\psi_f$ accordingly in the limit $M_f\to \infty$ so as to maintain some interactions).

At the  level of the equations of motion, in the limit $M_f\to \infty$ we obtain the massive gravity modified Einstein equation for $g\mn$, the free massless linearized Einstein equation for $\chi\mn$ which fully decouples and the equation of motion for all the matter fields $\psi_f$ on flat spacetime, (see also Ref.~\cite{Baccetti:2012bk}).

\subsubsection{(A)dS reference metric}

To consider massive gravity with an (A)dS reference metric as a limit of bi-gravity, we include a cosmological constant for the metric $f$ into~\eqref{Bigravity_Lagrangian}
\ba
\label{Lf_CC}
\L_{\rm CC, f}= - M_f^2 \int \d^4x \sqrt{-f} \Lambda_f \,.
\ea
There can also be in principle another cosmological constant living on top of the metric $g\mn$ but this can be included into the potential $\mathcal{U}(g,f)$. The background field equations of motion are then given by
\ba
M_f^2 G\mn[f] + \frac{m^2 \mpl^2}{4\sqrt{-g}} \(\frac{\delta}{\delta f^{\mu\nu}}\sqrt{-g}\, \mathcal{U}(g,f)\)&=&T\mn(\psi_f)-M_f^2\Lambda_f f\mn\\
\mpl^2 G\mn[g] + \frac{m^2 \mpl^2}{4\sqrt{-g}} \(\frac{\delta}{\delta g^{\mu\nu}}\sqrt{-g}\, \mathcal{U}(g,f)\)&=&T\mn(\psi_g)\,.
\ea
Taking now the limit $M_f\to \infty$ while keeping the cosmological constant $\Lambda_f$ fixed, the background solution for the metric $f\mn$ is nothing other than dS (or AdS depending on the sign of $\Lambda_f$). So we can now express the metric $f\mn$ as
\ba
f\mn=\gamma\mn+ \frac{1}{M_f}\chi\mn\,,
\ea
where $\gamma\mn$ is the dS metric with Hubble parameter $H=\sqrt{\Lambda_f/3}$. Taking the limit $M_f \to \infty$,   we recover massive gravity on (A)dS plus a completely decoupled massless spin-2 field $\chi\mn$,
\ba
\mathcal{L}_{\rm bi-gravity}- M_f^2 \int \d^4x \sqrt{-f} \Lambda_f\  \xrightarrow{M_f\to \infty}
\
\frac{\mpl^2}{2}\sqrt{-g}R+\frac{m^2}{4}\mathcal{U}(g,\gamma)
\\ \nn +\frac 12 \chi^{\mu\nu}\Ein^{\alpha\beta}\mn \chi_{\alpha\beta}\,,
\ea
where once again the scales $\mpl$ and $m$ are kept fixed in the limit $M_f\to \infty$.
$\gamma\mn$ now plays the role of a non-trivial reference metric for massive gravity. This corresponds to a theory of massive gravity on a more general reference metric as presented in~\cite{Hassan:2011tf}. Here again
the Lagrangian for massive gravity is given in~\eqref{dRGT_metric} with now $\mathcal{K}^{\mu}_{\nu}(g)=\delta^{\mu}_{\nu}-\sqrt{g^{\mu\alpha}\gamma_{\alpha\nu}}$. The massive gravity action remains fully non-linear in the limit $M_f\to \infty$ and is expressed solely in terms of the full metric $g\mn$ and the reference metric $\gamma\mn$, while the excitations $\chi\mn$ for the massless graviton remain dynamical but fully decouple from the massive sector.

\subsubsection{Arbitrary reference metric}
\label{sec:ArbitraryDL}

As is already clear from the previous discussion, to recover massive gravity on a non-trivial reference metric as a limit of bi-gravity, one needs to scale the Matter Lagrangian that couples to what will become the reference metric (say the metric $f$ for definiteness) in such a way that the Riemann curvature of $f$ remains finite in that decoupling limit. For a macroscopical description of the matter living on $f$ this is in principle always possible. For instance one can consider a point source  of mass $M_{\rm BH}$ living on the metric $f$. Then taking the limit $M_f, M_{\rm BH}\to \infty$ while keeping the ratio $M_{\rm BH}/M_f$ fixed, leads to a theory of massive gravity on a Schwarzschild reference metric and a decoupled massless graviton. However some care needs to be taken to see how this works when the dynamics of the matter sourcing $f$ is included.

As soon as the dynamics of the matter field is considered, one has to send the scale of that field to infinity so that it maintains some nonzero effect on $f$ in the limit $M_f\to  \infty$, \ie
 \ba
 \lim_{M_f\to \infty}\frac{1}{M_f^2}T^{\mu\nu} =  \lim_{M_f\to \infty}\frac{1}{\sqrt{-f}M_f^2}\frac{\delta \sqrt{-f} \L_{f}^{\rm (matter)}}{\delta f_{\mu\nu}} \to {\rm finite}\,.
 \ea
Nevertheless this can be achieved in such a way that the fluctuations of the matter fields remain finite and decouple in the limit $M_f \rightarrow \infty$. As an example suppose that the Lagrangian for the matter (for example a scalar field) sourcing the $f$ metric is
 \be
\L_{f}^{\rm (matter)} = \sqrt{-f} \left( - \frac{1}{2}f^{\mu\nu} \partial_{\mu} \chi \partial_{\nu} \chi - V_0 F\( \frac{\chi}{\lambda}\)  \right)
 \ee
 where $F(X)$ is an arbitrary dimensionless function of its argument. Then choosing $\chi$ to take the form
 \be
 \chi = M_f \bar \chi + \delta \chi \, ,
 \ee
  and rescaling $V_0 = M_f^2 \bar V_0 $ and $\lambda = M_f \bar \lambda$, then on taking the limit $M_f \rightarrow \infty$ keeping $\bar \chi$, $\delta \chi$, $\bar \lambda$ and $\bar V_0$ fixed, since
  \be
  \L_{f}^{\rm (matter)} \rightarrow M_f^2 \sqrt{-f} \left( - \frac{1}{2}f^{\mu\nu} \partial_{\mu} \bar \chi \partial_{\nu} \bar \chi - \bar V_0 F\( \frac{\bar \chi}{\bar \lambda}\)  \right)+ \text{fluctuations}\,,
  \ee
we find that the background stress energy blows up in such a way that $\frac{1}{M_f^2}T^{\mu\nu} $ remains finite and nontrivial, and in addition the background equations of motion for $\bar \chi$ remain well-defined and nontrivial in this limit,
 \be
 \Box_f \bar \chi = \frac{\bar V_0}{\bar \lambda} F'\( \frac{\bar \chi}{\bar \lambda}\) \, .
 \ee
This implies that even in the limit $M_f \rightarrow 0$, $f_{\mu\nu}$ can remain consistently as a nontrivial sourced metric which is a solution of some dynamical equations sourced by matter. In addition the action for the fluctuations $\delta \chi $ asymptotes to a free theory which is coupled only to the fluctuations of $f_{\mu\nu}$ which are themselves completely decoupled from the fluctuations of the metric $g$ and matter fields coupled to $g$.

As a result massive gravity with an arbitrary reference metric can be seen as a consistent limit of bi-gravity in which the additional degrees of freedom in the $f$ metric and matter that sources the background decouple.  Thus all solutions of massive gravity may be seen as $M_f \rightarrow \infty$ decoupling limits of solutions of bi-gravity. This will be discussed in more depth in Section~\ref{sec:DL bigravity}.  For an arbitrary reference metric which can be locally written as a small departures about Minkowski the decoupling limit is derived in Eq.~\eqref{DL arbitrary metric}.

Having derived massive gravity as a consistent decoupling limit of bi-gravity, we could of course do the same for any multi-metric theory. For instance out of $N$-interacting fields, we could take a limit so as to decouple one of the metrics, we then obtain the theory of $(N-1)$-interacting fields, all of which being massive and one decoupled massless spin-2 field.

\subsection{Decoupling limit of massive gravity}
\label{sec:DL MG}

We now turn to a different type of decoupling limit, which aims is to disentangle the dofs present in massive gravity itself and analyze the `irrelevant interactions' (in the usual EFT sense) that arise at the lowest possible scale. One could naively think that such interactions arise at the scale given by the graviton mass, but this is not so. In a generic theory of massive gravity with Fierz--Pauli at the linear level, the first irrelevant interactions typically arise at the scale $\Lambda_5=(m^4 \mpl)^{1/5}$. For the setups we have in mind, $m \ll \Lambda_5\ll \mpl$. But we shall see that interactions arising at such a low-energy scale are always pathological (reminiscent to the BD ghost~\cite{Creminelli:2005qk,Deffayet:2005ys}), and in ghost-free massive gravity the first (irrelevant)  interactions actually arise at the scale $\Lc=(m^3 \mpl)^{1/3}$.

We start by deriving the decoupling limit in the absence of vectors (helicity-1 modes) and then include them in the following section  \ref{sec:DL vectors}.
Since we are interested in the decoupling limit about flat spacetime, we look at the case where Minkowski is a vacuum solution to the equations of motion. This is the case in the absence of a cosmological constant and a tadpole and we thus focus on the case where $\alpha_0=\alpha_1=0$ in~\eqref{dRGT_metric}.

\subsubsection{Interaction scales}

In GR, the interactions of the helicity-2 mode arise at the very high energy scale, namely the Planck scale. In massive gravity a new scale enters and we expect some interactions to arise at a lower energy scale given by a geometric combination of the Planck scale and the graviton mass. The potential term $\mpl^2 m^2 \sqrt{-g}\L_n[\K[g,\eta]]$~\eqref{dRGT_metric} includes generic interactions between the canonically normalised helicity-0 ($\pi$), helicity-1 ($A_\mu$), and helicity-2 modes ($h\mn$) introduced in
\eqref{FP tot lin}
\ba
\label{operators}
\L_{j,k,\ell}&=&m^2 \mpl^2 \(\frac{h}{\mpl}\)^j\(\frac{\p A}{m \mpl}\)^{2k}\(\frac{\p^2 \pi}{m^2 \mpl}\)^\ell\nn \\
&=& \Lambda_{j,k, \ell}^{-4+(j+4k+3\ell)}\ h^{j}\(\p A\)^{2k}\(\p^2 \pi\)^\ell\,,
\ea
at the scale
\ba
\label{DLscale}
\Lambda_{j,k, \ell}=\(m^{2k+2\ell-2}\mpl^{j+2k+\ell-2}\)^{1/(j+4k+3\ell-4)}\,,
\ea
and
with $j,k,\ell \in \mathbb{N}$, and $j+2k+\ell > 2$.

Clearly the lowest interaction scale is $\Lambda_{j=0,k=0,\ell=3}\equiv\Lambda_5=(\mpl m^4)^{1/5}$ which arises for an operator of the form $(\p^2 \pi)^3$. If present such an interaction leads to an Ostrogradsky instability which is another manifestation of the BD ghost as identified in~\cite{Deffayet:2005ys}.

Even if that very interaction is absent there is actually an infinite set of dangerous interactions of the form $(\p^2 \pi)^\ell$ which arise at the scale $\Lambda_{j=0,k=0,\ell\ge3}$, with
\ba
\Lambda_5=(\mpl m^4)^{1/5}\le \(\Lambda_{j=0,k=0,\ell\ge3}\)< \Lc=(\mpl m^2)^{1/3}\,.
\ea
with $\Lambda_{j=0,k=0,\ell\to \infty}=\Lc$.

Any interaction with $j>0$ or $k>0$ automatically leads to a larger scale, so all the interactions arising at a scale between $\Lambda_5$ (inclusive) and $\Lambda_3$ are of the form $(\p^2 \pi)^\ell$ and carry an Ostrogradsky instability. For DGP we have already seen that there is no interactions at a scale below $\Lambda_3$. In what follows we show that same remains true for the ghost-free theory of massive gravity proposed in~\eqref{dRGT_metric}. To see this let us identify the interactions with $j=k=0$ and arbitrary power $\ell$ for $(\p^2 \pi)$.

\subsubsection{Operators below the scale $\Lc$}

We now express the potential term $\mpl^2 m^2 \sqrt{-g}\L_n[\K]$ introduced in~\eqref{dRGT_metric} using the metric in term of the helicity-0 mode, where we recall that the quantity $\K$ is defined in~\eqref{eqK},
as
$\K\mupn[g,\tilde f]=\delta\mupn -\big(\sqrt{g^{-1}\tilde f}\ \big)\mupn$, where $\tilde f$ is the `St\"uckelbergized' reference metric given in~\eqref{eta stuck}. Since we are interested in interactions without the helicity-2 and -1 modes ($j=k=0$), it is sufficient to follow the behaviour of the helicity-0 mode and so we have
\ba
\label{eta stuck hel0}
\left.\begin{array}{rcl}
\tilde f\mn \Big|_{h=A=0}&=& \eta\mn -\frac{2}{\mpl m^2}\Pi\mn+ \frac{1}{\mpl^2 m^4}\Pi^2\mn\\
g^{\mu\nu}\Big|_{h=0}&=&\eta^{\mu\nu}
\end{array}\right\}\hspace{15pt}\Longrightarrow \hspace{15pt}
\K\mupn|_{h=A=0}=\frac{\Pi\mupn}{\mpl m^2}\,,
\ea
with again $\Pi\mn=\p_\mu\p_\nu \pi$ and $\Pi^2\mn:=\eta^{\alpha\beta}\Pi_{\mu\alpha}\Pi_{\nu \beta}$.

As a result we infer that up to the scale $\Lc$ (excluded), the potential in~\eqref{dRGT_metric} is
\ba
\L_{\rm mass}&=&\frac{m^2\mpl^2}{4}\sqrt{-g}\sum_{n=2}^4\alpha_n \L_n[\K[g,\tilde f]]\Big|_{h=A=0} \\
 &=&  \frac{m^2\mpl^2}{4}\sum_{n=2}^4\alpha_n \L_n \left[\frac{\Pi\mn}{\mpl m^2}\right]\\
&=&\frac 14 \e^{\mu\nu\alpha\beta}\e_{\mu'\nu'\alpha'\beta'}\Big(
\frac{\alpha_2}{m^2} \delta^{\mu'}_\nu \delta^{\nu'}_\nu
+\frac{\alpha_3}{\mpl m^4} \delta^{\mu'}_\nu \Pi^{\nu'}_\nu
+\frac{\alpha_4}{\mpl^2 m^6} \Pi^{\mu'}_\nu \Pi^{\nu'}_\nu  \nn
\Big)\Pi^{\alpha'}_\alpha \Pi^{\beta'}_\beta\,,
\ea
where as mentioned earlier we focus  on the case without  a cosmological constant and tadpole \ie $\alpha_0=\alpha_1=0$.
All of these interactions are \emph{total derivatives}. So even though the ghost-free theory of massive gravity does in principle involve some interactions with higher derivatives of the form $(\p^2\pi)^\ell$ it does so in a very precise way so that all of these terms combine so as to give a total derivative and being harmless\epubtkFootnote{This is actually precisely the way ghost-free massive gravity was originally constructed in~\cite{deRham:2010iK,deRham:2010kj}.}.

As a result the potential term constructed proposed in Section~\ref{sec:ghost-free MG} (and derived from the deconstruction framework) is free of any interactions of the form $(\p^2 \pi)^\ell$. This means that the BD ghost as identified in the \stu language in~\cite{Deffayet:2005ys} is absent in this theory. However at this level, the BD ghost could still reappear through different operators at the scale $\Lc$ or higher.

\subsubsection{$\Lc$-decoupling limit}

Since there are no operators all the way up to the scale $\Lc$ (excluded), we can take the decoupling limit by sending $\mpl\to \infty$, $m\to 0$ and maintaining the scale $\Lc$ fixed.

The operators that arise at the scale $\Lc$ are the ones of the form~\eqref{operators} with either $j=1, k=0$ and arbitrary $\ell\ge 2$ or with $j=0, k=1$ and arbitrary $\ell\ge 1$. The second case scenario leads to vector interactions of the form $(\p A)^2 (\p^2 \pi)^\ell$ and  will be studied in the next subSection~\ref{sec:DL vectors}. For now we focus on the first kind of interactions of the form $h (\p^2 \pi)^\ell$,
\ba
\L_{\rm mass}^{\rm dec}=h\mn \bar X\mn\,,
\ea
with~\cite{deRham:2010kj} (see also refs.~\cite{deRham:2010iK} and~\cite{deRham:2011qq})
\ba
\bar X\mn&=&\frac{\delta}{\delta h\mn}
\L_{\rm mass}\Big|_{h=A=0} \\
&=&\frac{\mpl^2 m^2}{4}\frac{\delta}{\delta h\mnu}\(\sqrt{-g}
\sum_{n=2}^4\alpha_n \L_n[\K[g,\tilde f]]\)\Big|_{h=A=0}\,.\nn
\ea
Using the fact that
\ba
\frac{\delta \K^n}{\delta h^{\mu\nu}}\Bigg|_{h=A=0}=\frac n2 \(\Pi^{n-1}\mn-\Pi^{n}\mn\)\,,
\ea
we obtain
\ba
\bar X\mn = \frac {\Lc^3}8 \sum_{n=2}^4 \alpha_n \(\frac{4-n}{\Lc^{3n}}X^{(n)}\mn[\Pi]+\frac{n}{\Lc^{3(n-1)}}X\mn^{(n-1)}[\Pi]\)\,,
\ea
where the tensors $X^{(n)}\mn$ are constructed out of $\Pi\mn$, symbolically, $X^{(n)}\sim \Pi^{(n)}$ but in such a way that they are transverse and that their resulting equations of motion never involve more than two derivatives on each fields,
\ba
\label{X0}
X^{(0)}{}^\mu_{\, \mu'}[Q]&=&\ep^{\mu \nu \alpha \beta}\ep_{\mu' \nu \alpha \beta} \\
\label{X1}
X^{(1)}{}^\mu_{\, \mu'}[Q]&=&\ep^{\mu \nu \alpha \beta}\ep_{\mu' \nu' \alpha \beta} \ Q^{\nu'}_{\nu}\\
\label{X2}
X^{(2)}{}^\mu_{\, \mu'}[Q]&=&\ep^{\mu \nu \alpha \beta}\ep_{\mu' \nu' \alpha' \beta}  \ Q^{\nu'}_{\nu}\ Q^{\alpha'}_{\alpha}\\
\label{X3}
X^{(3)}{}^\mu_{\, \mu'}[Q]&=&\ep^{\mu \nu \alpha \beta}\ep_{\mu' \nu' \alpha' \beta'}\ Q^{\nu'}_{\nu}\ Q^{\alpha'}_{\alpha}Q^{\beta'}_\beta\\
\label{X4}
X^{(n\ge4)}{}^\mu_{\, \mu'}[Q]&=&0\,,
\ea
where we have included $X^{(0)}$ and $X^{(n\ge 4)}$ for completeness (these become relevant for instance in the context of bi-gravity). The generalization of these tensors to arbitrary dimensions is straightforward and in $d$-spacetime dimensions  there are $d$ such tensors, symbolically $X^{(n)}=\ep \ep \Pi^{n} \delta^{d-n-1}$ for $n=0,\cdots, d-1$.

Since we are dealing with the decoupling limit with $\mpl \to  \infty$ the metric is flat $g\mn = \eta\mn +\mpl^{-1}h\mn \to \eta\mn$ and all indices are raised and lowered with respect to the Minkowski metric.  These tensors $X^{(n)}\mn$ can be written more explicitly as follows
\ba
\label{X0 expl}
X^{(0)}\mn[Q]&=& 3!\eta\mn\\
\label{X1 expl}
X^{(1)}\mn[Q]&=&2!\([Q]\eta\mn-Q\mn\)\\
\label{X2 expl}
X^{(2)}\mn[Q]&=&([Q]^2-[Q^2])\eta\mn-2([Q]Q\mn-Q^2\mn)\\
\label{X3 expl}
X^{(3)}\mn[Q]&=&([Q]^3-3[Q][Q^2]+2[Q^3])\eta\mn\\
&&-3\([Q]^2Q\mn-2[Q]Q^2\mn-[Q^2]Q\mn+2Q^3\mn\)\nn\,.
\ea
Note that they also satisfy the recursive relation
\ba
X\mn^{(n)}=\frac{1}{4-n}\(-n \Pi^\alpha_{\, \mu}\delta^\beta_{\nu}+\Pi^{\alpha\beta}\eta\mn\)X^{(n-1)}\ab\,,
\ea
with $X^{(0)}\mn=3! \eta\mn$.

\subsubsection*{Decoupling limit}

From the expression of these tensors $X\mn$ in terms of the fully antisymmetric Levi-Cevita tensors, it is clear that the tensors $X\mn^{(n)}$ are transverse and that the equations of motion of $h^{\mu\nu}\bar X\mn$ with respect to both $h$ and $\pi$ never involve more than two derivatives. This decoupling limit is thus free of the Ostrogradsky instability which is the way the BD ghost would manifest itself in this language. This decoupling limit is actually free of any ghost-lie instability and the whole theory is free of the BD even beyond the decoupling limit as we shall see in depth in Section~\ref{sec:Abscence of ghost proof}.

Not only does the potential term proposed in~\eqref{dRGT_metric} remove any potential interactions of the form $(\p^2 \pi)^\ell$ which could have arisen at an energy between $\Lambda_5=(\mpl m^4)^{1/5}$ and $\Lc$ , but it also ensures that the interactions that arise at the scale $\Lc$ are healthy.

As already mentioned, in the decoupling limit $\mpl\to \infty$ the metric reduces to Minkowski and the standard Einstein--Hilbert term simply reduces to its linearized version. As a result, neglecting the vectors for now the full $\Lc$-decoupling limit of ghost-free massive gravity is given by
\ba
\label{DL MG1}
\L_{\Lc}&=&-\frac 14 h^{\mu\nu}\Ein^{\alpha\beta}\mn h\ab+\frac{1}{8}h\mnu \(2\alpha_2 X\mn^{(1)}+\frac{2\alpha_2+3\alpha_3}{\Lc^3}X\mn^{(2)}+\frac{\alpha_3+4\alpha_4}{\Lc^6}X^{(3)}\mn\)\\
&=&-\frac 14 h^{\mu\nu}\Ein^{\alpha\beta}\mn h\ab+h\mnu \sum_{n=1}^3 \frac{a_n}{\Lc^{3(n-1)}}X^{(n)}\mn\,,\nn
\ea
with $a_1=\alpha_2/4$, $a_2=(2\alpha_2+3\alpha_3)/8$ and $a_3=(\alpha_3+4\alpha_4)/8$ and the correct normalization should be $\alpha_2=1$.

\subsubsection*{Unmixing and Galileons}

As was already the case at the linearized level for the Fierz--Pauli theory (see Eqns.~\eqref{FP mixed} and~\eqref{FP tot lin}) the kinetic term for the helicity-0 mode appears mixed with the helicity-2 mode. It is thus convenient to diagonalize these two modes by performing the following shift,
\ba
\label{h diag}
h\mn = \tilde h\mn +\alpha_2\pi \eta\mn- \frac{2\alpha_2+3\alpha_3}{2\Lc^3} \p_\mu \pi \p_\nu \pi\,,
\ea
where the non-linear term has been included to unmix the coupling $h\mnu X^{(2)}\mn$,
leading to the following decoupling limit~\cite{deRham:2010iK}
\ba
\label{DL Lc}
\L_{\Lc} = -\frac 14 \Bigg[\tilde h^{\mu \nu}\hat {\mathcal{E}}^{\alpha \beta}\mn \tilde h\ab+\sum_{n=2}^5\frac{c_n}{\Lc^{3(n-2)}} \L^{(n)}_{\rm (Gal)}[\pi]-\frac{2(\alpha_3+4 \alpha_4)}{\Lc^6}\tilde h^{\mu\nu}X^{(3)}\mn\Bigg]\,,
\ea
where we introduced the Galileon Lagrangians $\L^{(n)}_{\rm (Gal)}[\pi]$ as defined in Ref.~\cite{Nicolis:2008in}
\ba
\L^{(n)}_{\rm (Gal)}[\pi]&=&\frac{1}{(6-n)!}(\p \pi)^2 \L_{n-2}[\Pi]\\
&=& -\frac{2}{n (5-n)!}\pi \L_{n-1}[\Pi]\,,
\ea
where the Lagrangians $\L_{n}[Q]=\ep \ep Q^n \delta^{4-n}$ for a tensor $Q\mupn$ are defined in (\ref{L0}-\ref{L4}), or more explicitly in (\ref{L0 X}-\ref{L4 X}), leading to the explicit form for the Galileon Lagrangians
\ba
\label{Gal 2}
\L^{(2)}_{\rm (Gal)}[\pi]&=&(\p \pi)^2\\
\label{Gal 3}
\L^{(3)}_{\rm (Gal)}[\pi]&=&(\p \pi)^2 [\Pi]\\
\label{Gal 4}
\L^{(4)}_{\rm (Gal)}[\pi]&=&(\p \pi)^2\([\Pi]^2-[\Pi^2]\)\\
\label{Gal 5}
\L^{(5)}_{\rm (Gal)}[\pi]&=&(\p \pi)^2\([\Pi]^3-3 [\Pi][\Pi^2]+2 [\Pi^3]\)\,,
\ea
and the coefficients $c_n$ are given in terms of the $\alpha_n$ as follows,
\ba
\label{coefficients Cn DL}
\begin{array}{lcl}
c_2=3 \alpha_2^2\,,& \qquad \qquad &
c_3=\frac 32 \alpha_2(2 \alpha_2+3 \alpha_3)\,,\\[5pt]
c_4=\frac 14 (4 \alpha_2^2+9\alpha_3^2+16\alpha_2 (\alpha_3+ \alpha_4))\,,& \qquad \qquad&
c_5=\frac 58(2\alpha_2+3 \alpha_3)(\alpha_3+4\alpha_4)\,.
\end{array}
\ea
Setting $\alpha_2=1$, we indeed recover the same normalization of $-3/4 (\p\pi)^2$ for the helicity-0 mode found in \eqref{FP tot lin}.

\subsubsection*{$X^{(3)}$-coupling}

In general, the last coupling $\tilde h^{\mu\nu}X^{(3)}\mn$ between the helicity-2 and helicity-0 mode cannot be removed by a local field redefinition. The non-local field redefinition
\ba
\tilde h\mn \to \tilde h\mn +G^{\rm massless}_{\mu\nu\alpha\beta}\, X^{(3)\, \alpha\beta}\,,
\ea
where $G^{\rm massless}_{\mu\nu\alpha\beta}$ is the propagator for a massless spin-2 field as defined in~\eqref{MasslessProp}, fully diagonalizes the helicity-0 and -2 mode at the price of introducing non-local interactions for $\pi$.

Note however that these non-local interactions do not hide any new degrees of freedom. Furthermore about some specific backgrounds, the field redefinition is local.  Indeed focusing on static and spherically symmetric configurations if we consider $\pi=\pi_0(r)$ and $\tilde h\mn$ given by
\ba
\tilde h\mn\d x^\mu \d x^\nu=-\psi(r)\d t^2+ \phi(r) \d r^2\,,
\ea
so that
\ba
\tilde h^{\mu\nu} X^{(3)}\mn=-\psi'(r)\pi_0'(r)^3\,.
\ea
The standard kinetic term for $\psi$ sets $\psi'(r)=\phi(r)/r$ as in GR and the $X^{(3)}$ coupling can be absorbed via the field redefinition, $\phi \to \bar  \phi -2(\alpha_3+4 \alpha_4)\pi'_0(r)^3/r \Lc^{-6}$, leading to the following new sextic interactions for $\pi$,
\ba
\tilde h\mnu X^{(3)}\mn \to - \frac{1}{r^2}\pi_0'(r)^6\,,
\ea
interestingly this new order-6 term satisfy all the relations of a Galileon interaction but cannot be expressed covariantly in a local way. See~\cite{Berezhiani:2008nr} for more details on spherically symmetric configurations with the $X^{(3)}$-coupling.

\subsubsection{Vector interactions in the $\Lc$-decoupling limit}
\label{sec:DL vectors}

As can be seen from the relation~\eqref{DLscale}, the scale associated with interactions mixing two helicity-1 fields with an arbitrary number of fields $\pi$, ($j=0, k=1$ and arbitrary $\ell$) is also $\Lambda_3$. So at that scale, there are actually an infinite number of interactions when including the mixing with between the helicity-1 and -0 modes (however as mentioned previously, since the vector field always appears quadratically it is always consistent to set them to zero as was performed previously).

The full decoupling limit including these interactions has been derived in Ref.~\cite{Ondo:2013wka}, (see also Ref.~\cite{Gabadadze:2013ria}) using the vielbein formulation of massive gravity as in~\eqref{dRGT_vielbein} and we review the formalism and the results in what follows.

In addition to the \stu fields associated with local covariance, in the vielbein formulation one also needs to introduce  6 additional \stu fields $\omega_{ab}$ associated to local Lorentz invariance, $\omega_{ab}=-\omega_{ba}$. These are non-dynamical since they never appear with derivatives, and can thus be treated as auxiliary fields which can be integrated. It is however useful to keep them in the decoupling limit action, so as to retain a closes-form expression. In terms of the Lorentz \stu fields, the full decoupling limit of massive gravity in four dimensions at the scale $\Lc$ is then (before diagonalization)~\cite{Ondo:2013wka}
\ba
\label{Full DL Lc}
\L^{(0)}_{\Lc}&=&-\frac 14h^{\mu \nu}\hat {\mathcal{E}}^{\alpha \beta}\mn h\ab
+\frac{1}{2}h\mnu \sum_{n=1}^3 \frac{a_n}{\Lc^{3(n-1)}}X^{(n)}\mn\\
&&+\frac{3\beta_1}{8}
 \delta^{\alpha\beta\gamma\delta}_{abcd}   \delta^a_\alpha\(  \delta^b_\beta F_{\ \gamma}^c \omega^d_{\  \delta}+2
[  \omega^b_{\  \beta} \omega^c_{\  \gamma}
  +\frac{1}{2}  \delta^b_{\beta}\omega^c_{\  \mu} \omega^{\mu}_{\  \gamma} ](\delta + \Pi)^d_{\delta}
 \) \nn \\
&&+ \frac{\beta_2}{8} \delta^{\alpha\beta\gamma\delta}_{abcd} \(\delta+\Pi\)^a_\alpha \left(  2 \delta^b_\beta F_{\ \gamma}^c \omega^d_{\  \delta}+
[  \omega^b_{\  \beta} \omega^c_{\  \gamma}
  +  \delta^b_{\beta}\omega^c_{\  \mu} \omega^{\mu}_{\  \gamma} ](\delta + \Pi)^d_{\delta}   \right)\nn \\
&&+  \frac{\beta_3}{48} \delta^{\alpha\beta\gamma\delta}_{abcd}  \(\delta+\Pi\)^a_\alpha\(\delta+\Pi\)^b_\beta \left(
 3  F_{\ \gamma}^c \omega^d_{\  \delta} +\omega^c_{\  \mu} \omega^{\mu}_{\  \gamma} (\delta + \Pi)^d_{\delta} \right) \nn \, ,
\ea
(the superscript $(0)$ indicates that this decoupling limit is taken with Minkowski as a reference metric),
with $F_{ab}=\p_a A_b-\p_b A_a$ and the coefficients $\beta_n$ are related to the $\alpha_n$ as in \eqref{beta<->alpha}.

The auxiliary Lorentz \stu fields carries all the non-linear mixing between the helicity-0 and -1 modes,
\ba
\omega_{ab}&=&\int_0^\infty \d u\, e^{-2 u}e^{-u \Pi_a^{a'}}F_{a'b'}e^{-u \Pi_b^{b'}}\\
&=&\sum_{n,m}\frac{(n+m)!}{2^{1+n+m} n! m!} (-1)^{n+m}\(\Pi^n\, F\, \Pi^m\)_{ab}\,.
\ea
In some special cases these sets of interactions can be resummed exactly, as was first performed in~\cite{deRham:2010tw}, (see also Refs.~\cite{Koyama:2011wx,Tasinato:2012ze}).

This decoupling limit includes non-linear combinations of the second-derivative tensor $\Pi\mn$ and the first derivative Maxwell tensor $F\mn$. Nevertheless, the structure of the interactions, always arising with a fully antisymmetric Levi-Cevita tensor is such that the equations of motion with respect to the helicity-2, -1 and -0 modes are manifestly second order. This is the same property as what happens for Galileon theories and ensures the absence of ghost in the full decoupling limit. When working beyond the decoupling limit, this property is no longer manifest, but as we shall see below, the \stu fields are no longer the correct representation of the physical degrees of freedom. Keeping the equations of motion second order in derivatives in the \stu fields is thus no longer the proper criteria for the absence of ghost. As we shall see below, the proper number of degrees of freedom is nonetheless maintained when working beyond the decoupling limit.

\subsubsection{Beyond the decoupling limit}
\label{sec:beyondDL}

\subsubsection*{Physical degrees of freedom}

In the previous section we have introduced four \stu fields $\phi^a$ which transform as scalar fields under coordinate transformation, so that the action of massive gravity is invariant under coordinate transformations. Furthermore the action is also invariant under global Lorentz transformations in the field space,
\ba
x^\mu \to x^\mu\,, \qquad g\mn \to g\mn\,, \hspace{10pt}{\rm and}\hspace{10pt}
\phi^a \to
\tilde \Lambda^a_{\, b} \phi^b\,.
\ea
In the DL, taking $\mpl \to \infty$, all fields are living on flat space-time, so in that limit, there is an additional global Lorentz symmetry acting this time on the space-time,
\ba
 x^\mu \to  \bar \Lambda\mupn \, x^\nu\,, \qquad h\mn \to \bar \Lambda^\alpha_{\, \mu} \bar \Lambda^\beta_{\, \nu} h\ab\,, \quad \text{and}\quad
\phi^a \to \phi^a\,.
\ea
The internal and space-time Lorentz symmetries are independent, (the internal one is always present while the space-time one is only there in the DL). In the DL we can identify both groups and work in the representation of the single group, so that the action is invariant under,
\ba
 x^\mu \to  \Lambda\mupn \, x^\nu\,, \qquad h\mn \to \Lambda^\alpha_{\, \mu} \Lambda^\beta_{\, \nu} h\ab\,, \quad \text{and}\quad
\phi^a \to \Lambda^a_{\, b}\phi^b\,.
\ea
The \stu fields $\phi^a$ then behave as Lorentz vectors under this identified group, and $\pi$ defined previously behaves as a Lorentz scalar. The helicity-0 mode of the graviton also behaves as a scalar in this limit, and $\pi$ captures the behaviour of the graviton helicity-0 mode. So in the DL limit, the right requirement for the absence of BD ghost is indeed the requirement that the equations of motion for $\pi$ remain at most second order (time) in derivative as was pointed out in~\cite{Deffayet:2005ys}, (see also~\cite{Creminelli:2005qk}). However beyond the DL, the helicity-0 mode of the graviton does not behave as a scalar field and neither does the $\pi$ in the split of the \stu fields. So beyond the DL there is no reason to anticipate that $\pi$ captures a whole degree of freedom, and it indeed, it does not. Beyond the DL, the equation of motion for $\pi$ will typically involve higher derivatives, but the correct requirement for the absence of ghost is different, as explained in Section~\ref{sec:stulanguage}. One should instead go back to the original four scalar \stu fields $\phi^a$ and check that out of these four fields only three of them be dynamical. This has been shown to be the case in Section~\ref{sec:stulanguage}. These three degrees of freedom, together with the two standard graviton polarizations then gives the correct five degrees of freedom and circumvent the BD ghost.

Recently much progress has been made in deriving the decoupling limit about arbitrary backgrounds, see Ref.~\cite{Kugo:2014hja}.

\subsubsection{Decoupling limit on (Anti) de Sitter}
\label{sec:DL on dS 2}

\subsubsection*{Linearized theory and Higuchi bound}

Before deriving the decoupling limit of massive gravity on (Anti) de Sitter,  we first need to analyze the linearized theory so as to infer the proper canonical normalization of the propagating dofs and the proper scaling in the decoupling limit, similarly as what was performed for massive gravity with flat reference metric. For simplicity we focus on $(3+1)$ dimensions here, and when relevant give the result in arbitrary dimensions. Linearized massive gravity on (A)dS was first derived in~\cite{Higuchi:1986py,Higuchi:1989gz}.
Since we are concerned with the decoupling limit of ghost-free massive gravity, we follow in this section the procedure presented in~\cite{deRham:2012kf}. We also focus on the dS case first before commenting on the extension to AdS.

At the linearized level about dS, ghost-free massive gravity reduces to the Fierz--Pauli action with  $g\mn = \g\mn + \tilde h\mn = \g\mn + h\mn / \mpl$, where $\gamma\mn$ is the dS metric with constant Hubble parameter $H_0$,
\ba
\L^{(2)}_{\rm MG, \, dS}=
-\frac {1}4 h^{\mu\nu}(\Ein_{\rm dS})^{\alpha\beta}\mn\, h_{\alpha\beta} - \frac {m^2}8  \g^{\mu\nu}\g^{\alpha \beta}\(H_{\mu\alpha}H_{\nu\beta}
-H\mn H_{\alpha \beta}\)\,,
\ea
where $H\mn$ is the tensor fluctuation as introduced in~\eqref{Hcov}, although now considered about the dS metric,
\ba
\label{cov H 2}
H\mn&=&h\mn+2\frac{\nabla_{(\mu} A_{\nu)}}{m}+2\frac{\Pi\mn}{m^2} \\
&&-\frac{1}{\mpl}
\left[\frac{\nabla_\mu A_{\alpha}}{m}+\frac{\Pi_{\mu\alpha}}{m^2}\right]
\left[\frac{\nabla_\nu A_{\beta}}{m}+\frac{\Pi_{\nu\beta}}{m^2}\right]\gamma^{\alpha\beta}\,,\nn
\ea
with $\Pi\mn=\nabla_\mu \nabla_\nu \pi$, $\nabla$ being the covariant derivative with respect to the dS metric $\g\mn$ and indices are raised and lowered with respect to this same metric. Similarly,  $\Ein_{\rm dS}$ is now the Lichnerowicz operator on de Sitter,
\ba
(\Ein_{\rm dS})^{\alpha\beta}\mn\, h_{\alpha\beta}
&=&-\frac 12 \Big[\Box  h\mn-2\nabla_{(\mu} \nabla_\alpha h^\alpha_{\nu)}+\nabla_\mu\nabla_\nu  h\\
&&
-\g \mn (\Box h-\nabla_\alpha\nabla_\beta  h^{\alpha\beta}) +6H_0^2 \(h\mn- \frac 12 h\g\mn\)\Big]\,.\nn
\ea
So at the linearized level and neglecting the vector fields, the helicity-0 and -2 mode of massive gravity on dS behave as
\ba
\label{L_2_ourway}
\L^{(2)}_{\rm MG, \, dS}&=&
-\frac {1}4 h^{\mu\nu}(\Ein_{\rm dS})^{\alpha\beta}\mn\, h_{\alpha\beta} - \frac{m^2}8
\(h\mn^2-h^2\)-\frac 18 F\mn^2\\
&&-\frac {1}2 h^{\mu\nu}\(\Pi\mn - [\Pi] \g\mn\)-\frac{1}{2m^2}\([\Pi^2]-[\Pi]^2\)\,.\nn
\ea
After integration by parts,
$[\Pi^2]=[\Pi]^2-3 H^2 (\p \pi)^2$.
The helicity-2 and -0 modes are thus diagonalized as in flat space-time by setting
$h\mn= \bar h\mn + \pi \g\mn$,
\ba
\label{quadratic}
\L^{(2)}_{\rm MG, \, dS}&=&
-\frac {1}4\bar h^{\mu\nu}(\Ein_{\rm dS})^{\alpha\beta}\mn\, \bar h_{\alpha\beta}-\frac{m^2}{8}\(\bar h\mn^2-\bar h^2\)-\frac 18 F\mn^2\\
&&-\frac{3}{4}\(1-2\(\frac{H}{m}\)^2\) \(\(\p \pi\)^2 - m^2 \bar h \pi -2 m^2 \pi^2\) \,.\nn
\ea


The most important difference from linearized massive gravity on Minkowski is that the properly canonically normalized helicity-0 mode is now instead
\ba
\phi= \sqrt{1-2\frac{H^2}{m^2}}\  \pi\,.
\ea
For a standard coupling of the form $\frac 1\mpl \pi T$, where $T$ is the trace of the stress-energy tensor, as we would infer from the coupling $\frac{1}{\mpl}h\mn T^{\mu\nu}$ after the shift $h\mn= \bar h\mn + \pi \g\mn$, this means that the properly normalized helicity-0 mode couples as
\ba
\L_{\rm helicity-0}^{\rm matter}= \frac{m^2}{\mpl \sqrt{m^2-2H^2} }\phi T\,,
\ea
and that coupling vanishes in the massless limit. This might suggest that in the massless limit $m\to 0$, the helicity-0 mode decouples, which would imply the absence of the standard  vDVZ discontinuity on (Anti) de Sitter~\cite{Kogan:2000uy,Porrati:2000cp}, unlike what was found on Minkowski, see Section~\ref{sec:vDVZ}, which confirms  the Newtonian approximation presented in~\cite{Deser:2001gz}.

While this observation is correct on AdS, in the dS one cannot take the massless limit without simultaneously sending $H\to 0$ at least the same rate. As a result, it would be incorrect to deduce that the helicity-0 mode decouples in the massless limit of massive gravity on dS.

To be more precise, the linearized action~\eqref{quadratic} is free from ghost and tachyons only if $m\equiv 0$ which corresponds to GR, or if $m^2>2 H^2$, which corresponds to the well-know Higuchi bound~\cite{Higuchi:1986py,Deser:2001wx}. In $d$ spacetime dimensions, the Higuchi bound is $m^2>(d-2)H^2$. In other words, on dS there is a forbidden range for the graviton mass, a theory with $0<m^2< 2H^2$ or with $m^2<0$ always excites at least one ghost degree of freedom. Notice that this ghost, (which we shall refer to as the Higuchi ghost from  now on) is distinct from the BD ghost which corresponded to an additional sixth degree of freedom. Here the theory propagates five dof (in four dimensions) and is thus free from the BD ghost (at least at this level), but at least one of the five dofs is a ghost. When $0<m^2< 2H^2$, the ghost is the helicity-0 mode, while for $m^2<0$, the ghost is he helicity-1 mode  (at quadratic order the helicity-1 mode comes in as $-\frac{m^2}{4}F\mn^2$). Furthermore when $m^2<0$, both the helicity-2 and -0 are also tachyonic, although this is arguably not necessarily a severe problem, especially not if the graviton mass is of the order of the Hubble parameter today, as it would take an amount of time comparable to the age of the Universe to see the effect of this tachyonic behavior. Finally the case, $m^2=2H^2$ (or $m^2=(d-2)H^2$ in $d$ spacetime dimensions), represents the Partially Massless case where the helicity-0 mode disappears. As we shall see in Section~\ref{sec:PM}, this is nothing other than a linear artefact and non-linearly the helicity-0 mode always reappears, so the PM case is infinitely strongly coupled and always pathological.

A summary of the different bounds is provided below as well as in Figure~\ref{Fig:Higuchi}:

\begin{itemize}
\item $m^2<0$: Helicity-1 modes are ghost, helicity-2 and -0 are tachyonic, \emph{sick theory}
\item $m^2=0$: \textbf{General Relativity}: two healthy (helicity-2) degrees of freedom, \textbf{healthy theory},
\item $0<m^2<2H^2$: One ``Higuchi ghost'' (helicity-0 mode) and four healthy degrees of freedom  (helicity-2 and -1 modes), \emph{sick theory},
\item $m^2=2 H^2$: \textbf{Partially Massless Gravity:} Four healthy degrees (helicity-2 and -1 modes), and one infinitely strongly coupled dof (helicity-0 mode), \emph{sick theory},
\item $m^2>2 H^2$: \textbf{Massive Gravity on dS:} Five healthy degrees of freedom, \emph{healthy theory}.
\end{itemize}

\epubtkImage{}{%
\begin{figure}[htb]
  \centerline{\includegraphics[width=13cm]{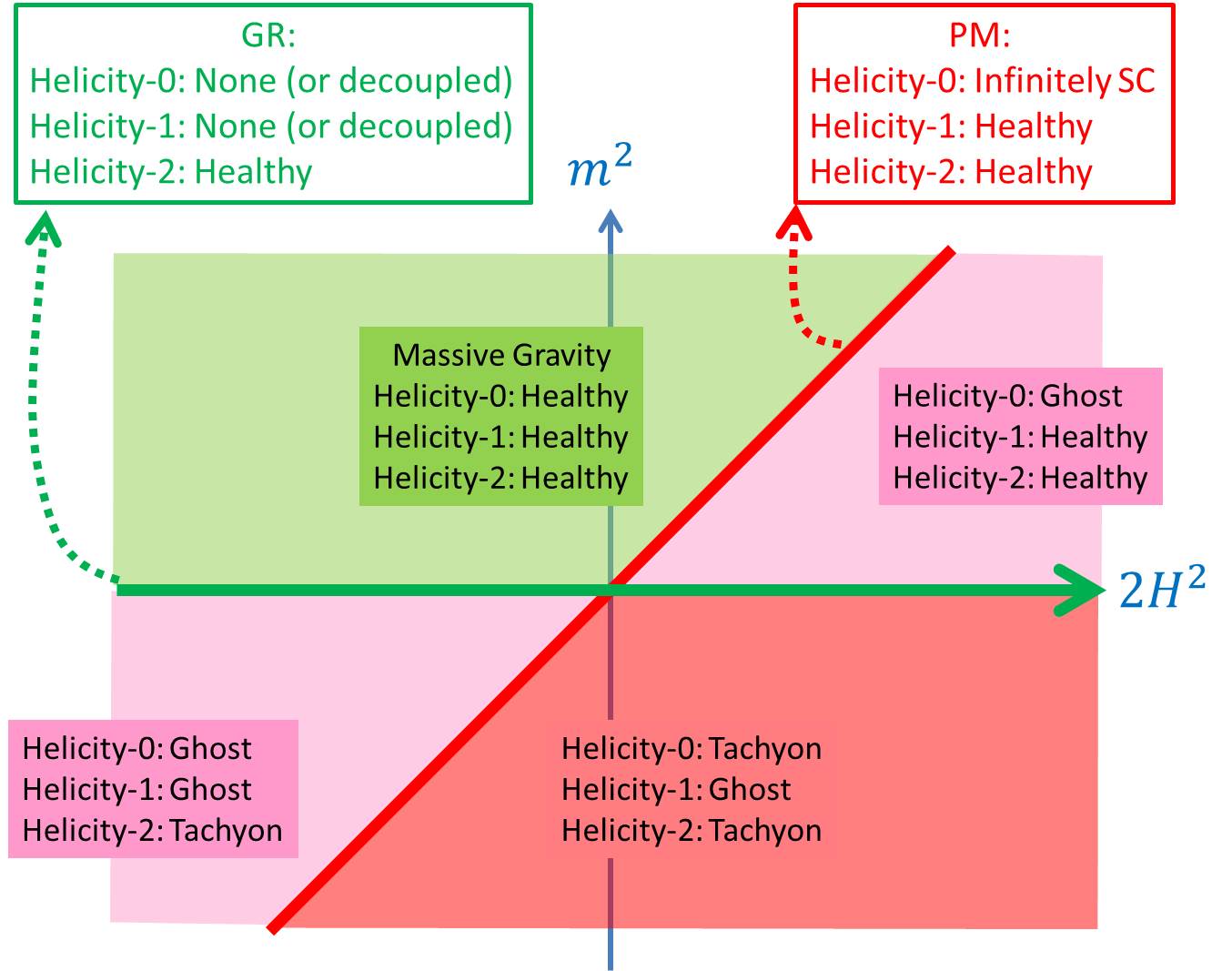}}
  \caption{Degrees of freedom for massive gravity on a maximally symmetric reference metric. The only theoretically allowed regions are the upper left Green region and the line $m=0$ corresponding to GR.}
  \label{Fig:Higuchi}
\end{figure}}

\subsubsection*{Massless and decoupling limit}

\begin{itemize}
\item As one can see from Figure~\ref{Fig:Higuchi}, in the case where $H^2<0$ (corresponding to massive gravity on AdS), one can take the massless limit $m\to 0$ while keeping the AdS length scale fixed in that limit. In that limit, the helicity-0 mode decouples from external matter sources and there is no vDVZ discontinuity. Notice however that the helicity-0 mode is nevertheless still strongly coupled at a low energy scale. \\
    When considering the decoupling limit $m\to 0$, $\mpl \to \infty$ of massive gravity on AdS, we have the choice on how we treat the scale $H$ in that limit. Keeping the AdS length scale fixed in that limit could lead to an interesting phenomenology in its own right, but is yet to be explored in depth.
\item In the dS case, the Higuchi forbidden region prevents us from taking the massless limit while keeping the scale $H$ fixed. As a result, the massless limit is only consistent if $H\to 0 $ simultaneously as $m\to 0$ and we thus recover the vDVZ discontinuity at the linear level in that limit.   \\
    When considering the decoupling limit $m\to 0$, $\mpl \to \infty$ of massive gravity on dS, we also have to send $H\to 0$. If $H/m\to 0$ in that  limit, we then recover the same decoupling limit as for massive gravity on Minkowski, and all the results of Section~\ref{sec:DL MG} apply. The case of interest is thus when the ratio $H/m$ remains fixed in the decoupling limit.
\end{itemize}

\subsubsection*{Decoupling limit}
\label{sec:DL on dS}

When taking the decoupling limit of massive gravity on dS, there are two additional contributions to take into account:
\begin{itemize}
\item First, as mentioned in Section~\ref{sec:beyondDL}, care needs to be applied to properly identify the helicity-0 mode on a curved background. In the case of (A)dS, the formalism was provided in Ref.~\cite{deRham:2012kf} by embedding a $d$-dimensional de Sitter spacetime  into a flat $(d+1)$-dimensional spacetime where the standard \stu trick could be applied. As a result the `covariant' fluctuation defined in~\eqref{Hcov} and used in~\eqref{cov H 2} needs to be generalized to (see Ref.~\cite{deRham:2012kf} for details)
    \ba
    \frac{1}{\mpl}H\mn &=& \frac{1}{\mpl}h\mn+\frac{2}{\Lc^3}\Pi\mn-\frac{1}{\Lc^6}\Pi^2\mn\\
    &&+\frac{1}{\Lc^3}\frac{H^2}{m^2}\((\p \pi)^2 (\g\mn-\frac{2}{\Lc^3}\Pi\mn)-\frac{1}{\Lc^6}\Pi_{\mu\alpha}\Pi_{\nu \beta}\p^\alpha \pi \p^\beta \pi\)\nn \\
    &&+H^2\frac{H^2}{m^2}\frac{(\p \pi)^4}{\Lc^9}+ \cdots\,.\nn
    \ea
Any corrections in the third line vanish in the decoupling limit and can thus be ignored, but the corrections  of order $H^2$ in the second line lead to new non-trivial contributions.
\item Second, as already encountered at the linearized level, what were total derivatives in Minkowski (for instance the combination $[\Pi^2]-[\Pi]^2$), now lead to new contributions on de Sitter. After integration by parts, $m^{-2}([\Pi^2]-[\Pi]^2)=m^{-2}R\mn \p^\mu \pi \p^\nu \pi=12 H^2/m^2 (\p \pi)^2$. This was the origin of the new kinetic structure for massive gravity on de Sitter and will have further effects in the decoupling limit when considering similar contributions from $\L_{3,4}(\Pi)$, where $\L_{3,4}$ are defined in (\ref{L3}, \ref{L4}) or more explicitly in (\ref{L3 X}, \ref{L4 X}).
\end{itemize}
Taking these two effects into account, we obtain the full decoupling limit for massive gravity on de Sitter,
\ba
\label{dL dS}
\L^{(\rm dS)}_{\Lc}=\L^{(0)}_{\Lc} +\frac{H^2}{m^2}\sum_{n=2}^5 \frac{\lambda_n}{\Lc^{3(n-1)}} \L^{(n)}_{\rm (Gal)}[\pi]\,,
\ea
where $\L^{(0)}_{\Lc}$ is the full Lagrangian obtained in the decoupling limit in Minkowski and  given in~\eqref{Full DL Lc}, and $\L^{(n)}_{\rm (Gal)}$ are the Galileon Lagrangians as encountered previously. Notice that while the ratio $H/m$ remains fixed,this decoupling limit is taken with $H,m \to 0$, so all the fields in~\eqref{dL dS} live on a Minkowski metric. The constant coefficients $\lambda_n$ depend on the free parameters of the ghost-free theory of massive gravity, for the theory~\eqref{dRGT_metric} with $\alpha_1=0$ and $\alpha_2=1$, we have
\ba
\lambda_2=\frac32\,, \hspace{10pt}\lambda_3=\frac 34\(1+2\alpha_3\)\,, \hspace{10pt} \lambda_4=\frac 14\(-1 +6 \alpha_4\)\,, \hspace{10pt}\lambda_5=-\frac 3{16}(\alpha_3+4\alpha_4)\,.
\ea
At this point we may perform the same field redefinition~\eqref{h diag} as in flat space and obtain the following semi-diagonalized decoupling limit,
\ba
\label{dL dS diag}
\L^{(\rm dS)}_{\Lc}&=&-\frac 14h^{\mu \nu}\hat {\mathcal{E}}^{\alpha \beta}\mn h\ab+\frac {\alpha_3+4\alpha_4}{8\Lc^9} h^{\mu\nu}X^{(3)}\mn +\sum_{n=2}^5\frac{\tilde c_n}{\Lc^{3(n-2)}}\L^{(n)}_{\rm (Gal)}[\pi]\\
&+&\text{Contributions from the helicity-1 modes}\nn
\,,
\ea
where the contributions from the helicity-1 modes are the same as the ones provided in~\eqref{Full DL Lc}, and the new coefficients $\tilde c_n=-c_n/4+H^2/m^2 \lambda_n$ cancel identically for $m^2=2H^2$, $\alpha_3=-1$ and $\alpha_4=-\alpha_3/4=1/4$, as pointed out in~\cite{deRham:2012kf}, and the same result holds for bi-gravity as pointed out in~\cite{Hassan:2012gz}. Interestingly for these specific parameters, the helicity-0 loses its kinetic term, and any self-mixing as well as any mixing with the helicity-2 mode. Nevertheless the mixing between the helicity-1 and -0 mode as presented in~\eqref{Full DL Lc} are still alive. There are no choices of parameters which would allow to remove the mixing with the helicity-1 mode and as a result, the helicity-0 mode generically reappears through that mixing. The loss of its kinetic term implies that the field is infinitely  strongly coupled on a configuration with zero vev for the helicity-1 mode and is thus an ill-defined theory. This was confirmed in various independent studies, see Refs.~\cite{Deser:2013uy,deRham:2013wv}.

\subsection{$\Lc$-Decoupling limit of bi-gravity}
\label{sec:DL bigravity}

We now proceed to derive the $\Lc$-decoupling limit of bi-gravity, and we will see how to recover the decoupling limit about any reference metric (including Minkowski and de Sitter) as special cases. As already seen in Section~\ref{sec:DL vectors}, the full DL is better formulated in the vielbein language, even though in that case \stu fields ought to be introduced for the broken diff and the broken Lorentz. Yet, this is a small price to pay, to keep the action in a much simpler form. We thus proceed in the rest of this section by deriving the $\Lc$-decoupling of bi-gravity and start in its vielbein formulation. We follow the derivation and formulation presented in~\cite{Fasiello:2013woa}. As previously, we focus on $(3+1)$-spacetime dimensions, although the whole formalism is trivially generalizable to arbitrary dimensions.

We start with the action~\eqref{bigravity} for bi-gravity, with the interaction
\ba
\label{Lfg bigravity}
\L_{g,f}&=&\frac{\mpl^2 m^2}{4} \int \d^4 x \sqrt{-g} \sum_{n=0}^4\alpha_n \L_n[\K[g,f]] \\
&=&
- \frac{ \mpl^2 m^2}{2} \ep_{abcd} \int  \Big[ \frac{\beta_0}{4!} e^a \wedge e^b \wedge e^c \wedge e^d  + \frac{\beta_1}{3!} f^a \wedge e^b \wedge e^c \wedge e^d   \\
&&+\frac{\beta_2}{2!2!} f^a \wedge f^b \wedge e^c \wedge e^d  + \frac{\beta_3}{3!} f^a \wedge f^b \wedge f^c \wedge e^d+ \frac{\beta_4}{4!} f^a \wedge f^b \wedge f^c \wedge f^d \Big]  \, ,\nn
\ea
where the relation between the $\alpha$'s and the $\beta$'s is given in~\eqref{beta<->alpha}.

We now introduce \stu fields $\phi^a=x^a-\chi^a$ for diffs and $\Lambda^a{}_b$ for the local Lorentz. 
In the case of massive gravity, there was no ambiguity in how to perform this `St\"uckelbergization' but in the case of bi-gravity, one can either `St\"uckelbergize the metric $f\mn$ or the metric $g\mn$.
In other words the broken diffs and local Lorentz symmetries can be restored by performing either one of the two replacements in~\eqref{Lfg bigravity},
\ba
\label{choice1}
f^a_{\mu} \to \tilde f^a_\mu=  \Lambda^a{}_b f^a_{c}(\phi(x)) \, \partial_{\mu} \phi^c\, .
\ea
or alternatively
\ba
e^a_{\mu} \to \tilde e^a_\mu=  \Lambda^a{}_b e^a_{c}(\phi(x)) \, \partial_{\mu} \phi^c\, .
\ea
For now we stick to the first choice~\eqref{choice1} but keep in mind that this freedom has deep consequences for the theory, and is at the origin of the duality presented in Section~\ref{sec:duality}.

Since we are interested in the decoupling limit, we now perform the following splits, (see Ref.~\cite{Ondo:2013wka} for more details),
\ba
e^a_{\mu} &=& \bar e^a_{\mu} + \frac{1}{2 \mpl } h^a_{\mu} \, , \qquad f^a_{\mu} = \bar e^a_{\mu} + \frac{1}{2 M_f } v^a_{\mu} \nonumber \\
\Lambda^a{}_{b} &=& e^{ \hat{\omega}^{a}{}_{ b}} = \delta^a{}_b + \hat{\omega}^{a}{}_{ b} + \frac{1}{2} \hat{\omega}^{a}{}_{c} \hat{\omega}^{c}{}_{b} + \cdots   \nn \\
\hat{\omega}^{a}_{\hspace{.2cm} b} &=& \frac{\omega^{a}_{\hspace{.2cm} b}}{m \mpl} \nn \\
\p_{\mu}\phi^a &=& \p_{\mu} \( x^a + \frac{A^a}{m \mpl} + \frac{ \p^{a}\pi}{ \Lc^3} \)
\ea
and perform the scaling or decoupling limit,
\ba
\label{Scaling 1}
\mpl   \rightarrow  \infty \,, \quad  M_f   \rightarrow  \infty \, ,\quad m \to  0
\ea
while keeping
\ba
\label{Scaling 2}
\Lc = (m^2\mpl)^{\frac{1}{3}} \to \text{constant}\,, \quad\, \mpl/M_f \to \text{ constant}\,, \\
 \quad   \text{and}  \, \quad \beta_n \to \text{constant}\,.\nn
\ea
Before performing any change of variables (any diagonalization), in addition to the kinetic term for quadratic $h$, $v$ and $A$, there are three contributions to the decoupling limit of bi-gravity:

\begin{enumerate}
\item[\ding{182}] Mixing of the helicity-0 mode with the helicity-1 mode $A_\mu$, as derived in~\eqref{Full DL Lc},
\item[\ding{183}] Mixing of the helicity-0 mode with the helicity-2 mode $h^a_\mu$, as derived in~\eqref{DL Lc},
\item[\ding{184}] Mixing of the helicity-0 mode with the new helicity-2 mode $v^a_\mu$,
\end{enumerate}

noticing that before field redefinitions, the helicity-0 mode do not self-interact (their self-interactions are constructed so as to be total derivatives).

As already explained in the previous section, the first contribution \ding{182} arising from the mixing between the helicity-0 and -1 modes is the same (in the decoupling limit) as what was obtained in Minkowski (and is independent of the coefficients $\beta_n$ or $\alpha_n$). This implies that the can be directly read of from the three last lines of~\eqref{Full DL Lc}. These contributions are the most complicated parts of the decoupling limit but remained unaffected by the dynamics of  $v$, \ie unaffected by the bi-gravity nature of the theory. This statement simply follows from scaling considerations. In the decoupling limit there cannot be any mixing between the helicity-1 and neither of the two helicity-2 modes. As a result, the helicity-1 modes only mix with themselves and the helicity-0 mode. Hence in the scaling limit (\ref{Scaling 1}, \ref{Scaling 2}) the helicity-1 decouples from the massless spin-2 field.

Furthermore, the first line of~\eqref{Full DL Lc} which corresponds to the dynamics of $h^a_\mu$ and the helicity-0 mode is also unaffected by the bi-gravity nature of the theory. Hence the second contribution \ding{183} is the also the same as previously derived. As a result,  the only new ingredient in bi-gravity is the mixing \ding{184} between the helicity-0 mode and the second helicity-2 mode $v^a_\mu$, given by a fixing of the form $h^{\mu\nu}X\mn$.

Unsurprisingly, these new contributions have the same form as \ding{183}, with three distinctions: First the way the coefficients enter in the expressions get modified ever so slightly ($\beta_1 \to \beta_1/3$ and $\beta_3\to 3 \beta_3$). Second, in the mass term the space-time index for $v^a_\mu$ ought to dressed with the \stu field,
\ba
v^a_\mu \to v^a_b \p_\mu \phi^b = v^a_b(\delta^b_\mu + \Pi^b_\mu/\Lc^3)\,.
\ea
Finally and most importantly, the helicity-2 field $v^\mu_a$ (which enters in the mass term) is now a function of the `St\"uckelbergized' coordinates $\phi^a$, which in the decoupling limit means that for the mass term
\ba
v^a_b=v^a_b[x^\mu+\p^\mu \pi/\Lc^3]\equiv v^a_b[\tilde x]\,.
\ea
These two effects do not need to be taken into account for the $v$ that enters in its standard curvature term as it is Lorentz and diff invariant.

Taking these three considerations into account, one obtains the decoupling limit for bi-gravity,
\ba
\label{Bigravity DL}
\L^{(\rm bi-gravity)}_{\Lc}&=&\L^{(0)}_{\Lc}-\frac 14v^{\mu \nu}[x]\hat {\mathcal{E}}^{\alpha \beta}\mn v\ab[ x]\\
\nn &-&\frac 12\frac{\mpl}{M_f} v^{\mu \beta}[\tilde x]\(\delta_\beta^\nu+\frac{\Pi_\beta^\nu}{\Lc^3}\)\sum_{n=0}^3 \frac{\tilde \beta_{n+1}}{\Lc^{3(n-1)}}X^{(n)}\mn[\Pi]\,,
\ea
with $\tilde \beta_n=\beta_n/(4-n)!(n-1)!$.
Modulo the non-trivial dependence on the coordinate $\tilde x=x +\p \pi/\Lc^3$, this is a remarkable simple
decoupling limit for bi-gravity. Out of this decoupling limit we can re-derive all the DL found previously very elegantly.

Notice as well the presence of a tadpole for $v$ if $\beta_1\ne 0$. When this tadpole vanishes (as well as the one for $h$), one can further take the limit $M_f\to \infty$ keeping all the other $\beta$'s fixed as well as $\Lc$, and recover straight away the decoupling limit of massive gravity on Minkowski found in~\eqref{Full DL Lc}, with a free and fully decoupled massless spin-2 field.

In the presence of a cosmological constant for both metrics (and thus a tadpole in this framework), we can also take the limit $M_f\to \infty$  and recover straight away the decoupling limit of massive gravity on (A)dS, as obtained in~\eqref{dL dS}.

This illustrates the strength of this generic decoupling limit for bi-gravity~\eqref{Bigravity DL}. In principle we could even go further and derive the decoupling limit of massive gravity on an arbitrary reference metric as performed in~\cite{Fasiello:2013woa}.
To obtain a general reference metric we first need to add an external source for $v_{\mu\nu}$ that generates a background for $\bar V_{\mu\nu}=M_f/\mpl \bar U\mn$. The reference metric is thus expressed in the local inertial frame as
\ba
f\mn&=&\eta\mn+ \frac{1}{M_f}\bar V\mn+\frac{1}{4M_f^2}\bar V_{\mu\alpha}\bar V_{\beta \nu}\eta^{\alpha\beta} +\frac{1}{M_f}v\mn+\mathcal{O}(M_f^{-2})\\
&=&\eta\mn+ \frac{1}{\mpl}\bar U\mn +\frac{1}{M_f}v\mn+\mathcal{O}(\mpl, M_f)^{-2}\,.
\ea
The fact that the metric $f$ looks like a perturbation away from Minkowski is related to the fact that the curvature needs to scale as $m^2$ in the decoupling limit in order to avoid the issues previously mentioned in the discussion of Section~\ref{sec:ArbitraryDL}.

We can then perform the scaling limit $M_f\to \infty$, while keeping the $\beta$'s  and the scale $\Lc=(\mpl m^2)^{1/3}$ fixed as well as the field $v\mn$ and the fixed tensor $\bar U\mn$. The decoupling limit is then simply given by
 \ba
\label{DL arbitrary metric}
\L^{(\rm \bar U)}_{\Lc}&=&\L^{(0)}_{\Lc}-\frac 12\bar U^{\mu \beta}[\tilde x]\(\delta_\beta^\nu+\frac{\Pi_\beta^\nu}{\Lc^3}\)\sum_{n=0}^3 \frac{\tilde \beta_{n+1}}{\Lc^{3(n-1)}}X^{(n)}\mn[\Pi]\\
&&-\frac 14v^{\mu \nu}\hat {\mathcal{E}}^{\alpha \beta}\mn v\ab\,,\nn
\ea
where the helicity-2 field $v$ fully decouples from the rest of the massive gravity sector on the first line which carries the other helicity-2 field as well as the helicity-1 and -0 modes. Notice that the general metric $\bar U$ has only an effect on the helicity-0 self-interactions, through the second term on the first line of~\eqref{DL arbitrary metric} (just as observed for the decoupling limit on AdS). These new interactions are ghost-free and look like Galileons for conformally flat $\bar U\mn= \lambda \eta\mn$, with $\lambda$ constant, but not in general. In particular the interactions found in~\eqref{DL arbitrary metric} would {\bf not} be the covariant Galileons found in~\cite{Deffayet:2009wt,Deffayet:2009mn,deRham:2010eu} (nor the ones found in~\cite{Gabadadze:2012tr}) for a generic metric.



\section{Extensions of Ghost-free Massive Gravity}
\label{sec:Extensions}

Massive gravity can be seen as a theory of a spin-2 field with the following free parameters in addition to the standard parameters of GR (\eg the cosmological constant, etc\ldots),
\begin{itemize}
  \setlength{\itemsep}{0pt}
   \setlength{\parskip}{0pt}
   \setlength{\parsep}{0pt}
\item Reference metric $f_{ab}$,
\item Graviton mass $m$,
\item $(d-2)$ dimensionless parameters $\alpha_n$ (or the $\beta$'s)\,.
\end{itemize}
As natural extensions of Massive gravity one can make any of these parameters dynamical. As already seen, the reference metric can be made dynamical leading to bi-gravity which in addition to massive spin-2 field carries a massless one as well.

Another natural extension is to promote the graviton mass $m$, or any of the free parameters $\alpha_n$ (or $\beta_n$) to a function of a new dynamical variable, say of an additional scalar field $\phi$. In principle the mass $m$ and the parameters $\alpha$'s can be thought as potentials for an arbitrary number of scalar  fields $m=m(\psi_j), \alpha_n=\alpha_n(\psi_j)$, and not necessarily the same fields for each one of them~\cite{Huang:2013mha}. So long as these functions are pure potentials and hide no kinetic terms for any new degree of freedom, the constraint analysis performed in Section~\ref{sec:Abscence of ghost proof} will go relatively unaffected, and the theory remains free from the BD ghost. This was shown explicitly for the mass-varying theory~\cite{Huang:2012pe,Hinterbichler:2013dv} (where the mass is promoted to a scalar function of a new single scalar field, $m=m(\phi)$, while the parameters $\alpha$ remain constant\epubtkFootnote{The non-renormalization theorem protects the parameters $\alpha$'s and the mass from acquiring large quantum corrections~\cite{deRham:2012ew,deRham:2013qqa} and it would be interesting to understand their implications in the case of a mass-varying gravity.}), as well as a general massive scalar-tensor theory~\cite{Huang:2013mha}, and for quasi-dilaton which allow for different couplings between the spin-2 and the scalar field, motivated by scale invariance. We review these models below, in Sections~\ref{sec:MassVarying} and \ref{sec:QuasiDilaton}.

Alternatively, rather than considering the parameters $m$ and $\alpha$ as arbitrary, one may set them to special values of special interest depending on the reference metric $f\mn$. Rather than an `extension' per se this is more special cases in the parameter space. The first obvious one is $m=0$ (for arbitrary reference metric and parameters $\alpha$), for which one recovers the theory of GR (so long as the spin-2 field couples to matter in a covariant way to start with). Alternatively, one may also sit on the Higuchi bound, (see Section~\ref{sec:DL on dS 2}) with the parameters $m^2=2H^2$, $\alpha_3=-1/3$ and $\alpha_4=1/12$ in four dimensions. This corresponds to the Partially Massless theory of gravity, which at the moment is pathological in its simplest realization and will be reviewed below, \ref{sec:PM}.

The coupling massive gravity to a DBI Galileon~\cite{deRham:2010eu} was considered in~\cite{Gabadadze:2012tr,Trodden:2012qe,Goon:2014ywa} leading to a generalized
 Galileon theory which maintains a Galileon symmetry on curved backgrounds.
This theory was shown to be free of any Ostrogradsky ghost in~\cite{Andrews:2013ora} and the cosmology was recently studied in~\cite{Hinterbichler:2013dv}  and perturbations in~\cite{Andrews:2013uca}.

Finally as other extensions to massive gravity, one can also consider  all the extensions applicable to GR. This includes the higher order Lovelock invariants in dimensions greater than four, as well as promoting the Einstein--Hilbert kinetic term to a function $f(R)$, which  is  equivalent to gravity with a scalar field. In the case of massive gravity this has been performed in~\cite{Cai:2013lqa} (see also~\cite{Bamba:2013fha,Kluson:2013yaa}), where the absence of BD ghost was proven via a constraint analysis, and the cosmology was explored (this was also  discussed in Section~\ref{sec:NoNewKin} and see also Section~\ref{sec:CosmoOtherProposals}). $f(R)$ extensions to bi-gravity were also derived in \cite{Nojiri:2012re,Nojiri:2012zu}.

Trace-anomaly driven inflation in bi-gravity was also explored in Ref.~\cite{Bamba:2014zra}.
Massless quantum effects can be taking into account by including the trace anomaly $\mathcal{T}_A$ given as \cite{Duff:1993wm}
\ba
\mathcal{T}_A= c_1(\frac 13 R^2-2R\mn^2+R_{\mu\nu\alpha\beta}^2+\frac 23 \Box R)+c_2 (R^2-4R\mn^2+R_{\mu\nu\alpha\beta}^2)+c_3 \Box R\,,
\ea
where $c_{1,2,3}$ are three constants depending on the field content (for instance the number of scalars, spinors, vectors, graviton etc.)
Including this trace anomaly to the bi-gravity de Sitter-like solutions were found which could represent a good model for anomaly-driven models of inflation.

\subsection{Mass-varying}
\label{sec:MassVarying}

The idea behind mass-varying gravity is to promote the graviton mass to a potential for an external scalar field $\psi$, $m \to m(\psi)$, which has its own dynamics~\cite{Huang:2012pe}, so that in four dimensions, the dRGT action for massive gravity gets promoted to
\ba
\L_{\rm Mass-Varying}=\frac{\mpl^2}{2}\int \d^4 x\sqrt{-g}\Bigg(R+\frac{m^2(\psi)}2\sum_{n=0}^4\alpha_n \L_n[\K]\\
-\frac 12 g^{\mu\nu}\p_\mu \psi \p_\nu \psi -W(\psi)\Bigg)\,,\nn
\ea
and the tensors $\K$ are given in~\eqref{eqK}.
This could also be performed for bi-gravity, where we would simply include the Einstein--Hilbert term for the metric $f\mn$. This formulation was then promoted not only to varying parameters $\alpha_{n} \to \alpha_{n}(\psi)$ but also to multiple fields $\psi_A$, with $A=1,\cdots, \mathcal{N}$ in~\cite{Huang:2013mha},
\ba
\L_{\rm Generalized\ MG}&=&\frac{\mpl^2}{2}\int \d^4 x\sqrt{-g}\Big[\Omega(\psi_A)R+\frac 12\sum_{n=0}^4\alpha_n(\psi_A) \L_n[\K]\\
&&-\frac 12 g^{\mu\nu}\p_\mu \psi_A \p_\nu \psi^A -W(\psi_A)\Big]\,.\nn
\ea
The absence of BD ghost in these theories were performed in~\cite{Huang:2012pe} and~\cite{Huang:2013mha} in unitary gauge, in the ADM language by means of a constraint analysis as formulated in Section~\ref{sec:ADMHamiltonian}. We recall that  in the absence of the scalar field $\psi$, the  primary second-class (Hamiltonian) constraint is given by
\ba
\mathcal{C}_0=\mathcal{R}_0(\gamma,p)+D^i_{\, j}n^j\mathcal{R}_i(\gamma,p)+m^2 \mathcal{U}_0(\gamma, n(\gamma,p))\approx 0\,.
\ea
In the case of a mass-varying theory of gravity, the entire argument remains the same, with the simple addition of the scalar field contribution,
\ba
\mathcal{C}_0^{\rm mass-varying}&=&\tilde{\mathcal{R}}_0(\gamma,p,\psi,p_\psi)+D^i_{\, j}n^j\tilde{\mathcal{R}}_i(\gamma,p,\psi,p_\psi)+m^2(\psi) \mathcal{U}_0(\gamma, n(\gamma,p))\nn \\
&\approx & 0\,,
\ea
where $p_\psi$ is the conjugate momentum associated with the scalar field $\psi$ and
\ba
\tilde{\mathcal{R}}_0(\gamma,p,\psi,p_\psi)&=&\mathcal{R}_0(\gamma,p)+\frac 12 \sqrt{\gamma}\p_i \psi \p^i \psi+\frac1{2\sqrt{\gamma}}p_\psi^2\\
\tilde{\mathcal{R}}_i(\gamma,p,\psi,p_\psi)&=&\mathcal{R}_i(\gamma,p)+p_\psi \p_i \psi \,.
\ea
Then the time-evolution of this primary constraint leads to a secondary constraint similarly as in Section~\ref{sec:ADMHamiltonian}. The expression for this secondary constraint is the same as in ~\eqref{secondary constraint} with a benign new contribution from the scalar field~\cite{Huang:2012pe}
\ba
\label{secondary constraint massvarying}
\tilde{\mathcal{C}}_2&=&\mathcal{C}_2+\frac{\p m^2(\psi)}{\p \psi}\left[\mathcal{U}_0\p_i \psi (\Nf n^i+\Nf^i)+\frac{\Nf}{\sqrt{\gamma}}\mathcal{U}_1p_\psi +\Nf\p_i \psi D^i_{\, k}n^k\right]\nn \\
&\approx& 0\,.
\ea
Then as in the normal fixed-mass case, the tertiary constraint is a constraint for the lapse and the system of constraint truncates leading to 5+1 physical degrees of freedom in four dimensions.
The same logic goes through for generalized massive gravity as explained in~\cite{Huang:2013mha}.

One of the important aspects of a mass-varying theory of massive gravity is that it allows more flexibility for the graviton mass. In the past the mass could have been much larger and could have lead to potential interesting features, be it for inflation (see for instance Refs.~\cite{Hinterbichler:2013dv,Lin:2013sja} and~\cite{Gumrukcuoglu:2012wt}), the Hartle--Hawking no-boundary proposal~\cite{Zhang:2012ap,Sasaki:2013nka,Zhang:2013pna}, or to avoid the Higuchi bound~\cite{Higuchi:1986py}, and yet be compatible with current bounds on the graviton mass. If the graviton mass is an effective description from higher dimensions it is also quite natural to imagine that the graviton mass would depend on some moduli.

\subsection{Quasi-dilaton}
\label{sec:QuasiDilaton}

The Planck scale $\mpl$, or Newton constant explicitly breaks scale invariance, but one can easily extend the theory of GR to a scale invariant one $\mpl \to \mpl e^{\lambda(x)}$ by including a dilaton scalar field $\lambda$ which naturally arises from string theory or from extra dimension compactification (see for instance~\cite{Damour:1994zq} and see Refs.~\cite{Piazza:2004df,Damour:1990tw,Gasperini:2001pc} for the role of a dilaton scalar field on cosmology).

When dealing with multi-gravity, one can extend the notion of conformal transformation to the global rescaling of the coordinate system of one metric with respect to that of another metric. In the case of massive gravity this amounts to considering the global rescaling of the reference coordinates with respect to the physical one. As already seen, the reference metric can  be promoted to a tensor with respect to transformations of the physical metric coordinates, by introducing four \stu fields $\phi^a$, $f\mn \to f_{ab}\p_\mu \phi^a\p_\nu \phi^b$.  Thus the theory can be made invariant under global rescaling of the reference metric if the reference metric is  promoted to a function of the \emph{quasi-dilaton} scalar field  $\sigma$,
\ba
f_{ab}\p_\mu \phi^a\p_\nu \phi^b \to e^{2\sigma/\mpl}f_{ab}\p_\mu \phi^a\p_\nu \phi^b\,.
\ea
This is the idea behind the Quasi-Dilaton theory of massive gravity proposed in Ref.~\cite{D'Amico:2012zv}. The theoretical consistency of this model was explored in~\cite{D'Amico:2012zv} and is reviewed below. The Vainshtein mechanism and the cosmology were also explored in~\cite{D'Amico:2012zv,D'Amico:2013kya} as well as in Refs.~\cite{Haghani:2013eya,Gannouji:2013rwa,DeFelice:2013tsa} and we review the cosmology in Section~\ref{sec:CosmoOtherProposals}. As we shall see in that section, one of the interests of Quasi-Dilaton massive gravity is the existence of spatially flat FLRW solutions, and particularly of self-accelerating solutions. Nevertheless such solutions have been shown to be strongly coupled within the region of interest~\cite{D'Amico:2013kya}, but an extension of that model was proposed in~\cite{DeFelice:2013tsa} and shown to be free from such issues.

Recently the decoupling limit of the original Quasi-Dilaton model was derived in \cite{Gabadadze:2014kaa}.
Interestingly a new self-accelerating solution was found in this model which admits no instability and all the modes are (sub)luminal for a given realistic set of parameters. The extension  of this solution to the full theory (beyond the decoupling limit) should provide for a consistent self-accelerating solution which is guaranteed to be stable (or with a harmless instability time scale of the order of the age of the Universe at least).

\subsubsection{Theory}

As already mentioned, the idea behind quasi-dilaton massive gravity (QMG) is to extend massive gravity to a theory which admits a new global symmetry. This is possible via the introduction of a quasi-dilaton scalar field $\sigma(x)$. The  action for QMG is thus given by
\ba
\label{QMG}
S_{\rm QMG}&=&\frac{\mpl^2}{2}\int \d^4x \sqrt{-g}\Big[R-\frac{\omega}{2\mpl^2}(\p \sigma)^2+\frac{m^2}{2}\sum_{n=0}^4\alpha_n \L_n[\tilde{\K}[g,\eta]]\Big]\\
&&+\int \d^4 x\sqrt{-g}\L_{\rm matter}(g, \psi)\,,\nn
\ea
where $\psi$ represent the matter fields, $g$ is the dynamical metric, and unless  specified otherwise all indices are raised and lowered with respect to $g$, and $R$ represents the scalar curvature with respect to $g$. The Lagrangians $\L_n$ were expressed in (\ref{L0}\,--\,\ref{L4}) or (\ref{L0 X}\,--\,\ref{L4 X}) and the tensor $\tilde K$ is given in terms of the \stu fields as
\ba
\tilde{\K}\mupn[g,\eta]=\delta\mupn-e^{\sigma/\mpl}\sqrt{g^{\mu\alpha}\p_\alpha \phi^a \p_\nu \phi^b \eta_{ab}}\,.
\ea
In the case of the QMG presented in~\cite{D'Amico:2012zv},  there is no cosmological constant nor tadpole ($\alpha_0=\alpha_1=0$) and $\alpha_2=1$.
This is a very special case of the generalized theory of massive gravity presented in~\cite{Huang:2013mha}, and the proof for the absence of BD ghost thus goes through in the same way. Here again the presence of the scalar field brings only minor  modifications to the Hamiltonian analysis in the ADM language as presented in Section~\ref{sec:MassVarying}, and so we do not reproduce the proof here. We simply note that the theory propagates six degrees of freedom in four dimensions and is manifestly free of any ghost on flat space time  provided that $\omega>1/6$. The key ingredient compared to mass-varying gravity or generalized massive gravity is the presence of a global rescaling symmetry which is both a space-time and internal transformation~\cite{D'Amico:2012zv},
\ba
x^\mu \to  e^{\xi}x^\mu, \quad g\mn \to e^{-2 \xi}g\mn, \quad
\sigma \to \sigma -\mpl \xi,  \quad \text{and}\quad \phi^a \to e^\xi \phi^a\,.
\ea
Notice that the matter action $\d^4 x\sqrt{-g}\L(g,\psi)$ breaks this symmetry, reason why it is called a `\emph{quasi}-dilaton'.

An  interesting feature of QMG is the fact that the decoupling limit leads to a bi-Galileon theory, one Galileon being the helicity-0 mode presented in Section~\ref{sec:DL MG}, and the other Galileon being the quasi-dilaton $\sigma$. Just as in massive gravity, there are no irrelevant operators arising at energy scale below $\Lc$, and at that scale the theory is given by
\ba
\label{QDil DL}
\L^{(\rm QMG)}_{\Lc}= \L^{(0)}_{\Lc}-\frac{\omega}{2}(\p \sigma)^2+ \frac12\sigma \sum_{n=1}^4\frac{(4-n)\alpha_n-(n+1)\alpha_{n+1}}{\Lc^{3(n-1)}}\L_n[\Pi] \,,
\ea
where the decoupling limit Lagrangian $\L^{(0)}_{\Lc}$ in the absence of the quasi-dilaton is given in~\eqref{Full DL Lc} and we recall that $\alpha_2=1$, $\alpha_1=0$, $\Pi\mupn=\p^\mu \p_\nu \pi$ and the Lagrangians $\L_n$ are expressed in (\ref{L1}-\ref{L4}) or (\ref{L1 X}-\ref{L4 X}). We see emerging a bi-Galileon theory for $\pi$ and $\sigma$, and thus the decoupling limit is manifestly ghost-free. We could then apply a similar argument as in Section~\ref{sec:ghostFreeBeyondDL} to infer the absence of BD ghost for the full theory based on this decoupling limit.
Up to integration by parts, the Lagrangian~\eqref{QDil DL} is invariant under both independent galilean transformation $\pi \to \pi + c+v_\mu x^\mu$ and $\sigma \to \sigma + \tilde c+\tilde v_\mu x^\mu$.

One of the relevance of this decoupling limit is that it makes the study of the Vainshtein mechanism more explicit. As we shall in what follows  (see Section~\ref{sec:Vainshtein}), the Galileon interactions are crucial for the Vainshtein mechanism to work.

Note that in~\eqref{QDil DL}, the interactions with the quasi-dilaton come in the combination $((4-n)\alpha_n-(n+1)\alpha_{n+1})$, while in $\L^{(0)}_{\Lc}$, the interactions between the helicity-0 and -2 modes come in the combination $((4-n)\alpha_n+(n+1)\alpha_{n+1})$. This implies that
in massive gravity, the interactions between the helicity-2 and -0 mode disappear in the special case where $\alpha_n=-(n+1)/(4-n)\alpha_{n+1}$ (this corresponds to the minimal model), and the Vainshtein mechanism is no longer active for spherically symmetric sources (see Refs.~\cite{Chkareuli:2011te,Berezhiani:2011mt,Berezhiani:2013dw,Berezhiani:2013dca,Renaux-Petel:2014pja}). In the case of QMG, the interactions with the quasi-dilaton survive in that specific case $\alpha_3=-4\alpha_4$, and a Vainshtein mechanism could still be feasible, although one might still need to consider non-asymptotically Minkowski configurations.

The cosmology of QMD was first discussed in~\cite{D'Amico:2012zv} where the existence of self-accelerating solutions was pointed out. This will be reviewed in the section on cosmology, see Section~\ref{sec:CosmoOtherProposals}. We now turn to the extended version of QMG recently proposed in Ref.~\cite{DeFelice:2013tsa}.

\subsubsection{Extended quasi-dilaton}

Keeping the same philosophy as the quasi-dilaton in mind, a simple but yet powerful extension was proposed in Ref.~\cite{DeFelice:2013tsa} and then further extended in~\cite{DeFelice:2013dua}, leading to interesting phenomenology and stable self-accelerating solutions. The phenomenology of this model was then further explored in~\cite{Bamba:2013aca}. The stability of the extended quasi-dilaton theory of massive gravity was explored in~\cite{Kluson:2013jea} and was proven to be ghost-free in~\cite{Mukohyama:2013raa}.

The key ingredient behind the extended quasi-dilaton theory of massive gravity (EMG) is to notice that two most important properties of QMG namely the absence of BD ghost and the existence of a global scaling symmetry are preserved if the covariantized reference metric is further generalized to include a disformal contribution of the form $\p_\mu \sigma \p_\nu \sigma$ (such a contribution to the reference metric can arise naturally from the brane-bending mode in higher dimensional braneworld models, see for instance~\cite{deRham:2010eu}).

The action for EMG then takes the same form as in~\eqref{QMG} with the tensor $\tilde{\K}$ promoted to
\ba
\label{tilde K}
\tilde{\K} \to \bar \K =  \mathbb{I}-e^{\sigma/\mpl}\sqrt{g^{-1}\bar f}\,,
\ea
with the tensor $\bar f\mn$ defined as
\ba
\label{fbar}
\bar f\mn=\p_\mu \phi^a \p_\nu \phi^b \eta_{ab}-\frac{\alpha_\sigma}{\mpl \Lc^3}e^{-2 \sigma/\mpl}\p_\mu \sigma \p_\nu \sigma\,,
\ea
where $\alpha_\sigma$ is a new coupling dimensionless constant (as mentioned in~\cite{DeFelice:2013tsa}, this coupling constant is expected to enjoy a non-renormalization theorem in the decoupling limit, and thus to receive quantum corrections which are always suppressed by at least $m^2/\Lc^2$). Furthermore this action can be generalized further by
\begin{itemize}
  \setlength{\itemsep}{0pt}
   \setlength{\parskip}{0pt}
   \setlength{\parsep}{0pt}
\item Considering different coupling constants for the $\bar \K$'s entering in $\L_2[\bar \K]$, $\L_3[\bar \K]$ and $\L_4[\bar \K]$.
\item One can also introduce what would be a cosmological constant for the metric $\bar f$, namely a new term of the form $\sqrt{-\bar f}e^{4 \sigma / \mpl}$.
\item General shift-symmetric Horndeski Lagrangians for the quasi-dilaton.
\end{itemize}
Even without these further generalizations, one can obtain self-accelerating solutions similarly as in the original QMG. For these self-accelerating solutions, the coupling constant $\alpha_\sigma$ does not enter the background equations of motion but plays a crucial role for the stability of the scalar perturbations on top of these solutions. This is one of the benefits of this extended quasi-dilaton theory of massive gravity.

\subsection{Partially massless}
\label{sec:PM}

\subsubsection{Motivations behind PM gravity}

The multiple proofs for the absence of BD ghost presented in Section~\ref{sec:Abscence of ghost proof} ensures that the ghost-free theory of massive gravity, (or dRGT) does not propagate more than five physical degrees of freedom in the graviton. For a generic finite mass $m$ the theory propagates exactly five degrees of freedom as can be shown from a linear analysis about a generic background. Yet one can ask whether there exists special points in parameter space where some of degrees of freedom decouple. General relativity, for which $m=0$ (and the other parameters $\alpha_n$ are finite) is one such example. In the massless limit of massive gravity the two helicity-1 modes and the helicity-0 mode decouple from the helicity-2 mode and we thus recover the theory of a massless spin-2 field corresponding to GR, and three decoupled degrees of freedom. The decoupling of the helicity-0 mode occurs via the Vainshtein mechanism\epubtkFootnote{Note that the Vainshtein mechanism does not occur for all parameters of the theory. In that case the massless limit does not reproduce GR.} as we shall see in Section~\ref{sec:Vainshtein}.

As seen in Section~\ref{sec:DL on dS 2}, when considering massive gravity on de Sitter as a reference metric, if the graviton mass is precisely $m^2=2H^2$, the helicity-0 mode disappears linearly as can be seen from the linearized Lagrangian~\eqref{quadratic}. The same occurs in any dimension when the graviton mass is tied to the de Sitter curvature by the relation $m^2=(d-2)H^2$. This special case is another point in parameter space where the helicity-0 mode could be decoupled, corresponding to a partially massless (PM) theory of gravity as first pointed out by Deser and Waldron~\cite{Deser:2001wx,Deser:2001us,Deser:2001xr}, (see also~\cite{Zinoviev:2001dt} for partially massless higher spin, and~\cite{Skvortsov:2006at} for related studies).

The absence of helicity-0 mode at the linearized level in PM is tied to the existence of a new scalar gauge symmetry at the linearized level when $m^2=2 H^2$ (or $(d-2)H^2$ in arbitrary dimensions), which is responsible for making the helicity-0 mode unphysical.  Indeed the action~\eqref{quadratic} is invariant under a special combination of a linearized diff and a conformal transformation~\cite{Deser:2001wx,Deser:2001us,Deser:2001xr},
\ba
\label{PM symmetry}
h\mn \to h\mn +\nabla_\mu \nabla_\nu \xi -(d-2)H^2 \xi \gamma\mn\,.
\ea
If a non-linear completion of PM gravity exist, then there must exist a non-linear completion of this symmetry which eliminates the helicity-0 mode to all orders. The existence of such a symmetry would lead to several outstanding features:
\begin{itemize}
\setlength{\itemsep}{0pt}
\setlength{\parskip}{0pt}
\setlength{\parsep}{0pt}
\item  It would protect the structure of the potential.
\item In the PM limit of massive gravity, the helicity-0 mode fully decouples from the helicity-2 mode and hence from external matter. As a consequence there is no Vainshtein mechanism that decouples the helicity-0 mode in the PM limit of massive gravity unlike in the massless limit. Rather the helicity-0 mode simply decouples without invoking any strong coupling effects and the theoretical and observational luggage that goes with it.
\item Last but not least, in PM gravity the symmetry underlying the theory is not diffeomorphism invariance but rather the one pointed out in~\eqref{PM symmetry}. This means that in PM gravity, an arbitrary cosmological constant does not satisfy the symmetry (unlike in GR). Rather the value of the cosmological constant is fixed by the gauge symmetry and is proportional to the graviton mass. As we shall see in Section~\ref{sec:NonRenorm} the graviton does not receive large quantum corrections (it is technically natural to set to small values).
    So if a PM theory of gravity existed it would have the potential to tackle the cosmological constant problem.
\end{itemize}
Crucially breaking of covariance implies that matter is no longer covariantly conserved. Instead the failure of energy conservation is proportional to the graviton mass,
\ba
\label{Matter-constraint}
\nabla_\mu \nabla_\nu T^{\mu\nu}=-\frac{m^2}{d-2}T\,,
\ea
which in practise is extremely small.

It is worth emphasizing that if a PM theory of gravity existed, it would be distinct from the minimal model of massive gravity where the non-linear interactions between the helicity-0 and -2 modes vanish in the decoupling limit but the helicity-0 mode is still fully present. PM gravity is also distinct from some specific branches of solutions found in Cosmology (see Section~\ref{sec:Cosmology}) on top of which the helicity-0 mode disappears. If a PM theory of gravity exists the helicity-0 mode would be fully absent of the whole theory and not only for some specific branches of solutions.

\subsubsection{The search for a PM theory of gravity}

\subsubsection*{A candidate for PM gravity:}

The previous considerations represent some strong motivations for finding a fully fledged theory of PM gravity (\ie beyond the linearized theory) and there has been many studies to find a non-linear realization of the PM symmetry. So far all these studies have in common to keep the kinetic term for gravity unchanged (\ie keeping the standard Einstein--Hilbert action, with a potential generalization to the Lovelock invariants~\cite{Hassan:2012rq}).

Under this assumption, it was shown in~\cite{Zinoviev:2006im,Joung:2012hz}, that while the linear level theory admits a symmetry in any dimensions, at the cubic level the PM symmetry only exists in $d=4$ spacetime dimensions, which could make the theory even more attractive.  It was also pointed out in~\cite{Deser:2004ji} that in four dimensions the theory is conformally invariant. Interestingly the restriction to four dimensions can be lifted in bi-gravity by including the Lovelock invariants~\cite{Hassan:2012rq}.

From the analysis in Section~\ref{sec:DL on dS 2} (see Ref.~\cite{deRham:2012kf}) one can see that the helicity-0 mode entirely disappears from the decoupling limit of ghost-free massive gravity, if one ignores the vectors and sets the parameters of the theory to $m^2=2H^2$, $\alpha_3=-1$ and $\alpha_4=1/4$ in four dimensions. The ghost-free theory of massive gravity with these parameters is thus a natural candidate for the PM theory of gravity. Following this analysis, it was also shown that bi-gravity with the same parameters for the interactions between the two metrics satisfies similar  properties~\cite{Hassan:2012gz}. Furthermore it was also shown in~\cite{deRham:2013wv} that the potential has to follow the same structure as that of ghost-free massive gravity to have a chance of being an acceptable candidate for PM gravity. In bi-gravity the same parameters as for massive gravity were considered as also being the natural candidate~\cite{Hassan:2012gz}.

\subsubsection*{Re-appearance of the Helicity-0 mode:}

Unfortunately, when analysing the interactions with the vector fields, it is clear from the decoupling limit~\eqref{Full DL Lc} that the helicity-0 mode reappears non-linearly through their couplings with the vector fields. These never cancel, not even in four dimensions and for no parameters of theory. So rather than being free from the helicity-0 mode, massive gravity with $m^2=(d-2)H^2$ has an infinitely strongly coupled helicity-0 mode and is thus  a sick theory. The absence of the helicity-0 mode is simple artefact of the linear theory.

As a result we can thus deduce that there is no theory of PM gravity. This result is consistent with many independent studies performed in the literature (see Refs.~\cite{Deser:2013uy,deRham:2013wv,Deser:2013bs,Deser:2013xb}).

\subsubsection*{Relaxing the assumptions:}

\begin{itemize}
\item One assumption behind this result is the form of the kinetic term for the helicity-2 mode, which is kept to the be Einstein--Hilbert term as in GR. A few studies have considered a generalization of that kinetic term to diffeomorphism-breaking ones~\cite{Folkerts:2011ev,Hinterbichler:2013eza} however further analysis~\cite{Kimura:2013ika,deRham:2013tfa} have shown that such interactions always lead to ghosts non-perturbatively. See Section~\ref{sec:NoNewKin} for further details.
\item Another potential way out is to consider the embedding of PM within bi-gravity or multi-gravity. Since bi-gravity is massive gravity and a decoupled massless spin-2 field in some limit it is unclear how bi-gravity could evade the results obtained in massive gravity but this approach has been explored in~\cite{Hassan:2012gz,Hassan:2012rq,Deser:2013gpa}.
\item The other assumptions are locality and Lorentz-invariance. It is well known that Lorentz-breaking theories of massive gravity can excite fewer than five degrees of freedom. This avenue is explored in Section~\ref{sec:LorentzViolating}.
\end{itemize}

To summarize there is to date no known non-linear PM symmetry which could project out the helicity-0 mode of the graviton while keeping the helicity-2 mode massive in a local and Lorentz invariant way.

\newpage

\section{Massive Gravity Field Theory}
\label{sec:MGFT}

\subsection{Vainshtein mechanism}
\label{sec:Vainshtein}

As seen earlier, in four dimensions a massless spin-2 field has five degrees of freedom, and there is no special PM case of gravity where the helicity-0 mode is unphysical while the graviton remains massive (or at least there is to date no known such theory). The helicity-0 mode couples to matter already at the linear level and this additional coupling leads to a extra force which is at the origin of the vDVZ discontinuity see in  Section~\ref{sec:vDVZ}. In this section we shall see how the non-linearities of the helicity-0 mode is responsible for a Vainshtein mechanism that screens the effect of this field in the vicinity of matter.

Since the Vainshtein mechanism relies strongly on non-linearites, this makes explicit solutions very hard to find. In most of the cases where the Vainshtein mechanism has been shown to work successfully, one assumes a static and spherically symmetric background source.
Already in that case the existence of consistent solutions which extrapolate from a well-behaved asymptotic behaviour at infinity to a screened solution close to the source are difficult to obtain numerically~\cite{Damour:2002gp} and were only recently unveiled~\cite{Babichev:2009jt,Babichev:2010jd} in the case of non-linear Fierz--Pauli gravity.

This review on massive gravity cannot do justice to all the ongoing work dedicated to the study of the Vainshtein mechanism (also sometimes called `kinetic Chameleon' as it relies on the kinetic interactions for the helicity-0 mode). In what follows we will give the general idea behind the Vainshtein mechanism starting from the decoupling limit of massive gravity and then show explicit solutions in the decoupling limit for static and spherically symmetric sources. Such an analysis is relevant for observational tests in the solar system  as well as for other astrophysical tests (such as binary pulsar timing), which we shall explore in Section~\ref{sec:pheno}. We refer to the following review on the Vainshtein mechanism for further details,
\cite{Babichev:2013usa} as well as to the following work~\cite{Deffayet:2008zz,Babichev:2009us,Chkareuli:2011te,Kaloper:2011qc,Babichev:2011iz,Gannouji:2011qz,Babichev:2012re,Kimura:2011dc,Hui:2012jb,Sbisa:2012zk,Hiramatsu:2012xj,Belikov:2012xp,Li:2013nua,Koyama:2013paa,Narikawa:2013pjr}.
Recently it was also shown that the Vainshtein mechanism works for bi-gravity, see Ref.~\cite{Babichev:2013pfa}.

We focus the rest of this section to the case of four space-time dimensions, although many of the results presented in what follows are well understood in arbitrary dimensions.

\subsubsection{Effective coupling to matter}
\label{sec:EffCoupling}

As already mentioned, the key ingredient behind the Vainshtein mechanism is the importance of interactions for the helicity-0 mode which we denote as $\pi$. From the decoupling limit analysis performed for massive gravity (see~\eqref{Full DL Lc}) and bi-gravity (see~\eqref{Bigravity DL}), we see that in some limit the helicity-0 mode $\pi$ behaves as a scalar field, which enjoys a special global symmetry
\ba
\pi \to \pi + c+v_\mu x^\mu\,,
\ea
and yet only carries two derivatives at the level of the equations of motion, (which as we have seen is another way to see the absence of BD ghost).

These types of interactions are very similar to the Galileon-type of
interactions introduced by Nicolis, Rattazzi and Trincherini in Ref.~\cite{Nicolis:2008in} as a generalization of the decoupling limit of DGP. For simplicity we shall focus most of the discussion on the Vainshtein mechanism with Galileons as a special example, and then mention in Section~\ref{sec:SSSMG} peculiarities that arise in the special case of massive gravity (see for instance Refs.~\cite{Berezhiani:2013dw,Berezhiani:2013dca}).

We thus start with a cubic Galileon theory
\ba
\label{CubicGal}
\L=-\frac12 (\p \pi)^2 -\frac{1}{\Lambda^3}(\p \pi)^2 \Box \pi +\frac{1}{\mpl}\pi T,,
\ea
where $T=T^\mu_{\, \mu}$ is the trace of the stress-energy tensor of external sources, and $\Lambda$ is the strong coupling scale of the theory. As seen earlier, in the case of massive gravity, $\Lambda=\Lc=(m^2\mpl)^{1/3}$.  This is actually precisely the way the helicity-0 mode enters in the decoupling limit of DGP~\cite{Luty:2003vm} as seen in Section~\ref{sec:DGP DL}. It is in that very context that the Vainshtein mechanism was first shown to work explicitly~\cite{Deffayet:2001uk}.

The essence of the Vainshtein mechanism is that close to a source,  the Galileon interactions dominate over the linear piece. We make use of this fact by splitting the source into a background contribution $T_0$ and a perturbation $\delta T$. The background source $T_0$ leads to a background profile $\pi_0$ for the field, and the response to the fluctuation $\delta T$ on top of this background is given by $\phi$, so that the total field is expressed as
\ba
\label{split}
\pi = \pi_0+\phi\,.
\ea
For a sufficiently large source (or as we shall see below if $T_0$ represents a static point-like source, then sufficiently close to the source), the non-linearities dominate and symbolically $\p^2 \pi_0 \gg \Lambda^3$.

We now follow the perturbations in the action~\eqref{CubicGal} and notice that the background configuration $\pi_0$ leads to a modified effective metric for the perturbations,
\ba
\label{Gal_perturbed}
\L^{(2)}=-\frac12 Z^{\mu\nu}(\pi_0)\p_\mu \phi \p_\nu \phi +\frac{1}{\mpl}\phi \delta T\,,
\ea
up to second order in perturbations, with the new effective metric $Z^{\mu\nu}$
\ba
Z^{\mu\nu}= \eta^{\mu\nu}+\frac{2}{\Lambda^3}X^{(1)\mu\nu}(\Pi_0)\,,
\ea
where the tensor $X^{(1)}$ is the same as that defined for massive gravity in~\eqref{X1} or in~\eqref{X1 expl}, so symbolically $Z$ is of the form $Z\sim 1+ \frac{\p^2\pi_0}{\Lambda^3}$.
One can generalize the initial action~\eqref{CubicGal} to arbitrary set of Galileon interactions
\ba
\label{Full Galileon}
\L=\pi \sum_{n=1}^4 \frac{c_{n+1}}{\Lambda^{3(n-1)}}\L_n[\Pi]\,,
\ea
with again $\Pi\mn=\p_\mu\p_\nu \pi$ and
where the scalars $\L_n$ have been defined in (\ref{L1}-\ref{L4}). The effective metric would then be of the form
\ba
\label{Z}
Z^{\mu\nu}(\pi_0)=\sum_{n=1}^4\frac{n(n+1)c_{n}}{\Lambda^{3(n-1)}} X^{(n-1)\mu\nu}(\Pi_0)\,,
\ea
where all the  tensors $X^{(n)}\mn$ are defined in (\ref{X0}-\ref{X4}). Notice that $\p_\mu Z^{\mu\nu}=0$ identically. For sufficiently large sources, the components of $Z$ are large, symbolically, $Z\sim \(\p^2\pi_0/\Lambda^3\)^n \gg 1$ for $n\ge 1$.

Canonically normalizing the fluctuations in~\eqref{Gal_perturbed}, we have symbolically,
\ba
\label{Canonical Norm}
\hat \phi = \sqrt{Z} \phi \,,
\ea
 assuming $Z^{\mu\nu} \sim Z \eta^{\mu\nu}$, which is not generally the case. Nevertheless this symbolic scaling is sufficient to get the essence of the idea. For a more explicit canonical normalization in specific configurations see Ref.~\cite{Nicolis:2008in}. As nicely explained in that reference, if $Z^{\mu\nu}$ is conformally flat, one should not only scale the field $\phi \to \hat \phi$  but also the space-like coordinates $x \to \hat x$ so at to obtain a standard canonically normalized field in the new system, $\int \d^4 \tilde x -\frac 12 (\partial_{\tilde x} \hat \phi)^2$.  For now we stick to the simple normalization~\eqref{Canonical Norm} as it is sufficient to see the essence of the Vainshtein mechanism. In terms of the canonically normalized field $\hat \phi$, the perturbed action~\eqref{Gal_perturbed} is then
\ba
\label{redressedL2}
\L^{(2)}=-\frac12(\p \hat \phi )^2 +\frac{1}{\mpl \sqrt{Z(\pi_0)}}\hat \phi \delta T\,,
\ea
which means that the coupling of the fluctuations to matter is medium dependent and can arise at a scale very different from the Planck scale. In particular, for a large background configuration, $\p^2 \pi_0 \gg \Lambda^3$ and $Z(\pi_0)\gg 1$, so the effective coupling scale to external matter is
\ba
M_{\rm eff}= \mpl \sqrt{Z}\gg \mpl\,,
\ea
and the coupling to matter is thus very suppressed. In massive gravity $\Lambda$ is related to the graviton mass, $\Lambda\sim m^{2/3}$, and so the effective coupling scale $M_{\rm eff} \to \infty$ as $m\to 0$, which shows how the helicity-0 mode characterized by $\pi$ decouples in the massless limit.

We now first review how the Vainshtein mechanism works more explicitly in a static and spherically symmetric configuration before applying it to other systems. Note that the Vainshtein mechanism relies on irrelevant operators. In a standard EFT this cannot be performed without going beyond the regime of validity of the EFT. In the context of Galileons and other very specific derivative theories, one can reorganize the EFT so that the operators considered can be large and yet remain within the regime of validity of the reorganized EFT. This will be discussed in more depth in what follows.

\subsubsection{Static and spherically symmetric configurations in Galileons}
\label{sec:SSSGalileon}

\subsubsection*{Suppression of the force}

We now consider a point like source
\ba
T_0=-M\delta^{(3)}(r)=-M \frac{\delta(r)}{4\pi r^2}\,,
\ea
where $M$ is the mass of the source localized at $r=0$. Since the source is static and spherically symmetric, we can focus on configurations which respect the same symmetry, $\pi_0=\pi_0(r)$. The background configuration for the field $\pi_0(r)$ in the case of the cubic Galileon~\eqref{CubicGal} satisfies the equation of motion~\cite{Nicolis:2004qq}
\ba
\label{DGP spherical}
\frac{1}{r^2}\p_r\left[r^3\(\frac{\pi_0'(r)}{r}+\frac{1}{\Lambda^3}\(\frac{\pi_0'(r)}{r}\)^2\)\right]=\frac{M}{4\pi \mpl}\frac{\delta(r)}{r^2}\,,
\ea
and so integrating both sides of the equation, we obtain an algebraic equation for $\pi_0'(r)$,
\ba
\label{pi_0 SSS}
\frac{\pi_0'(r)}{r}+\frac{1}{\Lambda^3}\(\frac{\pi_0'(r)}{r}\)^2=\frac{M}{\mpl}\frac{1}{4\pi r^3 }\,.
\ea
We can define the Vainshtein or strong coupling radius $r_*$ as
\ba
\label{r*}
r_*=\frac{1}{\Lambda}\(\frac{M}{4\pi\mpl}\)^{1/3}\,,
\ea
so that at large distances compared to that Vainshtein radius the linear term in~\eqref{DGP spherical} dominates while the interactions dominate at distances shorter than $r_*$,
\ba
&& {\rm for }\ r\gg r_*, \hspace{10pt} \pi_0'(r)\sim \frac{M}{4\pi\mpl}\frac{1}{r^2}\nn \\
\label{cubig Gal ST}
&& {\rm for }\ r\ll r_*, \hspace{10pt} \pi_0'(r)\sim \frac{M}{4\pi\mpl}\frac{1}{r_*^{3/2}r^{1/2}}\,.
\ea
So at large distances $r\gg r_*$ one recovers a Newton square law for the force mediated by $\pi$, and that fields mediates a force which is just a strong as standard gravity (\ie as the force mediated by the usual helicity-2 modes of the graviton). On shorter distances scales, \ie close to the localized source, the force mediated by the new field $\pi$ is much smaller than the standard gravitational one,
\ba
\label{forceSupCubicStatic}
\frac{F_{r\ll r_*}^{(\pi)}}{F_{\rm Newt}}\sim \(\frac{r}{r_*}\)^{3/2} \ll 1 \hspace{10pt}{\rm for}\hspace{10pt} r\ll r_\star\,.
\ea
In the case of the quartic Galileon (which typically arises in massive gravity), the force is even suppressed and goes as
\ba
\label{foceSupQuatric}
\frac{F_{r\ll r_*}^{({\rm quartic }\ \pi)}}{F_{\rm Newt}}\sim \(\frac{r}{r_*}\)^{2} \ll 1 \hspace{10pt}{\rm for}\hspace{10pt} r\ll r_\star\,.
\ea
For a Graviton mass of the order of the Hubble parameter today, \ie $\Lambda\sim (1000\mathrm{\ km})^{-1}$, then taking into account the mass of the Sun, the force at the position of the Earth is suppressed by 12 orders of magnitude compared to standard Newtoninan force in the case of the cubic Galileon and by 16 orders of magnitude in the quartic Galileon. This means that the extra force mediated by $\pi$ is utterly negligible compared to the standard force of gravity and deviations to GR are extremely small.

Considering the Earth-Moon system, the force mediated by $\pi$ at the surface of the Moon is suppressed by 13 orders of magnitude compared to the Newtonian one in the cubic Galileon. While small, this is still not far off from the possible detectability from the lunar laser ranging space experiment~\cite{Williams:2004qba}, as will be discussed further in what follows. Note that in the quartic Galileon, that force is suppressed instead by 17 orders of magnitude and is there again very negligible.

When applying this naive estimate~\eqref{forceSupCubicStatic} to the Hulse-Taylor system for instance, we would infer a suppression of 15 orders of magnitude compared to the standard GR results. As we shall see in what follows this estimate breaks down when the time evolution is not negligible. These points will be discussed in the phenomenology Section~\ref{sec:pheno}, but before considering these aspects we review in what follows different aspects of massive gravity from a field theory perspective, emphasizing the regime of validity of the theory as well as the quantum corrections that arise in such a theory and the emergence of superluminal propagation.

\subsubsection*{Perturbations}

We now consider perturbations riding on top of this background configuration for the Galileon field, $\pi= \pi_0(r)+\phi(x^\mu)$. As already derived in Section~\ref{sec:EffCoupling}, the perturbations $\phi$ see the effective space-dependent metric $Z^{\mu\nu}$ given in~\eqref{Z}. Focusing on the cubic Galileon for concreteness, the background solution for $\pi_0$ is given by~\eqref{pi_0 SSS}. In that case the effective metric is
\ba
Z^{\mu\nu}&=&\eta^{\mu\nu}+\frac{4}{\Lambda^3}\(\Box \pi_0\eta^{\mu\nu}-\p^\mu \p^\nu \pi_0\)\\
Z_{\mu\nu}\d x^\mu \d x^\nu&=& - \(1+\frac{4}{\Lambda^3}\(\frac{2\pi_0'(r)}{r}+\pi_0''(r)\)\)\d t^2 \\
&&+\(1+\frac{8\pi_0'(r)}{r\Lambda^3}\)\d r^2+\(1+\frac{4}{\Lambda^3}\(\frac{\pi_0'(r)}{r}+\pi_0''(r)\)\)r^2\d \Omega^2_2\nn\,,
\ea
so that close to the source, for $r\ll r_*$,
\ba
\label{Z Galileon SSS}
Z_{\mu\nu}\d x^\mu \d x^\nu&=&6\(\frac{r_*}{r}\)^{1/2}\(-\d t^2+\frac 43 \d r^2+\frac 13 r^2 \d \Omega^2_2\)+ \mathcal{O}(r_*/r)^0\,.
\ea
A few comments are in order:

\begin{itemize}
\item First we recover $Z\sim \sqrt{r_* / r} \gg 1$ for $r\ll r_*$, which is responsible for the redressing of the strong coupling scale as we shall see in~\eqref{Lambda*}. On the no-trivial background the new strong coupling scale is $\Lambda_*\sim \sqrt{Z}\Lambda \gg \Lambda$ for $r\ll r_*$. Similarly on top of this background the coupling to external matter no longer occurs at the Planck scale but rather at the scale $\sqrt{Z} \mpl \sim 10^7 \mpl$.
\item Second we see that within the regime of validity of the classical calculation, the modes propagating along the radial direction do so with a \emph{superluminal phase and group velocity} $c_r^2=4/3>1$ and the modes propagating in the orthoradial direction do so with a \emph{subluminal phase and group velocity} $c_\Omega^2=1/3$. This result occurs in any Galileon and multi-Galileon theory which exhibits the Vainshtein mechanism~\cite{Nicolis:2008in,deFromont:2013iwa,Garcia-Saenz:2013gya}. The subluminal velocity is not of great concern, not even for Cerenkov radiation since the coupling to other fields is so much suppressed, but the superluminal velocity has been source of many questions~\cite{Adams:2006sv}. It is definitely one of the biggest issues arising in these kinds of theories see Section~\ref{sec:SL}.
\end{itemize}

Before discussing about the biggest concerns of the theory namely the superluminalities and the low strong coupling scale we briefly present some subtleties that arise when considering Static and Spherically symmetric solutions in massive gravity as opposed to a generic Galileon theory.

\subsubsection{Static and spherically symmetric configurations in massive gravity}
\label{sec:SSSMG}

The Vainshtein mechanism was discussed directly in the context of massive gravity (rather than the Galileon larger family) in Refs.~\cite{Koyama:2011xz,Koyama:2011yg,Chkareuli:2011te,Sbisa:2012zk} and more recently in~\cite{Berezhiani:2013dw,Tasinato:2013rza,Berezhiani:2013dca}.  See also Refs.~\cite{Volkov:2013roa,Comelli:2011wq,Berezhiani:2008nr,Nieuwenhuizen:2011sq,Gruzinov:2011mm,Deffayet:2008zz,Babichev:2009us,Babichev:2009jt,Babichev:2010jd}
for other spherically symmetric solutions in massive gravity.

While the decoupling limit of massive gravity resembles that of a Galileon, it presents a few particularities which affects the precise realization of the Vainshtein mechanism:

\begin{itemize}
\item First if the parameters of the ghost-free theory of massive gravity are such that $\alpha_3+4 \alpha_4\ne 0$, there is a mixing $h^{\mu\nu}X^{(3)}\mn$ between the helicity-0 and -2 modes of the graviton that cannot be removed by a local field redefinition (unless we work in an special types of backgrounds). The effects of this coupling were explored in~\cite{Chkareuli:2011te,Berezhiani:2013dca} and it was shown that the theory does not exhibit any stable static and spherically symmetric configuration in presence of a localized point-like matter source. So in order to be phenomenologically viable, the theory of massive gravity needs to be tuned with $\alpha_3+4 \alpha_4=0$. Since these parameters do not get renormalized this is a \emph{tuning } and not a \emph{fine-tuning}.
\item When $\alpha_3+4 \alpha_4=0$ and the previous mixing $h^{\mu\nu}X^{(3)}\mn$ is absent, the decoupling limit of massive gravity resembles a specific quartic Galileon, where the coefficient of the cubic Galileon is related to quartic coefficient (and if one vanishes so does the other one),
  \ba
  \label{DL MG 3}
  \L_{\rm Helicity-0}&=&-\frac 34 (\p \pi)^2+\frac{3\alpha}{4\Lc^3}\L^{(3)}_{\rm (Gal)}[\pi]-\frac14\(\frac{\alpha}{\Lc^3}\)^2\L^{(4)}_{\rm (Gal)}[\pi]\\
  &&\nn+\frac{1}{\mpl}\(\pi T
  +\frac{\alpha}{\Lc^3}\p_\mu \pi\p_\nu \pi T^{\mu\nu}\)\,,
  \ea
  where we have set $\alpha_2=1$ and
  the Galileon Lagrangians $\L^{(3,4)}_{\rm (Gal)}[\pi]$ are given in \eqref{Gal 3} and \eqref{Gal 4}.
  Note that in this decoupling limit the graviton mass always enters in the combination $\alpha/\Lc^3$, with $\alpha = -(1+3/2\alpha_3)$. As a result this decoupling limit can never be used to directly probe the graviton mass itself but rather of the combination $\alpha/\Lc^3$~\cite{Berezhiani:2013dca}. Beyond the decoupling limit however the theory breaks the degeneracy between $\alpha$ and $m$.

  Not only is the cubic Galileon always present when the quartic Galileon is there, but one cannot prevent the new coupling to matter $\p_\mu \pi\p_\nu \p \pi T^{\mu\nu}$ which is typically absent in other Galileon theories.
\end{itemize}

The effect of the coupling $\p_\mu \pi\p_\nu \pi T^{\mu\nu}$ was explored in~\cite{Berezhiani:2013dw}. First it was shown that this coupling contributes to the definition of the kinetic term of $\pi$ and can lead to a ghost unless $\alpha>0$ so this restricts further the allowed region of parameter space for massive gravity. Furthermore even when $\alpha>0$, none of the static spherically symmetric which asymptote to $\pi \to 0$ at infinity (asymptotically flat solutions) extrapolate to a Vainshtein solution close to the source. Instead the Vainshtein solution near the source extrapolate to cosmological solutions at infinity which is independent of the source
\ba
\pi_0(r) & \to & \frac{3+\sqrt{3}}{4}\frac{\Lc^3}{\alpha} r^2 \qquad \text{for}\quad r\gg r_* \\
\pi_0(r) & \to & \(\frac{\Lc^3}{\alpha}\)^{2/3} \(\frac{M}{4\pi \mpl}\)^{1/3} \qquad \text{for}\quad r\ll r_*\,.
\ea
If $\pi$ was a scalar field in its own right such an asymptotic condition would not be acceptable. However in massive gravity $\pi$ is the helicity-0 mode of the gravity and its effect always enters from the \stu combination $\p_\mu \p_\nu \pi$, which goes to a constant at infinity. Furthermore this result is only derived in the decoupling limit, but in the fully fledged theory of massive gravity, the graviton mass kicks in at the distance scale $\ell\sim m^{-1}$ and suppresses any effect at these scales.

Interestingly when performing the perturbation analysis on this solution, the modes along all directions are subluminal, unlike what was found for the Galileon in~\eqref{Z Galileon SSS}. It is yet unclear whether this is an accident to this specific solution or if this is something generic in \emph{consistent} solutions of massive gravity.

\subsection{Validity of the EFT}
\label{sec:EFT}

The Vainshtein mechanism presented previously relies crucially on interactions which are important at a low energy scale $\Lambda \ll \mpl$. These interactions are operators of dimension larger than four, for instance the cubic Galileon $(\p \pi)^2 \Box \pi$ is a dimension-7 operator and the quartic Galileon is a dimension-10 operator. The same can be seen directly within massive gravity. In the decoupling limit~\eqref{DL MG1}, the terms $h^{\mu\nu}X^{(2,3)}\mn$ are respectively dimension-7 and-10 operators. These operators are thus irrelevant from a traditional EFT viewpoint and the theory is hence not renormalizable.

This comes as no surprise, since gravity itself is not renormalizable and there is thus no reason to expect massive gravity nor its decoupling limit to be renormalizable. However for the Vainshtein mechanism to be successful in massive gravity, we are required to work within a regime where these operators dominate over the marginal ones (\ie over the standard kinetic term $(\p \pi)^2$ in the strongly coupled region where $\p^2 \pi \gg \Lambda^3$). It is therefore natural to wonder whether or not one can ever use the effective field description within the strong coupling region without going outside the regime of validity of the theory.

The answer to this question relies on two essential features:

\begin{enumerate}
\item First, as we shall see in what follows, the Galileon interactions or the interactions that arise in the decoupling limit of massive gravity and which are essential for the Vainshtein mechanism do not get renormalized within the decoupling limit (they enjoy a non-renormalization theorem which we review in what follows).
\item The non-renormalization theorem together with the shift and Galileon symmetry implies that only higher operators of the form $\(\p^\ell \pi\)^m$, with $\ell,m \ge 2$ are generated by quantum corrections. These operators differ from the Galileon operators in that they always generate terms that more than two derivatives on the field at the level of the equation of motion (or they always have two or more derivatives per field at the level of the action).
\end{enumerate}

This means that there exists a regime of interest for the theory, for which the operators generated by quantum corrections are irrelevant (non-important compared to the Galileon interactions). Within the strong coupling region, the field itself can take large values, $\pi \sim \Lambda$, $\p \pi \sim \Lambda^2$, $\p^2 \pi \sim \Lambda^3$, and one can still rely on the Galileon interactions and take no other operator into account so long as any further derivative of the field is suppressed, $\p^{n} \pi \ll \Lambda^{n+1}$ for any $n\ge 3$.

This is similar to the situation in DBI scalar field models, where the field operator itself and its velocity is considered to be large $\pi \sim \Lambda$ and $\p \pi \sim \Lambda^2$, but the field acceleration and any higher derivatives are suppressed $\p^n \pi \ll \Lambda^{n+1}$ for $n\ge 2$ (see~\cite{deRham:2010eu}). In other words, the Effective Field expansion should be reorganized so that operators which do not give equations of motion with more than two derivatives (\ie Galileon interactions) are considered to be large and ought to be treated as the \emph{relevant operators}, while all other interactions (which lead to terms in the equations of motion with more than two derivatives) are treated as \emph{irrelevant corrections} in the effective field theory language.

Finally, as mentioned previously, the Vainshtein mechanism itself changes the canonical scale and thus the scale at which the fluctuations become strongly coupled. On top of a background configuration, interactions do not arise at the scale $\Lambda$ but rather at the rescaled strong coupling scale $\Lambda_*= \sqrt{Z}\Lambda$, where $Z$ is expressed in~\eqref{Z}. In the strong coupling region, $Z\gg1$ and so $\Lambda_*\gg \Lambda$. The higher interactions for fluctuations on top of the background configuration are hence much smaller than expected and their quantum corrections are therefore suppressed.

When taking the cubic Galileon and considering the strong coupling effect from a static and spherically symmetric source then
\ba
\label{Lambda*}
\Lambda_*\sim \sqrt{Z} \Lambda\sim \sqrt{\frac{\pi_0'(r)}{r \Lambda^3}}\Lambda\,,
\ea
where the profile for the cubic Galileon in the strong coupling region is given in~\eqref{cubig Gal ST}. If the source is considered to be the Earth, then at the surface of the Earth this gives
\ba
\Lambda_*\sim \(\frac{M}{\mpl}\frac{1}{(r\Lambda)^3}\)\Lambda \sim 10^7 \Lambda \sim {\rm cm}^{-1}\,,
\ea
taking $\Lambda\sim(1000\mathrm{\ km})^{-1}$, which would be the scale $\Lc$ in massive gravity for a graviton mass of the order of the Hubble parameter today. In the quartic Galileon this enhancement in the strong coupling scale does not work as well in the purely static and spherically symmetric case~\cite{Burrage:2012ja} however considering a more realistic scenario and taking the smallest breaking of the spherical symmetry into account (for instance the Earth dipole) leads to a comparable result of a few cm~\cite{Berezhiani:2013dca}. Notice that this is the redressed strong coupling scale when taking into consideration only the effect of the Earth. When getting to these smaller distance scales, all the other matter sources surrounding whichever experiment or scattering process needs to be accounted for and this pushes the redressed strong coupling scale even higher~\cite{Berezhiani:2013dca}.

\subsection{Non-renormalization}
\label{sec:NonRenorm}

The \emph{non-renormalization} theorem mentioned above states that within a Galileon theory the Galileon operators themselves do not get renormalized.
This was originally understood within the context of the cubic Galileon in the procedure established in~\cite{Nicolis:2004qq} and is easily generalizable to all the Galileons~\cite{Nicolis:2008in}. In what follows, we review the essence of non-renormalization theorem within the context of massive gravity as derived in~\cite{deRham:2012ew}.

Let us start with the decoupling limit of massive gravity~\eqref{DL MG1} in the absence of vector modes (the Vainshtein mechanism presented previously does not rely on these modes and it thus consistent for the purpose of this discussion to ignore them). This decoupling limit is a very special scalar-tensor theory on flat spacetime
\ba
\label{DL 2}
\L_{\Lc}=-\frac 14h^{\mu \nu}\hat {\mathcal{E}}^{\alpha \beta}\mn h\ab-\frac 14 h^{\mu\nu}\sum_{n=1}^3 \frac{c_n}{\Lc^{3(n-1)}}X^{(n)}\mn\,,
\ea
where the coefficients $c_n$ are given in \eqref{coefficients Cn DL} and the tensors $X^{(n)}$ are given in (\ref{X1}\,--\,\ref{X3}) or (\ref{X0 expl}\,--\,\ref{X3 expl}).
The theory described by~\eqref{DL 2} (including the two interactions $h X^{(2,3)}$) enjoys two kinds of symmetries: a gauge symmetry for $h\mn$ (linearized diffeomorphism) $h\mn\to h\mn +\p_{(\mu}\xi_{\nu)}$ and a global shift and Galilean symmetry for $\pi$, $\pi \to \pi + c + v_\mu x^\mu$. Notice that unlike in a pure Galileon theory, here the global symmetry for $\pi$ is an exact symmetry of the Lagrangian (not a symmetry up to boundary terms).
This means that the quantum corrections generated by this theory ought to preserve the same kinds of symmetries.

The non-renormalization theorem follows simply from the antisymmetric structure of the interactions~\eqref{X2} and \eqref{X3}.
Let us consider the contributions of the vertices
\ba
V_2&=&h^{\mu\nu}X^{(2)}\mn = h^{\mu\nu}\ep^{\mu\alpha\beta\gamma}\ep^{\nu\alpha'\beta'}_{\phantom{\nu\alpha'\beta'}\gamma}\p_\alpha \p_{\alpha'}\pi \p_\beta \p_{\beta'}\pi\\
V_3&=&h^{\mu\nu}X^{(3)}\mn = h^{\mu\nu}\ep^{\mu\alpha\beta\gamma}\ep^{\nu\alpha'\beta'\gamma'}\p_\alpha \p_{\alpha'}\pi \p_\beta \p_{\beta'}\pi \p_\gamma \p_{\gamma'}\pi
\ea
to an arbitrary diagram. If all the external legs of this diagram are $\pi$ fields then it follows immediately that the contribution of the process goes as $(\p^2 \pi)^n$ or with more derivatives and is thus not an operator which was originally present in~\eqref{DL 2}. So let us consider the case where a vertex (say $V_3$) contributes to the diagram with a spin-2 external leg of momentum $p_\mu$. The contribution from that vertex to the whole diagram is given by
\ba
i\mathcal{M}_{V_3}&\propto& i \int \frac{\d^4k}{(2\pi)^4}\frac{\d^4q}{(2\pi)^4}\mathcal{G}_{k}\mathcal{G}_{q}\mathcal{G}_{p-k-q}\\
&& \times \Big[
\epsilon^{*\mu\nu}\ep_{\mu}^{\phantom{\mu}\alpha\beta\gamma}\ep_{\nu}^{\phantom{\nu}\alpha'\beta'\gamma'}
k_\alpha k_{\alpha'} q_\beta q_{\beta'}(p-k-q)_\gamma (p-k-q)_{\gamma'}\Big]\nn \\
&\propto& i \epsilon^{*\mu\nu}\ep_{\mu}^{\phantom{\mu}\alpha\beta\gamma}\ep_{\nu}^{\phantom{\nu}\alpha'\beta'\gamma'}
p_\gamma p_{\gamma'}
\int \frac{\d^4k}{(2\pi)^4}\frac{\d^4q}{(2\pi)^4}\mathcal{G}_{k}\mathcal{G}_{q}\mathcal{G}_{p-k-q}
k_\alpha k_{\alpha'} q_\beta q_{\beta'}\,,\nn
\ea
where $\epsilon^{*\mu\nu}$ is the polarization of the spin-2 external leg and $\mathcal{G}_k$ is the Feynman propagator for the $\pi$-particle, $\mathcal{G}_k=i(k^2-i \ep)^{-1}$. This contribution is quadratic in the momentum of the external spin-2 field $p_\gamma p_{\gamma'}$, which means that in position space it has to involve at least two derivatives in $h\mn$ (there could be more derivatives arising from the integral over the propagator $\mathcal{G}_{p-k-q}$ inside the loops). The same result holds when inserting a $V_2$ vertex as explained in~\cite{deRham:2012ew}. As a result any diagram in this theory can only generate terms of the form $(\p^2 h)^\ell (\p^2 \pi)^m$, or terms with even more derivatives. As a result the operators presented in~\eqref{DL 2} or in the decoupling limit of massive gravity are not renormalized. This means that within the decoupling limit the scale $\Lambda$ does not get renormalized, and it can be set to an arbitrarily small value (compared to the Planck scale) without running issues. The same holds for the other parameter $c_2$ or $c_3$.

When working beyond the decoupling limit, we expect operators of the form $h^2 (\p^2\pi)^n$ to spoil this non-renormalization theorem. However these operators are $\mpl$ suppressed, and so they lead to quantum corrections which are themselves $\mpl$ suppressed. This means that the quantum corrections to the graviton mass suppressed as well~\cite{deRham:2012ew}
\ba
\label{delta m}
\delta m^2 \lesssim m^2 \(\frac{m}{\mpl}\)^{2/3}\,.
\ea
This result is crucial for the theory. It implies that a small graviton mass is technically natural.

\subsection{Quantum corrections beyond the decoupling limit}

As already emphasized, the consistency of massive gravity relies crucially on a very specific set of allowed interactions summarized in Section~\ref{sec:dRGTsummary}. Unlike for GR, these interactions are not protected by any (known) symmetry and we thus expect quantum corrections to destabilize this structure. Depending on the scale at which these quantum corrections kick in, this could lead to a ghost at an unacceptably low scale.

Furthermore as discussed previously, the mass of the graviton itself is subject to quantum corrections, and for the theory to be viable the graviton mass ought to be tuned to extremely small values. This tuning would be technically unnatural if the graviton mass large quantum corrections.

We first summarize the results found so far in the literature before providing further details
\begin{enumerate}
  \setlength{\parskip}{0pt}
  \setlength{\parsep}{0pt}
\item \textbf{Destabilization of the potential:}\\
At one-loop, matter fields do not destabilize the structure of the potential. Graviton loops on the hand do lead to new operators which do not belong to the ghost-free family of interactions presented in (\ref{L0}\,--\,\ref{L4}), however they are irrelevant below the Planck scale.
\item \textbf{Technically natural graviton mass:}\\
As already seen in~\eqref{delta m}, the quantum corrections for the graviton mass are suppressed by the graviton mass itself, $\delta m^2 \lesssim m^2 (m/\mpl)^{2/3}$ this result is confirmed at one-loop beyond the decoupling limit and as result a small graviton mass is technically natural.
\end{enumerate}

\subsubsection{Matter loops}

The essence of these arguments go as follows: Consider a `covariant' coupling to matter, $\L_{\rm matter}(g\mn, \psi_i)$, for any species $\psi_i$ be it a scalar, a vector, or a fermion (in which case the coupling has to be performed in the vielbein formulation of gravity, see~\eqref{SDirac}).

At one loop, virtual matter fields do not mix with the virtual graviton. As a result as far as matter loops are concerned, they are `unaware' of the graviton mass, and only lead to quantum corrections which are already present in GR and respect diffeomorphism invariance. So the only potential term (\ie operator with no derivatives on the metric fluctuation) it can lead to is the cosmological constant.

This result was confirmed at the level of the one-loop effective action in~\cite{deRham:2013qqa} where it was shown that a field of mass $M$ leads to a running of the cosmological constant $\delta \Lambda_{\rm CC}\sim M^4$. This result is of course well-known and is at the origin of the old cosmological constant problem~\cite{Weinberg:1988cp}. The key element in the context of massive gravity is that this cosmological constant does not lead to any ghost and no new operators are generated from matter loops, at the one-loop level (and this independently of the regularization scheme used, be it dimensional regularization, cutoff regularization, or other.) At higher loops we expect virtual matter fields and graviton to mix and effect on the structure of the potential still remains to be explored.

\subsubsection{Graviton loops}

When considering virtual gravitons running in the loops, the theory does receive quantum corrections which do not respect the ghost-free structure of the potential. These are of course suppressed by the Planck scale and the graviton mass and so in dimensional regularization,  we generate new operators of the form\footnote{This result has been checked explicitly in Ref.~\cite{deRham:2013qqa} using dimensional regularization or following the log divergences.
Taking power law divergences seriously would also allow for a scalings of the form $(\Lambda_{\rm Cutoff}^4/\mpl^n) h^n$, which are no longer suppressed by the mass scale $m$ (although the mass scale $m$ would never enter with negative powers at one loop.) However it is well known that power law divergences cannot be trusted as they depend on the measure of the path integral and can lead to erroneous results in cases where the higher energy theory is known. See Ref.~\cite{Burgess:1992gx} and references therein for known examples and an instructive discussion on the use and abuses of power law divergences.}
\ba
\label{QC}
\L_{\rm QC}^{\rm (potential)}\sim \frac{m^4}{\mpl^n} \ h^n\,,
\ea
with $n\ge 2$, and
where $m$ is the graviton mass, and the contractions of $h$ do not obey the structure presented in (\ref{L0}-\ref{L4}). In a normal effective field theory this is not an issue as such operators are clearly irrelevant below the Planck scale. However for massive gravity, the situation is more subtle.

As see in Section~\ref{sec:Vainshtein} (see also Section~\ref{sec:EFT}), massive gravity is phenomenologically viable only if it has an active Vainshtein mechanism which screens the effect of the helicity-0 mode in the vicinity of dense environments. This Vainshtein mechanisms relies on having a large background for the helicity-0 mode, $\pi=\pi_0+\delta \pi$ with $\p^2 \pi_0 \gg \Lc^3=m^2\mpl$, which in unitary gauge implies $h=h_0+\delta h$, with $h_0\gg \mpl$.

To mimic this effect, we consider a given background for $h=h_0\gg \mpl$. Perturbing the new operators~\eqref{QC} about this background leads to a contribution at quadratic order for the perturbations $\delta h$ which does not satisfy the Fierz--Pauli structure,
\ba
\L_{\rm QC}^{(2)}\sim \frac{m^4\, h_0^{n-2}}{\mpl^n}\ \delta h^2\, .
\ea
In terms of the helicity-0 mode $\pi$, considering $\delta h \sim \p^2 \pi/m^2$ this leads to higher derivative interactions
\ba
\label{QC background}
\L_{\rm QC}^{(2)}\sim \frac{h_0^{n-2}}{\mpl^n}\ \(\p^2 \pi\)^2\,,
\ea
which revive the BD ghost at the scale
$m^2_{\rm ghost}\sim h_0^2 (\mpl/h_0)^n$. $m^2_{\rm ghost}$. The mass of the ghost can be made arbitrarily small, (smaller than $\Lc$) by taking $n\gg1$ and $h_0\gtrsim \mpl$ as is needed for the Vainshtein mechanism. In itself this would be a disaster for the theory as it means precisely in the regime where we need the Vainshtein mechanism to work, a ghost appears at an arbitrarily small scale and we can no longer trust the theory.

The resolution to this issue lies within the Vainshtein mechanism itself and its implementation not only at the classical level as was done to estimate the mass of the ghost in~\eqref{QC background} but also within the calculation of the quantum corrections themselves. To take the Vainshtein mechanism consistently into account one needs to consider the effective action \emph{redressed} by the interactions themselves (as was performed at the classical level for instance in~\eqref{redressedL2}).

This redressing was taken into at the level of the one-loop effective action in Ref.~\cite{deRham:2013qqa} and it was shown that when resumed, the large background configuration has the effect of further suppressing the quantum corrections so that the mass the ghost never reaches below the Planck scale even when $h_0 \ll \mpl$.
To be more precise~\eqref{QC background} is only one term in an infinite order expansion in $h_0$. Resuming these terms leads rather to contribution of the form (symbolically)
\ba
\label{QC background resumed}
\L_{\rm QC}^{(2)}\sim \frac{1}{1+\frac{h_0}{\mpl}}\frac{1}{\mpl^2}\ \(\p^2 \pi\)^2\,,
\ea
so that the effective scale at which this operator is relevant is well above the Planck scale when $h_0\gtrsim \mpl$ and is at the Planck scale when working in the weak-field regime $h_0 \lesssim \mpl$. Notice that $h_0\sim-\mpl$ corresponds to a physical singularity in massive gravity (see~\cite{Berezhiani:2011mt}), and the theory would break down at that point anyways, irrespectively of the ghost.

As a result at the one loop level the quantum corrections destabilize the structure of the potential but in a way which is irrelevant below the Planck scale.

\subsection{Strong coupling scale vs cutoff}
\label{sec:StrongCoupling}

Whether it is to compute the Vainshtein mechanism or quantum corrections to massive gravity, it is crucial to realize that the scale $\Lambda=(m^2\mpl)^{1/3}$ (denoted as $\Lambda$ in what follows) is \emph{not} necessarily the cutoff of the theory.

The cutoff of a theory corresponds to the scale at which the given theory breaks down and new physics is required to describe nature. For GR the cutoff is the Planck scale. For massive gravity the cutoff could potentially be below the Planck scale, but is likely well above the scale $\Lambda$, and the redressed scale $\Lambda_*$ computed in section~\eqref{Lambda*}. Instead $\Lambda$ (or $\Lambda_*$ on some backgrounds) is the strong-coupling scale of the theory.

When hitting the scale $\Lambda$ or $\Lambda_*$ perturbativity breaks down (in the standard field representation of the theory), which means that in that representation loops ought to be taken into account to derive the correct physical results at these scales. However it does not necessarily mean that new physics should be taken into account. The fact that tree-level calculations do not account for the full results does in no way imply that theory itself breaks down at these scales, only that perturbation theory breaks down.

Massive gravity is of course not the only theory whose strong coupling scale departs from its cutoff. See for instance Ref.~\cite{Aydemir:2012nz} for other examples in chiral theory, or in gravity coupled to many species. To get more intuition on these types of theories and on the distinction between strong coupling scale and cutoff, consider a large number $N\gg 1$ of scalar fields coupled to gravity. In that case the effective strong coupling scale seen by these scalars is $M_{\rm eff}=\mpl/\sqrt{N}\ll \mpl$, while the cutoff of the theory is still $\mpl$ (the scale at which new physics enters in GR is independent of the number of species living in GR).

The philosophy behind~\cite{Aydemir:2012nz} is precisely analogous to the distinction between the strong coupling scale and the cutoff (onset of new physics) that arises in massive gravity, and summarizing the results of~\cite{Aydemir:2012nz} would not make justice of their work, instead we quote the abstract and encourage the reader to refer to that article for further details:

\begin{quote}
``In effective field theories it is common to identify the onset of new physics with the violation of tree-level unitarity. However, we show that this is parametrically incorrect in the case of chiral perturbation theory, and is probably theoretically incorrect in general. In the chiral theory, we explore perturbative unitarity violation as a function of the number of colors and the number of flavors, holding the scale of the ``new physics'' (\ie QCD) fixed. This demonstrates that the onset of new physics is parametrically uncorrelated with tree-unitarity violation. When the latter scale is lower than that of new physics, the effective theory must heal its unitarity violation itself, which is expected because the field theory satisfies the requirements of unitarity. (\dots) A similar example can be seen in the case of general relativity coupled to multiple matter fields, where iteration of the vacuum polarization diagram restores unitarity. We present arguments that suggest the correct identification should be connected to the onset of inelasticity rather than unitarity violation.''~\cite{Aydemir:2012nz}.
\end{quote}

\subsection{Superluminalities and (a)causality}
\label{sec:SL}

Besides the presence of a low strong coupling scale in massive gravity (which is a requirement for the Vainshtein mechanism, and is thus not a feature that should necessarily try to avoid), another point of concern is the possibility to have superluminal propagation. This statements requires a qualification and to avoid any confusion, we shall first review the distinction between \emph{phase} velocity, \emph{group} velocity, \emph{signal} velocity and \emph{front} velocity and their different implications. We follow the same description as in~\cite{Milonni} and~\cite{Brillouin} and refer to these books and references therein for further details.

\begin{enumerate}
\item \textbf{Phase Velocity:} For a wave of constant frequency, the phase velocity is the speed at which the peaks of the oscillations propagate. For a wave~\cite{Brillouin}
  \ba
  f(t,x)=A \sin (\omega t- k x)= A\sin \(\omega \(t-\frac{x}{v_{\rm phase}}\)\)\,,
  \ea
  the phase velocity $v_{\rm phase}$ is given by
  \ba
  v_{\rm phase}=\frac \omega k\,.
  \ea
\item \textbf{Group Velocity:} If the amplitude of the signal varies, then the group velocity represents the speed at which the modulation or envelop of the signal propagates. In a medium where the phase velocity is constant and does not depend on frequency, the phase and the group velocity are the same. More generally in a medium with dispersion relation $\omega(k)$, the group velocity is
  \ba
  v_{\rm group}=\frac{\p \omega(k)}{\p k}\,.
  \ea
  We are familiar with the notion that the phase velocity can be larger than speed of light $c$ (in this review we use units where $c=1$.) Similarly, it has been known for now almost a century that
  \begin{center}
  \begin{minipage}{10cm}
  \textit{``(...) the group velocity could exceed $c$ in a spectral region of an anomalous dispersion''}~\cite{Milonni}\,.
  \end{minipage}
  \end{center}
While being a source of concern at first, it is now well-understood not to be in any conflict with the theory of general (or special) relativity and not to be the source of any acausality. The resolution lies in the fact that the group velocity does not represent the speed at which \emph{new} information is transmitted. That speed is instead refer as the \emph{front} velocity as we shall see below.
\item \textbf{Signal Velocity} ``{\it yields the arrival of the main signal, with intensities of the order of magnitude of the input signal}''~\cite{Brillouin}. 
  Nowadays it is common to define the signal velocity as the velocity from the part of the pulse which has reached at least half the maximum intensity. However as mentioned in~\cite{Milonni}, this notion of speed rather is arbitrary and some known physical systems can exhibit a signal velocity larger than $c$.
\item \textbf{Front Velocity:} Physically, the front velocity represents the speed of the front of a disturbance, or in other words ``{\it Front velocity (...) correspond[s] to the speed at which the very first, extremely small (perhaps invisible) vibrations will occur.}''~\cite{Brillouin}. 

  The front velocity is thus the speed at which the very first piece of information of the first ``forerunner" propagates once a front or a ``{\it sudden discontinuous turn-on of a field}'' is turned on~\cite{Milonni}.

  ``{\it The front is defined as a surface beyond which, at a given instant in time the medium is completely at rest}''~\cite{Brillouin}, 
  \ba
  f(t,x)=\theta(t)\sin(\omega t- k x)\,,
  \ea
  where $\theta(t)$ is the Heaviside step function.

  In practise the front velocity is the large $k$ (high frequency) limit of the phase velocity.
\end{enumerate}
The distinction between these four types  of velocities in presented in Figure~\ref{Fig:PhaseGroupVelocity}. They are important to keep in mind and especially to be distinguished when it comes to \emph{superluminal} propagation. \textbf{Superluminal phase, group and signal velocities} have been observed and measured experimentally in different physical systems and yet cause no contradiction with special relativity nor do they signal acausalities. See Ref.~\cite{Hollowood:2007kt} for an enlightening discussion of the case of QED in curved spacetime.

The \textbf{front velocity} on the other hand,  is the real `measure' of the speed of propagation of new information, and the front velocity is always (and should always be) (sub)luminal.
As shown in \cite{Shore:2007um}, ``{\it the `speed of light' relevant for causality is $v_{\rm ph}(\infty)$,
i.e. the high-frequency limit of the phase velocity. Determining this requires a knowledge
of the UV completion of the quantum field theory.}" In other words there is no sense in computing a classical version of the front velocity since quantum corrections always dominate.

When it comes to the presence of superluminalities in massive gravity and theories of Galileons this distinction is crucial. We first summarize the current state of the situation in the context of both Galileons and massive gravity and then give further details and examples in what follows:

\begin{itemize}
\item In Galileons theories the presence of superluminal group velocity has been established for all the parameters which exhibit an active Vainshtein mechanism. These are present in spherically symmetric configurations near massive sources as well as in self-sourced plane waves and other configurations for which no special kind of matter is required.
\item Since massive gravity reduces to a specific Galileon theory in some limit we expect the same result to be true there well and to yield solutions with superluminal group velocity. However to date no fully consistent solution has yet been found in massive gravity which exhibits superluminal group velocity (let alone superluminal front velocity which would be the real signal of acausality). Only local configurations have been found with superluminal group velocity or finite frequency phase velocity but it has not been proven that these are stable global solutions. Actually in all the cases where this has been checked explicitely so far, these local configurations have been shown not to be part of global stable solutions.
\end{itemize}

It is also worth noting that the potential existence of superluminal propagation is not restricted to theories which break the gauge symmetry. For instance massless spin-3/2 are also known to propagate superluminal modes on some non-trivial backgrounds~\cite{Henneaux:2013vca}.

\epubtkImage{}{%
\begin{figure}[htb]
  \centerline{
    \includegraphics[width=7.5cm]{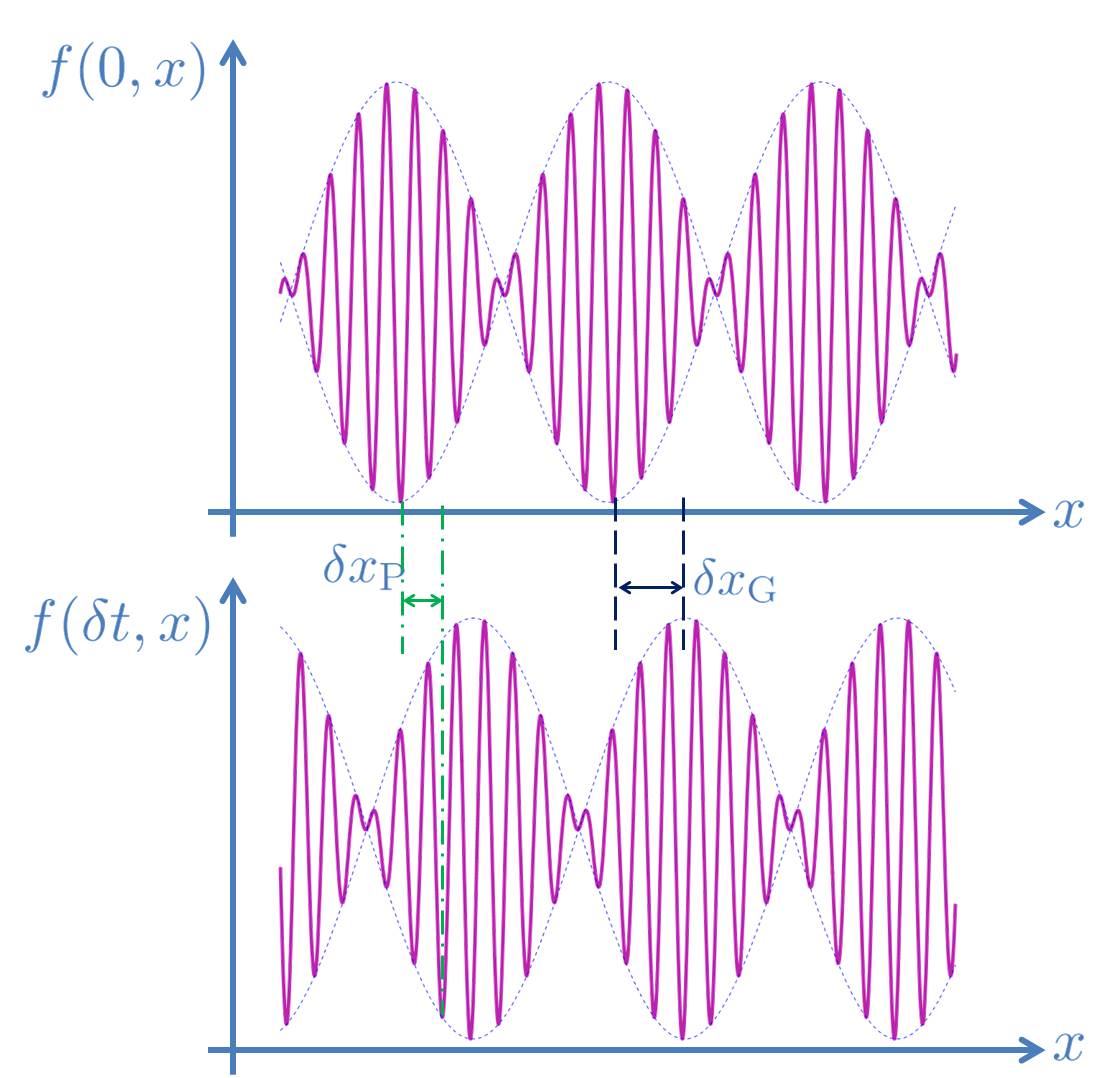}\quad
    \includegraphics[width=7.5cm]{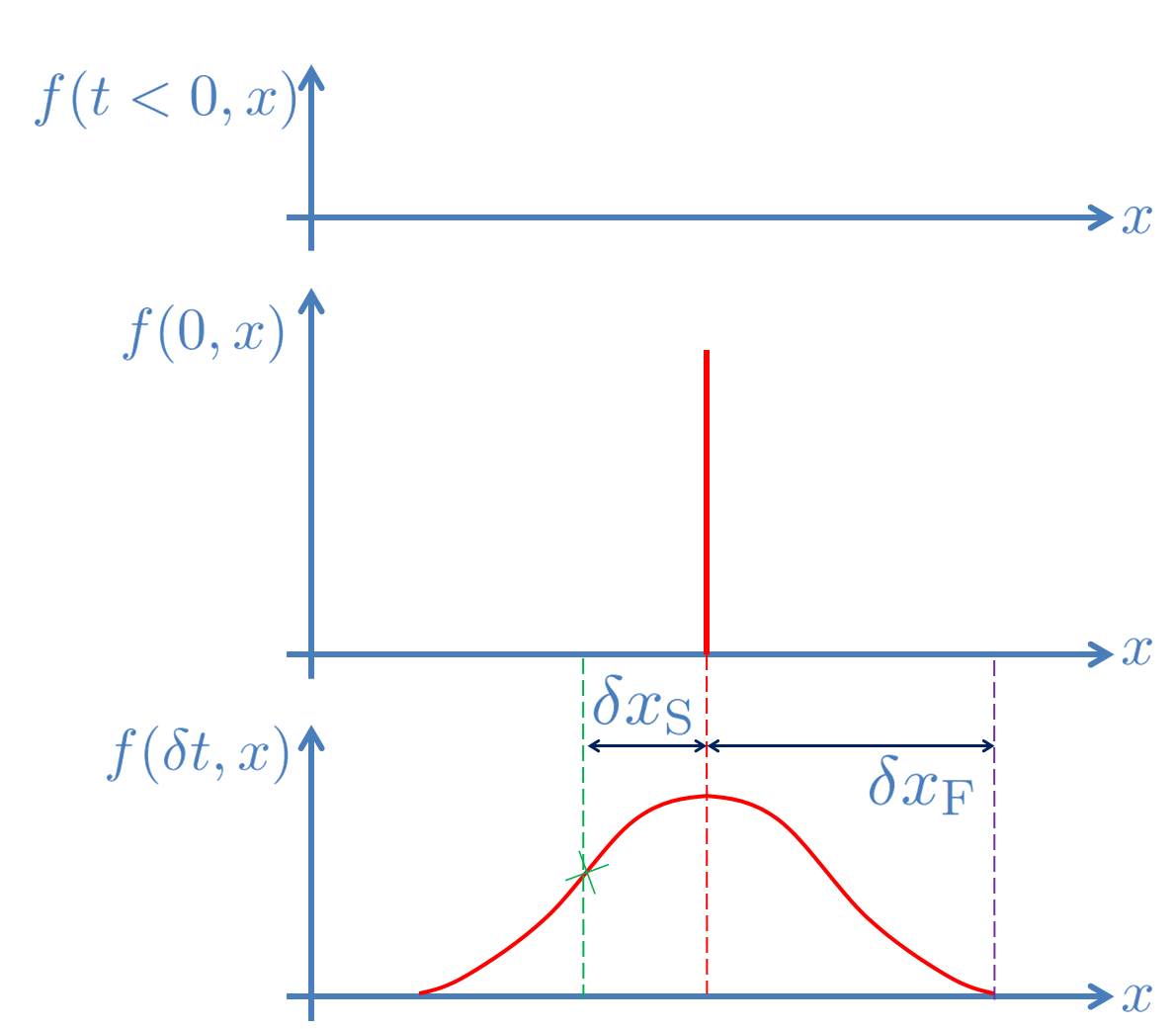}
  }
  \caption{Difference between phase, group, signal and front velocities. At $t=\delta t$, the phase and group velocities are represented on the left and given respectively by $v_{\rm phase}= \delta x_P/\delta t$ and $v_{\rm group}= \delta x_G/\delta t$ (in the limit $\delta t \to 0$.) The signal and front velocity represented on the right are given by $v_{\rm signal}=\delta x_S/\delta t$ (where $\delta x_S$ is the point where at least half the intensity of the original signal is reached.) The front velocity is given by $v_{\rm front}= \delta x_F/\delta t$.}
  \label{Fig:PhaseGroupVelocity}
\end{figure}}

\subsubsection{Superluminalities in Galileons}
\label{sec:SLGal}

Superluminalities in Galileon and other closely related theories have been pointed out in several studies for more a while~\cite{Nicolis:2008in,Adams:2006sv,Goon:2010xh,Evslin:2011vh,Curtright:2012gx,deFromont:2013iwa,Garcia-Saenz:2013gya}. Note also that Ref.~\cite{Hinterbichler:2009kq} was the first work to point out the existence of superluminal propagation in the higher-dimensional picture of DGP rather than in its purely four-dimensional decoupling limit. See also Refs.~\cite{Creminelli:2010ba,Creminelli:2012my,Hinterbichler:2012fr,Hinterbichler:2012yn,Easson:2013bda,Elder:2013gya} for related discussions on super- versus sub- luminal propagation in conformal Galileon and other DBI-related models.

The physical interpretation of these superluminal propagations was studied in other non-Galileon models in~\cite{Dubovsky:2005xd,Babichev:2007dw} and see~\cite{Dvali:2012zc,Vikman:2012bx} for their potential connection with classicalization~\cite{Dvali:2010bf,Dvali:2010jz,Dvali:2011nj,Alberte:2012is}.

In all the examples found so far what has been pointed out is the existence of a superluminal group velocity, which is the regime inspected is the same as the phase velocity. As we will be seen below (see Section~\ref{sec:duality}), in the one example where we can compute the phase velocity for momenta at which loops ought to be taken into account, we find (thanks to a dual description) that the corresponding front velocity is exactly luminal even though the low-energy group velocity is superluminal. This is no indications that all Galileon theories are causal but it comes to show how a specific Galileon theory which exhibits superluminal group velocity in some regime is dual to a causal theory.

In most of the cases considered superluminal propagation was identified in a spherically symmetric setting in the vicinity of a localized mass as was presented in Section~\ref{sec:SSSGalileon}. To convince the reader that these superluminalities are independent of the coupling to matter, we show here how superluminal propagation can already occur in the vacuum in any Galileon theories without even the need of any external matter.

Consider an arbitrary quintic Galileon
\ba
\L=\pi \sum_{n=1}^4 \frac{c_{n+1}}{\Lambda^{3(n-1)}} \L_n(\Pi)\,,
\ea
where the $\L_n$ are given in~\eqref{L1}-\eqref{L4} and we choose the canonical normalization $c_2=1/12$. One can check that any plane-wave configuration of the form
\ba
\label{pi_0 plane wave}
\pi_0(x^\mu)=F(x^1-t)\,,
\ea
is a solution of the vacuum equations of motion for any arbitrary function $F$,
\ba
\sum_{n=1}^4\frac{(n+1)c_n}{\Lambda^{3(n-1)}}\L_n(\Pi_0)=0\,,
\ea
with $\Pi_{0\, \mu\nu}=\p_\mu\p_\nu \pi_0$,
since $\L_n(\p_\mu\p_\nu \pi_0)=0$ for any $n\ge1$ for a plane-wave of the form~\eqref{pi_0 plane wave}.

Now considering perturbations riding on top of the plane-wave, $\pi(x^\mu)=\pi_0(t,x^1)+\delta \pi(x^\mu)$, these perturbations see an effective background-dependent metric similarly as in Section~\ref{sec:EffCoupling} and have the linearized equation of motion
\ba
Z^{\mu \nu}(\pi_0) \p_\mu \p_\nu \delta \pi=0\,,
\ea
with $Z^{\mu\nu}$ given in~\eqref{Z}
\ba
Z^{\mu\nu}(\pi_0)&=& \sum_{n=0}^3\frac{(n+1)(n+2)c_{n+2}}{\Lambda^{3 n}}X^{(n)\, \mu\nu}(\Pi_0)\\
&=&\left[\eta^{\mu\nu}-\frac{12c_3}{\Lambda^3}F''(x^1-t)(\delta^\mu_0+\delta^\mu_1)(\delta^\nu_0+\delta^\nu_1)\right]\,.
\ea
A perturbation traveling along the direction $x^1$ has a velocity $v$ which satisfies
\ba
Z^{00}v^2+2Z^{01}v+Z^{11}=0\,.
\ea
So depending on wether the perturbation travels with or against the flow of the plane wave, it will have a velocity $v$ given by
\ba
v=-1\qquad \text{or}\quad v=\frac{1-\frac{12c_3}{\Lambda^3}F''(x^1-t)}{1+\frac{12c_3}{\Lambda^3}F''(x^1-t)}\,.
\ea
So a plane wave which admits\epubtkFootnote{If $c_3=0$, we can easily generalize the background solution to find other configurations that admit a superluminal propagation.}
\ba
12c_3 F''<-\Lambda^3\,,
\ea
the perturbation propagates with a superluminal velocity. However this velocity corresponds to the group velocity and in order to infer whether or not there is any acausality we need to derive the front velocity, which is the large momentum limit of the phase velocity. The derivation presented here presents a tree-level calculation and to compute the large momentum limit one would need to include loop corrections. This is especially important as $12c_3 F''\to -\Lambda^3$ as the theory becomes (infinitely) strongly coupled at that point~\cite{Burrage:2011cr}. So far no computation has properly taken these quantum effects into account, and the (a)causality of Galileons theories is yet to determined.

\subsubsection{Superluminalities in massive gravity}

The existence of superluminal propagation directly in massive gravity has been pointed out in many references in the literature~\cite{Burrage:2011cr,Gruzinov:2011sq,Deser:2012qx,Deser:2013eua} (see also~\cite{Yu:2013owa} for another nice discussion). Unfortunately none of these studies have qualified the type of velocity which exhibits superluminal propagation. On closer inspection it appears that there again for all the cases cited the superluminal propagation has so far always been computed classically without taking into account quantum corrections. These results are thus always valid for the low frequency group velocity but never for the front velocity which requires a fully fledged calculation beyond the tree-level classical approximation \cite{Shore:2007um}.

Furthermore while it is very likely that massive gravity admits superluminal propagation, to date there is no known consistent solution of massive gravity which has been shown to admit superluminal (even of group) velocity. We review the arguments in favor of superluminal propagation in what follows together with their limitations. Notice as well that while a Galileon theory typically admits superluminal propagation on top of static and spherically symmetric Vainshtein solutions as presented in Section~\ref{sec:SSSGalileon}, this is not the case for massive gravity see Section~\ref{sec:SSSMG} and \cite{Berezhiani:2013dw}.

\begin{enumerate}
\item \textbf{\textit{Argument:} Some background solutions of massive gravity admit superluminal propagation.} \\
\textbf{\textit{Limitation of the argument:} the solutions inspected were not physical.}\\
Ref.~\cite{Gruzinov:2011sq} was the first work to point out the presence of superluminal group velocity in the full theory of massive gravity rather than in its Galileon decoupling limit. These superluminal modes ride on top of a solution which is unfortunately unrealistic for different reasons. First the solution itself is unstable. Second the solution has no rest frame (if seen as a perfect fluid) or one would need to perform a superluminal boost to bring the solution to its rest frame. Finally, to exist, such a solution should be sourced by a matter source with complex eigenvalues~\cite{deRham:2011pt}. As a result the solution cannot be trusted in the first place, and so neither can the superluminal propagation of fluctuations about it.
\item \textbf{\textit{Argument:} Some background solutions of the decoupling limit of massive gravity admit superluminal propagation.} \\
\textbf{\textit{Limitation of the argument:} the solutions were only found in a finite region of space and time.}\\
In Ref.~\cite{Burrage:2011cr} superluminal propagation was found in the decoupling limit of massive gravity. These solutions do not require any special kind of matter, however the background has only be solved locally and it has not (yet) been shown whether or not they could extrapolate to sensible and stable asymptotic solutions.
\item \textbf{\textit{Argument:} There are some exact solutions of massive gravity for which the determinant of the kinetic matrix vanishes thus massive gravity is acausal.} \\
\textbf{\textit{Limitation of the argument:} misuse of the characteristics analysis -- what has really been identified is the absence of BD ghost.}\\
Ref.~\cite{Deser:2012qx} presented some solutions which appeared to admit some \emph{instantaneous modes} in the full theory of massive gravity. Unfortunately the results presented in~\cite{Deser:2012qx} were due to a misuse of the characteristics analysis.

  The confusion in the characteristics analysis arises from the very constraint that eliminates the BD ghost. The existence of such a constraint was discussed in length in many different formulations in Section~\ref{sec:Abscence of ghost proof} and it is precisely what makes ghost-free (or dRGT) massive gravity special and theoretically viable. Due to the presence of this constraint, the characteristics analysis should be performed \emph{after} solving for the constraints and not before \cite{Izumi:2013poa}.

  In~\cite{Deser:2012qx} it was pointed out that the determinant of the time kinetic matrix vanished in ghost-free massive gravity \emph{before} solving for the constraint. This result was then interpreted as the propagation of instantaneous modes and it was further argued that the theory was then acausal. This result is simply an artefact of not properly taking into account the constraint and performing a characteristics analysis on a set of modes which are not all dynamical (since two phase space variables are constrained by the primary and secondary constrains~\cite{Hassan:2011hr,Hassan:2011ea}). In other word it is precisely what would--have--been the BD ghost which is responsible for canceling the determinant of the time kinetic matrix. This does not mean that the BD ghost propagates instantaneously but rather that the BD ghost is not present in that theory, which is the very point of the theory.

  One can show that the determinant of the time kinetic matrix in general does not vanish when computing it {\bf after} solving for the constraints.  In summary the results presented in~\cite{Deser:2012qx} cannot be used to deduce the causality of the theory or absence thereof.\vspace{10pt}

\item \textbf{\textit{Argument:} Massive gravity admits shock wave solutions which admit superluminal and instantaneous modes.} \\
\textbf{\textit{Limitation of the argument:} These configurations lie beyond the regime of validity of the classical theory.}\\
  Shock wave local solutions on top of which the fluctuations are superluminal were found in~\cite{Deser:2013eua}.  Furthermore a characteristic analysis reveals the possibility for spacelike hypersurfaces to be characteristic. While interesting, such configurations lie beyond the regime of validity of the classical theory and quantum corrections ought to be included.

  Having said that, it is likely that the characteristic analysis performed in \cite{Deser:2013eua} and then in \cite{Deser:2013qza} would give the same results had it been performed on regular solutions\footnote{We thank the authors of \cite{Deser:2013eua,Deser:2013qza} for pointing this out.}. This point is discussed below.\vspace{10pt}

\item \textbf{\textit{Argument:} The characteristic analysis shows that some field configurations of massive gravity admit superluminal propagation and the possibility for spacelike hypersurfaces to be characteristic.} \\
\textbf{\textit{Limitation of the argument:} Same as point 2. Putting this limitation aside this result is certainly correct {\it classically} and in complete agreement with previous results presented in the literature (see point 2 where local solutions were given).}\\
Even though the characteristic analysis presented in~\cite{Deser:2013eua} used shock wave local configurations, it is also  valid for smooth wave solutions which would be within the regime of validity of the theory. In \cite{Deser:2013qza} the characteristic analysis for a shock wave was presented again and it was argued that CTCs were {\it likely} to exist.

To better see the essence behind the general characteristic analysis argument, let us look at the (simpler yet representative) case of a Proca field with an additional quartic interaction as explored in \cite{Ong:2013qja,Velo:1970ur},
\ba
\label{Proca lambda}
\L=-\frac 14 F\mn^2-\frac 12 m^2 A^\mu A_\mu-\frac 14 \lambda (A^\mu A_\mu)^2\,.
\ea
The idea behind the characteristic analysis is to ``{\it replace the highest derivative terms $\p^N A$ by $k^N \tilde A$}"~\cite{Ong:2013qja} so that one of the equations of motion is
\ba
 \left[(m^2+\lambda A^\nu A_\nu)k^\alpha k_\alpha+2\lambda (A^\nu k_\nu )^2 \right]k^\mu \tilde A_\mu =0  \,.
\ea
When $\lambda\ne 0$, one can solve this equation maintaining  $k^\mu \tilde A_\mu \ne 0$.
Then there are certainly field configurations for which the normal to the characteristic surface is timelike and thus the mode with $k^\mu \tilde A_\mu \ne 0$ can propagate superluminally in this Proca field theory. However as we shall see below this very combination $\mathcal{Z}=\left[(m^2+\lambda A^\nu A_\nu)k^\alpha k_\alpha+2\lambda (A^\nu k_\nu )^2 \right]=0$ with $k^\mu$ timelike (say $k^\mu=(1,0,0,0)$) is the coefficient of the time-like kinetic term of the helicity-0 mode. So one can never have $\left[(m^2+\lambda A^\nu A_\nu)k^\alpha k_\alpha+2\lambda (A^\nu k_\nu )^2 \right]=0$ with $k^\mu=(1,0,0,0)$ (or any timelike direction) without automatically having an infinitely strongly helicity-0 mode and thus automatically going beyond the regime of validity of the theory (see Ref.~\cite{Burrage:2011cr} for more details.)

To see this more precisely, let us perform the characteristic analysis in the \stu language.  An analysis performed in unitary gauge is of course perfectly acceptable, but to connect with previous work in Galileons and in massive gravity the \stu formalism is useful.

In the \stu language, $A_\mu \to A_\mu +m^{-1}\p_\mu \pi$, keeping track of the terms quadratic in $\pi$, we have
\ba
\L_\pi^{(2)}&=&-\frac 12 Z^{\mu\nu}\p_\mu \pi \p_\nu \pi\,,\\
{\rm with}\hspace{10pt} Z^{\mu\nu}[A_\mu]&=& \eta^{\mu\nu}+\frac{\lambda}{m^2} A^2 \eta^{\mu\nu}+2 \frac{\lambda}{m^2}A^\mu A^\nu\,.
\ea
It is now clear that the combination found in the characteristic analysis $\mathcal{Z}$ is nothing other than
\ba
\mathcal{Z}\equiv Z^{\mu\nu}k_\mu k_\nu\,,
\ea
where $Z^{\mu\nu}$ is the kinetic matrix of the helicity-0 mode. Thus a configuration with $\mathcal{Z}=0$ with $k^\mu=(1,0,0,0)$ implies that the $Z^{00}$ component of helicity-0 mode kinetic matrix vanishes. This means that the conjugate momentum associated to $\pi$ cannot be solved  for in this time-slicing, or that the helicity-0 mode is infinitely strongly coupled.

This result should sound familiar as it echoes what has already been shown to happen in the decoupling limit of massive gravity, or here of the Proca field theory (see \cite{Babichev:2007dw,Vikman:2007sj} for related discussions in that case). Considering the decoupling limit of \eqref{Proca lambda} with $m\to 0$ and $\hat \lambda = \lambda/m^4 \to $ const, we obtain a decoupled massless gauge field and a scalar field,
\ba
\L_{\rm DL} = -\frac 14 F\mn^2 -\frac 12 (\p \pi)^2-\frac{\hat \lambda}{4} (\p \pi)^4\,.
\ea
For fluctuations about a given background configuration $\pi=\pi_0(x)+\delta \pi$, the fluctuations see an effective metric $\tilde Z^{\mu\nu}(\pi_0)$ given by
\ba
\tilde Z^{\mu\nu}(\pi_0)=\(1+\hat \lambda(\p \pi_0)^2\) \eta^{\mu\nu}+2\hat \lambda \p^\mu \pi_0 \p^\nu \pi_0\,.
\ea
Of course unsurprisingly, we find $\tilde  Z^{\mu\nu}(\pi_0)\equiv m^{-2} Z^{\mu\nu}[m^{-1}\p_\mu \pi_0]$. The fact that we can find superluminal or instantaneous propagation in the characteristic analysis is equivalent to the statement that in the decoupling limit there exists classical field configurations for $\pi_0$ for which the fluctuations propagate superluminally (or even instantaneously). Thus the results of the characteristic analysis are in agreement with previous results in the decoupling limit as was pointed out for instance in \cite{Adams:2006sv,Nicolis:2008in,Burrage:2011cr}.

Once again, if one starts with a field configuration where the kinetic matrix is well defined, one cannot reach a region where one of the eigenvalues of $Z^{\mu\nu}$ crosses zero without going beyond the regime of validity of the theory as described in~\cite{Burrage:2011cr}.  See also Refs.~\cite{Hollowood:2007kt,Shore:2007um} for the use of the characteristic analysis and its relation to (micro-)causality.
\end{enumerate}

  The presence of instantaneous modes in some (self-accelerating) solutions of massive gravity was actually pointed out from the very beginning. See Refs.~\cite{deRham:2010tw} and~\cite{Koyama:2011wx} for an analysis of self-accelerating solutions in the decoupling limit, and~\cite{DeFelice:2012mx} for self-accelerating solutions in the full theory (see also~\cite{Gratia:2012wt} for a complementary analysis of self-accelerating solutions.) All these analysis had already found instantaneous modes on some self-accelerating branches of massive gravity. However as pointed out in all these analysis, the real question is to establish whether or not these solutions lie within the regime of validity of the EFT, and whether one could reach such solutions with a finite amount of energy and while remaining within the regime of validity of the EFT.

    This aspect connects with Hawking's chronology protection argument which is already in effect in GR~\cite{Hawking:1991nk,Hawking:1991pk}, (see also~\cite{Visser:1992py} and~\cite{Visser:1995cc} for a comprehensive review). This argument can be extended to Galileon theories and to massive gravity as was shown in Ref.~\cite{Burrage:2011cr}.

 It was pointed out in~\cite{Burrage:2011cr} and in many other preceding works
 that there exists local backgrounds in Galileon theories and in massive gravity which admit superluminal and instantaneous propagation. (As already mentioned, in point 2. above in massive gravity it is however unclear whether these localized backgrounds admit stable and consistent global realizations). The worry with superluminal propagation is that it could imply the presence of  CTCs (closed timelike curves). However when `cranking up' the background sufficiently so as to reach a solution which would admit CTCs, the Galileon or the helicity-0 mode of the graviton becomes inevitably infinitely strongly coupled. This means that the effective field theory used breaks down and the background becomes unstable with arbitrarily fast decay time before any CTC can ever be formed.\\

{\bf Summary:} Several analyses have confirmed the existence of local configurations admiting superluminalities in massive gravity.
 At this point, we leave it to the reader's discretion to decide whether the existence of local classical  configurations which admit superluminalities and instantaneous propagation means that the theory should be discarded. We  bear in mind the following considerations:
\begin{itemize}
\item No stable global solutions have been found with the same properties.
\item No CTCs  can been constructed within the regime of validity of the theory.
As shown in Ref.~\cite{Burrage:2011cr}  CTCs constructed with these configurations always lie beyond the regime of validity of the theory. Indeed in order to create a CTC, a mode needs to become instantaneous.  As soon as a mode becomes instantaneous, the regime of validity of the classical theory is null and classical considerations are thus obsolete.
\item Finally and most importantly, all the results presented so far for Galileons and Massive Gravity (including the ones summarized here), rely on classical configurations. As was explained at the beginning of this section causality is determined by the front velocity for which classical considerations break down. Therefore no classical calculations can ever prove or disprove the (a)causality of a theory.
\end{itemize}

 \subsubsection{Superluminalities vs Boulware--Deser ghost vs Vainshtein}

 We finish by addressing what would be an interesting connection between the presence of superluminalities and the very constraint of massive gravity which removes the BD ghost  which was pointed out in \cite{Deser:2012qx,Deser:2013eua,Deser:2013qza}. Actually one can show that the presence of local configurations which admits superluminalities is generic to any theories of massive gravity, including DGP, Cascading gravity, non-Fierz-Pauli massive gravity and even other braneworld models and is not specific to the presence of a constraint which removes the BD ghost. For instance consider a theory of massive gravity for which the cubic interactions about flat spacetime different than that of the ghost-free model of massive gravity. Then as shown in section \ref{sec:BDghost} (for instance Eqns.~\eqref{FP1} or~\ref{FP2}, see also \cite{Creminelli:2005qk,Deffayet:2005ys}) the decoupling limit analysis leads to terms of the form
 \ba
 \L_{{\rm FP},\, \pi}=-\frac 12 (\p \pi)^2 + \frac{4}{\mpl m^4} \([\Pi][\Pi^2]-[\Pi^3]\)\,.
 \ea
As we have shown earlier, results from this decoupling limit are in full agreement with a characteristic analysis.

The plane wave solutions provided in \eqref{pi_0 plane wave} is still a vacuum solution in this case.  Following the same analysis as that provided in section \ref{sec:SLGal}, one can easily find modes propagating with superluminal group and phase velocity for appropriate choices of functions $F(x^1-t)$ (while keeping within the regime of validity of the theory.)

Alternatively, let us look a background configuration $\bar \pi$ with  $\bar \Pi=\p^2 \bar \pi$.  Without loss of generality at any point $x$ one can diagonalize the matrix $\bar \Pi$. Focusing on a mode traveling along the $x^1$ direction with momentum $k_\mu=\(k_0,k_1,0,0\)$, we find the dispersion relation
\ba
\label{eqk}
\(k_0^2-k_1^2\)+\frac{16}{\mpl m^4} \(k_0-k_1\)^2 \left[k_1^2\(\bar \Pi^0_{\ 0}+\bar \Pi^2_{\ 2}+\bar \Pi^3_{\ 3}\)+k_0^2\(\bar \Pi^1_{\ 1}+\bar \Pi^2_{\ 2}+\bar \Pi^3_{\ 3}\)
+2 k_0 k_1 \bar \Pi^\mu_{\ \mu}\right]\qquad \\=0\,.\nn
\ea
The presence of higher power in $k$ is nothing else but the signal of the BD ghost about generic backgrounds where $\bar \Pi\ne 0$. Performing a characteristic analysis at this point would focus on the higher powers in $k$ which are intrinsic to the ghost. One can follow instead the non-ghost mode which is already present even when $\bar \Pi= 0$. To follow this mode it is therefore sufficient to perform a perturbative analysis in $k$.  \eqref{eqk} can always be solved for $k_0=k_1$ as well as  for
\ba
k_0=-k_1 +\frac{32}{\mpl m^4}k_1^3\(\bar\Pi^0_{\ 0}-\bar\Pi^1_{\ 1}\)+\mathcal{O}\(\frac{k_1^5 \bar \Pi^2}{m^8\mpl}\)\,.
\ea
We can therefore always find a configuration for which $k_0^2>k_1^2$ at least perturbatively which is sufficient to imply the existence of superluminalities. Even if this calculation was performed perturbatively, it still implies the presence classical  superluminalities like in the previous analysis of Galileon theories or ghost-free massive gravity.

As a result the presence of local solutions in massive gravity which admit superluminalities is not connected to the constraint that removes the BD ghost.  Rather it is likely that the presence of superluminalities could be tied to the Vainshtein mechanism (with flat asymptotic boundary conditions), which as we have seen is crucial for these types of theories (see Refs.~\cite{Adams:2006sv,Hinterbichler:2009kq} and~\cite{deFromont:2013iwa} for a possible connection.) More recently the presence of superluminalities has also been connected to the idea of classicalization which is tied to the Vainshtein mechanism \cite{Dvali:2012zc,Vikman:2012bx}. It is possible that the only way these superluminalities could make sense is through this idea of classicalization. Needless to say this is very much speculative at the moment. Perhaps the Galileon dualities presented below could help understanding these open questions.

\subsection{Galileon duality}
\label{sec:duality}

The low strong coupling scale and the presence of superluminalities raises the question of how to understand the theory beyond the redressed strong coupling scale, and whether or not the superluminalities are present in the front velocity.

A non-trivial map between the conformal Galileon and the DBI conformal Galileon was recently presented in~\cite{Creminelli:2013fxa} (see also~\cite{Bellucci:2002ji}). The conformal Galileon side admits superluminal propagation while the DBI side of the map is luminal. Since both sides are related by a `simple' field redefinition which does not change the physics, and cannot change the causality of the theory, this suggests that the superluminalities encountered in that example must be in the group velocity rather than the front velocity.

Recently another Galileon duality was proposed in~\cite{Curtright:2012gx} and~\cite{deRham:2013hsa} by use of simple Legendre transform. First encountered within the decoupling limit of bi-gravity~\cite{Fasiello:2013woa} the duality can be seen as being related to the freedom in how to introduce the \stu fields. However the duality survives independently from bi-gravity and could be significant in the context of massive gravity.

To illustrate this duality we start with a full Galileon in $d$ dimensions as in~\eqref{Full Galileon}
\ba
\label{Full Galileon 2}
S=\int \d^d x \(\pi \sum_{n=1}^{d} \frac{c_{n+1}}{\Lambda^{3(n-1)}}\L_n[\Pi]\)\,,
\ea
and perform the field redefinition
\ba
\pi(x) \qquad&\to& \qquad\rho(\tilde x)= - \pi(x) -\frac 1{2\Lambda^3} \(\frac{\p \pi(x)}{\p x^\mu} \)^2\\
x^\mu \qquad&\to& \qquad\tilde x^\mu = x^\mu +\eta^{\mu\nu}\frac{1}{\Lambda^3}\frac{\p \pi(x)}{\p x^\nu}\,,
\ea
This transformation is fully invertible without requiring any inverse of derivatives,
\ba
\rho(\tilde x) \qquad&\to& \qquad\pi(x)= - \rho(\tilde x) -\frac 1{2\Lambda^3} \(\frac{\p \rho(\tilde x)}{\p \tilde x^\mu} \)^2\\
\tilde x^\mu \qquad&\to& \qquad x^\mu = \tilde x^\mu +\eta^{\mu\nu}\frac{1}{\Lambda^3}\frac{\p \rho(\tilde x)}{\p \tilde x^\nu} \,,
\ea
so the field transformation is not non-local (at least not in the traditional sense) and does not hide degrees of freedom.

In terms of the dual field $\rho(\tilde x)$, the Galileon theory~\eqref{Full Galileon 2} is nothing other than another Galileon with different coefficients,
\ba
\label{Dual Galileon}
S=\int \d^d \tilde x \(\rho(\tilde x) \sum_{n=1}^{d} \frac{p_{n+1}}{\Lambda^{3(n-1)}}\L_n[\Sigma]\)\,,
\ea
with $\Sigma\mn=\p^2 \rho(\tilde x)/ \p \tilde x^\mu\p \tilde x^\nu$ and the new coefficients are given by~\cite{deRham:2013hsa}
\ba
p_n=\frac 1n \sum_{k=2}^{d+1}(-1)^kc_k\frac{k(d-k+1)!}{(n-k)! (d-n+1)!}\,.
\ea
This duality thus maps a Galileon to another Galileon theory with different coefficients. In particular this means that the free theory $c_{n>2}=0$ maps to another non-trivial $(d+1)^{\rm th}$ order Galileon theory with $p_n\ne 0$ for any $2\le n \le d+1$. This dual Galileon theory admits superluminal propagation precisely in the same way as was pointed out on the spherically symmetric configurations of Section~\ref{sec:SSSGalileon} or on the plane wave solutions of Section~\ref{sec:SLGal}. Yet this non-trivial Galileon is dual to a free theory which is causal and luminal by definition.

What was computed in these examples for a non-trivial Galileon theory (and in all the examples known so far in the literature) is only the tree-level group velocity valid till the (redressed) strong coupling scale of the theory. Once hitting the (redressed) strong coupling scale the loops need to be included. In the dual free theory however there are no loops to account for, and thus the result of luminal velocity in that free theory is valid at all scale and has to match the front velocity. This is strongly suggestive that the front velocity in that example of non-trivial Galileon theory is luminal and the theory is causal even though it exhibits a superluminal group velocity.

It is clear at this point that a deeper understanding of this class of theories is required. We expect this will be the subject of further studies. In the rest of this review we focus on some phenomenological aspects of massive gravity before presenting other theories of massive gravity.

\newpage
\part{Phenomenological Aspects of Ghost-free Massive Gravity}
\label{part:pheno}

\section{Phenomenology}
\label{sec:pheno}

Below we summarize some of the phenomenology of massive gravity and DGP. Many other interesting results have been derived in the literature, including the implication for the very Early Universe. For instance false vacuum decay and the Hartle--Hawking no-boundary proposal was studied in the context of massive gravity in~\cite{Zhang:2012ap,Sasaki:2013nka,Zhang:2013pna} where it was shown that the graviton mass could increase the rate. The implications of massive gravity to the cyclic Universe were also studied in Ref.~\cite{Cai:2012ag} with a regular bounce.

\subsection{Gravitational waves}

\subsubsection{Speed of propagation}

If the photon had a mass it would no longer propagate at `the speed of light', but at a lower speed. For the photon its speed of propagation is known with such an accuracy in so many different media that it can be used to put the most stringent constraints on the photon mass to~\cite{Beringer:1900zz} $m_{\gamma}< 10^{-18}\mathrm{\ eV}$.
In the rest of this review we will adopt the viewpoint that the photon is massless and that light does indeed propagate at the `speed of light'.

The earliest bounds on the graviton mass were based on the same idea. As described in~\cite{Will:2005va}, (see also~\cite{Maggiore:1999vm}), if the graviton had a mass, gravitational waves would propagate at a speed different than that of light, $v_g^2=1-m^2/E^2$ (assuming a speed of light $c=1$). This different velocity between the light and gravitational waves would manifest itself in observations of supernovae.
Assuming the emission of a gravitational wave with frequency larger than the graviton mass, this could lead to a bound on the graviton mass of $m< 10^{-23}\mathrm{\ eV}$ considering a frequency of 100~Hz and a supernovae located 200~Mpc away~\cite{Will:2005va} (assuming that the photon propagates at the speed of light).

Alternatively, another way to test the speed of gravitational waves and bound the graviton mass without relying on any assumptions on the photon is through the observation of inspiralling compact objects which allows to derive the frequency-dependence of GWs. The detection of GWs in Advanced LIGO could then bound the graviton mass potentially all the way down to $m< 10^{-29}\mathrm{\ eV}$~\cite{Will:2005va,Will:1997bb,Berti:2004bd}.

The graviton mass is also relevant for the production of primordial gravitational waves during inflation. Following the analysis of~\cite{Gumrukcuoglu:2012wt} it was shown that the graviton mass opens up the production of gravitational waves during inflation with a sharp peak with a height and position which depend on the graviton mass. See also \cite{Mohseni:2011vv} for the study of exact plane wave solutions in massive gravity.

Nevertheless, these bounds on the graviton mass are relatively weak compared to the typical value of $m\sim 10^{-30}-10^{-33}\mathrm{\ eV}$ considered till now in this review. The reason for this is because these bounds do not take into account the effects arising from the additional polarization in the gravitational waves which would be present if the graviton had a mass in a Lorentz-invariant theory. For the photon, if it had a mass, the additional polarization would decouple and would therefore be irrelevant (this is related to the absence of vDVZ discontinuity at the classical level for a Proca theory.) In massive gravity however the helicity-0 mode of the graviton couples to matter. As we shall see below, the bounds on the graviton mass inferred from the absence of fifth forces are typically much more stringent.

\subsubsection{Additional polarizations}
\label{sec:AddPol}

One of the predictions of GR is the existence of gravitational waves (GW) with two transverse independent polarizations.

While GWs have not been directly detected via interferometer yet, they have been detected through the spin-down of binary pulsar systems~\cite{Hulse:1974eb,Taylor:1989sw,Weisberg:2004hi}. This detection via binary pulsars does not count as a direct detection, but it matches expectations from GWs with such an accuracy, and for now so many different systems of different relative masses that it seems unlikely that the spin-down could be due to something different than the emission of GWs.

\begin{figure}[]
\begin{center}
\includegraphics[width=13cm]{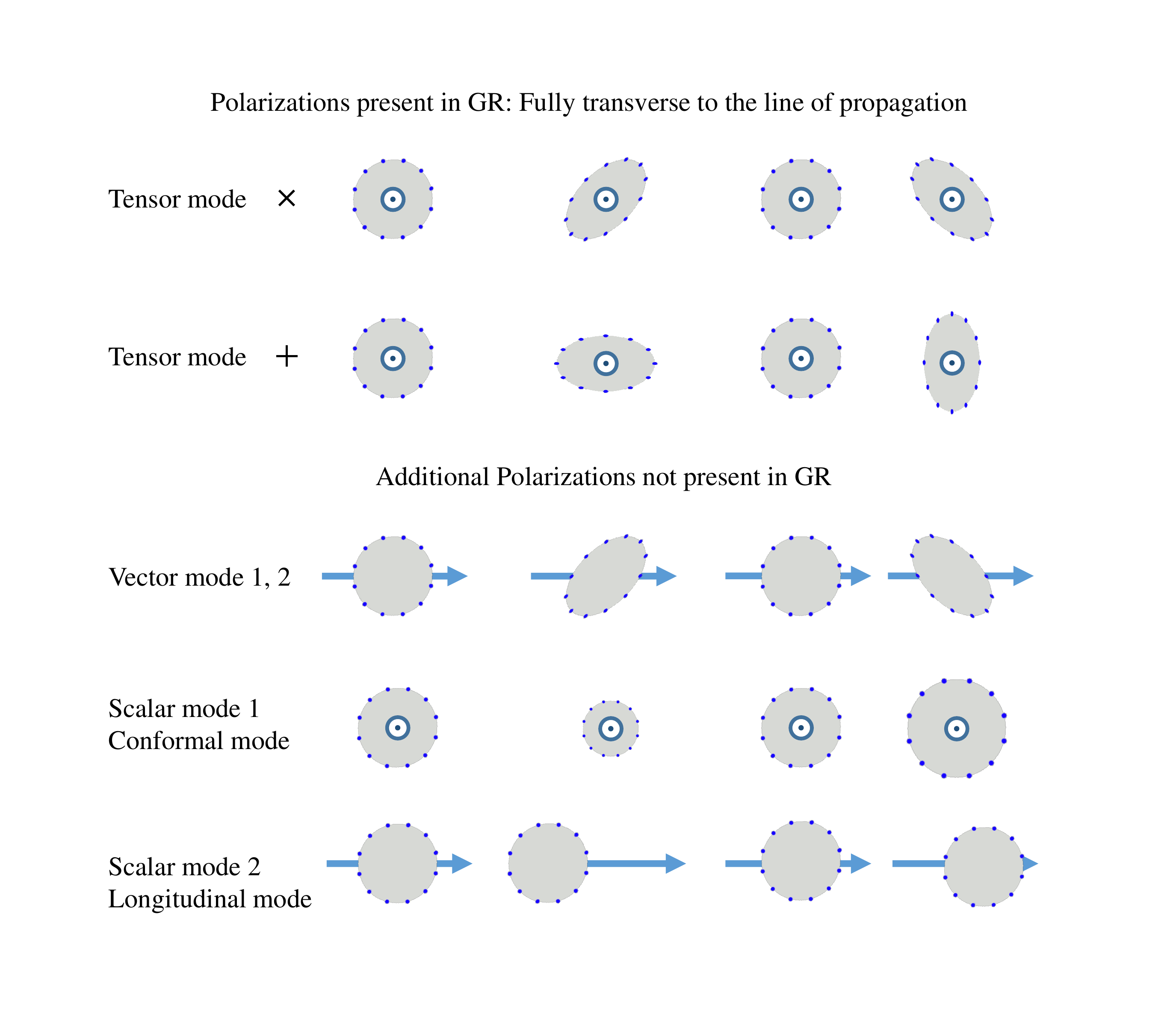}
%
\caption{\label{CdR:fig:GWs}Polarizations of Gravitational Waves in General Relativity and potential additional polarizations in modified gravity.}
\end{center}      
\end{figure}

In a modified theory of gravity, one could expect a total of up to six polarizations for the GWs as seen in Fig.~\ref{CdR:fig:GWs}.

As emphasized in the first part of this review, and particularly in Section~\ref{sec:BDghost}, the sixth excitation, namely the longitudinal one, represents a ghost degree of freedom. Thus if that mode is observed it cannot be arising from a Lorentz-invariant massive graviton. Its presence could be linked for instance to new scalar degrees of freedom which are independent from the graviton itself. In massive gravity, only five polarizations are expected.
Notice however that the helicity-1 mode does not couple directly to matter or external sources, so it is unlikely that GWs with polarizations which mix the transverse and longitudinal directions would be produced in a natural process.

Furthermore, any physical process which is expected to produce GWs would include very dense sources where the Vainshtein mechanism will thus be expected to be active and screen the effect of the helicity-0 mode. As a result the excitation of the breathing mode is expected to be suppressed in any theory of massive gravity which includes an active Vainshtein mechanism.

So while one could in principle expect up to six polarizations for GWs in a modified theory of gravity, in massive gravity only the two helicity-2 polarizations are expected to be produced in a potentially observable amount by interferometers like advanced-LIGO~\cite{Harry:2010zz}. To summarize, in ghost-free massive gravity or DGP we expect the following:

\begin{itemize}
\item The helicity-2 modes are produced in the same way as in GR and would be indistinguishable if they travel distances smaller than the graviton Compton wavelength
\item The helicity-1 modes are not produced
\item The breathing or conformal mode is produced but suppressed by the Vainshtein mechanism and so the magnitude of this mode is suppressed compared to the helicity-2 polarization by many orders of magnitudes.
\item The longitudinal mode does not exist in a ghost-free theory of massive gravity. If such a mode is observed it must be arise from another field independent from the graviton.
\end{itemize}

We will also discuss the implications for indirect detection of GWs via binary pulsar spin-down in Section~\ref{sec:Pulsars}.
We will see that already in these setups the radiation in the breathing mode is suppressed by 8 orders of magnitude compared to that in the helicity-2 mode. In more relativistic systems such as black hole mergers, this suppression will be even bigger as the Vainshtein mechanism is stronger in these cases, and so we do not expect to see the helicity-0 mode component of a GW emitted by such systems.

To summarize, while additional polarizations are present in massive gravity, we do not expect to be able to observe them in current interferometers. However these additional polarizations, and in particular the breathing mode can have larger effects on solar system tests of gravity (see Section~\ref{sec:SolarSystem}) as well as for weak lensing (see Section~\ref{sec:Lensing}), as we review in what follows. They also have important implications for black holes as we discuss in Section~\ref{sec:BH} and in cosmology in Section~\ref{sec:Cosmology}.

\subsection{Solar system}
\label{sec:SolarSystem}

A lot of the phenomenology of massive gravity can be derived from its decoupling limit where it resembles a Galileon theory. Since the Galileon was first encountered in DGP most of the phenomenology was first derived for that model. The extension to massive gravity is usually relatively straightforward with a few subtleties which we mention at the end. We start by reviewing the phenomenology assuming a cubic Galileon decoupling limit, which is directly applicable for DGP and then extend to the quartic Galileon and ghost-free massive gravity.

Within the context of DGP a lot of its phenomenology within the solar system was derived in~\cite{Lue:2002sw,Lue:2005ya} using the full higher-dimensional picture as well as in~\cite{Dvali:2002vf}. In these work the effect from the helicity-0 mode in the advanced of the perihelion were computed explicitly. In particular in~\cite{Dvali:2002vf} it was shown how an infrared modification of gravity could have an effect on small solar system scales and in particular on the Moon. In what follows we review their approach.

Consider a point source of mass $M$ localized at $r=0$.
In GR (or rather Newtonian gravity as it is a sufficient approximation), the gravitational potential mediated by the point source is
\ba
\Psi(r)=-h_{00} =- \frac{M}{4\pi \mpl}\frac{1}{r}=-\frac{1}{\mpl}\frac{r_S}{r}\,,
\ea
where $r_S$ is the Schwarzschild radius associated with the source. Now in a theory of massive gravity the helicity-0 mode of the graviton also contributes to the gravitational potential with an additional amount $\delta \Psi$. As seen in Section~\ref{sec:Vainshtein}, when the Vainshtein mechanism is active the contribution from the helicity-0 mode is very much suppressed. $\delta \Psi \ll \Psi$ but measurements in the Solar system are reaching such a level of accuracy than even a small deviation $\delta \Psi$ could in principle be observable~\cite{Williams:2004qba}.

In the decoupling limit of DGP, matter fields couple to the following perturbed metric
\ba
h\mn^{\rm DGP}= h\mn^{\rm Einstein} + \pi_0 \eta\mn\,,
\ea
where $\pi$ is the helicity-0 mode of the graviton (up to some dimensionless numerical factors which we have set to unity). In massive gravity, matter couples to the following metric (see the discussion in Section~\ref{sec:SSSMG} and~\eqref{DL MG 3}),
\ba
h\mn^{\rm massive\ gravity}= h\mn^{\rm Einstein} + \pi_0 \eta\mn+\frac{\alpha}{\Lc^3}\p_\mu\pi_0 \p_\nu \pi_0\,.
\ea
The deviation $\delta \Psi$ to the gravitational potential is thus given by
\ba
\delta \Psi=- \pi_0\,,
\ea
(notice that in the static and spherically symmetric case $\p_\mu\pi \p_\nu \pi$ leads to no correction to the gravitational potential).

Following~\cite{Dvali:2002vf} we define as $\e$ the fractional change in the gravitational potential
\ba
\e(r)= \frac{\delta \Psi}{\Psi}= \frac{\pi_0(r)}{\mpl}\frac{r}{r_S}\,.
\ea
This change in the Newtonian force implies a change in the motion of a test particle (for instance the Moon) within that gravitational field of the localized mass $M$ (of for instance the Earth) as compared to GR. For elliptical orbits this leads to an additional angular precession of the perihelion due to the force mediated by the helicity-0 mode on top of that of GR.
The additional advanced of the perihelion per orbit is given in terms of $\e$ as
\ba
\delta\phi = \pi R_0 \frac{\d}{\d r}\(r^2 \frac{\d }{\d r}(r^{-1}\e)\)\Big|_{R_0}\,,
\ea
where $R_0$ is the mean orbit radius,
(notice the $\pi$ in that expression is the standard value $\pi=3.14\ldots$ nothing to do with the helicity-0 mode).

\subsubsection*{DGP and cubic Galileon}

In the decoupling limit of DGP (cubic Galileon) $\pi$ was given in~\eqref{cubig Gal ST} and $r^{-1}\e\sim (r/r_*^3)^{1/2}$ where $r_*$ is the strong coupling radius derived in~\eqref{r*}, $r_*=\Lc^{-1}(M/4\pi\mpl)^{1/3}$ leading to an anomalous advance of the perihelion
\ba
\delta \phi\sim \frac{3\pi}{4}\(\frac{r}{r_*}\)^{3/2}\,.
\ea
When the graviton mass goes to zero $\Lc\to \infty$ and the departure from GR goes to zero, this is another example of how the Vainshtein mechanism arises. Interestingly it was pointed out in~\cite{Lue:2002sw,Lue:2005ya} that in DGP the sign of this anomalous angle depends on whether on the branch studied (self-accelerating branch - or normal branch).

For the Earth-Moon system, taking $\Lc^3(m^2\mpl)$ with $m\sim H_0\sim 10^{-33}\mathrm{\ eV}$, this leads to an anomalous precision of the order of~\cite{Dvali:2002vf}
\ba
\delta \phi \sim 10^{-12}\ {\rm rad/orbit}\,,
\ea
which is just on the edge of the level of accuracy currently reached by the lunar laser ranging experiment~\cite{Williams:2004qba} (for instance the accuracy quoted for the effective variation of the Gravitational constant is $(4\pm 9)\times 10^{-13}$/year $\sim (0.5 \pm 1)\times 10^{-11}$/orbit).

As pointed out in~\cite{Dvali:2002vf} and~\cite{Lue:2002sw,Lue:2005ya} the effect could be bigger for the advance of the perihelion of Mars around the Sun, but at the moment the accuracy is slightly less. \\

\subsubsection*{Massive gravity and quartic Galileon:}

As already mentioned in Section~\ref{sec:SSSGalileon}, the Vainshtein mechanism is typically much stronger\epubtkFootnote{This is not to say that perturbations and/or perturbativity do not break down earlier in the quartic Galileon, see for instance Section~\ref{QuarticGalileon} below as well as~\cite{Burrage:2012ja,Berezhiani:2013dw}, which is another sign that the Vainshtein mechanism works better in that case.} in the spherically symmetric configuration of the quartic Galileon and thus in massive gravity (see for instance the suppression of the force given in~\eqref{foceSupQuatric}). Using the same values as before for a quartic Galileon we obtain
\ba
\delta \phi\sim 2\pi \(\frac{r}{r_*}\)^{2}\sim 10^{-16}\, / \text{orbit}\,.
\ea
Furthermore, in massive gravity the parameter that enters this relation is not directly the graviton but rather the graviton mass weighted with the coefficient $\alpha=-(1+3/2\alpha_3)$ which depends on the cubic potential term $\L_3$, assuming that $\alpha_4=-\alpha_3/4$, (see Section~\ref{sec:SSSMG} for more precision)
\ba
\delta \phi\sim 10^{-16}\(\frac{1}{\alpha}\frac{m^2}{(10^{-33}\mathrm{\ eV})^2}\)^{2/3} / \text{orbit}\,,
\ea
This is typically very far from observations unless we are very close to the minimal model.\epubtkFootnote{The minimal model does not have a Vainshtein mechanism~\cite{Renaux-Petel:2014pja} in the static and spherically symmetric configuration so in the limit $\alpha_3\to -1/3$, or equivalently $\alpha\to 0$, we indeed expect an order one correction.}

\subsection{Lensing}
\label{sec:Lensing}

As mentioned previously, one peculiarity of massive gravity not found in DGP nor in a typical Galileon theory (unless we derive the Galileons from a higher-dimensional brane picture~\cite{deRham:2010eu}) is the new disformal coupling to matter of the form $\p_\mu\pi\p_\nu\pi T^{\mu\nu}$ which means that the helicity-0 mode also couples to conformal matter.

In the vacuum, for a static and spherically symmetric configuration the coupling $\p_\mu\pi\p_\nu\pi T^{\mu\nu}$ plays no role. So to the level at which we are working when deriving the Vainshtein mechanism about a point-like mass this additional coupling to matter does not affect the background configuration of the field (see~\cite{deRham:2012ew} for a discussion outside the vacuum, taking into account for the instance the effect of the Earth atmosphere). However it does affect this disformal coupling does affect the effect metric seen by perturbed sources on top of this configuration. This could have some implications for structure formation is to the best of our knowledge have not been fully explored yet, and does affect the bending of light. This effect was pointed out in~\cite{Wyman:2011mp} and the effects to gravitational lensing were explored. We review the key results in what follows and refer to~\cite{Wyman:2011mp} for further discussions (see also~\cite{Sjors:2011iv}).

In GR, the relevant potential for lensing is $\Phi_L= \frac 12 \(\Phi - \Psi\)\sim 2\(r_S/r\)$, where we use the same notation as before, $h_{00}=\Psi$ and $h_{ij}=\Phi \delta_{ij}$. A conformal coupling of the form $\pi \eta\mn$ does not affect this lensing potential but the disformal coupling $\alpha/\Lc^3 \mpl \p_\mu \pi \p_\mu \pi T^{\mu\nu}$ leads to a new contribution $\delta \Phi$ given by
\ba
\delta \Phi=\frac{\alpha}{\Lc^3\mpl}\pi_0'(r)^2\,.
\ea
[Note we use a different notation that in~\cite{Wyman:2011mp}, here $\alpha=1+3\alpha_3$.] This new contribution to the lensing potential leads to an anomalous fractional lensing of
\ba
\mathcal{R}=\frac{\frac 12 \delta \Phi}{\Phi_L}\sim \frac{r}{4r_S} \(\frac{\Lc}{\alpha^{1/3} \mpl}\)\(\frac{M}{4\pi \mpl}\)^{2/3}\,.
\ea
For the bending of light about the Sun, this leads to an effect of the order of
\ba
\mathcal{R}\sim 10^{-11} \(\frac{1}{\alpha}\frac{m^2}{(10^{-33}{\rm eV})^2}\)^{1/3}\,,
\ea
which is utterly negligible. Note that this is a tree-level calculation. When getting at these distances loops ought to be taken into account as well.

At the level of galaxies or clusters of galaxy, the effect might be more tangible. The reason for that is that for the mass of a galaxy, the associated strong coupling radius is not much larger than the galaxy itself and thus at the edge of a galaxy these effects could be stronger. These effects were investigated in~\cite{Wyman:2011mp} where it was shown a few percent effect on the tangential shear caused by the helicity-0 mode of the graviton or of a disformal Galileon considering a Navarro--Frenk--White halo profile, for some parameters of the theory. Interestingly the effect peaks at some specific radius which is the same for any halo when measured in units of the viral radius. Even though the effect is small this peak could provide a smoking gun for such modifications of gravity.

Recently another analysis was performed in Ref.~\cite{Narikawa:2013pjr} where the possibility to
testing theories of modified gravity exhibiting the Vainshtein mechanism against
observations of cluster lensing was explored. In such theories, like in massive gravity, the second derivative of the field can be large at the transition between the screened and unscreened region, leading to observational signatures in cluster lensing.

\subsection{Pulsars}
\label{sec:Pulsars}

One of the main predictions of massive gravity is the presence of new polarizations for GWs. While these new polarization might not be detectable in GW interferometers as explained in Section~\ref{sec:AddPol}, we could still expect them to lead to detectable effects in the binary pulsar systems whose spin-down is in extremely good agreement with GR. In this section we thus consider the power emitted in the helicity-0 mode of the graviton in a binary-pulsar system. We use the effective action approach derived by Goldberger and Rothstein in~\cite{Goldberger:2004jt} and start with the decoupling limit of DGP before exploring that of ghost-free massive gravity and discussing the subtleties that arise in that case.  We mainly focus on the monopole and quadrupole radiation although the whole formalism can be derived for any multipoles We follow the derivation of Refs.~\cite{deRham:2012fw,deRham:2012fg}, see also Refs.~\cite{Chu:2012kz,Andrews:2013qva} for related studies.

In order to account for the Vainshtein mechanism into account we perform a similar background-perturbation split as was performed in Section~\ref{sec:Vainshtein}. The source is thus split as $T=T_0+\delta T$ where $T_0$ is a static and spherically source representing the total mass localized at the center of mass and $\delta T$ captures the motion of the companions with respect to the center of mass.

This matter profile leads to a profile for the helicity-0 mode (here mimicked as a cubic Galileon which is the case for DGP) as in~\eqref{split} as
$\pi=\pi_0(r)+\phi$,  where the background $\pi_0(r)$ has the same static and spherical symmetry as $T_0$ and so has the same profile as in Section~\ref{sec:SSSGalileon}.


The background configuration $\pi_0(r)$ of the field was derived in~\eqref{pi_0 SSS} where $M$ accounts in this case for the total mass of both companions and $r$ is the distance to the center of mass.
Following the same procedure, the fluctuation $\phi$ then follows a modified Klein--Gordon equation
\ba
Z(\pi_0) \p^2_x \phi(x)=0\,,
\ea
where the Vainshtein mechanism is fully encoded in the background dependent prefactor  $Z(\pi_0)\sim 1+ \p^2 \pi_0/\Lambda^3$ and $Z(\pi_0)\gg 1$ in the vicinity of the binary pulsar system (well within the strong coupling radius defined in~\eqref{r*}.)

Expanding the field in spherical  harmonics the mode functions satisfy
\ba
Z(\pi_0) \p_x^2\left[u_\ell (r) Y_{\ell m}(\Omega)e^{- i \omega t}\right]=0\,,
\ea
where the modes are normalized so as to satisfy the standard normalization in the WKB region, for $r\gg \omega^{-1}$.

The total power emitted via the field $\pi$ is given by the sum over these mode functions,
\ba
P^{(\pi)}=\sum_{\ell=0}^\infty P^{(\pi)}_\ell=\sum_{\ell=0}^\infty\sum_{m=-\ell}^\ell\sum_{n\ge0}(n \Omega_P)\Big|
\frac{1}{\mpl T_P}\int_0^{T_P}\d^4xu_\ell(r)Y_{\ell,m} e^{-i n \Omega_P t} \delta T \Big|^2\,,
\ea
where $T_P$ is the orbital period of the binary system and $\Omega_P=2\pi/T_p$ is the corresponding angular velocity. $P^{(\pi)}_0$ is the power emitted in the monopole, $P^{(\pi)}_1$ in the dipole $P^{(\pi)}_2$ in the quadrupole of the field $\pi$ uniquely, etc\ldots in addition to the standard power emitted in the helicity-2 quadrupole channel of GR.

Without the Vainshtein mechanism, the mode functions would be the same as for a standard free-field in flat space-time, $u_\ell\sim\frac{1}{r \sqrt{\pi \omega}}\, \cos(\omega r)$ and the power emitted in the monopole would be larger than that emitted in GR, which would be clearly ruled out by observations. The Vainshtein mechanism is thus crucial here as well for the viability of DGP or ghost-free massive gravity.

\subsubsection*{Monopole}

Taking the prefactor $Z(\pi_0)$ into account, the zero mode for the monopole is given instead by
\ba
u_0(r)\sim \frac{1}{(\omega r_*^3)^{1/4}}\(1-\frac{(\omega r)^2}{4}+\cdots\)\,,
\ea
in the strong coupling regime $r\ll \omega^{-1}\ll r_*$ which is the region where the radiation would be emitted.
As a result, the power emitted in the monopole channel through the field $\pi$ is given by~\cite{deRham:2012fw}
\ba
P^{(\pi)}_{0}= \kappa \frac{(\Omega_P \bar r)^4}{(\Omega_P r_*)^{3/2}}\frac{\mathcal{M}^2}{\mpl^2}\Omega_P^2\,,
\ea
where $\mathcal{M}$ is the reduced mass and $\bar r$ is the semi-major axis of the orbit and $\kappa$ is a numerical prefactor of order 1 which depends on the eccentricity of the orbit.

This is to be compared with the Peters--Mathews formula for the power emitted in GR (in the helicity-2 modes) in the quadrupole~\cite{Peters:1963ux},
\ba
P^{\rm (Peters-Mathews)}_2=\tilde \kappa\, (\Omega_P \bar r)^4 \frac{\tilde{\mathcal{M}}^2}{\mpl^2}\Omega_P^2\,,
\ea
where $\tilde \kappa$ is again a different numerical prefactor which depends on the eccentricity of the orbit, and $\tilde{\mathcal{M}}$ is a different combination of the companion masses, when both masses are the same (as is almost the case for the Hulse--Taylor pulsar), $\mathcal{M}=\tilde{\mathcal{M}}$.

We see that the radiation in the monopole is suppressed by a factor of $(\Omega_P r_*)^{-3/2}$ compared with the GR result. For the Hulse--Taylor pulsar this is a suppression of 10 orders of magnitudes which is completely unobservable (at best the precision of the GR result is of 3 orders of magnitude).

Notice however that the suppression is far less than what was naively anticipated from the static approximation in Section~\ref{sec:SSSGalileon}.

The same analysis can be performed for the dipole emission with an even larger suppression of about 19 orders of magnitude compared the Peters--Mathews formula.

\subsubsection*{Quadrupole}

The quadrupole emission in the field $\pi$ is slightly larger than the monopole. The reason  is that energy conservation makes the non-relativistic limit of the monopole radiation irrelevant and one needs to  take the first relativistic correction into account to emit in that channel. This is not so for the quadrupole as it does not correspond to the charge associated with any Noether current even in the non-relativistic limit.

In the non-relativistic limit, the mode function for the quadrupole is
simply \newline $u_2(r)\sim (\omega r)^{3/2}/(\omega r_*^3)^{1/4}$ yielding a quadrupole emission
\ba
P^{(\pi)}_2 = \bar \kappa \frac{(\Omega_P \bar r)^3}{(\Omega_P r_*)^{3/2}}\frac{\bar{\mathcal{M}}^2}{\mpl^2}\Omega_P^2\,,
\ea
where $\bar \kappa$ is another numerical factor which depends on the eccentricity of the orbit and $\bar{\mathcal{M}}$ another reduced mass.
The Vainshtein suppression in the quadrupole is $(\Omega_P r_*)^{-3/2}(\Omega \bar r)^{-1}\sim 10^{-8}$ for the Hulse--Taylor pulsar, and is thus well below the limit of being detectable.

\subsubsection*{Quartic Galileon}
\label{QuarticGalileon}

When extending the analysis to more general Galileons or to massive gravity which includes a quartic Galileon, we expect a priori by following the analysis of Section~\ref{sec:SSSGalileon}, to find a stronger Vainshtein suppression. This result is indeed correct when considering the power radiated in only one multipole. For instance in a quartic Galileon, the power emitted in the field $\pi$ via the quadrupole channel is suppressed by 12 orders  of magnitude compared the GR emission.

However this estimation does not account for the fact that there could be many multipoles contributing with the same strength in a quartic Galileon theory~\cite{deRham:2012fg}.

In a quartic Galileon theory, the effective metric in the strong coupling radius for a static and spherically symmetric background is
\ba
Z\mn \d x^\mu \d x^\nu\sim \(\frac{\pi_0'}{\Lambda^3 r}\)^2 \(- \d t^2 +\d r^2 +r^2_*\d \Omega^2\)\,,
\ea
the fact that the angular direction is not suppressed by $r^2$ but rather by a constant $r_*^2$ implies that the multipoles are no longer suppressed by additional powers of velocity as is the case in GR or in the cubic Galileon. This implies that many multipoles contribute with the same strength, yielding a potentially large results. This is a sign that perturbation theory is not under control on top of this static and spherically symmetric background and one should really consider a more realistic background which will resume some of these contributions.

In situations where there is a large hierarchy between the mass of the two objects (which is the case for instance within the solar system), perturbation theory can be seen to remain under control and the power emitted in the quartic Galileon is completely negligible.

\subsection{Black holes}
\label{sec:BH}

As in any gravitational theory, the existence and properties of black holes are crucially important for probing the non-perturbative aspects of gravity. The celebrated black hole theorems of GR play a significant role in guiding understanding of non-perturbative aspects of quantum gravity. Furthermore the phenomenology of black holes is becoming increasingly important as understanding of astrophysical black holes increases.

Massive gravity and its extensions certainly exhibit black hole solutions and if the Vainshtein mechanism is successful then we would expect solutions which look arbitrary close to the Schwarzschild and Kerr solutions of GR. However, as in the case of cosmological solutions, the situation is more complicated due to the absence of a unique static spherically symmetric solution that arises from the existence of additional degrees of freedom, and also the existence of other branches of solutions which may or may not be physical. There are a handful of known exact solutions in massive gravity~\cite{Nieuwenhuizen:2011sq,Koyama:2011xz,Koyama:2011yg,Gruzinov:2011mm,Comelli:2011wq,Berezhiani:2011mt,Volkov:2012wp,Cai:2012db,Volkov:2013roa,Tasinato:2013rza,Arraut:2013bqa,Kodama:2013rea}, but the most interesting and physically relevant solutions probably correspond to the generic case where exact analytic solutions cannot be obtained. A recent review of black hole solutions in bi-gravity and massive gravity is given in~\cite{Volkov:2013roa}.

An interesting effect was recently found in the context of bi-gravity in Ref.~\cite{Babichev:2013una}. In that case the Schwarzschild solutions were shown to be unstable (with a Gregory-Laflamme type of instability \cite{Gregory:1993vy,Gregory:1994bj}) at a scale dictated by the graviton mass, i.e. the instability rate is of the order of the age of the Universe. See also Ref.~\cite{Babichev:2014oua} where the analysis was generalized to the non-bidiagonal. In this more general situation, spherically symmetric perturbations were also found but generically no instabilities.

Since all black holes solutions of massive gravity arise as decoupling limits $M_f \rightarrow \infty$ of solutions in bi-gravity\epubtkFootnote{In taking this limit, it is crucial that the second metric $f_{\mu\nu}$ be written in a locally inertial coordinate system, \ie  a system which is locally Minkowski. Failure to do this will lead to the erroneous conclusion that massive gravity on Minkowski is not a limit of bi-gravity.}, we can consider from the outset the bi-gravity solutions and consider the massive gravity limit after the fact. Let us consider then the bi-gravity action expressed as
\ba
S &=& \frac{\mpl^2}{2} \int \d^4 x \sqrt{-g} R[g]+\frac{M_f^2}{2} \int \d^4 x \sqrt{-f} R[f]  \\
&+&\frac{m^2 M_{\rm eff}^2}{4}\int \d^4 x \sqrt{-g} \sum_{n} \frac{\beta_n}{n!} {\L}_n(\sqrt{\mathbb{X}}) + \text{Matter}\nn\,,
\ea
where $M_{\rm eff}^{-2}= \mpl^{-2}+M_f^{-2}$. Here the definition is such that in the limit $M_f \rightarrow \infty$ the $\beta_n$'s correspond the usual expressions in massive gravity. We may imagine matter coupled to both metrics although to take the massive gravity limit we should imagine black holes formed from matter which exclusively couples to the $g$ metric.

One immediate consequence of working with bi-gravity, is that since the $g$ metric is sourced by polynomials of $\sqrt{\mathbb{X}}=\sqrt{g^{-1}f}$ whereas the $f$ metric is sourced by polynomials of $\sqrt{f^{-1}g}$. We thus require that $\mathbb{X}$ is invertible away from curvature singularities. This is equivalent to saying that the eigenvalues of $g^{-1}f$ and $f^{-1}g$ should not pass through zero away from a curvature singularity. This in turn means that if one metric is diagonal and admits a horizon, the second metric if it is diagonal must admit a horizon at the same place, \ie  two diagonal metrics have common horizons. This is a generic observation that is valid for any theory with more than one metric~\cite{Deffayet:2011rh} regardless of the field equations. Equivalently this implies that if $f$ is a diagonal metric without horizons, \eg Minkowski spacetime, then the metric for a black hole must be non-diagonal when working in unitary gauge. This is consistent with the known exact solutions. For certain solutions it may be possible by means of introducing \stu fields to put both metrics in diagonal form, due to the \stu fields absorbing the off-diagonal terms. However for the generic solution we would expect that at least one metric to be non-diagonal even with \stu fields present.

Working with a static spherically symmetric ansatz for both metrics we find in general that bi-gravity admits Schwarzschild-(anti) de Sitter type metrics of the form (see~\cite{Volkov:2013roa} for a review)
\ba
&& \d s_g^2 = -D(r) \d t^2 + \frac{1}{D(r)} \d r ^2+ r^2 \d \Omega^2 \, , \\
&& \d s_f^2=-\Delta(U) \d T^2+ \frac{1}{\Delta(U)} \d U ^2+ U^2 \d \Omega^2 \, ,
\ea
where
\ba
&&D(r) = 1- \frac{2M}{8 \pi \mpl^2 r}- \frac{1}{3}\Lambda_g r^2 \, , \\
&& \Delta(U) = 1- \frac{1}{3}\Lambda_f U^2 \, ,
\ea
are the familiar metric functions for de Sitter and Schwarzschild de Sitter.

The $f$-metric coordinates are related to those of the $g$ metric by (in other words the profiles of the \stu fields)
\be
 U =u r \, , \quad \,  T=u t - u \int \frac{D(r)-\Delta(U)}{D(r)\Delta(U)} \d r \, ,
\ea
where the constant $u$ is given by
\be
u=- \frac{2 \beta_2}{\beta_3}\pm  \frac{1}{\beta_3} \sqrt{4 \beta_2^2-6 \beta_1 \beta_3} \, .
\ee
Finally the two effective cosmological constants that arise from the mass terms are
\ba
\Lambda_g = -\frac{m^2 M_{\rm eff}^2}{\mpl^2}\left( 6 \beta_0 + 2 \beta_1 u + \frac{1}{2} \beta_2 u^2 \right) \, ,\\
\Lambda_f = -\frac{m^2 M_{\rm eff}^2}{M_f^2 u^2 }\left( \frac{1}{2} \beta_2 + \frac{1}{2} \beta_3 u + \frac{1}{4} \beta_4 u^2 \right) \, .
\ea
In this form we see that in the limit $M_f \rightarrow \infty$ we have $\Lambda_f \rightarrow 0$ and $M_{\rm eff} \rightarrow \mpl$ and then these solutions match onto the known exact black holes solutions in massive gravity in the absence of charge~\cite{Nieuwenhuizen:2011sq,Koyama:2011xz,Koyama:2011yg,Gruzinov:2011mm,Comelli:2011wq,Berezhiani:2011mt,Volkov:2012wp,Volkov:2013roa,Tasinato:2013rza,Arraut:2013bqa,Kodama:2013rea}. Note in particular that for every set of $\beta_n$'s there are two branches of solutions determined by the two possible values of $u$.

These solutions describe black holes sourced by matter minimally coupled to metric $g$ with mass $M$. An obvious generalization is to assume that the matter couples to both metrics, with effective masses $M_1$ and $M_2$ so that
\ba
&&D(r) = 1- \frac{2M_1}{8 \pi \mpl^2 r}- \frac{1}{3}\Lambda_g r^2 \, , \\
&& \Delta(U) = 1 -\frac{2M_2}{8 \pi M_f^2 U}- \frac{1}{3}\Lambda_f U^2 \, .
\ea
Although these are exact solutions, not all of them are stable for all values and ranges of parameters and in certain cases it is found that the quadratic kinetic term for various fluctuations vanishes indicating a linearization instability which means these are not good vacuum solutions. On the other hand neither are these the most general black hole like solutions; the general case requires numerical analysis to solve the equations which is a subject of ongoing work (see, e.g.,~\cite{Volkov:2012wp}). We note only that in~\cite{Volkov:2012wp} a distinct class of solutions is obtained numerically in bi-gravity for which the two metrics take the diagonal form
\ba
&& \d s_g^2 = -Q(r)^2 \d t^2 + \frac{1}{N(r)^2} \d r ^2+ r^2 \d \Omega^2 \, , \\
&& \d s_f^2=-A(r)^2 \d T^2+ \frac{U'(r)^2}{Y(r)^2} \d r ^2+ U(r)^2 \d \Omega^2 \, ,
\ea
where $Q,N,A,Y,U$ are five functions of radius that are numerically obtained solutions of five differential equations. According to the previous arguments about diagonal metrics~\cite{Deffayet:2011rh} these solutions do not correspond to black holes in the massive gravity on Minkowski limit $M_f \rightarrow \infty$, however the limit $M_f \rightarrow \infty$ can be taken and they correspond to black hole solutions in a theory of massive gravity in which the reference metric is Schwarzschild (--\,de Sitter or anti-de Sitter). The arguments of~\cite{Deffayet:2011rh} are then evaded since the reference metric itself admits a horizon.

%
%
%
%
%
%
%
%
%
%
%
%
%
%
%
%
%
%
%
%
%
%
%
%
%

\section{Cosmology}
\label{sec:Cosmology}

One of the principal motivations for considering massive theories of gravity is their potential to address, or at least provide a new perspective on, the issue of cosmic acceleration as already discussed in Section~\ref{sec:HigherDim Scenarios}. Adding a mass for the graviton keeps physics at small scales largely equivalent to GR because of the Vainshtein mechanism. However it inevitably modifies gravity in at large distances, \ie in the Infrared. This modification of gravity is thus most significant for sources which are long wavelength. The cosmological constant is the most infrared source possible since it is build entirely out of zero momentum modes and for this reason we may hope that the nature of a cosmological constant in a theory of massive gravity or similar infrared modification is changed.

There have been two principal ideas for how massive theories of gravity could be useful for addressing the cosmological constant. On the one hand by weakening gravity in the infrared, they may weaken the sensitivity of the dynamics to an already existing large cosmological constant. This is the idea behind screening or degravitating solutions~\cite{Dvali:2002fz,Dvali:2002pe,ArkaniHamed:2002fu,Dvali:2007kt} (see Section~\ref{sec:degravitation}). The second idea is that a condensate of massive gravitons could form which act as a source for self-acceleration, potentially explaining the current cosmic acceleration without the need to introduce a non-zero cosmological constant (as in the case of the DGP model~\cite{Deffayet:2000uy,Deffayet:2001pu}, see Section~\ref{sec:DGP Selfacceleration}). This idea does not address the `old cosmological constant problem'~\cite{Weinberg:1988cp} but rather assumes that some other symmetry, or mechanism exists which ensures the vacuum energy vanishes.  Given this, massive theories of gravity could potential provide an explanation for the currently small, and hence technically unnatural value of the cosmological constant, by tying it to the small, technically natural, value of the graviton mass.

Thus the idea of screening/degravitation and self-acceleration are logically opposites to each other, but there is some evidence that both can be achieved in massive theories of gravity. This evidence is provided by the decoupling limit of massive gravity to which we review first. We then go on to discuss attempts to find exact solutions in massive gravity and its various extensions.

\subsection{Cosmology in the decoupling limit}

A great deal of understanding about the cosmological solutions in massive gravity theories can be learned from considering the `decoupling limit' of massive gravity discussed in Section~\ref{sec:DL MG}. The idea here is to recognize that locally, \ie in the vicinity of a point, any FLRW geometry can be expressed as a small perturbation about Minkowski spacetime (about $\vec x=0$) with the perturbation expansion being good for distances small relative to the curvature radius of the geometry:
\ba
\d s^2 &=& - \left[ 1- (\dot H + H^2) \vec x{} ^2\right] \d t^2+ \left[ 1- \frac{1}{2} H^2 \vec x{}^2\right] \d \vec x^2\\
&=& \left( \eta_{\mu\nu}+ \frac{1}{\mpl} h_{\mu\nu}^{\rm FLRW}\right) \d x^{\mu} \d x^{\nu} \, .
\ea
In the decoupling limit $\mpl \rightarrow \infty$, $m \rightarrow 0$ we keep the canonically normalized metric perturbation $h_{\mu\nu}$ fixed. Thus the decoupling limit corresponds to keeping $H^2 \mpl$ and $\dot H \mpl$ fixed, or equivalently $H^2/m^2$ and $\dot H/m^2$ fixed. Despite the fact that $H \rightarrow 0 $ vanishes in this limit, the analogue of the Friedmann equation remains nontrivial if we also scale the energy density such that $\rho/\mpl$ remains finite. Because of this fact it is possible to analyze the modification to the Friedmann equation in the decoupling limit\epubtkFootnote{In the context of DGP, the Friedmann equation was derived in Section~\ref{sec:DGP Friedmann} from the full five-dimensional picture, but one would have obtained the correct result if derived instead from the decoupling limit. The reason is the main modification of the Friedmann equation arises from the presence of the helicity-0 mode which is already captured in the decoupling limit. }.

The generic form for the helicity-0 mode which preserves isotropy near $\vec x=0$ is
\ba
\pi = A(t) + B(t) \vec x^2+ \dots \,.
\ea
In the specific case where $\dot B(t)=0$ this also preserves homogeneity in a theory in which the Galileon symmetry is exact, as in massive gravity, since a translation in $\vec x$ corresponds to a Galileon transformation of $\pi$ which leaves invariant the combination $\partial_{\mu} \partial_{\nu} \pi$. In Ref.~\cite{deRham:2010tw} this ansatz was used to derive the existence of both self-accelerating and screening solutions.

\subsubsection*{Friedmann equation in the decoupling limit}

We start with the decoupling limit Lagrangian given in~\eqref{DL MG1}. Following the same notation as in Ref.~\cite{deRham:2010tw} we set $a_n=-c_n/2$, where the coefficients $c_n$ are given in terms of the $\alpha_n$'s in \eqref{coefficients Cn DL}.
The self-accelerating branch of solutions then corresponds to the ansatz
\ba
\pi = \frac{1}{2} q_0 \Lambda_3^3 x_{\mu} x^{\mu}+ \phi \, \\
h_{\mu \nu}= -\frac{1}{2} H_{\rm dS}^2 x_{\mu} x^{\mu}+ \chi_{\mu\nu} \, \\
T_{\mu\nu}= - \lambda \eta_{\mu\nu} + \tau_{\mu\nu} \, ,
\ea
where $\pi,\chi$ and $\tau$ correspond to the fluctuations about the background solution.

For this ansatz, the background equations of motion reduce to
\ba
\label{eq.HdS}
H_{\rm dS}\(a_1 + 2 a_2 q_0+ 3 a_3 q_0^2\) =0 \\
\label{eq.FriedmanDL}
H_{\rm dS}^2 = \frac{\lambda}{3 \mpl^2} + \frac{2 \Lambda_3^3}{\mpl} (a_1 q_0+a_2 q_0^2 + a_3 q_0^3) \,.
\ea
In the `self-accelerating branch' when $H_{\rm dS}\ne 0$,  the first constraint can be used to infer  $q_0$ and the second one corresponds to  the effective Friedmann equation. We see that even in the absence of a cosmological constant $\lambda=0$, for generic coefficients we have a constant $H_{\rm dS}$ solution which corresponds to a self-accelerating de Sitter solution.

The stability of these solutions can be analyzed by looking at the Lagrangian for the quadratic fluctuations
\ba
{\cal L}^{(2)} = - \frac{1}{4} \chi^{\mu \nu}\Ein\mn\abu \chi_{\alpha \beta}+ \frac{6 H_{\rm dS}^2 \mpl}{\Lambda_3^3} ( a_2+ 3 a_3 q_0) \phi \Box \phi+ \frac{1}{\mpl} \chi^{\mu\nu} \tau_{\mu\nu} \,.
\ea
Thus we see that the helicity zero mode is stable provided that
\ba
( a_2+ 3 a_3 q_0) >0 \, .
\ea
However, these solutions exhibit a peculiarity. To this order the helicity-0 mode fluctuations do not couple to the matter perturbations (there is no kinetic mixing between $\pi$  and $\chi\mn$). This means that there is no Vainshtein effect, but at the same time there is no vDVZ discontinuity for the Vainshtein effect to resolve!

\subsubsection*{Screening solution}

Another way to solve the system of equations~\eqref{eq.HdS} and~\eqref{eq.FriedmanDL} is to consider instead flat solutions $H_{\rm dS}=0$. Then~\eqref{eq.HdS} is trivially satisfied and we see the existence of a `screening solution' in the Friedmann equation~\eqref{eq.FriedmanDL}, which can accommodate a cosmological constant without any acceleration. This occurs when the helicity-0 mode `absorbs' the contribution from the cosmological constant $\lambda$, and the background configuration for $\pi$ parametrized by $q_0$ satisfies
\ba
(a_1 q_{0}+a_2 q_{0}^2 + a_3 q_{0}^3)=-\frac{\lambda}{6 \mpl \Lambda_3^3} \, .
\ea
Perturbations about this screened configuration then behave as
\ba
{\cal L}^{(2)} = - \frac{1}{2} \chi^{\mu \nu}\Ein_{\mu\nu}\abu\chi_{\alpha \beta}+ \frac{3}{2}\phi \Box \phi+ \frac{1}{\mpl} (\chi^{\mu\nu} + \eta^{\mu\nu} \phi)\tau_{\mu\nu} \, .
\ea
In this case the perturbations are stable, and the Vainshtein mechanism is present which is necessary to resolve the vDVZ discontinuity. Furthermore since the background contribution to the metric perturbation vanishes $h_{\mu\nu}=0$, they correspond to Minkowski solutions which are sourced by a nonzero cosmological constant. In the case where $a_3=0$ these solutions only exist if $\lambda < \mpl \Lambda_3^3 \frac{3a_1^2}{2a_2}$. In the case where $a_3\ne 0$ there is no upper bound on the cosmological constant which can be screened via this mechanism.

In this branch of solution, the strong coupling scale for fluctuations on top of this configuration becomes of the same order of magnitude as that of the screened cosmological constant. For a large cosmological constant the strong coupling scale becomes to large and the helicity-0 mode would thus not be sufficiently Vainshtein screened.

Thus while these solutions seem to indicate positively that there are self-screening solutions which can accommodate a continuous range of values for the cosmological constant and still remain flat, the range is too small to significantly change the Old Cosmological Constant problem. Nevertheless the considerable difficulty in attacking the old cosmological constant problem means that these solutions deserve further attention as they also provide a proof of principle on how Weinberg's no go could be evaded~\cite{Weinberg:1988cp}. We emphasize that what prevents a large cosmological constant from being screened is not an issue in the theoretical tuning but rather an observational bound, so this is already a step forward.

These two classes of solutions are both maximally symmetric. However, the general cosmological solution is isotropic but inhomogeneous. This is due to the fact that a nontrivial time dependence for the matter source will inevitably source $B(t)$, and as soon as $\dot B \neq 0$ the solutions are inhomogeneous. In fact as we now explain in general the full nonlinear solution is inevitably inhomogeneous due to the existence of a no-go theorem against spatially flat and closed FLRW solutions.

\subsection{FLRW solutions in the full theory}

\subsubsection{Absence of flat/closed FLRW solutions}

A nontrivial consequence of the fact that diffeomorphism invariance is broken in massive gravity is that there are no spatially flat or closed FLRW solutions~\cite{DAmico:2011jj}. This result follows from the different nature of the Hamiltonian constraint. For instance, choosing a spatially flat form for the metric $\d s^2 = - \d t^2+ a(t)^2 \d \vec x^2$, the mini-superspace Lagrangian takes the schematic form \be {\cal L} = -3 \mpl^2 \frac{a \dot a^2}{N} + F_1(a) + F_2(a) N \, .
\ee
Consistency of the constraint equation obtained from varying with respect to $N$ and the acceleration equation for $\ddot a$ implies
\ba
\frac{\partial F_1(a)}{\partial t}=\dot a \frac{\partial F_1(a)}{\partial a} =0 \, .
\ea
In GR since $F_1(a)=0$ there is no analogue of this equation. In the present case this equation can be solved either by imposing $\dot a=0$ which implies the absence of any dynamic FLRW solutions, or by solving $\frac{\partial F_1(a)}{\partial a}=0$ for fixed $a$ which implies the same thing. Thus there are no nontrivial spatially flat FLRW solutions in massive gravity in which the reference metric is Minkowski. The result extends also to spatially closed cosmological solutions. As a result different alternatives have been explored in the literature to study the cosmology of massive gravity. See Figure~\ref{Fig:NoFRW} for a summary of these different approaches.

\epubtkImage{}{%
\begin{figure}[htb]
  \centerline{\includegraphics[width=\textwidth]{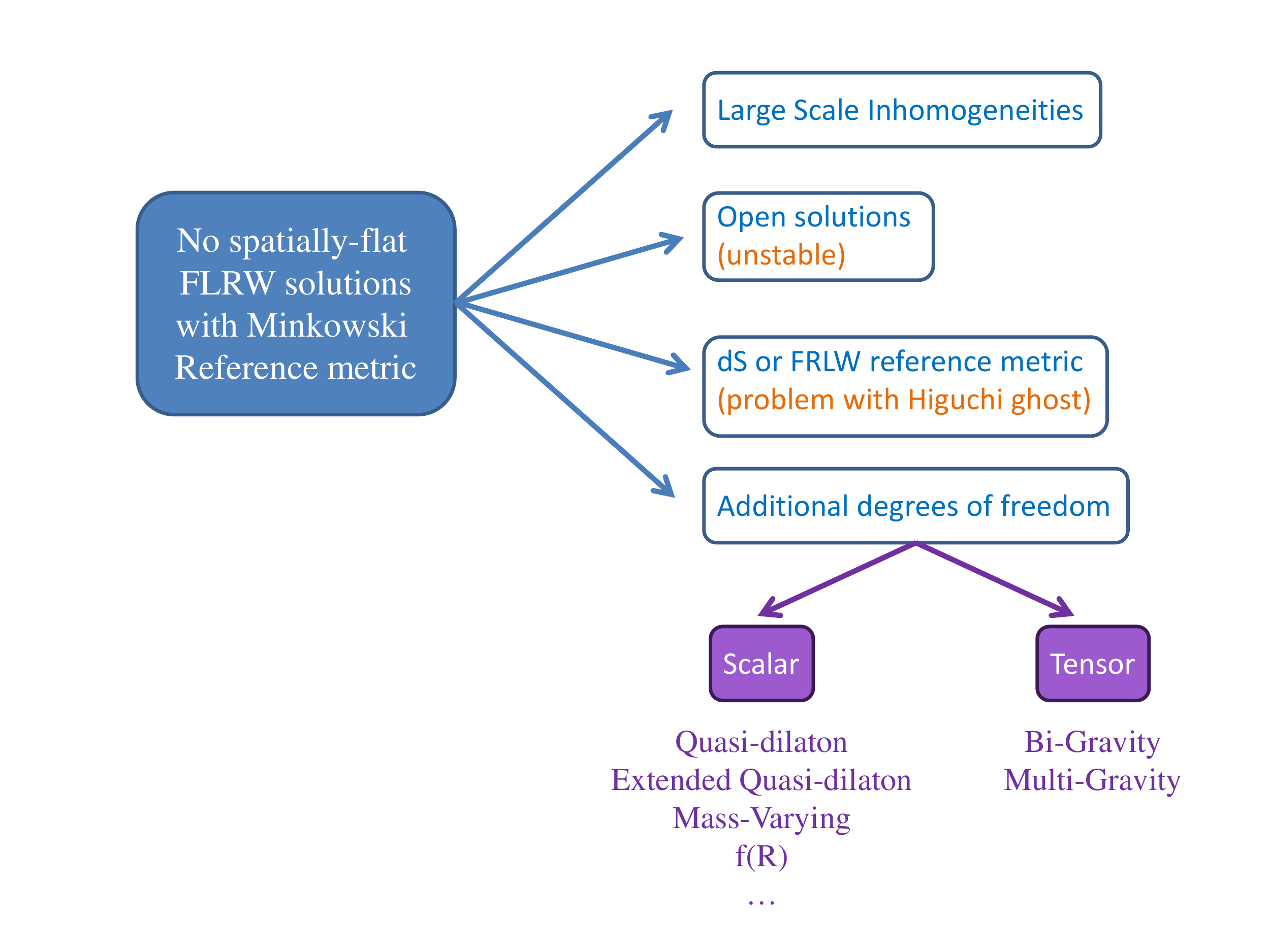}}
  \caption{Alternative ways in deriving the cosmology in massive gravity.}
\label{Fig:NoFRW}
\end{figure}}

\subsubsection{Open FLRW solutions}
\label{sec:openFRW}

While the previous argument rules out the possibility of spatially flat and closed FLRW solutions, open ones are allowed~\cite{Gumrukcuoglu:2011ew}. To see this we make the ansatz
$\d s^2 = - \d t^2+ a(t)^2 \d \Omega_{H^3}^2$
where $\d \Omega_{H^3}^2$ expressed in the form
\ba
\d \Omega_{H^3}^2= \d \vec x^2 - |k| \frac{( \vec x . \d \vec x)^2}{(1+ |k| \vec x^2)} = \frac{\d r^2}{1+|k| r^2} + r^2 \d \Omega_{S^2}^2 \, ,
\ea
is the metric on a hyperbolic space, and express the reference metric in terms of \stu fields
$\tilde f_{\mu\nu} \d x^{\mu} \d x^{\nu} = \eta_{ab} \partial_{\mu} \phi^a\partial_{\nu} \phi^b$ with
\ba
\phi^0 = f(t) \sqrt{1+ |k| \vec x^2} \, , \\
\phi^i = \sqrt{|k|} f(t) x^i \, .
\ea
then the mini-superspace Lagrangian of~\eqref{dRGT_metric} takes the form
\ba
\L_{\rm mGR} &=&-3\mpl^2 |k|N a -3 \mpl^2 \frac{a \dot a^2}{N}
+3 m^2\mpl^2 \Big[2 a^2 \mathcal{X}(2 N a - \dot f a - N \sqrt{|k| }f \nn \\
&+&   \alpha_3 a^2 \mathcal{X}^2 (4 N a - 3\dot f a - N \sqrt{|k| }f )
+ 4 \alpha_4 a^3 \mathcal{X}^3(N- \dot f) \Big] \, , \nn
\ea
with $\mathcal{X}=1- \frac{ \sqrt{|k|} f}{a}$.
In this case the analogue additional constraint imposed by consistency of the Friedmann and acceleration (Raychaudhuri) equation is
\be
  a\mathcal{X}\left( \(3- \frac{2 \sqrt{|k|} f}{a} \) +\frac32 \alpha_3 \(3- \frac{ \sqrt{|k|} f}{a} \) \mathcal{X} +6\alpha_4 \mathcal{X}^2\right)=0 \, .
\ea
The solution for which $\mathcal{X}=0 $ is essentially Minkowski spacetime in the open slicing, and is thus uninteresting as a cosmology.

Focusing on the other branch and assuming $\mathcal{X}\ne 0$, the general solution determines $f(t)$ in terms of $a(t)$ takes the form $f(t) = \frac{1}{\sqrt{|k|}}u a(t)$ where $u$ is a constant determined by the quadratic equation
\ba
 3- 2 u + \frac 32 \alpha_3 \(3- u \) (1-u) +6\alpha_4 (1-u)^2=0 \, .
 \ea
The resulting Friedmann equation is  then
\be
3\mpl^2H^2 - \frac{3 \mpl^2 |k|}{a^2} = \rho +2 m^2 \rho_m \, ,
\ee
where
\be
 \rho_m=- (1-u)\left( 3\(2-u \)
+  \frac 32 \alpha_3 \(4- u\) (1-u)+6\alpha_4 (1-u)^2 \right)  \, .
\ee
Despite the positive existence of open FLRW solutions in massive gravity, there remain problems of either strong coupling (due to absence of quadratic kinetic terms for physical degrees of freedom) or other instabilities which essentially rule out the physical relevance of these FLRW solutions~\cite{Gumrukcuoglu:2011zh,DeFelice:2012mx,Vakili:2012tm}.

\subsection{Inhomogenous/anisotropic cosmological solutions}

As pointed out in~\cite{DAmico:2011jj} the absence of FLRW solutions in massive gravity should not be viewed as an observational flaw of the theory. On the contrary the Vainshtein mechanism guarantees that there exist inhomogeneous cosmological solutions which approximate the normal FLRW solutions of GR as closely as desired in the limit $m \rightarrow 0$. Rather, it is the existence of a new physical length scale $1/m$ in massive gravity, which cause the dynamics to be inhomogeneous at cosmological scales. If this scale $1/m$ is comparable to or larger than the current Hubble radius, then the effects of these inhomogeneities would only become apparent today, with the universe locally appearing as homogenous for most of its history in the local patch which we observe.

One way to understand how the Vainshtein mechanism recovers the prediction of homogeneity and isotropy is to work in the formulation of massive gravity in which the \stu fields are turned on. In this formulation, the \stu fields can exhibit order unity inhomogeneities with the metric remaining approximately homogeneous. Matter which couples only to the metric will perceive an effectively homogenous and anisotropic universe, and only through interaction with the Vainshtein suppressed additional scalar and vector degrees of freedom would it be possible to perceive the inhomogeneities. This is achieved because the metric is sourced by the \stu fields through terms in the equations of motion which are suppressed by $m^2$. Thus as long as $R \gg m^2$ the metric remains effectively homogeneous and isotropic despite the existence of no-go theorems against exact homogeneity and isotropy.

In this regard a whole range of exact solutions have been studied exhibiting these properties~\cite{Koyama:2011wx,Volkov:2011an,Koyama:2011xz,Chamseddine:2011bu,Gratia:2012wt,Kobayashi:2012fz,Tasinato:2012ze,Wyman:2012iw,Volkov:2012zb,Khosravi:2013axa,Volkov:2013roa,Gratia:2013gka,DeFelice:2013bxa,DeFelice:2013awa,DeFelice:2012mx,Tasinato:2013rza,PhysRevD.88.063006}. A generalization of some of these solutions was presented in Ref.~\cite{Motohashi:2012jd} and Ref.~\cite{Gratia:2013uza}.
In particular we note that in~\cite{Volkov:2012cf,Volkov:2012zb} the most general exact solution of massive gravity is obtained in which the metric is homogenous and isotropic with the \stu fields inhomogeneous. These solutions exist because the effective contribution to the stress energy tensor from the mass term (\ie viewing the mass term corrections as a modification to the energy density) remains homogenous and isotropic despite the fact that it is build out of \stu fields which are themselves inhomogeneous.

Let us briefly discuss how these solutions are obtained\epubtkFootnote{See~\cite{Volkov:2013roa} for a recent review and more details. The convention on the parameters $b_n$ there is related to our $\beta_n$'s here via $b_n = - \frac{1}{4} (4-n)! \beta_n$.}. As we have already discussed all solutions of massive gravity can be seen as $M_f \rightarrow \infty$ decoupling limits of bi-gravity. Therefore we may consider the case of inhomogeneous solutions in bi-gravity and the solutions of massive gravity can always be derived as a limit of these bi-gravity solutions. We thus begin with the action
\ba
S &=& \frac{\mpl^2}{2} \int \d^4 x \sqrt{-g} R[g]+\frac{M_f^2}{2} \int \d^4 x \sqrt{-f} R[f]  \\
&+&\frac{m^2 M_{\rm eff}^2}{4}\int \d^4 x \sqrt{-g} \sum_{n} \frac{\beta_n}{n!} {\L}_n(\sqrt{\mathbb{X}}) + \text{Matter}\nn\,,
\ea
where $M_{\rm eff}^{-2}= \mpl^{-2}+M_f^{-2}$ and $\sqrt{\mathbb{X}}=\sqrt{g^{-1}f}$ and we may imagine matter coupled to both $f$ and $g$ but for simplicity let us imagine matter is either minimally coupled to $g$ or it is minimally coupled to $f$.

\subsubsection{Special isotropic and inhomogeneous solutions}

Although it is possible to find solutions in which the two metrics are proportional to each other $f_{\mu\nu}= C^2 g_{\mu\nu}$~\cite{Volkov:2013roa}, these solutions require in addition that the stress energies of matter sourcing $f$ and $g$ are proportional to one another. This is clearly too restrictive a condition to be phenomenologically interesting. A more general and physically realistic assumption is to suppose that both metrics are isotropic but not necessarily homogenous.
This is covered by the ansatz
\ba
\d s^2_g &=& - Q(t,r)^2 \d t^2+ N(t,r)^2 \d r^2 + R^2 \d^2 \Omega_{S^2} \, ,\\
\d s^2_f &=& - (a(t,r) Q(t,r)^2 \d t+ c(t,r) N(t,r) \d r)^2\\
&& + ( c(t,r) Q(t,r) \d t - b(t,r)n(t,r) N(t,r) \d r)^2   \nn + u(t,r)^2 R^2 \d^2 \Omega_{S^2} \, ,
\ea
and $\d^2 \Omega_{S^2}$ is the metric on a unit 2-sphere. To put the $g$ metric in diagonal form we have made use of the one copy of overall diff invariance present in bi-gravity. To distinguish from the bi-diagonal case we shall assume that $c(t,r) \neq 0$. The bi-diagonal case allows for homogenous and isotropic solutions for both metrics which will be dealt with in Section~\ref{sec:bigravityFRW}. The square root may be easily taken to give
\ba
\sqrt{\mathbb{X}} = \(\begin{array}{cccc}
a & c N/Q & 0 & 0\\
-cN/Q & b & 0 & 0\\
0 & 0 & u & 0\\
0 & 0 & 0 & u
\end{array}\)\,,
\ea
which can easily be used to determine the contribution of the mass terms to the equations of motion for $f$ and $g$. This leads to a set of partial differential equations for $Q,R,N,n,c,b$ which in general require numerical analysis. As in GR, due to the presence of constraints associated with diffeomorphism invariance, and the Hamiltonian constraint for the massive graviton, several of these equations will be first order in time-derivatives. This simplifies matters somewhat but not sufficiently to make analytic progress. Analytic progress can be made however by making additional more restrictive assumptions, at the cost of potentially losing the most physically interesting solutions.

\subsubsection*{Effective cosmological constant}

For instance, from the above form we may determine that the effective contribution to the stress energy tensor sourcing $g$ arising from the mass term is of the form
\ba
T_{\rm mass}{}^0_{\ r} = -m^2 \frac{M_{\rm eff}^2}{\mpl^2} \frac{cN}{Q} \left( \frac{3}{2} \beta_1+ \beta_2 u + \frac{1}{4} \beta_3 u^2 \right) \, .
\ea
If we make the admittedly restrictive assumption that the metric $g$ is of the FLRW form or is static, then this requires that $T^0{}_r$=0 which for $c \neq 0$ implies
\ba
\frac{3}{2} \beta_1+ \beta_2 u + \frac{1}{4} \beta_3 u^2 =0\,.
\ea
This should be viewed as an equation for $u(t,r)$ whose solution is
\ba
u(t,r) = u=- \frac{2 \beta_2}{\beta_3}\pm  \frac{1}{\beta_3} \sqrt{4 \beta_2^2-6 \beta_1 \beta_3} \, .
\ea
Then conservation of energy imposes further
\ba
T_{\rm mass}{}^0_{\ 0}-T_{\rm mass}{}^\theta_{\ \theta}\hspace{-5pt}&=&\hspace{-5pt} -m^2 \frac{M_{\rm eff}^2}{\mpl^2} \left( \frac{1}{2} \beta_2 + \frac{1}{4} \beta_3 u\right)\left( (u-a)(u-b)+c^2\right)\nn \\
&=&0 \, ,
\ea
since $u$ is already fixed we should view this generically as an equation for $c(t,r)$ in terms of $a(t,r)$ and $b(t,r)$
 \ba
 (u-a)(u-b)+c^2=0 \, .
 \ea
With these assumptions the contribution of the mass term to the effective stress energy tensor sourcing each metric becomes equivalent to a cosmological constant for each metric $T_{\rm mass}{}^{\mu}{}_{\nu}(g) = - \Lambda_g \delta^{\mu}{}_{\nu}$ and $T_{\rm mass}{}^{\mu}{}_{\nu}(f) = - \Lambda_f \delta^{\mu}{}_{\nu}$ with
\ba
\Lambda_g &=& -\frac{m^2 M_{\rm eff}^2}{\mpl^2}\left( 6 \beta_0 + 2 \beta_1 u + \frac{1}{2} \beta_2 u^2 \right) \, ,\\
\Lambda_f &=& -\frac{m^2 M_{\rm eff}^2}{M_f^2 u^2 }\left( \frac{1}{2} \beta_2 + \frac{1}{2} \beta_3 u + \frac{1}{4} \beta_4 u^2 \right) \, .
\ea
Thus all of the potential dynamics of the mass term is reduced to an effective cosmological constant. Let us stress again that this rather special fact is dependent on the rather restrictive assumptions imposed on the metric $g$ and that we certainly do not expect this to be the case for the most general time-dependent, isotropic, inhomogeneous solution.

\subsubsection*{Massive gravity limit}

As usual we can take the $M_f \rightarrow 0$ limit to recover solutions for massive gravity on Minkowski (if $\Lambda_f \rightarrow 0$) or more generally if the scaling of the parameters $\beta_n$ is chosen so that $\Lambda_f$ and $\left( 6 \beta_0 + 2 \beta_1 u + \frac{1}{2} \beta_2 u^2 \right)$ and hence $\Lambda_g$ remains finite in the limit then these will give rise to solutions for massive gravity for which the reference metric is any Einstein space for which
\be
G_{\mu\nu}(f) = - \Lambda_f f_{\mu\nu} \, .
\ee
For example this includes the interesting cases of de Sitter and anti-de Sitter reference metrics.

Thus for example, assuming no additional matter couples to the $f$ metric, both bi-gravity and massive gravity on a fixed reference metric admit exact cosmological solutions for which the $f$ metric is de Sitter or anti-de Sitter
\ba
\d s^2_g &=& - \d t^2+a(t)^2 \left(\frac{\d r^2}{1-k r^2} + r^2 \d \Omega^2 \right) \\
\d s^2_f &=& - \Delta(U) \d T^2+\left(\frac{\d U^2}{\Delta(U)} + U^2 \d \Omega^2 \right)\,,
\ea
where $\Delta(U)=1 -\frac{\Lambda_f}{3}U^2$, and the scale factor $a(t)$ satisfies
\be
3 \mpl^2 \left( H^2 + \frac{k}{a^2}\right)= \Lambda_g +\rho_M(t)\,,
\ee
where $\rho_M(t)$ is the energy density of matter minimally coupled to $g$, $H = \dot a/a$, and $U,T$ can be expressed as a function of $r$ and $t$ and comparing with the previous representation $U = u r$. The one remaining undetermined function is $T(t,r)$ and this is determined by the constraint that $(u-a)(u-b)+c^2=0$ and the conversion relations
\ba
\sqrt{\Delta(U)} \d T &=&  a(t,r) \d t+ c(t,r) \frac{1}{\sqrt{1-k r^2}} \d r \, ,\\
\frac{1}{\sqrt{\Delta(U)}} \d U&=& c(t,r)  \d t - b(t,r)n(t,r) \frac{1}{\sqrt{1-k r^2}} \d r \, ,\\
 U(t,r) &=& u r \, ,
\ea
which determine $b(t,r)$ and $c(t,r)$ in terms of $\dot T$ and $T'$.
These relations are difficult to solve exactly, but if we consider the special case $\Lambda_f=0$ which corresponds in particular to massive gravity on Minkowski then the solution is
\be
T(t,r) = q \int^t \d t \frac{1}{\dot a} + \left( \frac{u^2}{4 q}+q r^2\right) a \, ,
\ee
where $q$ is an integration constant.

In particular in the open universe case $\Lambda_f=0$, $k=0$, $q=u$, $T = u a \sqrt{1+|k| r^2}$, we recover the open universe solution of massive gravity considered in Section~\ref{sec:openFRW} where for comparison $f(t) =\frac{1}{\sqrt{|k|}} u a(t)$, $\phi^0(t,r) = T(t,r)$ and $\phi^r=U(t,r)$.

\subsubsection{General anisotropic and inhomogeneous solutions}

Let us reiterate again that there are a large class of inhomogeneous but isotropic cosmological solutions for which the effective Friedmann equation for the $g$ metric is the same as in GR with just the addition of a cosmological constant which depends on the graviton mass parameters. However these are not the most general solutions, and as we have already discussed many of the exact solutions of this form considered so far have been found to be unstable, in particular through the absence of kinetic terms for degrees of freedom which implies infinite strong coupling. However all the exact solutions arise from making a strong restriction on one or the other of the metrics which is not expected to be the case in general. Thus the search for the `correct' cosmological solution of massive gravity and bi-gravity will almost certainly require a numerical solution of the general equations for $Q,R,N,n,c,b$, and their stability.

Closely related to this, we may consider solutions which maintain homogeneity, but are an\-iso\-tropic~\cite{Gumrukcuoglu:2012aa,Maeda:2013bha,DeFelice:2013awa}. In~\cite{Maeda:2013bha} the general Bianchi class A cosmological solutions in bi-gravity are studied. There it is shown that the generic anisotropic cosmological solution in bi-gravity asymptotes to a self-accelerating solution, with an acceleration determined by the mass terms, but with an anisotropy that falls of less rapidly than in GR. In particular the anisotropic contribution to the effective energy density redshifts like non-relativistic matter. In~\cite{Gumrukcuoglu:2012aa,DeFelice:2013awa} it is found that if the reference metric is made to be of an anisotropic FLRW form, then for a range of parameters and initial conditions stable ghost free cosmological solutions can be found.

These analyses are ongoing and it has been uncovered that certain classes of exact solutions exhibit strong coupling instabilities due to vanishing kinetic terms and related pathologies. However this simply indicates that these solutions are not good semi-classical backgrounds. The general inhomogeneous cosmological solution (for which the metric is also inhomogeneous) is not known at present, and it is unlikely it will be possible to obtain it exactly. Thus it is at present unclear what are the precise nonlinear completions of the stable inhomogeneous cosmological solutions that can be found in the decoupling limit. Thus the understanding of the cosmology of massive gravity should be regarded as very much work in progress, at present it is unclear what semi-classical solutions of massive gravity  are the most relevant for connecting with our observed cosmological evolution.

\subsection{Massive gravity on FLRW and bi-gravity}

\subsubsection{FLRW reference metric}

One straightforward extension of the massive gravity framework is to allow for modifications to the reference metric, either by making it cosmological or by extending to bi-gravity (or multi-gravity). In the former case, the no-go theorem is immediately avoided since if the reference metric is itself an FLRW geometry, there can no longer be any obstruction to finding FLRW geometries.

The case of massive gravity with a spatially flat FLRW reference metric was worked out in~\cite{Fasiello:2012rw} where it was found that if using the convention for which the massive gravity Lagrangian is~\eqref{LmGR} with the potential given in terms of the coefficient $\beta's$ as in~\eqref{mGR2},
then the Friedmann equation takes the form
\ba
H^2 = -\sum_{n=0}^3 \frac{3(4-n) m^2 \beta_n}{2 n!} \left(\frac{b}{a} \right)^n + \frac{1}{3 \mpl^2} \rho \,.
\ea
Here the dynamical and reference metrics in the form
\ba
 g_{\mu\nu}\d x^{\mu} \d x^{\nu} &=& - \d t^2+ a(t)^2 \d \vec x^2 \, , \\
 f_{\mu\nu}\d x^{\mu} \d x^{\nu} &=& - M^2 \d t^2+ b(t)^2 \d \vec x^2 \,
\ea
and the Hubble constants are related by
\be
\sum_{n=0}^2 \frac{(3-n)\beta_{n+1}}{2 n!} \left( \frac{b}{a}\right)^{n+1} \left( \frac{H}{b}- \frac{H_f}{a}\right) =0\, .
\ee
Ensuring a nonzero ghost-free kinetic term in the vector sector requires us to always solve this equation with \ba
\frac{H}{H_f}= \frac{b}{a}\,,
\ea
so that the Friedmann equation takes the form
\ba
H^2 = -\sum_{n=0}^3 \frac{3(4-n) m^2 \beta_n}{2 n!} \left(\frac{H}{H_f} \right)^n + \frac{1}{3 \mpl^2} \rho \,,
\ea
where $H_f$ is the Hubble parameter for the reference metric.
By itself this Friedmann equation looks healthy in the sense that it admits FLRW solutions that can be made as close as desired to the usual solutions of GR.

However in practice the generalization of the Higuchi consideration~\cite{Higuchi:1986py} to this case leads to an unacceptable bound (see Section~\ref{sec:DL on dS}).

It is a straightforward consequence of the representation theory for the de Sitter group that a unitary massive spin-2 representation only exists in four dimensions for $m^2 \ge 2 H^2$ as was the case in de Sitter. Although this result only holds for linearized fluctuations around de Sitter, its origin as a bound comes from the requirement that the kinetic term for the helicity zero mode is positive, \ie  the absence of ghosts in the scalar perturbations sector. In particular the kinetic term for the helicity-0 mode $\pi$ takes the form \ba
{\cal L}_{\rm helicity-0}  \propto -m^2 (m^2 - 2H^2) (\partial \pi)^2 \, .
\ea
Thus there should exist an appropriate generalization of this bound for any cosmological solution of nonlinear massive gravity for which there an FLRW reference metric.

This generalized bound was worked out in~\cite{Fasiello:2012rw} and takes the form
\ba
-\frac{m^2}{4 \mpl^2 } \frac{H}{H_f} \left[3 \beta_1 + 4 \beta_2 \frac{H}{H_f} + \beta_3 \frac{H^2}{H_f^2}\right] \ge 2 H^2 \, .
\ea
Again by itself this equation is easy to satisfy. However combined with the Friedmann equation we see that the two equations are generically in conflict if in addition we require that the massive gravity corrections to the Friedmann equation are small for most of the history of the universe, \ie during radiation and matter domination
\ba
-\sum_{n=0}^3 \frac{3 (4-n)m^2 \beta_n}{2 n!} \left(\frac{H}{H_f} \right)^n \le H^2
\ea
This phenomenological requirement essentially rules out the applicability of FLRW cosmological solutions in massive gravity with an FLRW reference metric.

This latter problem which is severe for massive gravity with dS or FLRW reference metrics\epubtkFootnote{Notice that this is not an issue in massive gravity with a flat reference metric since the analogue Friedmann equation does not even exist.}, gets resolved in bi-gravity extensions, at least for a finite regime of parameters.

\subsubsection{Bi-gravity}
\label{sec:bigravityFRW}

Cosmological solutions in bi-gravity have been considered in~\cite{Volkov:2011an,vonStrauss:2011mq,Comelli:2011zm,Comelli:2012db,Akrami:2012vf,Volkov:2012cf,Volkov:2013roa,Akrami:2013pna,Akrami:2013ffa,Berg:2012kn}. We keep the same notation as previously and consider the action for bi-gravity as in~\eqref{bigravity} (in terms of the $\beta$'s where the conversion between the $\beta$'s and the $\alpha$'s is given in~\eqref{beta<->alpha})
\ba
\L_{\rm bi-gravity}=\frac{\mpl^2}{2} \sqrt{-g} R[g]+\frac{M_f^2}{2} \sqrt{-f} R[f]\nn \\
+\frac{m^2\mpl^2}{4}\sum_{n=0}^4\frac{\beta_n}{n!}\L_n[\sqrt{\mathbb{X}}]+\L_{\rm matter}[g,\psi_i]\,,
\ea
 assuming that matter only couples to the $g$ metric.  Then the two Friedmann equations for each Hubble parameter take the respective form
\ba
\label{Friedmann Hg}
H^2 &=& -\sum_{n=0}^3 \frac{3(4-n) m^2 \beta_n}{2 n!} \left(\frac{H}{H_f} \right)^n + \frac{1}{3 \mpl^2} \rho\\
\label{Friedmann Hf}
H^2_f&=&-\frac{\mpl^2}{M_f^2} \left[  \sum_{n=0}^3 \frac{3 m^2 \beta_{n+1}}{2 n!} \left(\frac{H}{H_f} \right)^{n-3} \right] \, .
\ea
Crucially the generalization of the Higuchi bound now becomes
\ba
-\frac{m^2}{4} \frac{H}{H_f} \left[ 3\beta_1 + 4 \beta_2 \frac{H}{H_f} + \beta_3 \frac{H^2}{H_f^2}\right] \left[ 1+ \left(\frac{H_f \mpl}{H M_f} \right)^2\right] \ge 2 H^2 \, .
\ea
The important new feature is the last term in square brackets. Although this tends to unity in the limit $M_f \rightarrow \infty$, which is consistent with the massive gravity result, for finite $M_f$ it opens a new regime where the bound is satisfied by having $\left(\frac{H_f \mpl}{H M_f} \right)^2 \gg 1$ (notice that in our convention the $\beta$'s are typically negative). One may show~\cite{Fasiello:2013woa} that it is straightforward to find solutions of both Friedmann equations which are consistent with the Higuchi bound over the entire history of the universe. For example, choosing the parameters $\beta_2= \beta_3=0$ and solving for $H_f$ the effective Friedmann equation for the metric which matter couples to is
\ba
\label{Friedmann bigravity}
H^2 = \frac{1}{6 \mpl^2} \left(\rho(a) + \sqrt{\rho(a)^2 + \frac{12m^4 \mpl^6}{M_f^2}} \right)
\ea
and the generalization of the Higuchi bound is
\ba
\left(1+ \frac{16 M_f^2}{3\mpl^2 \beta_1^2} \frac{H^4}{m^4} \right) > 0 \, .
\ea
which is trivially satisfied at all times. More generally there is an open set of such solutions. The observationally viability of the self-accelerating branch of these models has been considered in~\cite{Akrami:2012vf,Akrami:2013pna} with generally positive results. Growth histories of the bi-gravity cosmological solutions have been considered in~\cite{Berg:2012kn}. However while avoiding the Higuchi bound indicates absence of ghosts, it has been argued that these solutions may admit gradient instabilities in their cosmological perturbations~\cite{Comelli:2012db}.

We should stress again that just as in massive gravity, the absence of FLRW solutions should not be viewed as an inconsistency of the theory with observations, also in bi-gravity these solutions may not necessarily be the ones of most relevance for connecting with observations. It is only that they are the most straightforward to obtain analytically.  Thus cosmological solutions in bi-gravity, just as in massive gravity, should very much be viewed as a work in progress.

\subsection{Other proposals for cosmological solutions}
\label{sec:CosmoOtherProposals}


Finally, we may note that more serious modifications the massive gravity framework have been considered in order to allow for FLRW solutions. These include mass-varying gravity and the quasi-dilaton models~\cite{D'Amico:2012zv,D'Amico:2013kya}. In~\cite{Gumrukcuoglu:2013nza} it was shown that mass-varying gravity and the quasi-dilaton model could allow for stable cosmological solutions but for the original quasi-dilaton theory the self-accelerating solutions are always unstable. On the other hand the generalizations of the quasi-dilaton~\cite{DeFelice:2013dua,DeFelice:2013tsa} appears to allow stable cosmological solutions.

In addition one can find cosmological solutions in non-Lorentz invariant versions of massive gravity~\cite{Comelli:2013tja} (and~\cite{Comelli:2012vz,Comelli:2013txa,Comelli:2013paa}). We can also allow the mass to become dependent on a field~\cite{Wu:2013ii,Leon:2013qh}, extend to multiple metrics/vierbeins~\cite{Tamanini:2013xia}, extensions with $f(R)$ terms either in massive gravity~\cite{Cai:2013lqa} or in bi-gravity~\cite{Nojiri:2012re,Nojiri:2012zu} which leads to interesting self-accelerating solutions. Alternatively one can consider other extensions to the form of the mass terms by coupling massive gravity to the DBI Galileons~\cite{Gabadadze:2012tr,Andrews:2013ora,Andrews:2013uca,Hinterbichler:2013dv}.

As an example, we present here the cosmology of the extension of the quasi-dilaton model considered in~\cite{DeFelice:2013tsa} where the reference metric $\bar f\mn$ is given in~\eqref{fbar} and depends explicitly on the dynamical quasi-dilaton field $\sigma$.

The action takes the familiar form with an additional kinetic term introduced for the quasi-dilaton which respects the global symmetry
\ba
S &=& \mpl^2 \int \d^4 x \sqrt{-g} \left[\frac{1}{2} R- \Lambda - \frac{\omega}{2 \mpl^2} (\partial \sigma)^2  \right. \nn \\ && \left. +\frac{m^2}{4} \(\L_2[\tilde{\K}] +\alpha_3\L_3[\tilde{\K}] +\alpha_4 \L_4[\tilde{\K}] \) \right]  \, ,
\ea
where the tensor $\tilde{\K}$ is given in~\eqref{tilde K}.

The background ansatz is taken as
\ba
\d s^2 &=& - N^2(t) \d t^2 + a(t)^2 \d \vec x{}^2 \, ,  \\
\phi^0 &=& \phi^0(t) \, , \\
\phi^i &=& x^i \, ,\\
\sigma &=& \sigma(t) \, ,
\ea
so that
\ba
&& f_{\mu\nu} \d x^{\mu} \d x^{\nu} = - n(t)^2 \d t^2 + \d \vec x{}^2 \, ,\\
&& n(t)^2 = (\dot \phi^0(t))^2+ \frac{\alpha_{\sigma}}{\mpl^2 m^2} \dot \sigma^2 \, .
\ea
The equation that for normal massive gravity forbids FLRW solutions follows from varying with respect to $\phi^0$ and takes the form
\ba
\partial_t \left( a^4 X(1-X) J \right) =0 \, ,
\ea
where
\ba
X = \frac{e^{\sigma/\mpl}}{a} \qquad \text{and}\qquad
J = 3 + \frac{9}{2}(1-X) \alpha_3 + 6(1-X)^2 \alpha_4 \, .
\ea
As the universe expands $ X(1-X) J \sim 1/a^4$ which for one branch of solutions implies $J \rightarrow 0$ which determines a fixed constant asymptotic value of $X$ from $J=0$.
In this asymptotic limit the effective Friedmann equation becomes
\ba
\left( \frac{3}{2} - \omega \right) H^2 = \Lambda + \Lambda_X
\ea
where
\ba
\Lambda_X= m^2 (X-1)\left[ 6-3X+\frac{3}{2}(X-4)(X-1) \alpha_3 +6 (X-1)^2 \alpha_4\right] \,,
\ea
defines an effective cosmological constant which gives rise to self-acceleration even when $\Lambda=0$ (for $\omega<6$).

The analysis of~\cite{DeFelice:2013tsa} shows that these self-accelerating cosmological solutions are ghost free provided that
\ba
0< \omega < 6\, , \qquad X^2 < \frac{\alpha_{\sigma} H^2}{m_g^2} < r^2 X^2 \ee where \be r= 1+ \frac{\omega H^2}{m^2 X^2 (\frac{3}{2}\alpha_3 (X-1)-2)} \, .
\ea
In particular this implies that $\alpha_{\sigma}>0$ which demonstrates that the original quasi-dilaton model~\cite{D'Amico:2012zv,D'Amico:2012pi} has a scalar (Higuchi type) ghost. The analysis of~\cite{DeFelice:2013dua} confirms these properties in a more general extension of this model.

\newpage

\part{Other Theories of  Massive Gravity}
\label{Part:OtherMG}

\section{New Massive Gravity}
\label{sec:NMG}

\subsection{Formulation}

Independently of the formal development of massive gravity in four dimensions described above, there has been interest in constructing a purely three dimensional theory of massive gravity. Three dimensions are special for the following reason: for a massless graviton in three dimensions there are no propagating degrees of freedom. This follows simply by counting, a symmetric tensor in 3 dimensions has 6 components. A massless graviton must admit a diffeomorphism symmetry which renders 3 of the degrees of freedom pure gauge, and the remaining 3 are non-dynamical due to the associated first class constraints. On the contrary a massive graviton in 3 dimensions has the same number of degrees of freedom as a massless graviton in four dimensions, namely 2. Combining these two facts together, in 3 dimensions it should be possible to construct a diffeomorphism invariant theory of massive gravity. The usual massless graviton implied by diffeomorphism invariance is absent and only the massive degree of freedom remains.

A diffeomorphism and parity invariant theory in three dimensions was given in~\cite{Bergshoeff:2009hq} and referred to as `new massive gravity' (NMG). In its original formulation the action is taken to be
\ba
S_{\rm NMG} = \frac{1}{\kappa^2} \int \d^3 x \sqrt{-g} \left[ \sigma R+ \frac{1}{m^2} \(R_{\mu\nu}R^{\mu \nu}- \frac{3}{8}R^2\)\right] \,,
\ea
where
$\kappa^2 = 1/M_3$ defines the three dimensional Planck mass, $\sigma= \pm 1$ and $m$ is the mass of the graviton. In this form the action is manifestly diffeomorphism invariant and constructed entirely out of the metric $g_{\mu\nu}$. However to see that it really describes a massive graviton, it is helpful to introduce an auxiliary field $f_{\mu \nu}$ which will see below also admits an interpretation as a metric, to give a quasi-bi-gravity formulation
\ba
\label{NMG 2}
S_{\rm NMG}= M_3  \int \d^3 x \sqrt{-g} \left[ \sigma R- q^{\mu\nu}G_{\mu\nu}- \frac{1}{4} m^2 (q_{\mu\nu} q^{\mu\nu}-q^2)\right] \,.
\ea
The kinetic term for $q\mn$ appears from the mixing with $G\mn$.
Although this is not a true bi-gravity theory, since there is no direct Einstein--Hilbert term for $q_{\mu\nu}$, we shall see below that it is a well-defined decoupling limit of a bi-gravity theory, and for this reason it makes sense to think of $q_{\mu\nu}$ as effectively a metric degree of freedom. In this form we see that the special form of $R\mn^2-3/8R^2$ was designed so that $q_{\mu\nu}$ has the Fierz--Pauli mass term. It is now straightforward to see that this corresponds to a theory of massive gravity by perturbing around Minkowski spacetime.
Defining
\ba
g_{\mu\nu} =\eta_{\mu\nu} + \frac{1}{\sqrt{M_3}} h_{\mu\nu} \, ,
\ea
and perturbing to quadratic order in $h_{\mu\nu}$ and $q_{\mu\nu}$ we have
\ba
S_{{2}} = M_3  \int \d^3 x  \left[ -\frac{\sigma}{2} h^{\mu\nu} \Ein\mn\abu h\ab
- q^{\mu\nu} \Ein\mn\abu  h\ab- \frac{1}{4} m^2 (q_{\mu\nu} q^{\mu\nu}-q^2)\right] \, .
\ea
Finally diagonalizing as $h_{\mu\nu} = \tilde h_{\mu\nu} - \sigma q_{\mu\nu}$ we obtain
\ba
S_{{2}} = M_3  \int \d^3 x  \left[ -\frac{\sigma}{2} \tilde h^{\mu\nu} \Ein\mn\abu  \tilde h\ab +\frac{\sigma}{2} q^{\mu\nu} \Ein\mn\abu  q\ab- \frac{1}{4} m^2 (q_{\mu\nu} q^{\mu\nu}-q^2)\right] \, ,
\ea
which is manifestly a decoupled massless graviton and massive graviton. Crucially however we see that the kinetic terms of each have the opposite sign. Since only the degrees of freedom of the massive graviton $q_{\mu\nu}$ are propagating, unitarity when coupled to other sources forces us to choose $\sigma=-1$. The apparently ghostly massless graviton does not lead to any unitarity violation, at least in perturbation theory, as there is no massless pole in the propagator.
The stability of the vacua was further shown in different gauges in Ref.~\cite{Ghodsi:2012fg}.

\subsection{Absence of Boulware--Deser ghost}

The auxiliary field formulation of new massive gravity is also useful for understanding the absence of the BD ghost~\cite{deRham:2011ca}.
Setting $\sigma=-1$ as imposed previously and working with the formulation~\eqref{NMG 2},
we can introduce new vector and scalar degrees of freedom as follows
\ba
q_{\mu\nu} = \frac{1}{\sqrt{M_3}}\bar q_{\mu\nu} + \nabla_{\mu} V_{\nu}+ \nabla_{\nu} V_{\mu} \,,
\ea
with
\ba
V_{\mu}= \frac{1}{\sqrt{M_3 m}}A_{\mu}+ \frac{\nabla_{\mu} \pi}{\sqrt{M_3}m^2} \,,
\ea
where the factors of $\sqrt{M_3}$ and $m$ are chosen for canonical normalization. $A_\mu$ represents the helicity-1 mode which carries 1 degree of freedom and $\pi$ the helicity-0 mode that carries 1 degree freedom. These two modes carries all the dynamical fields.

Introducing new fields in this way also introduced new symmetries. Specifically there is a $U(1)$ symmetry
\ba
\pi \rightarrow \pi + m \chi \, , \quad A_{\mu} \rightarrow A_{\mu} - \chi \, ,
\ea
and a linear diffeomorphism symmetry
\ba
\bar q_{\mu\nu} \rightarrow \bar q_{\mu\nu}+ \nabla_{\mu} \chi_{\nu}+  \nabla_{\nu} \chi_{\mu} \, , \quad A_{\mu} \rightarrow A_{\mu} - \sqrt{m} \chi_{\mu} \, .
\ea
Substituting in the action, integrating by parts and using the Bianchi identity $\nabla_{\mu}G^{\mu}_{\nu}=0$ we obtain
\ba S_{\rm NMG}&=&  \int \d^3 x \sqrt{-g} \Bigg[ -M_3 R-  \sqrt{M_3}\bar q^{\mu\nu}G_{\mu\nu}  \\
&-& \frac{1}{4}  \Big((m \bar q_{\mu\nu}+ \nabla_{\mu} A_{\nu}+\nabla_{\nu} A_{\mu} +\frac{2}{m} \nabla_{\mu} \nabla_{\nu} \pi)^2\nn \\
&-&(m \bar q+ 2 \nabla A+\frac{2}{m} \Box \pi)^2 \Big)\Bigg]  \nn \,.
\ea
Although this action contains apparently higher order terms due to its dependence on $\nabla_{\mu} \nabla_{\nu} \pi$, this dependence is Galileon-like in that the equations of motion for all fields are second order. For instance the naively dangerous combination \be (\nabla_{\mu} \nabla_{\nu} \pi)^2- (\Box \pi)^2 \ee is up to a boundary term equivalent to $R_{\mu\nu} \nabla^{\mu} \nabla^{\nu}\pi$. In~\cite{deRham:2011ca} it is shown that the resulting equations of motion of all fields are second order due to these special Fierz--Pauli combinations.

As a result of the introduction of the new gauge symmetries, we straightforwardly count the number of non-perturbative degrees of freedom. The total number of fields are 16: 6 from $g_{\mu\nu}$, 6 from $q_{\mu\nu}$, 3 from $A_{\mu}$ and one from $\pi$. The total number of gauge symmetries are 7: 3 from diffeomorphisms, 3 from linear diffeomorphisms and 1 from the $U(1)$. Thus the total number of degrees of freedom are $16 - 7 \, (\text{gauge})-7 \, (\text{constraint}) =2$ which agrees with the linearized analysis. An independent argument leading to the same result is given in~\cite{Hohm:2012vh} where NMG including its topologically massive extension (see below) are presented in Hamiltonian form using Einstein--Cartan language (see also~\cite{Deser:2009hb}).

\subsection{Decoupling limit of new massive gravity}

The formalism of the previous section is also useful for deriving the decoupling limit of NMG which as in the higher dimensional case, determines the leading interactions for the helicity-0 mode. The decoupling limit~\cite{deRham:2011ca} is defined as the limit \be
M_3 \rightarrow \infty \, , \quad  m \rightarrow 0 \, \quad \Lambda_{5/2} = (\sqrt{M_3} m^2)^{2/5}= \text{fixed} \, .
\ee
As usual the metric is scaled as
\ba
g_{\mu\nu} = \eta_{\mu \nu}+ \frac{1}{\sqrt{M_3}} h_{\mu\nu} \, ,
\ea
and in the action
\ba
S_{\rm NMG}&=&  \int \d^3 x \sqrt{-g} \Bigg[ -M_3 R-  \sqrt{M_3}\bar q^{\mu\nu}G_{\mu\nu}  \\
&-&  \frac{1}{4}  \Big((m \bar q_{\mu\nu}+ \nabla_{\mu} A_{\nu}+\nabla_{\nu} A_{\mu} +\frac{2}{m} \nabla_{\mu} \nabla_{\nu} \pi)^2\nn\\
&-&(m \bar q+ 2 \nabla A+\frac{2}{m} \Box \pi)^2 \Big)\Bigg]  \nn \,,
\ea
the normalizations have been chosen so that we keep
$A_{\mu}$ and $\pi$ fixed in the limit.
We readily find
\ba
S_{\rm dec} &=& \int \d^3 x \Bigg[ +\frac{1}{2} h^{\mu\nu} \Ein\mn\abu h\ab
- \bar q^{\mu\nu} \Ein\mn\abu h\ab-\bar q^{\mu\nu}(\partial_{\mu} \partial_{\nu}\pi -\eta_{\mu\nu} \Box \pi )\nn\\ &&- \frac{1}{4} F_{\mu\nu}F^{\mu\nu}
+   \frac{1}{\Lambda_{5/2}^{5/2}} \Ein^{\mu \nu}_{\alpha \beta} h^{\alpha \beta} (\partial_{\mu} \pi \partial_{\nu}\pi - \eta_{\mu\nu} (\partial \pi)^2)\Bigg] \, ,
\ea
where all raising and lowering is understood with respect to the 3 dimensional Minkowski metric.
Performing the field redefinition $h_{\mu\nu} = 2 \pi \eta_{\mu\nu} + \tilde h_{\mu\nu}+ \bar q_{\mu\nu}$ we finally obtain
\ba
S_{\rm dec} &=& \int \d^3 x \Bigg[ +\frac{1}{2} \tilde h^{\mu\nu} \Ein\mn\abu \tilde h\ab
- \frac{1}{2} \bar q^{\mu\nu} \Ein\mn\abu \bar q\ab   \nn \\
&& - \frac{1}{4} F_{\mu\nu}F^{\mu\nu}   -\frac{1}{2} (\partial \pi)^2 - \frac{1}{2 \Lambda_{5/2}^{5/2}} (\partial \pi)^2 \Box \pi \Bigg] \, .
\ea
Thus we see that in the decoupling limit, NMG becomes equivalent to two massless gravitons which have no degrees of freedom, one massless spin-1 particle which has one degree of freedom, and one scalar $\pi$ which has a cubic Galileon interaction. This confirms that the strong coupling scale for NMG is $\Lambda_{5/2}$.

The decoupling limit clarifies one crucial aspect of NMG. It has been suggested that NMG could be power counting renormalizable following previous arguments for topological massive gravity~\cite{Deser:1990bj} due to the softer nature of divergences in three-dimensional and the existence of a dimensionless combination of the Planck mass and the graviton mass. This is in fact clearly not the case since the above cubic interaction is a non-renormalizable operator and dominates the Feynman diagrams leading to perturbative unitarity violation at the strong coupling scale $\Lambda_{5/2}$ (see Section~\ref{sec:StrongCoupling} for further discussion on the distinction between the breakdown of perturbative unitarity and the breakdown of the theory).

\subsection{Connection with bi-gravity}

The existence of the NMG theory at first sight appears to be something of an anomaly that cannot be reproduced in higher dimensions. There also does not at first sight seem to be any obvious connection with the diffeomorphism breaking ghost-free massive gravity model  (or dRGT) and multi-gravity extensions. However in~\cite{Paulos:2012xe} it was shown that NMG, and certain extensions to it, could all be obtained as scaling limits of the same 3 dimensional bi-gravity models that are consistent with ghost-free massive gravity in a different decoupling limit. As we already mentioned, the key to seeing this is the auxiliary formulation where the tensor $f_{\mu\nu}$ is related to the missing extra metric of the bi-gravity theory.

Starting with the 3 dimensional version of bi-gravity~\cite{Hassan:2011zd} in the form
\ba
S =  \int \d^3 x \left[\frac{M_g}{2} \sqrt{-g} R[g]+\frac{M_f}{2} \sqrt{-f} R[f] - m^2 \U[g,f] \right] \, ,
\ea
where the bi-gravity potential takes the standard form in terms of characteristic polynomials similarly as in~\eqref{U}
\ba
\U[g,f] = -\frac{M_{\rm eff}}{4}\sum_{n=0}^3 \alpha_n \L_n(\cal K) \,,
\ea
and $\K$ is given in~\eqref{eqK} in terms of the two dynamical metrics $g$ and $f$.
The scale $M_{\rm eff}$  is defined as
$M_{\rm eff}^{-1}= M_g^{-1}+ M_f^{-1}$.
The idea is to define a scaling limit~\cite{Paulos:2012xe} as follows
\ba
M_f \rightarrow + \infty
\ea
keeping $M_3=-(M_g+M_f)$ fixed and keeping $q_{\mu\nu}$ fixed in the definition
\ba
f_{\mu\nu}= g_{\mu\nu}- \frac{M_3}{M_f}q_{\mu\nu} \, .
\ea
Since ${\cal K}_{\mu\nu} \rightarrow \frac{M_3}{2M_f} q_{\mu\nu}$ then we have in the limit
\ba
 S =  \int \d^3 x \Bigg[-\frac{M_3}2 \sqrt{-g} R[g] - \frac{M_3}{2} \, q^{\mu\nu}G_{\mu\nu}(g)
+ \frac{m^2 M_{\rm eff}}{4} \sum_{n=0}^3 \alpha_n (-1)^n \left( \frac{M_3}{M_f} \right)^n
\L_n(q) \Bigg] \nn
\ea
which prompts the definition of a new set of coefficients
\ba
c_n = -\frac{(-1)^n}{2M_3}\bar{M} \alpha_n  \left( \frac{M_3}{M_f} \right)^n \,,
\ea
so that
\ba
S = M_3 \int \d^3 x \left[-\sqrt{-g} R[g] -  \, q^{\mu\nu}G_{\mu\nu}(g) - m^2 \sum_{n=0}^3 c_n {\cal L}_n(q) \right] \, .
\ea
Since this theory is obtained as a scaling limit of the ghost-free bi-gravity action, it is guaranteed to be free from the BD ghost. We see that in the case $c_2=1/4$, $c_3=c_4=0$ we obtain the auxiliary field formulation of NMG, justifying the connection between the auxiliary field $q_{\mu\nu}$ and the bi-gravity metric $f_{\mu\nu}$.

\subsection{3D massive gravity extensions}

The generic form of the auxiliary field formulation of NMG derived above~\cite{Paulos:2012xe}
\ba
\label{NMGgeneral}
S = M_3 \int \d^3 x \left[-\sqrt{-g} R[g] -  \, q^{\mu\nu}G_{\mu\nu}(g) - m^2 \sum_{n=0}^3 c_n {\cal L}_n(q) \right] \, ,
\ea
demonstrates that there exists a two parameter family extensions of NMG determined by nonzero coefficients for $c_3$ and $c_4$. The purely metric formulation for the generic case can be determined by integrating out the auxiliary field $q_{\mu\nu}$. The equation of motion for $q_{\mu\nu}$ is given symbolically
\ba
-G - m^2 \sum_{n=1}^3 n c_n \, \epsilon \, \epsilon \, q^{n-1} g^{3-n} =0 \, .
\ee
This is a quadratic equation for the tensor $q_{\mu\nu}$. Together these two additional degrees of freedom give the cubic curvature~\cite{Sinha:2010ai} and Born--Infeld extension NMG~\cite{Gullu:2010pc}. Although additional higher derivative corrections have been proposed based on consistency with the holographic c-theorem~\cite{Paulos:2010ke}, the above connection suggests that Eq.~(\ref{NMGgeneral}) is the most general set of interactions allowed in NMG which are free from the BD ghost.

In the specific case of the Born--Infeld extension~\cite{Gullu:2010pc} the action is
\ba
S_{\rm B.I} = 4m^2 M_3 \int \d^3 x \left[\sqrt{-g} - \sqrt{- \det[g_{\mu\nu} - \frac{1}{m^2} G_{\mu\nu}]} \right] \, .
\ea
It is straightforward to show that on expanding the square root to second order in $1/m^2$ we recover the original NMG action. The specific case of the Born--Infeld extension of NMG, also has a surprising role as a counterterm in the AdS\sub{4} holographic renormalization group~\cite{Jatkar:2011ue}. The significance of this relation is unclear at present.

\subsection{Other 3D theories}

\subsubsection{Topological massive gravity}

In four dimensions the massive spin two representations of the Poincar\'e group must come in positive and negative helicity pairs. By contrast in three dimensions the positive and negative helicity states are completely independent. Thus while a parity preserving theory of massive gravity in 3 dimensions will contain two propagating degrees of freedom, it seems possible in principle for there to exist an interacting theory for one of the helicity modes alone. What is certainly possible is that one can give different interactions to the two helicity modes. Such a theory necessarily breaks parity, and was found in~\cite{Deser:1981wh,Deser:1982vy}. This theory is known as `topologically massive gravity' (TMG) and is described by the Einstein--Hilbert action, with cosmological constant, supplemented by a term constructed entirely out of the connection (hence the name topological)
\ba
S = \frac{M_3}{2} \int \d^3 x  \sqrt{-g }(R-2 \Lambda) + \frac{1}{4\mu} \epsilon^{\lambda \mu\nu } \Gamma^{\rho}_{\lambda \sigma} \left[ \partial_{\mu} \Gamma^{\sigma}_{\rho \nu}+ \frac{2}{3} \Gamma^{\sigma}_{\mu \tau}\Gamma^{\tau}_{\nu \rho} \right] \,.
\ea
The new interaction is a gravitational Chern--Simons term and is responsible for the parity breaking. More generally this action may be supplemented to the NMG Lagrangian interactions and so the TMG can be viewed as a special case of the full extended parity violating NMG.

The equations of motion for topologically massive gravity take the form
\ba
 G_{\mu\nu} + \Lambda g_{\mu\nu} + \frac{1}{\mu} C_{\mu\nu}=0 \, ,
\ea
where $C_{\mu\nu}$ is the Cotton tensor which is given by
\ba
 C_{\mu \nu} = \epsilon_{\mu}{}^{\alpha \beta} \nabla_{\alpha} (R_{\beta \nu}- \frac{1}{4} g_{\beta \nu} R) \,.
\ea
Einstein metrics for which $G_{\mu\nu}= - \Lambda g_{\mu \nu}$ remain as a subspace of general set of vacuum solutions. In the case where the cosmological constant is negative $\Lambda  = -1/\ell^2$ we can use the correspondence of Brown and Henneaux~\cite{Brown:1986nw} to map the theory of gravity on an asymptotically AdS$_3$ space to a 2D CFT living at the boundary.

The AdS/CFT in the context of Topological massive gravity was also studied in Ref.~\cite{Skenderis:2009nt}.

\subsubsection{Supergravity extensions}

As with any gravitational theory it is natural to ask whether extensions exist which exhibit local supersymmetry, \ie supergravity. A supersymmetric extension to topologically massive gravity was given in~\cite{Deser:1982sw}. An $N=1$ supergravity extension of NMG including the topologically massive gravity terms was given in~\cite{Andringa:2009yc} and further generalized in~\cite{Bergshoeff:2009aq}. The construction requires the introduction of an `auxiliary' bosonic scalar field $S$ so that the form of the action is
\ba
S &=& \frac{1}{\kappa^2} \int \d^3 x \sqrt{-g} \left[ M{\cal L}_C + \sigma {\cal L}_{E.H.}+ \frac{1}{m^2}{\cal L}_K + \frac{1}{8 \tilde m^2} {\cal L}_{R^2}+ \frac{1}{\hat m^2}{\cal L}_{S^4}+ \frac{1}{\hat \mu} {\cal L}_{S^3}\right]  \nn \\
&&+ \int \d^3 x \frac{1}{\mu}{\cal L}_{\rm top}\,,
\ea
where
\ba
&&{\cal L}_C = S + \text{fermions} \\ && {\cal L}_{\rm E.H.} = R-2S^2+ \text{fermions} \\
&& {\cal L}_K= K-\frac{1}{2}S^2 R - \frac{3}{2} S^4 + \text{fermions} \\ && {\cal L}_{R^2}= -16 \left[ (\partial S)^2- \frac{9}{4} (S^2 + \frac{1}{6}R)^2 \right] + \text{fermions} \\ && {\cal L}_{S^4} = S^4 + \frac{3}{10} RS^2 + \text{fermions} \\ && {\cal L}_{S^3} = S^3 + \frac{1}{2} RS+ \text{fermions} \\ && {\cal L}_{\rm top} = \frac{1}{4} \epsilon^{\lambda \mu\nu } \Gamma^{\rho}_{\lambda \sigma} \left[ \partial_{\mu} \Gamma^{\sigma}_{\rho \nu}+ \frac{2}{3} \Gamma^{\sigma}_{\mu \tau}\Gamma^{\tau}_{\nu \rho} \right]+ \text{fermions} \,.
\ea
The fermion terms complete each term in the Lagrangian into an independent supersymmetric invariant. In other words supersymmetry alone places no further restrictions on the parameters in the theory. It can be shown that the theory admits supersymmetric AdS vacua~\cite{Andringa:2009yc,Bergshoeff:2009aq}. The extensions of this supergravity theory to larger numbers of supersymmetries is considered at the linearized level in~\cite{Bergshoeff:2010ui}.

Moreover, $N=2$ supergravity extensions of TMG were recently constructed in Ref.~\cite{Kuzenko:2013uya} and its $N=3$ and $N=4$ supergravity extensions in Ref.~\cite{Kuzenko:2014jra}.

\subsubsection{Critical gravity}

Finally, let us comment on a special case of three dimensional gravity known as log gravity~\cite{Bergshoeff:2011ri} or critical gravity in analogy with the general dimension case~\cite{Lu:2011zk,Deser:2011xc,Alishahiha:2011yb}. For a special choice of parameters of the theory, there is a degeneracy in the equations of motion for the two degrees of freedom leading to the fact that one of the modes of the theory becomes a `logarithmic' mode.

Indeed, at the special point $\mu \ell  = 1$, (where $\ell$ is the AdS length scale, $\Lambda=-1/\ell^2$), known as the `chiral point' the left-moving (in the language of the boundary CFT) excitations of the theory become pure gauge and it has been argued that the theory then becomes purely an interacting theory for the right moving graviton~\cite{Carlip:2008jk}. In Ref.~\cite{Li:2008dq} it was earlier argue that there was no massive graviton excitations at the critical point $\mu\ell=1$, however Ref.~\cite{Carlip:2008jk} found one massive graviton excitation for every finite and non-zero value of $\mu\ell$, including at the critical point $\mu\ell=1$.

This case was further analysed in  \cite{Grumiller:2008qz}, see also Ref.~\cite{Grumiller:2013at} for a recent review.  It was shown that the degeneration of the massive graviton mode with the left moving boundary graviton leads to logarithmic excitations.

To be more precise, starting with the auxiliary formulation of NMG with a cosmological constant $\lambda m^2$
\ba
S_{\rm NMG}= M_3  \int \d^3 x \sqrt{-g} \left[ \sigma R- 2 \lambda m^2 - q^{\mu\nu}G_{\mu\nu}
- \frac{1}{4} m^2 (q_{\mu\nu} q^{\mu\nu}-q^2)\right]\,,
\ea
we can look for AdS vacuum solutions for which the associated cosmological constant $\Lambda=-1/\ell^2$ in $G_{\mu\nu}= -\Lambda g_{\mu\nu}$ is not the same as $\lambda m^2$. The relation between the two is set by the vacuum equations to be
\ba
-\frac{1}{4m^2} \Lambda^2- \Lambda \sigma + \lambda m^2 =0 \, ,
\ea
which generically has two solutions. Perturbing the action to quadratic order around this vacuum solution we have
\ba
S_{{2}} = M_3  \int \d^3 x  \left[ -\frac{\bar \sigma}{2} h^{\mu\nu} {\cal G}_{\mu\nu}  - q^{\mu\nu} {\cal G}_{\mu\nu} - \frac{1}{4} m^2 (q_{\mu\nu} q^{\mu\nu}-q^2)\right] \, .
\ea
where
\ba
{\cal G}_{\mu\nu}(h) = \Ein\mn\abu h\ab- 2\Lambda h_{\mu\nu}+ \Lambda \bar g_{\mu\nu} h
\ea
and
\ba
\bar \sigma = \sigma - \frac{\Lambda}{3m^2}
\ea
where we raise and lower the indices with respect to the background AdS metric $\bar g_{\mu\nu}$.

As usual it is apparent that this theory describes one massless graviton (with no propagating degrees of freedom) and one massive one whose mass is given by $M^2 = - m^2 \bar \sigma$. However by choosing $\bar \sigma =0$ the massive mode becomes degenerate with the existing massless one.

In this case the action is
\ba
S_{{2}} = M_3  \int \d^3 x  \left[  - q^{\mu\nu} {\cal G}_{\mu\nu} - \frac{1}{4} m^2 (q_{\mu\nu} q^{\mu\nu}-q^2)\right] \, ,
\ea
and varying with respect to $h_{\mu\nu}$ and $q_{\mu\nu}$ we obtain the equations of motion \ba && {\cal G}_{\mu\nu}(q)=0 \,  , \\ &&{\cal G}_{\mu\nu}(h)+ \frac{1}{2} (q_{\mu\nu} - \bar g_{\mu\nu} q)=0 \, .
\ea
Choosing the gauge $\nabla^{\mu}h_{\mu\nu}- \nabla_{\nu}h=0$ the equations of motion imply $h=0$ and the resulting equation of motion for $h_{\mu\nu}$ takes the form \be \label{logeq} \left[ \Box - 2 \Lambda \right]^2 h_{\mu\nu} =0  \, .
\ee
It is this factorization of the equations of motion into a square of an operator that is characteristic of the critical/log gravity theories. Although the equation of motion is solved by the usual massless models for which $\left[ \Box - 2 \Lambda \right] h_{\mu\nu} =0$, there are additional logarithmic modes which do not solve this equation but do solve Eq.~(\ref{logeq}). These are so-called because they behave logarithmically in $\rho$ asymptotically when the AdS metric is put in the form $\d \rho^2= \ell^2 ( - \cosh(\rho)^2 \d \tau^2+ \d \rho^2 + \sinh( \rho)^2 \d \theta^2)$.  The presence of these log modes was shown to remain beyond the linear regime, see Ref.~\cite{Grumiller:2008pr}.

Based on this result as well as on the finiteness and conservation of the stress tensor and on the emergence of a Jordan cell structure in the Hamiltonian,  the correspondence to a logarithmic CFT was conjectured in Ref.~\cite{Grumiller:2008qz}, where the to be dual log CFTs representations have degeneracies in the spectrum of scaling dimensions.

Strong indications for this correspondence appeared in many different ways.
First, consistent boundary conditions which allow the log modes were provided in Ref.~\cite{Grumiller:2008es}, were it was shown that in addition to the Brown-Henneaux boundary conditions one could also consider more general ones. These boundary conditions were further explored in \cite{Henneaux:2009pw,Maloney:2009ck} where it was shown that the stress-energy tensor for these boundary conditions are finite and not chiral, giving another indication that the theory could be dual to a logarithmic CFT.

Then specific correlator functions were computed and compared. Ref.~\cite{Skenderis:2009nt} checked the 2-point correlators and Ref.~\cite{Grumiller:2009mw} the 3-point ones.
A similar analysis was also performed within the context of NMG in Ref.~\cite{Grumiller:2009sn}
where the 2-point correlators were computed at the chiral point and shown to behave as those of a logarithmic CFT.

Further checks for this AdS/log CFT  include the 1-loop partition function as computed in Ref.~\cite{Gaberdiel:2010xv}. See also  Ref.~\cite{Grumiller:2013at}  for a review of other checks.

It has been shown however that ultimately these theories are non-unitary due to the fact that there is a non-zero inner product between the log modes and the normal models and the inability to construct a positive definite norm on the Hilbert space~\cite{Porrati:2011ku}.

\subsection{Black holes and other exact solutions}

A great deal of physics can be learned from studying exact solutions, in particular those corresponding to black hole geometries. Black holes are also important probes of the non-perturbative aspects of gravitational theories. We briefly review here the types of exact solutions obtained in the literature.

In the case of topologically massive gravity, a one-parameter family of extensions to the BTZ black hole have been obtained in~\cite{Garbarz:2008qn}. In the case of NMG as well as the usual BTZ black holes obtained in the presence of a negative cosmological constant there are in a addition a class of warped AdS$_3$ black holes~\cite{Clement:2009gq} whose metric takes the form \be \d s^2 = - \beta^2 \frac{\rho^2-\rho_0^2}{r^2} \d t^2 +r^2 \left( \d \phi - \frac{\rho+(1-\beta^2) \omega}{r^2} \d t\right)^2+ \frac{1}{\beta^2 \zeta^2} \frac{\d \rho^2}{\rho^2- \rho_0^2} \, , \ee where the radial coordinate $r$ is given by \be
r^2 = \rho^2 + 2\omega \rho + \omega^2 (1- \beta^2) + \frac{\beta^2 \rho_0^2}{1-\beta^2} \, .
\ee
and the parameters $\beta$ and $\zeta$ are determined in terms of the graviton mass $m$ and the cosmological constant $\Lambda$ by \be
\beta^2 = \frac{9-21 \Lambda/m^2 \mp 2 \sqrt{3 (5+7 \Lambda/m^2}}{4(1-\Lambda/m^2)} \,, \quad \zeta^{-2} = \frac{21-4\beta^2}{8 m^2} \, .
\ee
This metric exhibits two horizons at $\rho = \pm \rho_0$, if $\beta^2\ge 0$ and $\rho_0$ is real. Absence of closed timelike curves requires that $\beta^2\le 1$. This puts the allowed range on the values of $\Lambda$ to be \be
-\frac{35 m^2}{289} \le \Lambda \le \frac{m^2}{21} \, .
\ee
AdS waves, extensions of plane (pp) waves anti-de Sitter spacetime have been considered in~\cite{AyonBeato:2009yq}. Further work on extensions to black hole solutions, including charged black hole solutions can be found in~\cite{Oliva:2009ip,Clement:2009ka,Giribet:2009qz,Ahmedov:2010uk,Ahmedov:2011yd,Kwon:2011ey,Perez:2011qp,Ghodsi:2010ev}. We note in particular the existence of a class of Lifshitz black holes~\cite{AyonBeato:2009nh} which exhibit the Lifshitz anisotropic scale symmetry \be t \rightarrow \lambda^z t \, , \quad \vec x \rightarrow \lambda \vec x \, , \ee where $z$ is the dynamical critical exponent. As an example for $z=3$ the following Lifshitz black hole can be found~\cite{AyonBeato:2009nh}
\ba
\d s^2 = - \frac{r^6}{\ell^6} \left( 1- \frac{M l^2}{r^2} \right) \d t^2 + \frac{\d r^2}{\left( \frac{r^2}{\ell^2}-M \right)} + r^2 \d \phi^2 \, .
\ea
This metric has a curvature singularity at $r=0$ and a horizon at $r_+ = \ell \sqrt{M}$. The Lifshitz symmetry is preserved if we scale $t \rightarrow \lambda^3 t$, $x \rightarrow \lambda x$, $r \rightarrow \lambda^{-1} r$ and in addition we scale the black hole mass as $M \rightarrow \lambda^{-2} M$. The metric should be contrasted with the normal BTZ black hole which corresponds to $z=1$
\ba
\d s^2 = - \frac{r^2}{\ell^2} \left( 1- \frac{M \ell^2}{r^2} \right) \d t^2 + \frac{\d r^2}{\left( \frac{r^2}{\ell^2}-M \right)} + r^2 \d \phi^2 \, .
\ea
Exact solutions for charged Black Holes were also derived in Ref.~\cite{Ghodsi:2010gk} and an exact, non-stationary solution of TMG and NMG  with the asymptotic charges
of a BTZ black hole was find in \cite{Flory:2013ssa}.  This exact solution was shown to admit a timelike singularity. Other exact asymptotically AdS-like solutions were found in Ref.~\cite{Ghodsi:2011ua}.

\subsection{New massive gravity holography}

One of the most interesting avenues of exploration for NMG has been in the context of Maldacena's AdS/CFT correspondence~\cite{Maldacena:1997re}. According to this correspondence, NMG with a cosmological constant chosen so that there are asymptotically anti-de Sitter solutions is dual to a conformal field theory (CFT). This has been considered in~\cite{Bergshoeff:2009aq,Liu:2009bk,Liu:2009kc} where it was found that the requirements of bulk unitarity actually lead to a negative central charge.

The argument for this proceeds from the identification of the central charge of the dual two dimensional field theory with the entropy of a black hole in the bulk using Cardy's formula.  The entropy of the black hole is given by~\cite{Kraus:2005vz}
\ba
S = \frac{A_{\rm BTZ}}{4 G_3} \Omega
\ea
where $G_3$ is the 3-dimensional Newton constant and $\Omega = \frac{2 G_3 }{3 \ell} c$ where $l$ is the AdS radius and $c$ is the central charge. This formula is such that $c=1$ for pure Einstein--Hilbert gravity with a negative cosmological constant.

A universal formula for this central charge has been obtained as is given by
\ba
c = \frac{\ell}{2 G_3} g_{\mu\nu} \frac{\partial {\cal L}}{\partial R_{\mu\nu}} \,.
\ea
This result essentially follows from using the Wald entropy formula~\cite{Wald:1993nt} for a higher derivative gravity theory and identifying this with the central change through the Cardy formula. Applying this argument for new massive gravity we obtain~\cite{Bergshoeff:2009aq}
\ba
c = \frac{3 \ell}{2 G_3} \left( \sigma + \frac{1}{2 m^2 \ell^2} \right) \, .
\ea
Since $\sigma=-1$ is required for bulk unitarity, we must choose $m^2>0$ to have a chance of getting $c$ positive. Then we are led to conclude that the central charge is only positive if
\ba
\Lambda = - \frac{1}{\ell^2} <-2 m^2 \, .
\ea
However, unitarity in the bulk requires $m^2 > -\Lambda/2$ and this excludes this possibility. We are thus led to conclude that NMG cannot be unitary both in the bulk and in the dual CFT.
This failure to maintain both bulk and boundary unitarity can be resolved by a modification of NMG to a full bi-gravity model, namely Zwei-Dreibein gravity to which we turn next.

\subsection{Zwei-dreibein gravity}

As we have seen there is a conflict in NMG between unitarity in the bulk, \ie  the requirement that the massive gravitons are not ghosts, and unitarity in dual CFT as required by the positivity of the central charge. This conflict may be resolved however by replacing NMG with the 3 dimensional bi-gravity extension of ghost-free massive gravity  that we have already discussed. In particular if we work in the Einstein--Cartan formulation in 3 dimensions, then the metric is replaced by a `dreibein' and since this is a bi-gravity model, we need two `dreibeins'. This gives us the Zwei-dreibein gravity~\cite{Bergshoeff:2013xma}.

In the notation of~\cite{Bergshoeff:2013xma} the Lagrangian is given by
\ba
{\cal L} &=& - \sigma M_1 e_a R^a(e) -M_2  f_a R^a(f) - \frac{1}{6} m^2 M_1 \alpha_1 \epsilon_{abc} e^a e^b e^c \nn \\
&& -\frac{1}{6} m^2 M_2 \alpha_2 \epsilon_{abc} f^a f^b f^c
+ \frac{1}{2} m^2 M_{12} \epsilon_{abc} ( \beta_1 e^a e^b f^c+ \beta_2 e^a f^b f^c)
\ea
where we have suppressed the wedge products $e^3= e \wedge e \wedge e$, $R^a(e)$ is Lorentz vector valued curvature two-form for the spin-connection associated with the dreibein $e$ and $R^a(f)$ that associated with the dreibein $f$. Since we are in three dimensions, the spin-connection can be written as a Lorentz vector dualizing with the Levi-Civita symbol $\omega^a= \epsilon^{abc} \omega_{bc}$.
This is nothing other than the vierbein representation of bi-gravity with the usual ghost-free (dRGT) mass terms. As we have already discussed, NMG and its various extensions arise in appropriate scaling limits.

A computation of the central charge following the same procedure was given in~\cite{Bergshoeff:2013xma} with the result that
\ba
c = 12 \pi \ell (\sigma M_1 + \gamma M_2) \, .
\ea
Defining the parameter $\gamma$ via the relation
\ba
(\alpha_2 (\sigma M_1 + M_2) + \beta_2 M_2) \gamma^2 + 2 (M_2 \beta_1 - \sigma M_1 \beta_2) \gamma \nn\\
- \sigma ( \alpha_1 ( \sigma M_1 + M_2) + \beta_1 M_1) =0 \, ,
\ea
then bulk unitarity requires  $\gamma/(\sigma M_1 + \gamma M_2) <0 $. In order to have $c>0$ we thus need $ \gamma <0$ which in turn implies $\sigma >1$ (since $M_1$ and $M_2$ are defined as positive). The absence of tachyons in the AdS vacuum requires $\beta_1 + \gamma \beta_2>0$, and this assumes a real solution for $\gamma$ for a negative $\Lambda$.
 There are an open set of such solutions to these conditions, which shows that the conditions for unitarity are not finely tuned. For example in~\cite{Bergshoeff:2013xma} it is shown that there is an open set of solutions which are close to the special case $M_1 = M_2$, $\beta_1 = \beta_2=1$, $\gamma=1$ and $\alpha_1=\alpha_2 = 3/2+ \frac{1}{\ell^2 m^2}$. This result is not in contradiction with the scaling limit that reproduces NMG, because this scaling limit requires the choice $\sigma=-1$ which is in contradiction with positive central charge.

These results potentially have an impact on the higher dimensional case. We see that in three dimensions we potentially have a diffeomorphism invariant theory of massive gravity (\ie bi-gravity) which at least for AdS solutions exhibits unitarity both in the bulk and in the boundary CFT for a finite range of parameters in the theory. However these bi-gravity models are easily extended into all dimensions as we have already discussed and it is similarly easy to find AdS solutions which exhibit bulk unitarity. It would be extremely interesting to see if the associated dual CFTs are also unitary thus providing a potential holographic description of generalized theories of massive gravity.


\section{Lorentz-Violating Massive Gravity}
\label{sec:LorentzViolating}

%
%

\subsection{\textit{SO(3)}-invariant mass terms}

The entire analysis performed so far is based on assuming Lorentz invariance.  In what follows we briefly review a few other potentially viable theories of massive gravity where Lorentz invariance is broken and their respective cosmology.

Prior to the formulation of the ghost-free theory of massive gravity, it was believed that no Lorentz invariant theories of massive gravity could evade the BD ghost and Lorentz-violating theories were thus the best hope and we refer to~\cite{Rubakov:2008nh} for a thorough review on the field. A thorough analysis of Lorentz-violating theories of massive gravity was performed in~\cite{Dubovsky:2004sg} and more recently in~\cite{Comelli:2013txa}. See also Refs.~\cite{Gabadadze:2004iv,Blas:2007zz,CuadrosMelgar:2011yw} for other complementary studies.
Since this field has been reviewed in~\cite{Rubakov:2008nh} we only summarize the key results in this section (see also~\cite{Blas:2010hb} for a more recent review on many developments in Lorentz violating theories.)
See also Ref.~\cite{Lin:2013sja} for an interesting spontaneous breaking of Lorentz invariance in ghost-free massive gravity using three scalar fields, and Ref.~\cite{Lin:2013aha} for a $SO(3)$-invariant ghost-free theory of massive gravity which can be formulated with three \stu scalar fields and propagating five degrees of freedom.

In most theories of Lorentz-violating massive gravity, the $SO(3,1)$ Poincar\'e group is broken down to a $SO(3)$ rotation group. This implies the presence of a preferred time. Preferred-frame effects are however strongly constrained by solar system tests~\cite{Will:2005va} as well as pulsar tests~\cite{Bell:1995jz}, see also~\cite{Yagi:2013qpa,Yagi:2013ava} for more recent and even tighter constraints.

At the linearized level the general mass term which satisfies this rotation symmetry is
\ba
\label{SO3 mass term}
\L_{{\rm SO}(3)\ {\rm mass}}= \frac 18 \(m_0^2 h_{00}^2+2 m_1^2 h_{0i}^2-m_2^2 h_{ij}^2+m_3^2 h^i_i h^j_j-2 m_4^2 h_{00}h^i_i\)\,,
\ea
where space indices are raised and lowered with respect to the flat spatial metric $\delta_{ij}$.
This extends the Lorentz invariant mass term presented in~\eqref{general mass term}. In the rest of this section we will establish the analogue of the Fierz--Pauli mass term~\eqref{FPmass} in this Lorentz violating case and establish the conditions on the different mass parameters $m_{0,1,2,3,4}$.

We note that Lorentz invariance is restored when $m_1=m_2$, $m_3=m_4$ and $m_0^2=-m_1^2+m_3^2$. The Fierz--Pauli structure then further fixes $m_1=m_3$ implying $m_0=0$ which is precisely what ensures the presence of a constraint and the absence of BD ghost (at least at the linearized level).

Out of these 5 mass parameters some of them have a direct physical meaning~\cite{Dubovsky:2004sg,Rubakov:2004eb,Rubakov:2008nh}
\begin{itemize}
\item The parameter $m_2$ is the one that represents the mass of the helicity-2 mode. As a result we should impose $m_2^2\ge 0$ to avoid tachyon-like instabilities. Although we should bear in mind that if that mass parameter is of the order of the Hubble parameter today  $m_2\simeq 10^{-33}\mathrm{\ eV}$, then such an instability would not be problematic.

\item The parameter $m_1$ is the one responsible for turning on a kinetic term for the two helicity-1 modes. Since $m_1=m_2$ in a Lorentz-invariant theory of massive gravity, the helicity-1 mode cannot be turned off ($m_1=0$) while maintaining the graviton massive ($m_2\ne 0$). This is a standard result of Lorentz invariant massive gravity seen so far where the helicity-1 mode is always present. For Lorentz breaking theories the theory is quite different and one can easily switch off at the linearized level the helicity-1 modes in a theory of Lorentz-breaking massive gravity. The absence of a ghost in the helicity-1 mode requires $m_1^2\ge 0$.

\item If $m_0\ne 0$ and $m_1\ne 0$ and $m_4\ne 0$ then two scalar degrees of freedom are present already at the linear level about flat space-time and one of these is always a ghost. The absence of ghost requires either $m_0=0$ or $m_1=0$ or finally $m_4=0$ and $m_2=m_3$.

    In the last scenario where $m_4=0$ and $m_2=m_3$, the scalar degree of freedom loses its gradient terms at the linear level which means that this mode is infinitely strongly coupled unless no gradient appears fully non-linearly either.

    The case $m_0$ has an interesting phenomenology as will be described below. While it propagates five degrees of freedom about Minkowski it avoids the vDVZ discontinuity in an interesting way.

    Finally the case $m_1=0$ (including when $m_0=0$) will be discussed in more detail in what follows.  It is free of both scalar (and vector) degrees of freedom at the linear level about Minkowski  and thus evades the vDVZ discontinuity in a straightforward way.

\item The analogue of the Higuchi bound was investigated in~\cite{Blas:2009my}. In de Sitter with constant curvature $H$, the generalized Higuchi bound is
 \ba
m_4^4+2 H^2\(3(m_3^2-m_4^2)-m_2^2\)> m_4^2(m_1^2-m_4^2)\qquad \text{if}\quad m_0=0\,,
\ea
while if instead $m_1=0$  then no scalar degree of freedom are propagating on de Sitter either so there is no analogue of the Higuchi bound (a scalar starts propagating on FLRW solutions but it does not lead to an equivalent Higuchi bound either. However the absence of tachyon and gradient instabilities do impose some conditions between the different mass parameters).
\end{itemize}

As shown in the case of the Fierz--Pauli mass term and its non-linear extension, one of the most natural way to follow the physical degrees of freedom and their health is to restore the broken symmetry with the appropriate number of \stu fields.

In Section~\ref{sec:nonlinearStuck} we reviewed how to restore the broken diffeomorphism invariance using four \stu fields $\phi^a$ using the relation~\eqref{stuH}. When Lorentz invariance is broken the \stu trick has to be performed slightly differently. Performing an ADM decomposition which is appropriate for the type of Lorentz breaking we are considering, we can use for \stu scalar fields $\Phi=\Phi^0$ and $\Phi^i$, $i=1,\cdots,3$ to define the following four-dimensional scalar, vector and tensors~\cite{Comelli:2013txa}
\ba
\label{Lorentz break n}
n&=&(-g^{\mu\nu}\p_\mu\Phi\p_\nu \Phi)^{-1/2}\\
\label{Lorentz break nmu}
n_\mu &=& n\, \p_\mu \Phi\\
\label{Lorentz break Y}
Y\mupn&=& g^{\mu \alpha}\p_\alpha \Phi^i \p_\nu \Phi^j \delta_{ij}\\
\label{Lorentz break G}
\Gamma\mupn &=& Y\mupn+n^\mu n^\alpha \p_\alpha \Phi^i \p_\nu \Phi^j \delta_{ij}\,.
\ea
$n$ can be thought of as the `St\"uckelbergized' version of the lapse and $\Gamma\mupn$ as that of the spatial metric.

In the Lorentz-invariant case we are stuck with the combination $X\mupn=Y\mupn-n^{-2}g^{\mu\alpha}n_\alpha n_\nu$, but this combination can be broken here and the mass term can depend separately on $n, n_\mu$ and $Y$. This allows for new mass terms. In~\cite{Comelli:2013txa} this framework was derived and used to find new mass terms  that exhibit five degrees of freedom. This formalism was also developed in~\cite{Dubovsky:2004sg} and used to derive new mass terms that also have fewer degrees of freedom. We review both cases in what follows.

\subsection{Phase $m_1=0$}
\label{sec:m1=0}

\subsubsection{Degrees of freedom on Minkowski}

As already mentioned the helicity-1 mode have no kinetic term at the linear level on Minkowski if $m_1=0$. Furthermore it turns out that the field $v$ in~\eqref{scalarPert} is the Lagrange multiplier which removes the BD ghost (as opposed to the field $\psi$ in the case $m_0=0$ presented previously). It imposes the constraint $\dot \tau=0$ which in turns implies $\tau=0$. Using this constraint back in the action one can check that there remains no time derivatives on any of the scalar fields which means that there are no propagating helicity-0 mode on Minkowski either~\cite{Dubovsky:2004sg,Rubakov:2004eb,Rubakov:2008nh}. So in the case where $m_1=0$ there are only 2 modes propagating in the graviton on Minkowski, the 2 helicity-2 modes as in GR.

In this case the absence of the ghost can be seen to follow from the presence of a residual symmetry on flat space~\cite{Dubovsky:2004sg,Rubakov:2008nh}
\ba
x^i\to x^i + \xi^i(t)\,,
\ea
for three arbitrary functions $\xi^i(t)$.
In the \stu language this implies the following internal symmetry
\ba
\Phi^i\to \Phi^i+ \xi^i(\Phi)\,.
\ea

To maintain this symmetry non-linearly the mass term should be a function of $n$ and $\Gamma\mupn$~\cite{Rubakov:2008nh}
\ba
\label{m1 mass}
\L_{{\rm mass}}= -\frac{m^2\mpl^2}{8}\sqrt{-g}\, F\(n, \Gamma\mupn\)\,.
\ea

The absence of helicity-1 and -0 modes while keeping the helicity-2 mode massive makes this Lorentz violating theory of gravity especially attractive. Its cosmology  was explored in~\cite{Dubovsky:2005dw}
and it turns out that this theory of massive gravity could be a candidate for Cold Dark Matter as shown in~\cite{Dubovsky:2004ud}.

Moreover explicit black hole solutions were presented in~\cite{Rubakov:2008nh} where it was shown that in this theory of massive gravity black holes have hair and the \stu fields (in the \stu formulation of the theory) do affect the solution. This result is tightly linked to the fact that this theory of massive gravity admits instantaneous interactions which is generic to any action of the form~\eqref{m1 mass}.

\subsubsection{Non-perturbative degrees of freedom}

Perturbations on more general FLRW backgrounds were then considered more recently in~\cite{Blas:2009my}. Unlike in Minkowski, scalar perturbations on curved backgrounds are shown to behave in a similar way for the cases $m_1=0$ and $m_0=0$. However as we shall see below the case $m_0=0$ propagates five degrees of freedom including a helicity-0 mode that behaves as a scalar it follows  that on generic backgrounds the theory with $m_1=0$ also propagates a helicity-0 mode. The helicity-0 mode is thus infinitely strongly coupled when considered perturbatively about Minkowski.

\subsection{General massive gravity ($m_0=0$)}
\label{sec:m0=0}

In~\cite{Comelli:2013txa} the most general mass term which extends~\eqref{SO3 mass term} non-linearly was considered. It can be written using the \stu variables  defined in~\eqref{Lorentz break n},~\eqref{Lorentz break nmu} and~\eqref{Lorentz break Y},
\ba
\label{SO3 mass term nonlinear}
\L_{{\rm SO}(3)\ {\rm mass}}= -\frac{m^2\mpl^2}{8}\sqrt{-g}\, V\(n, n_\mu, Y\mupn\)\,.
\ea
Generalizing the Hamiltonian analysis for this mass term and requiring the propagation of five degrees of freedom about any background led to the Lorentz-invariant ghost-free theory of massive gravity presented in Section~\ref{sec:ghost-free MG} as well as two new theories of Lorentz breaking massive gravity.

All of these cases ensures the absence of  BD ghost by having $m_0=0$. The case where the BD ghost is projected thanks to the requirement $m_1=0$ is discussed in Section~\ref{sec:m1=0}.

\subsubsection{First explicit Lorentz-breaking example with five dofs}

The first explicit realization of a consistent nonlinear Lorentz breaking model is as follows~\cite{Comelli:2013txa}
\ba
\label{V1}
V_1\(n, n_\mu, Y\mupn\)=n^{-1}\Big[n+\zeta(\Gamma)\Big]U(\tilde{\K})+n^{-1}\mathcal{C}(\Gamma) \, ,
\ea
with
\ba
\tilde{\K}\mupn=\(\Gamma^{\mu\alpha}-\frac{n^2 n^\mu n^\alpha}{\Big[n+\zeta(\Gamma)\Big]^2}\)
\p_\alpha \Phi^i \p_\nu \Phi^j \delta_{ij} \, ,
\ea
and where $U$, $C$ and $\zeta$ are scalar functions.

The fact that several independent functions enter the mass term will be of great interest for cosmology as one of these functions (namely $\mathcal{C}$) can be used to satisfy the Bianchi identity while the other function can be used for an appropriate cosmological history.

The special case $\zeta=0$ is what is referred to as the `minimal model' and was investigated in~\cite{Comelli:2013paa}. In unitary gauge, this minimal model is simply \ba
\L^{\rm (minimal)}_{{\rm SO}(3)\ {\rm mass}}= -\frac{m^2\mpl^2}{8}\sqrt{-g}\(U(g^{ik}\delta_{kj})+N^{-1}\mathcal{C}(\gamma^{ik}\delta_{ij})\)\,,
\ea
where $\gamma_{ij}$ is the spatial part of the metric and $g^{ij}=\gamma^{ij}-N^{-2}N^i N^j$, where $N$ is the lapse and $N^i$ the shift.

This minimal model is of special interest as both the primary and secondary second-class constraints that remove the sixth degree of freedom can be found explicitly and on the constraint surface the contribution of the mass term to the  Hamiltonian is
\ba
H\propto\mpl^2 m^2 \int \d^3x \sqrt{\gamma}\mathcal{C}(\gamma^{ik}\delta_{ij})\,,
\ea
where the overall factor is positive so the Hamiltonian is positive definite as long as the function $\mathcal{C}$ is positive.

At the linearized level about Minkowski (which is a vacuum solution) this theory can be parameterized in terms of the  mass scales introduces in~\eqref{SO3 mass term} with $m_0=0$, so the BD ghost is projected out in a way similar as in ghost-free massive gravity.

Interestingly if $\mathcal{C}=0$ this theory corresponds to $m_1=0$ (in addition to $m_0=0$) which as seen earlier the helicity-1 mode is absent at the linearized level. However they survive non-linearly and so the case $\mathcal{C}=0$ is infinitely strongly coupled.

\subsubsection{Second example of Lorentz-breaking with five dofs}

Another example of Lorentz breaking $SO(3)$ invariant theory of massive gravity was provided in~\cite{Comelli:2013txa}. In that case the \stu language is not particularly illuminating and we simply give the form of the mass term in unitary gauge ($\Phi=t$ and $\Phi^i=x^i$),
\ba
V_2&=&\frac{c_1}{2}\left[\vec{N}^T\(N \, \mathbb{I}+\mathbb{M}\)^{-1}\(\mathbb{F}+N^{-1}\mathbb{M} \mathbb{F}\) \(N \, \mathbb{I}+\mathbb{M}\)^{-1} \vec{N}
\right]\\
&&+\mathcal{C}+N^{-1}\tilde{\mathcal{C}}\,.\nn
\ea
where $\mathbb{F}=\{f_{ij}\}$ is the spatial part of the reference metric (for a Minkowski reference metric $f_{ij}=\delta_ij{}$),  $c_1$ is a constant and $\mathcal{C}$, $\tilde{\mathcal{C}}$ are functions of the spatial metric $\gamma_{ij}$ while $\mathbb{M}$ is a rank-3 matrix which depends on $\gamma^{ik}f_{kj}$.

Interestingly $\tilde{\mathcal{C}}$ does not enter the Hamiltonian on the constraint surface. The contribution of this mass term to the on-shell Hamitonian is~\cite{Comelli:2013txa}
\ba
H\propto\mpl^2 m^2 \int \d^3x \sqrt{\gamma}\left[
-\frac{c_1}{2} \vec{N}^T\(N \, \mathbb{I}+\mathbb{M}\)^{-1} \mathbb{F} \mathbb{M} \(N \, \mathbb{I}+\mathbb{M}\)^{-1} \vec{N}+\mathcal{C}
\right]\,, \hspace{10pt}
\ea
with a positive coefficient, which implies that $\mathcal{C}$ should be bounded from below

\subsubsection{Absence of vDVZ and strong coupling scale}

Unlike in the Lorentz-invariant case,  the kinetic term for the \stu fields does not only arise from the mixing with the helicity-2 mode.

When looking at perturbations about Minkowski and focusing on the   scalar modes we can follow the analysis of~\cite{Rubakov:2004eb},
\ba
\label{scalarPert}
\d s^2=-(1-\psi)\d t^2 +2 \p_i v \d x^i \d t+\(\delta_{ij}+ \tau \delta_{ij}+\p_i \p_j \sigma\)\d x^i \d x^j\,,
\ea
when $m_0=0$ $\psi$ plays the role of the Lagrange multiplier for the primary constraint imposing
\ba
\sigma=\(\frac{2}{m_4^2}-\frac{3}{\nabla}\)\tau\,,
\ea
where $\nabla$ is the three-dimensional Laplacian. The secondary constraint then imposes the relation
\ba
v=\frac{2}{m_1^2}\dot \tau\,,
\ea
where dots represent derivatives with respect to the time.
Using these relations for $v$ and $\sigma$ we obtain the Lagrangian for the remaining scalar mode (the helicity-0 mode) $\tau$~\cite{Rubakov:2004eb,Dubovsky:2004sg},
\ba
\label{L tau}
\L_\tau&=&\frac{\mpl^2}{4}\Bigg(\left[\(\frac{4}{m_4^2}-\frac{4}{m_1^2}\)\nabla \tau  -3 \tau\right]\ddot \tau-2\frac{m_2^2-m_3^2}{m_4^4}(\nabla \tau)^2\\
&+&\(4 \frac{m_2^2}{m_4^2}-1\)\tau \nabla \tau -3m_2^2 \tau^2
\Bigg)\,.
\ea
In terms of power counting this means that the Lagrangian includes terms of the form $\mpl^2 m^2 \nabla \phi \ddot \phi$  arising from the term going as $(m_4^{-2}-m_1^{-2})\nabla \tau \ddot \tau$ (where $\phi$ designates the helicity-0 mode which  includes a combination of $\sigma$ and $v$). Such terms are not present in the Lorentz-invariant Fierz--Pauli case and its non-linear ghost-free extension since $m_4=m_1$ in that case, and they play a crucial role in this Lorentz violating setting.

Indeed in the small mass limit these terms $\mpl^2 m^2 \nabla \phi \ddot \phi$ dominate over the ones that go as $\mpl^2 m^4 \phi \ddot \phi$ (\ie the  ones present in the Lorentz invariant case). This means that in the small mass limit, the correct canonical normalization of the helicity-0 mode $\phi$ is not of the form $\hat \phi = \phi /\mpl m^2$ but rather $\hat \phi = \phi/\mpl m \sqrt{\nabla}$, which is crucial in determining the strong coupling scale and the absence of vDVZ discontinuity:
\begin{itemize}
\item The new canonical normalization implies a much larger strong coupling scale that goes as $\Lambda_2=(\mpl m)^{1/2}$ rather than $\Lc=(\mpl m^2)^{1/3}$ as is the case in DGP and ghost-free massive gravity.
\item Furthermore in the massless limit the coupling of the helicity-0 mode to the tensor vanishes fasters than some of the Lorentz-violating kinetic interactions in~\eqref{L tau} (which is scales as $m \hat h \p^2 \hat \phi$). This means that one can take the massless limit $m\to 0$ in such a way that the coupling to the helicity-2 mode disappears and so does the coupling of the helicity-0 mode to matter (since this coupling arises after de-mixing of the helicity-0 and -2 modes). This implies the absence of vDVZ discontinuity in this Lorentz-violating theory despite the presence of five degrees of freedom.
\end{itemize}

The absence  of vDVZ discontinuity and the larger strong coupling scale $\Lambda_2$ makes this theory more tractable at small mass scales.
We emphasize however that the absence of vDVZ discontinuity does prevent some sort of Vainshtein mechanism to still come into play since the theory is still strongly coupled at the scale $\Lambda_2\ll\mpl$. This is similar to what happens for the Lorentz-invariant ghost-free theory of massive gravity on AdS (see Section~\ref{sec:DL on dS 2} and \cite{deRham:2012kf}). Interestingly however the same redressing of the strong coupling scale as in DGP or ghost-free massive gravity was explored in~\cite{Comelli:2013tja} where it was shown that in the vicinity of a localized mass, the strong coupling scale gets redressed in such a way that the weak field approximation remains valid till the Schwarzschild radius of the mass, \ie exactly as in GR.

In these theories, bounds on the graviton comes from the exponential decay in the Yukawa potential which switches gravity off at the graviton's Compton wavelength, so the Compton wavelength ought to be larger than the largest gravitational bound states which are of about 5~Mpc, putting a bound on the graviton mass of  $m\lesssim 10^{-30}\mathrm{\ eV}$  in which case $\Lambda_2\sim\(10^{-4}{\rm mm}\)^{-1}$~\cite{Comelli:2013paa,Goldhaber:1974wg}.

\subsubsection{Cosmology of general massive gravity}

The cosmology of general massive gravity was recently studied in~\cite{Comelli:2013tja} and we summarize their results in what follows.

In Section~\ref{sec:Cosmology} we showed how the Bianchi identity in ghost-free massive gravity prevents the existence of spatially flat FLRW solutions. The situation is similar in general Lorentz violating theories of massive gravity unless the function $\mathcal{C}$ in~\eqref{V1} is chosen so as to satisfy the following relation when the shift and $n^i$ vanish~\cite{Comelli:2013tja}
\ba
H \(\mathcal{C}'-\frac 12 \mathcal{C}\)=0\,.
\ea
Choosing a function $\mathcal{C}$  which satisfies the appropriate condition to allow for FLRW  solutions, the Friedmann equation then depends entirely on the function $U(\tilde{\K})$ also defined in~\eqref{V1}. In this case the graviton potential ~\eqref{SO3 mass term nonlinear} acts as an effective `dark fluid' with respective energy density and pressure dictated by the function $U$~\cite{Comelli:2013tja}
\ba
\rho_{\rm eff}=\frac{m^2}4 U(\tilde{\K})\qquad p_{\rm eff}=\frac{m^2}4 (2U'(\tilde{\K})-U(\tilde{\K}))\,,
\ea
leading to an effective phantom-like behaviour when $2U'/U<0$.

This solution is stable and healthy  as long as the second derivative of $\mathcal{C}$ satisfies some conditions which can easily be accommodated for appropriate functions $\mathcal{C}$ and $U$.

Expanding $U$ in terms of the scale factor for late time $U=\sum_{n\ge 0}\bar{U}_n (a-1)^n$ one can use
 CMB and BAO data from~\cite{Ade:2013zuv} to put constraints on the first terms of that series~\cite{Comelli:2013tja}
\ba
 \frac{\bar{U}_1}{\bar{U}_0}=0.12\pm2.1 \qquad \text{and}\qquad
 \frac{\bar{U}_2}{\bar{U}_0}<2\pm3 \qquad \text{at}\qquad 95\%\ \mathrm{C.L.}
\ea
Focusing instead on early time cosmology BBN data can similarly be used to constrains the function $U$, see~\cite{Comelli:2013tja} for more details.

\section{Non-local massive gravity}

The ghost-free theory of massive gravity proposed in Section~\ref{sec:ghost-free MG} as well as the Lorentz-violating theories of the previous section require an auxiliary metric. new massive gravity on the other hand can be formulated in a way which requires no mention of an auxiliary metric. Note however that all of these theories do break one copy of diffeomorphism invariance, and this occurs in bi-gravity as well and in the zwei dreibein extension of new massive gravity.

One of the motivations of non-local theories of massive gravity is to formulate the theory without any reference metric.\epubtkFootnote{Notice that even if massive gravity is formulated without the need of a reference metric, this does not change the fact that one copy of diffeomorphism invariance in broken leading to additional degrees of freedom as is the case in new massive gravity.} This is the main idea behind the non-local theory of massive gravity introduced in~\cite{Jaccard:2013gla}.\epubtkFootnote{See also~\cite{Khoury:2006fg,Deser:2007jk,Barvinsky:2011hd,Barvinsky:2011rk,Barvinsky:2012ts} for other ghost-free non-local modifications of gravity, but where the graviton is massless.}

Starting with the linearized equation about flat space-time of the Fierz--Pauli theory
\ba
\delta G\mn -\frac 12 m^2\(h\mn- h\eta\mn\)= 8 \pi G T\mn\,,
\ea
where $\delta G\mn = \Ein^{\alpha \beta}\mn h\ab$ is the linearized Einstein tensor, this modified Einstein equation can be `covariantized' so as to be valid about for any background metric. The linearized Einstein tensor $\delta G\mn$ gets immediately covariantized to the full Einstein tensor $G\mn$. The mass term on the other hand is more subtle and involves non-local operators. Its covariantization can take different forms, and the ones considered in the literature which do not involve a reference metric are
\ba
\label{cov mass term}
\frac 12 \(h\mn- h\eta\mn\) \longrightarrow \left\{\begin{array}{ll}
\(\Box_g^{-1}G\mn\)^T  &  \qquad \text{Ref.~\cite{Jaccard:2013gla}}\\
\frac 38 \(g\mn \Box_g^{-1}R\)^T  &\qquad  \text{Refs.~\cite{Maggiore:2013mea,Foffa:2013sma,Foffa:2013vma}}
\end{array}
\right.\,,
\ea
where $\Box_g$ is the covariant d'Alembertian $\Box_g=g^{\mu\nu}\nabla_\mu \nabla_\nu$ and $\Box_g^{-1}$ represents the retarded propagator.
One could also consider  a linear combination of both possibilities. Furthermore any of these terms could also be implemented by additional terms that vanish on flat space, but one should take great care in ensuring that they do not propagate additional degrees of freedom (and ghosts).

Following~\cite{Jaccard:2013gla} we use the notation where $T$ designates the transverse part of a tensor. For any tensor $S\mn$,
\ba
S\mn=S\mn^T+\nabla_{(\mu}S_{\nu)}\,,
\ea
with $\nabla^\mu S^T\mn=0$.
In flat space we can infer the relation~\cite{Jaccard:2013gla}
\ba
S\mn^T=S\mn-\frac{2}{\Box} \p_{(\mu} \p^\alpha S_{\nu) \alpha}+\frac{1}{\Box^2}\p_\mu \p_\nu \p^\alpha \p^\beta S\ab\,.
\ea

The theory propagates what looks like a ghost-like instability irrespectively of the exact formulation chosen in~\eqref{cov mass term}. However it was recently argued that the would-be ghost is not a radiative degree of freedom and therefore does not lead to any vacuum decay. It remains an open question of whether the would be ghost can be avoided in the full nonlinear theory.

The cosmology of this model was studied in~\cite{Maggiore:2013mea,Foffa:2013vma}. The new contribution~\eqref{cov mass term} in the Einstein equation can play the role of dark energy. Taking the second formulation of~\eqref{cov mass term} and setting the  graviton mass to $m\simeq 0.67 H_0$ where $H_0$ is the Hubble parameter today reproduces the observed amount of dark energy. The mass term acts as a dark fluid with effective time-dependent equation of state $\omega_{\rm eff}(a)\simeq -1.04 -0.02(1-a)$ where $a$ is the scale factor, and is thus phantom-like.

Since this theory is formulated at the level of the equations of motion and not at the level of the action and since  it includes non-local operators
 it ought to be thought as an effective classical theory. These equations of motion should not be used to get some insight on the quantum nature of the theory nor on its quantum stability. New physics would kick in when quantum corrections ought to be taken into account. It remains an open question at the moment of how to embed nonlocal massive gravity into a consistent quantum effective field theory.

Notice however that an action principle was proposed in Ref.~\cite{Modesto:2013jea}, (focusing on four dimensions),
\ba
S= \int \d^4 x \sqrt{-g}\left[\frac{\mpl^2}{2}R+\bar \lambda
+\frac{\mpl^2}{2M^2} R\mn h\(-\frac{\Box}{M^2}\)G^{\mu\nu} \right]\,,
\ea
where the function $h$ is defined as
\ba
h(z)=\frac{\Box +m^2}{\Box}\frac{1}{z}|p_s(z)|e^{\frac 12 \Gamma(0,p_s^2(z))+\frac 12 \gamma_E}\,,
\ea
where $\gamma_E=0.577216$ is the Euler's constant, $\Gamma(b,z)=\int_z^\infty t^{b-1} e^{-t}$ is the incomplete gamma function, $s$ is a integer $s>3$ and $p_s(z)$ is a real polynomial of rank $s$.
Upon deriving the equations of motion we recover the non-local massive gravity Einstein equation presented above \cite{Modesto:2013jea},
\ba
G\mn+\frac{m^2}{\Box}G\mn=\frac{\mpl^2}{2}T\mn\,,
\ea
up to order $R^2$ corrections. We point out however that in this action derivation principle the operator $\Box^{-1}$ likely correspond to a symmetrized Green's function, while in \eqref{cov mass term} causality requires $\Box^{-1}$ to represent the retarded one.

We stress however that this theory should be considered as a classical theory uniquely and not be quantized. It is an interesting question of whether or not the ghost reappears when considering quantum fluctuations like the ones that seed any cosmological perturbations.

\newpage

\section{Outlook}

The past decade has witnessed a revival of interest in massive gravity as a potential alternative to GR. The original theoretical obstacles that came in the way of deriving a consistent theory of massive gravity have now been overcome but with them comes a new set of challenges which will be decisive in establishing the viability of such theories. The presence of a low strong coupling scale on which the Vainshtein mechanism rely, has opened the door to a new way to think about these types of effective field theories. At the moment it is yet unclear whether these types of theories could lead to an alternative to UV completion. The superluminalities that also arise in many cases with the Vainshtein  mechanism should also be understood in more depth. At the moment its real implications are not well understood and no case of true acausality has been shown to be present within the regime of validity of the theory. Finally, the difficulty in finding fully-fledged cosmological and black holes solutions in many of these theories (both in ghost-free massive gravity and bi-gravity, and in other extensions or related models such as cascading gravity) makes their full phenomenology still evasive. Nevertheless the well understood decoupling limits of these models can be used to say a great deal about phenomenology without going into the complications of the full theories. These represent many open questions in massive gravity which reflect the fact that the field is yet extremely young and many developments are still in progress.

\section*{Acknowledgments}

CdR wishes to thank Denis Comelli, Matteo Fasiello, Gregory Gabadadze, Daniel Grumiller, Kurt Hinterbichler, Andrew Matas, Shinji Mukohyama, Nicholas Ondo, Luigi Pilo and especially Andrew Tolley for useful discussions. CdR is supported by Department of Energy grant DE-SC0009946.

\newpage

\bibliography{refs}

\end{document}